\documentclass[useAMS,usenatbib]{mn2e}

\usepackage{rotating, graphicx}
\usepackage{subfig}
\usepackage{pdflscape}
\usepackage{natbib}
\usepackage{amsmath}
\usepackage{courier}
\usepackage{color}
\usepackage{float}
\restylefloat{table}
\usepackage{footnote}
\usepackage[T1]{fontenc}
\usepackage{aecompl}
\usepackage{amssymb}
\usepackage{rotating}

\title[A correlation between ionisation and cloud mass-loss rate]{Connecting the dots: a correlation between ionising radiation and cloud mass-loss rate traced by optical integral field spectroscopy}
\author[A.~F. McLeod et al.]{A.~F. McLeod$^{1,2}$\thanks{E-mail: amcleod@eso.org}, M. Gritschneder$^{2}$, J.~E. Dale$^{2,3}$, A. Ginsburg$^{1}$, P. D. Klaassen$^{4}$,  \newauthor J.~C. Mottram$^{5}$, T. Preibisch$^{2}$, S. Ramsay$^{1}$, M. Reiter$^{6}$ and L. Testi$^{1,7}$
\\
$^{1}$European Southern Observatory, Karl-Schwarzschild-Str. 2, 85748 Garching bei M{\"u}nchen, Germany \\
$^{2}$Universit\"ats-Sternwarte, Ludwig-Maximilians-Universit\"at M\"unchen, Scheinerstr.~1, 81679 M\"unchen, Germany\\
$^{3}$Excellence Cluster `Universe', Boltzmannstr. 2, 85748 Garching bei M{\"u}nchen, Germany\\
$^{4}$UK Astronomy Technology Center, Royal Observatory Edinburgh, Blackford Hill, Edinburgh EH9 3HJ, UK\\
$^{5}$Max Planck Institute for Astronomy, K{\"o}nigstuhl 17, 69117 Heidelberg\\
$^{6}$University of Michigan Department of Astronomy, 1085 S. University, Ann Arbor, MI 48109\\
$^{7}$INAF/Osservatorio Astrofisico of Arcetri, Largo E. Fermi, 5, 50125 Firenze, Italy}
\begin{document}

\date{Accepted . Received ; in original form}

\pagerange{\pageref{firstpage}--\pageref{lastpage}} \pubyear{2014}

\maketitle

\label{firstpage}

\begin{abstract}
We present an analysis of the effect of feedback from O- and B-type stars with data from the integral field spectrograph MUSE mounted on the Very Large Telescope of pillar-like structures in the Carina Nebular Complex, one of the most massive star-forming regions in the Galaxy. For the observed pillars, we compute gas electron densities and temperatures maps, produce integrated line and velocity maps of the ionised gas, study the ionisation fronts at the pillar tips, analyse the properties of the single regions, and detect two ionised jets originating from two distinct pillar tips. For each pillar tip we determine the incident ionising photon flux $Q_\mathrm{0,pil}$ originating from the nearby massive O- and B-type stars and compute the mass-loss rate $\dot{M}$ of the pillar tips due to photo-evaporation caused by the incident ionising radiation. We combine the results of the Carina data set with archival MUSE data of a pillar in NGC 3603 and with previously published MUSE data of the Pillars of Creation in M16, and with a total of 10 analysed pillars, find tight correlations between the ionising photon flux and the electron density, the electron density and the distance from the ionising sources, and the ionising photon flux and the mass-loss rate. The combined MUSE data sets of pillars in regions with different physical conditions and stellar content therefore yield an empirical quantification of the feedback effects of ionising radiation. In agreement with models, we find that $\dot{M}\propto Q_\mathrm{0,pil}^{1/2}$.
\end{abstract}

\begin{keywords}
\textsc{HII} regions, (ISM): jets and outflows, (ISM:) individual: NGC 3372, NGC 3603, M\,16 
\end{keywords}

\section{Introduction}

Feedback from massive stars is a crucial factor in the baryon cycle of galaxies. It regulates how they turn their gas reservoir into stars and therefore plays a fundamental role in galaxy evolution \citep{sommerlarsen03}. However, its exact role is one of the main missing ingredients to connect the observed galaxy population to $\Lambda$CDM cosmology (e.g.~\citealt{vogel14}, \citealt{schaye15}). On smaller, molecular cloud-scales (1-10 pc), massive stars (which generally form in a clustered environment, rather than in isolation, \citealt{beuther07}, \citealt{bressert10}) profoundly affect their immediate environment: stellar winds, photo-ionising radiation and supernova events are responsible for inflating \textsc{HII} regions and shell-like structures, which both suppress and enhance star formation by clearing away the molecular gas but also locally compressing it to create high-density regions and therefore possible sites of new star formation \citep{walch14}.

Feedback from massive stars is also thought to be responsible for exposing pillar-like structures and globules from surrounding gas and dust (\citealt{mellema06}, \citealt{arthur11}, \citealt{tremblin12}). However, the relative importance of the different pillar formation scenarios is debated: is a pre-existing dense structure exposed by stellar feedback, or does feedback collect and compress the molecular gas to form overdensities which subsequently form pillars via instabilities (\citealt{gritsch10}, \citealt{tremblin12})? Notwithstanding the contentious formation mechanism, these structures are formed throughout star forming regions, both observed (\citealt{hester96}, \citealt{klaassen14}) and simulated (\citealt{gritsch10}, \citealt{dale12}), some host forming stars at their tips which are sometimes known to launch bipolar jets (\citealt{smith10}, \citealt{reiter13}), and they are found to photo-evaporate under the influence of the nearby massive stars (\citealt{west13}, \citealt{M16}). The mechanism of photo-evaporation is explained by the strong ionising radiation from the nearby O- and B-type stars impinging on the matter of the pillar tips: the pressure of the pillar material increases and it is photo-ionised, leading to a photo-evaporative flow of matter streaming away from the pillar surface \citep{hester96}. Despite the fact that the photo-evaporative effect has been observed in many previous works, and the connection between the mass-loss rate and the flux from the ionising stars has been analytically described (e.g. \citealt{bertoldi89}, \citealt{bertoldi90}, \citealt{lefloch94}, \citealt{mellema98}), no direct observational tests of the theories have been performed.

In a previous study, we targeted the iconic Pillars of Creation in M\,16 (\citealt{M16}, henceforth referred to as MC15) with the integral field unit (IFU) MUSE mounted on the Very Large Telescope (VLT). In projection, these pillars are situated $\sim$ 2 pc south-west of the massive cluster NGC 6611, which is responsible for the radiation field which is photo-evaporating the nearby pillars. For these structures, we computed the mass-loss rate due to photo-evaporation by making use of the simultaneous imaging and spectroscopic capabilities of MUSE, which allowed us to determine both the morphology and the kinematics of the ionised gas by covering all the ionised emission lines in the 4650 \AA\ - 9300 \AA\ range covered by MUSE. Together with their molecular mass of about 200 $M_{\odot}$ \citep{white99}, we estimated a lifetime of $\sim$ 3 Myr for the pillars in M\,16.

However interesting for the particular case of the M\,16 region, the results obtained for the Pillars of Creation alone are not enough to analyse and quantify ionising feedback in general. To do this, the interplay between the ionising star cluster and the affected (photo-evaporating) pillar-like structures needs to be analysed in a representative sample of regions to test for the existence of a relationship between the two. In this paper, we exploit a unique data set of pillar-like structures in the Carina star-forming region observed with the IFU MUSE. In combination with archival MUSE data of a pillar in NGC 3603 and the previously published M\,16 data, we attempt a first quantitative analysis of ionising feedback in high-mass star-forming regions by comparing the photo-evaporative effect in these different regions. 

The Carina Nebula Complex (CNC) is a rich, nearby (2.3 kpc, \citealt{smith06}) and well studied star forming region, with tens of young star clusters, O- and B-type stars and many pillar-like structures that surround the central part of the Complex \citep{smith08}. Because of this, it is the ideal region to study the stellar populations of young, massive clusters, as well as their feedback on the surrounding molecular clouds. It hosts about 65 O-type stars as well as three WNL stars\footnote{Late-type Wolf-Rayet stars with nitrogen-dominated spectra that show hydrogen lines.}, which together emit a total ionising photon luminosity of about 10$^{51}$ photons s$^{-1}$ \citep{smith06b}. Despite the high degree of feedback and the presence of already exposed star clusters, as well as the evolved massive LBV $\eta$ Car, the CNC is also known to host vigorous ongoing star formation as well as a few $10^{4}M_{\odot}$ in dense, cold clouds which have not yet formed stars \citep{preibisch12}. The massive stellar content of the CNC is predominantly found in the three main young clusters, namely Trumpler 14, 15 and 16, which together account for $\sim$~93\% of the total ionising flux. Tr 14 and Tr 16 are located toward the center of the giant Carina HII region and have ages of about 1-2 Myr \citep{hur12}. While Tr 14 hosts $\sim$ 10 O-stars (ranging from O2 I to O6.5 V) and has a very compact spatial configuration with a core radius of just 0.15 pc and a central stellar density of the order of 10$^{4}$ M$_{\odot}$ pc$^{-2}$, Tr 16 hosts $\sim$18 O-stars, a WNL star, and is a loose open cluster and consists of several subclusters (\citealt{feigelson11}, \citealt{wolk11}). By contrast, Tr 15 only hosts about 6 O-stars, none of them of spectral type earlier than O8, and is older than Tr 14 and 16, with an age between 5 and 10 Myr \citep{wang11}. The structure and properties of the dense, cool clouds in the CNC was recently studied in new detail by deep sub-mm mapping observations with LABOCA at the APEX telescope \citep{CNC-Laboca} and by the $70 - 500\,\mu$m far-infrared observations with the \textit{Herschel} space observatory \citep{CNC-Herschel1,CNC-Herschel2}. These observations showed that most of the dense gas in the CNC is concentrated in numerous pillar-like cloud structures, which are shaped by the irradiation of nearby high-mass stars.

The multitude of pillars, globules and outflows in the CNC have been thoroughly analysed in \cite{hartigan15}. These authors used both broad- and narrowband near infrared (I, K, Br$\gamma$ and H$_{2}$) and optical (H$\alpha$, [SII] and [OIII]) data to detect previously unknown candidate jets (to complement and complete the already well known population of Herbig-Haro objects from \citealt{smith10}) and identify the main irradiated surfaces between the \textsc{HII} region and the surrounding molecular clouds. In this very exhaustive study, \citeauthor{hartigan15} analyse 63 different regions within the CNC, each containing one or more pillars or irradiated globules. The MUSE CNC data set presented in this paper covers 5 of the \citeauthor{hartigan15} regions and contains 5 distinct pillars as well as one irradiated globule. The regions were chosen to maximise the number of covered pillars, sample these from a variety of different locations within the CNC, and minimise the needed amount of telescope time. This unique data set, in combination with MUSE observations of pillars in M\,16 (MC15) and NGC 3603, offers for the first time the possibility of analysing both the morphology and the kinematics of the ionised gas with the unique combination of angular resolution and spectral coverage offered by MUSE. With this data, we analyse the ionising feedback from the massive stars in the CNC by computing and comparing the ionising photon flux and photo-evaporation rate for a total of 10 pillars, and we deliver an observational quantification of the effects of ionising feedback.

The observations and data reduction are discussed in Section \ref{obs}. The analysis in this paper is divided into four main topics: a study of the behaviour of the various detected emission lines at the pillar/ambient matter interfaces (Section \ref{profiles}), the computation and discussion of the physical parameters of the observed regions (Section \ref{physparams}), the connection between the ionising massive stars and the photo-evaporation of the surrounding pillars (Section \ref{main}), and the discussion of two detected bipolar jets originating from the pillar tips (Section \ref{jetsection}). Conclusions are present in Section \ref{conclusions}.

\section{Observations and data reduction}\label{obs}
\subsection{Carina}
\begin{figure*}
\includegraphics[scale=0.5]{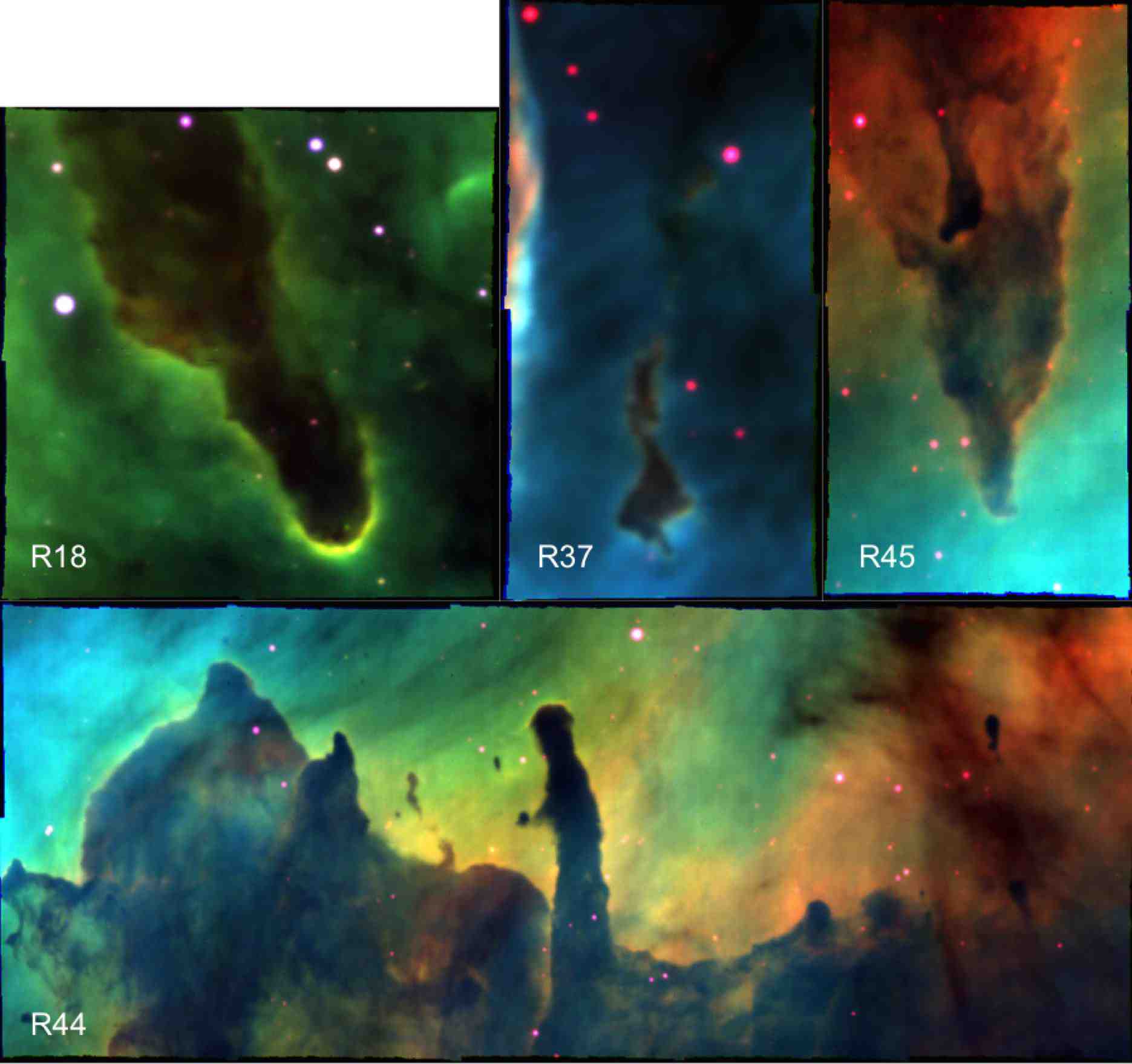}
\caption{Three-color composites of integrated line maps of the four observed regions in Carina (R18, R37, R45 and R44). Red is [SII]$\lambda$6717, green is H$\alpha$ and blue is [OIII]$\lambda$5007. The images are not continuum-subtracted. See text Section \ref{obs} for further details.}
\label{rgb}
\end{figure*}

\begin{figure*}
\includegraphics[scale=0.6]{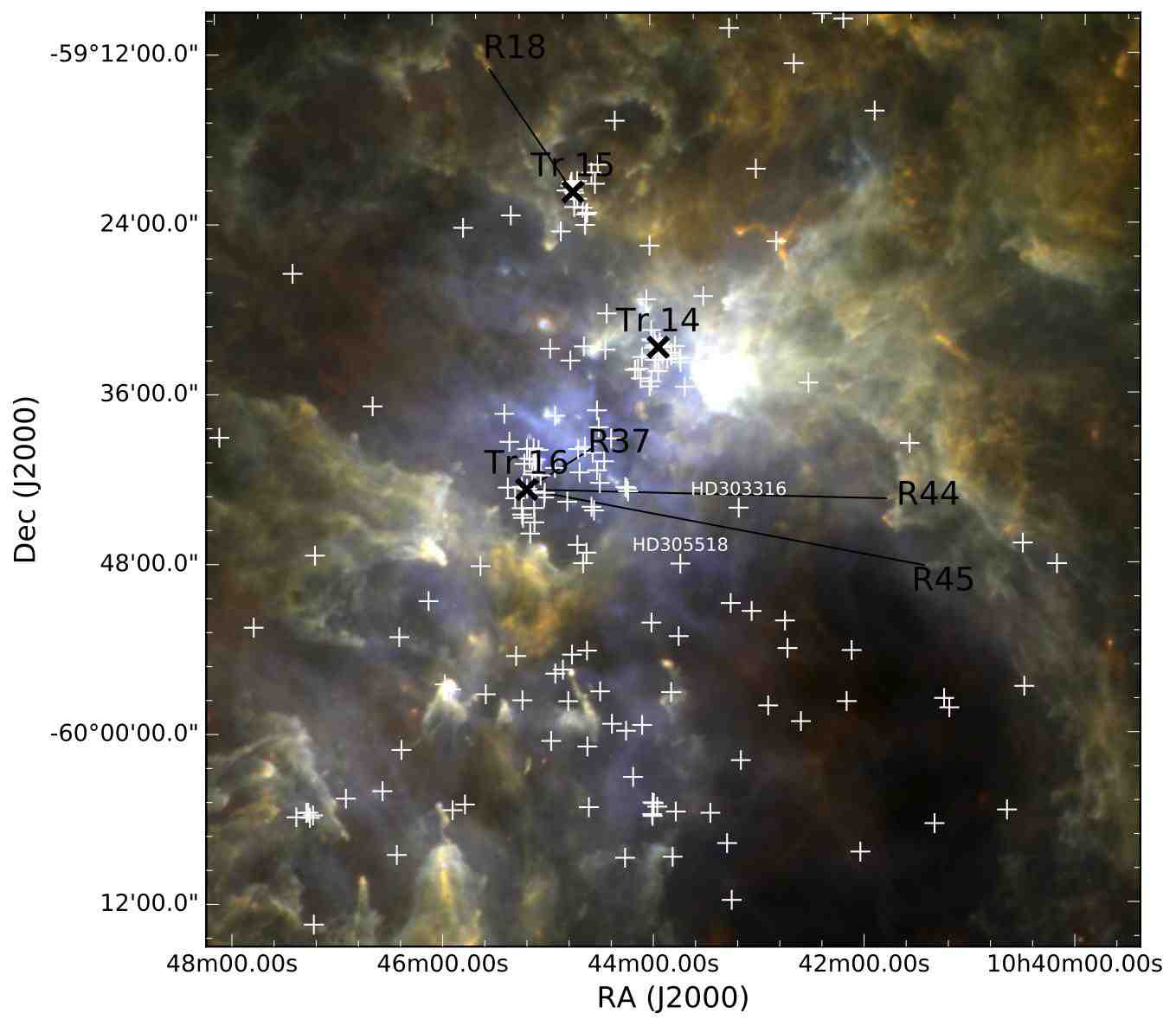}
\caption{Three-color \emph{Herschel} composite (red = 250 $\mu$m, green = 160 $\mu$m, blue = 70 $\mu$m). The 4 observed regions in Carina (R18, R37, R44 and R45) as well as the three most massive clusters (Tr 14, 15 and 16) are marked in black, O- and B-stars from \protect\cite{gagne11} are marked with white crosses, the projected distances from each region to the main nearby clusters are traced with the black lines, and two O-stars (HD 303316 and HD 305518) are highlighted, see text Section \ref{clusters}.}
\label{orient}
\end{figure*}

We observed a sample of pillar-like structures in the Carina Nebula Complex with the IFU MUSE under the program 096.C-0574(A) (PI McLeod). The observing program targeted five regions sampled from \cite{hartigan15}, who present near-infrared images from NEWFIRM and optical images from MOSAIC observations\footnote{the NOAO Extremely Wide-Field Infrared Imager and the NOAO/KPNO Mosaic Wide Field Imager, thus both wide-field imagers covering a much larger area than the HST and Spitzer surveys from \cite{smith10}, \cite{smith10b} and \cite{povich11}.} of pillars, globules and jets in the Carina star-forming region. As already mentioned, the five regions from \citeauthor{hartigan15}, Areas 18, 22, 37, 44 and 45 (for simplicity named R18, R22, R37, R44 and R45 henceforth in this paper) were selected based on two criteria: the presence of pillar-like structures and their different locations within the CNC (to sample different feedback environments and conditions). Only four of the five target regions are discussed in this work, as the observations of R22 showed that this region is too heavily extincted to identify the pillars, and therefore not suited for the analysis in this work. RGB composites of the four discussed regions (R18, R37, R44 and R45) are shown in Fig.~\ref{rgb}. The colours correspond to [SII]$\lambda$6717 (red), H$\alpha$ (green) and [OIII]$\lambda$5007 (blue, the images shown in this figure are not continuum-subtracted). Fig.~\ref{orient} shows the location of the four regions with respect to the three main CNC clusters (black crosses) and the location of known O- and B-type stars from \cite{gagne11} (white crosses). Two O-stars (HD 303316 and HD 305518 are highlighted, as discussed in Section \ref{clusters}).

With these observations, we sample a total of 5 pillar-like structures (R18, R45 and three distinct pillars in R44, where we refer to the pillars as R44 P1, R44 P2 and R44 P3 as indicated by the slit positions in Fig.\ref{slits}) as well as one detached globule (R37) in the relative vicinity of two of the three most massive clusters in the CNC, Tr 15 and Tr 16. As will be discussed later (see Section \ref{conclusions}), a dedicated MUSE data set covering the massive pillars around Tr 14, as well as the cluster stars themselves, has been recently observed (VLT program 097.C-0137(A), PI McLeod) and will be presented in a forthcoming publication.  

The total amount of telescope time awarded to this program was 5.5 hours, and the data was taken during the period from October 31st, 2015 to January 17th, 2016. The central coordinates of each observed region, as well as the date of observation and number of mosaic pointings per region are listed in Table \ref{pointings}. The exposure time for all observations was 90 seconds, and each pointing was observed three times following a 90$^{\circ}$ dither pattern, a method generally used to minimise instrument artefacts and proven to be successful with our MUSE science verification data of pillar-like structures in M\,16 (MC15). The data reduction was carried out in the ESO \textsc{esorex} environment of the MUSE pipeline \citep{pipeline} and using standard calibrations. Each pointing was observed in the nominal wavelength range of MUSE (4650-9300 \AA, resolving power R = 2000-4000) and exploiting the instrument's Wide Field Mode with a field of view of 1x1 arcminutes. At the distance of Carina, the seeing-limited resolution of the MUSE observations of 0.2 arcsec is about 0.002 pc.

\begin{table*}
\begin{center}
\caption{MUSE observations of pillar-like structures in the Carina Region. Coordinates are taken from \protect\cite{hartigan15}.}
\begin{tabular}{lcccc}
\hline
\hline
Region & R.A. (J2000) & Dec (J2000) & Observing date & No. of mosaic pointings \\
R18 & 10:45:31.98 & -59:12:23 & 1/2 Nov. 2015 & 2 \\
R37 & 10:44:31.44 & -59:39:21 & 31 Oct./1 Nov. 2015 & 2 \\
R44 & 10:41:39.21 & -59:43:33 & 30 Nov./1 Dec. 2015 & 10 \\
 & & & \& 10/1 Dec. 2015 & \\
R45 & 10:41:20.69 & -59:48:23 & 15/16 Nov. 2015 & 2 \\ 
\hline
\end{tabular}
\label{pointings}
\end{center}
\end{table*}

\subsection{NGC 3603 and M\,16}\label{ngc3603}
We compare the observations of the Carina pillars to similar structures in two other regions covered by MUSE, namely the Pillars of Creation in the Eagle Nebula (M\,16), as well as a pillar south-west of the massive cluster in NGC 3603. Both the MUSE M\,16 and NGC 3603 data were taken during the instrument's science verification run in June 2014. The M\,16 data set (program 60.A-9309(A), PI McLeod) is presented and analysed in MC15,  which gives details of the observational setup and data reduction. The NGC 3603 and M16 pillars were included to sample pillars not only across various star-forming regions, but also in different ionising conditions.

\begin{figure}
\includegraphics[scale=0.35]{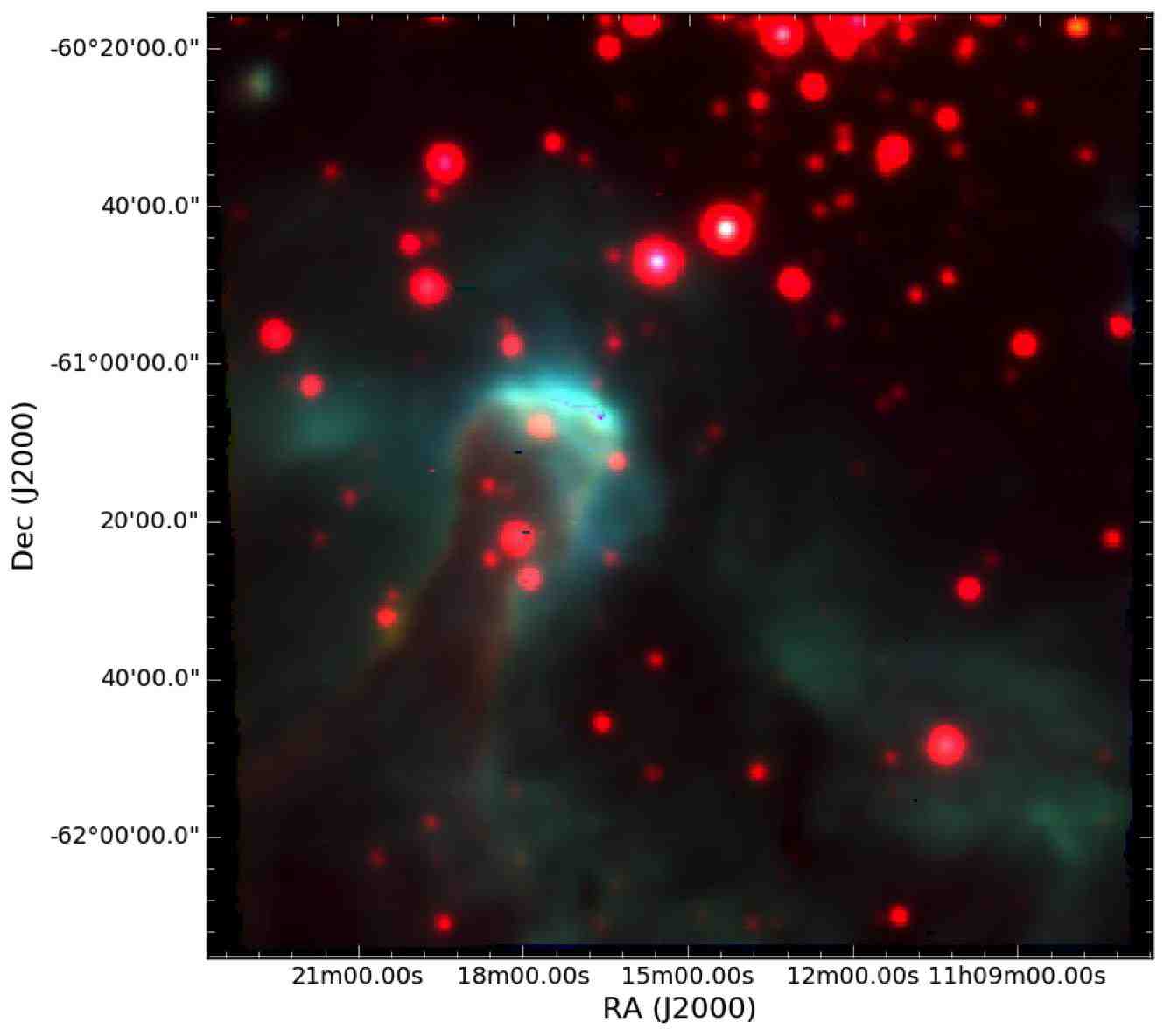}
\caption{RGB composite of the pillar in NGC3603, red is [SII]$\lambda$6717, green is H$\alpha$ and blue is [OIII]$\lambda$5007 (as for Fig.~\ref{rgb} the integrated emission line maps are not continuum-subtracted). See text Section \ref{ngc3603}.}
\label{ngc3603fig}
\end{figure}

The massive cluster in NGC 3603, in projection only about 1 pc away from the observed pillar, is one of the most massive clusters in the Milky Way with an age of 1-2 Myr (\citealt{sung04}, \citealt{harayama08}) and mass estimates ranging from $\sim$ 10$^{4}$ M$_{\odot}$ \citep{harayama08} to about 1.8$\times$10$^{4}$ M$_{\odot}$ \citep{rochau10}. The exposed cluster is surrounded by molecular cloud structures, among which are two prominent pillars (e.g. \citealt{west13}), which are under the influence of the large number of O-type stars ($>$ 30, \citealt{melena08}) that dominate the feedback in the region. It is the ideal region to analyse one of the more extreme Galactic feedback environments, as the bolometric luminosity of the HD 97950 cluster\footnote{We use HD 97950 as the designation for the massive cluster in the NGC 3603 region, although it originally refers to the star HD 97950, which could be resolved into a multiple system in the centre of a massive cluster in \cite{melena08}.} is about 100 times greater than that of the Orion cluster, and about 10\% that of the massive 30 Doradus cluster in the Large Magellanic Cloud. With MUSE, the pillar south-east of the cluster was covered, and an RGB composite of [SII], H$\alpha$ and [OIII] is shown in Fig.~\ref{ngc3603fig}.

The archival NGC 3603 data set (program 60.A-9344(A)) consists of a single pointing covered with four different exposure times, namely 10, 60, 300 and 2100 seconds. For the purpose of this analysis, the 10 and 60 second exposures did not deliver the necessary S/N, and the main ionised emission lines are saturated in the 2100 seconds exposure. We therefore reduced the 300s data cube in the same manner as described in in the previous subsection. Unfortunately however, the NGC 3603 data was not taken with a 90$^{\circ}$ rotation dither pattern as was done for M\,16 and Carina, and the resulting images are therefore heavily contaminated by instrument artefacts, visible as a striped pattern most noticeable in the velocity maps (see e.~g. Figure \ref{velmaps}d).

\section{Analysis}
This work is oriented toward ionising feedback from massive stars and star clusters on the nearby pillar-like structures. The aim is to find a link between the feedback-driving stars and the feedback-affected surrounding molecular clouds by analysing the pillars pointing back towards the ionising star clusters. The analysis consists of four parts which are then combined to construct a clear picture of ionising feedback: we analyse the intensity of the ionised emission lines towards the pillars (Section \ref{profiles}); we determine the physical parameters for the different conditions present in each region (Section \ref{physparams}); we determine the mass-loss rate due to photo-evaporation of the pillars (Section \ref{main}); and we relate the presence and morphology of ionised jets at the pillar tips to the feedback conditions of the regions (Section \ref{jetsection}). The novelty of this analysis consist in the exploitation of a uniform IFU data set of pillars in different star forming regions and under different feedback conditions, which allows a simultaneous study of the physical parameters and the kinematics.

\subsection{Emission line intensity profiles}\label{profiles}

\begin{figure*}
\mbox{
\subfloat[]{\includegraphics[scale=0.32]{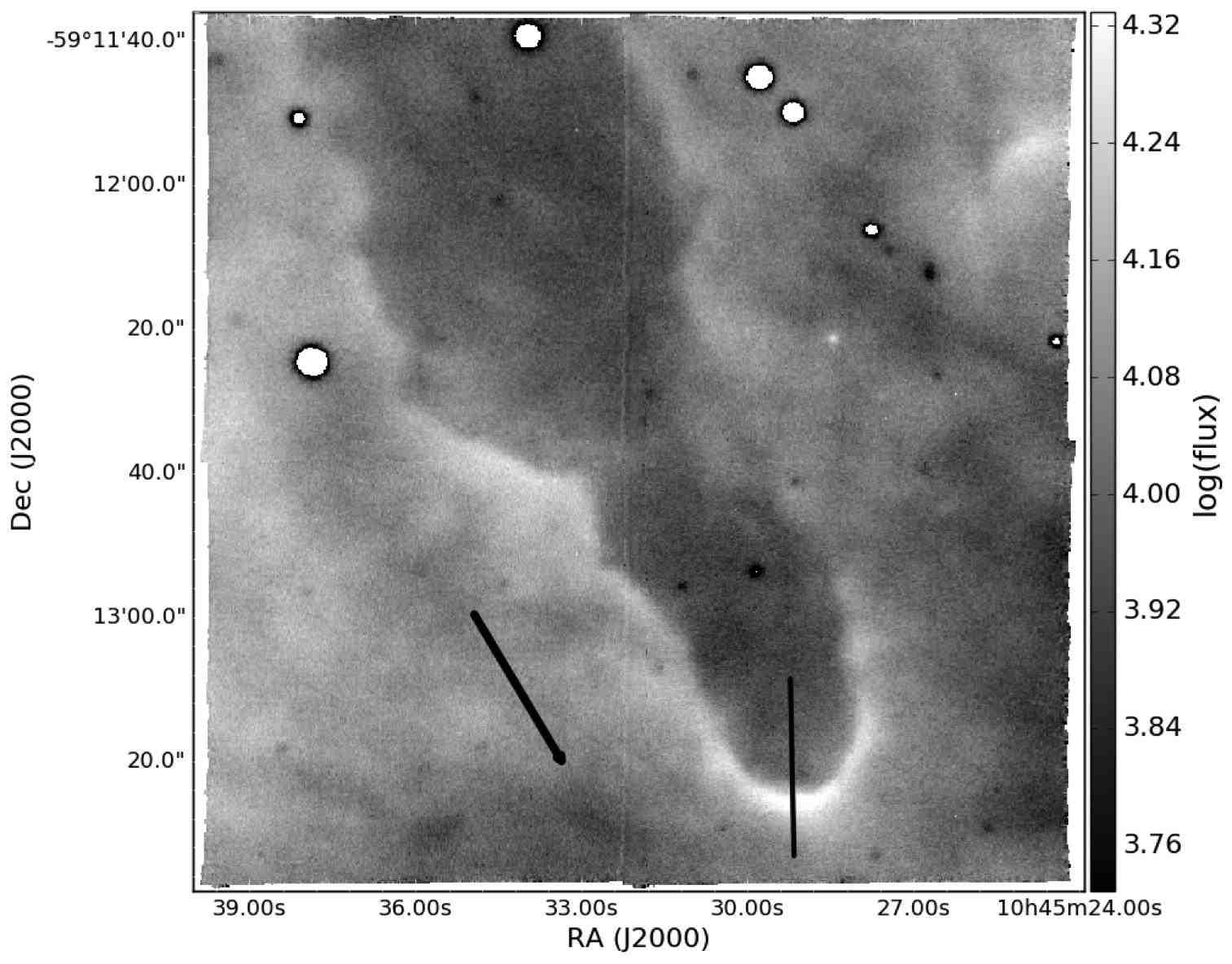}}}
\mbox{
\subfloat[]{\includegraphics[scale=0.42]{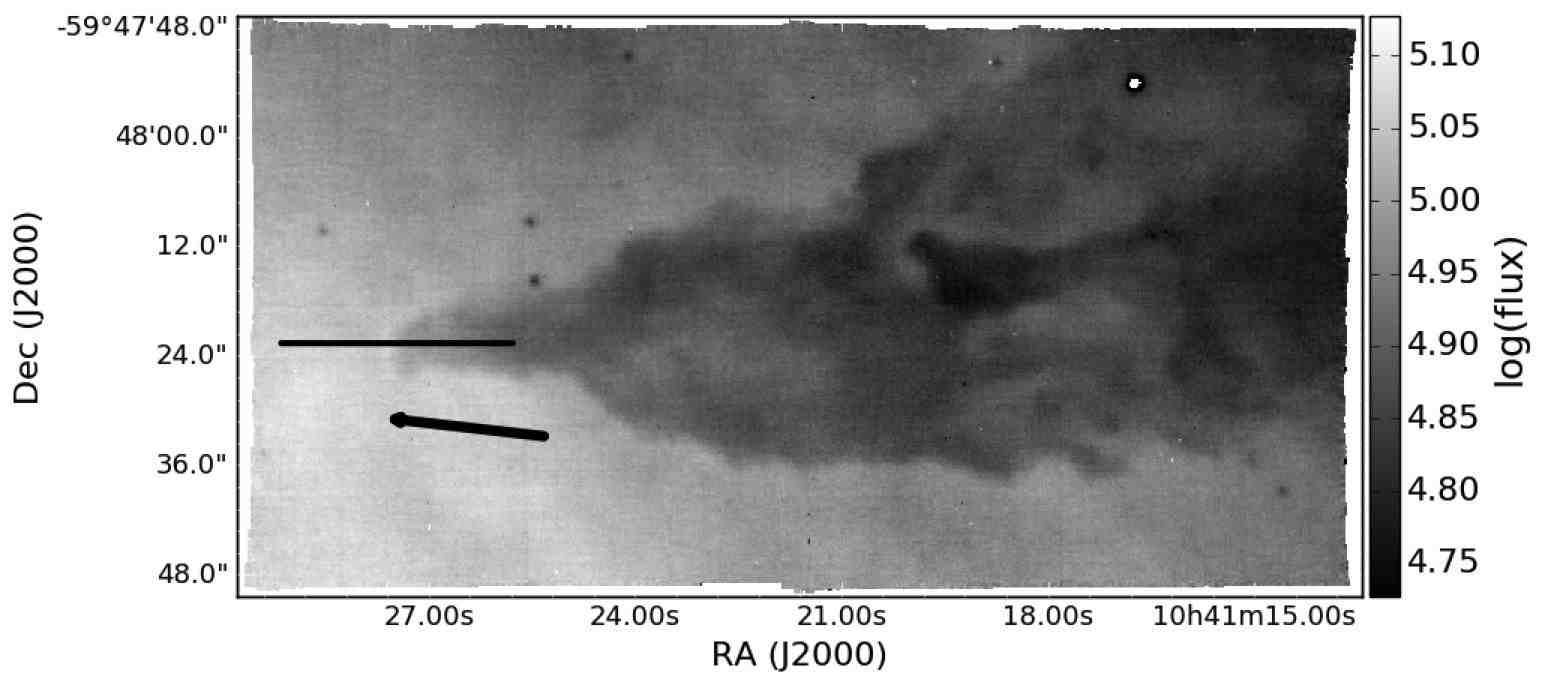}}}
\mbox{
\subfloat[]{\includegraphics[scale=0.44]{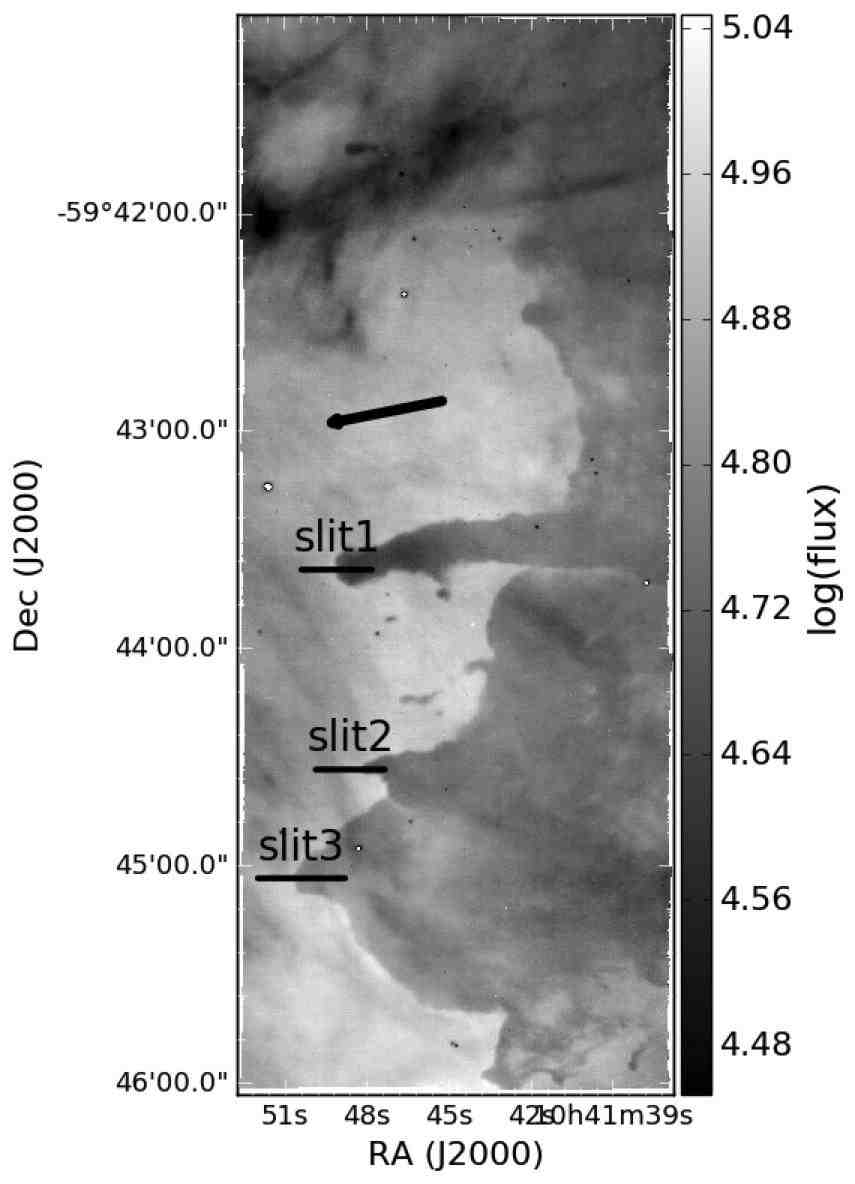}}
\subfloat[]{\includegraphics[scale=0.44]{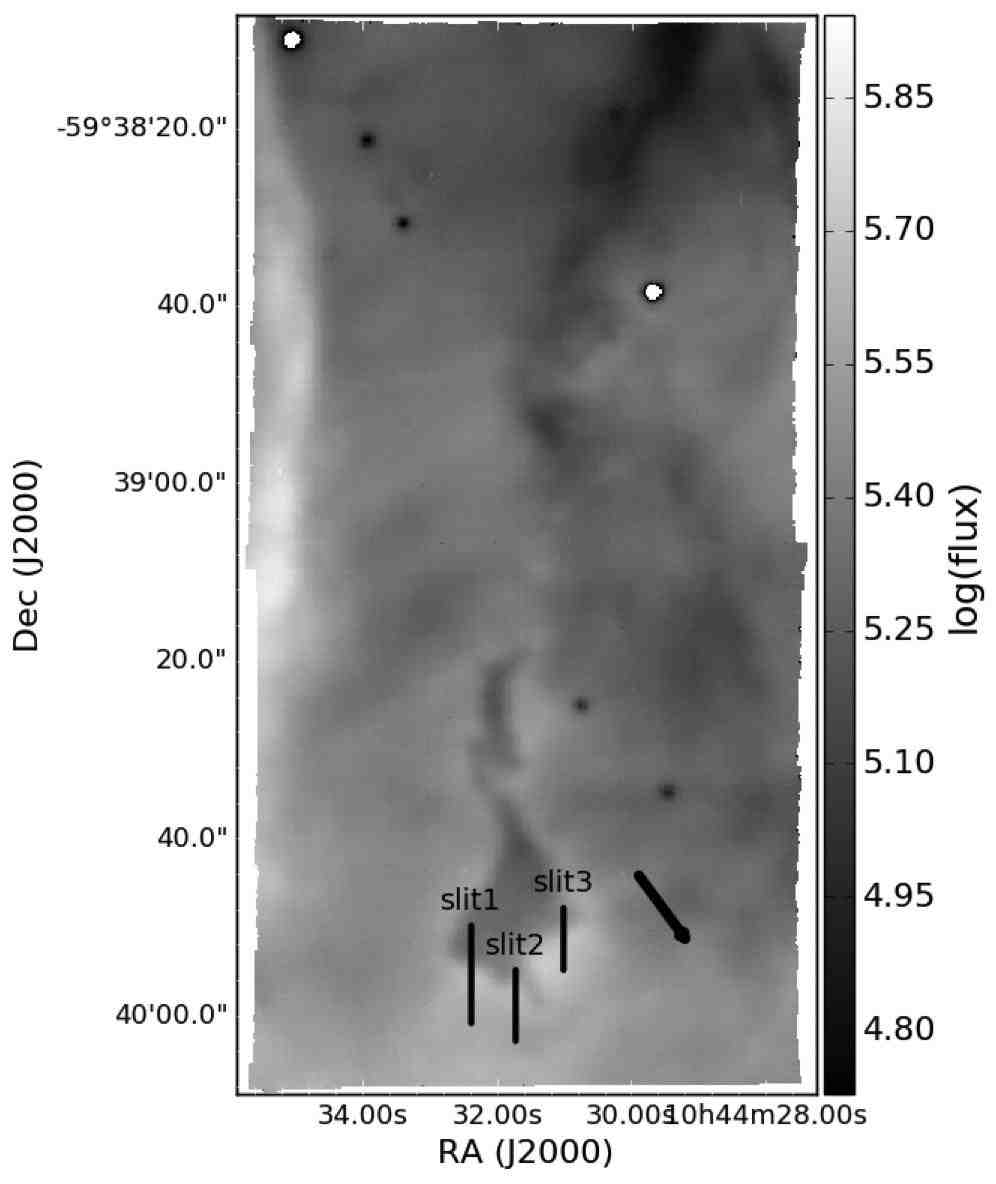}}}
\caption{Continuum-subtracted H$\alpha$ integrated intensity maps of the four Carina regions: R18 (panel (a)), R45 (panel (b)), R44 (panel (c)) and R37 (panel (d)). The black lines indicate the positions of the slits used to compute the line intensity profiles (see Section \ref{profiles}). The flux is measured in 10$^{-20}$ erg s$^{-1}$ cm$^{-2}$ \AA$^{-1}$. Black arrows indicate the direction towards the main ionising sources for each pillar.}
\label{slits}
\end{figure*}

Figure \ref{slits} shows continuum-subtracted integrated H$\alpha$ line maps (all integration intervals span $\pm$ 3 \AA\ around the central emission line wavelength) of all the discussed Carina pillars, while emission line maps of the other main lines used in this analysis (H$\beta$, [OIII]$\lambda$4959,5007, [SII]$\lambda$6717,31, [SIII]$\lambda$9069, [NII]$\lambda$6548,84, [OI]$\lambda$6300, [OII]$\lambda$7320,30) are shown in the Appendix in Figures \ref{maps1} to \ref{maps10}. From the emission line maps, the ionisation fronts at the illuminated pillar tips are clearly visible, and in the case of R44-P3 and R45 the left and right sides of the pillars are also being illuminated (see Fig.~\ref{rgb}), most probably by the nearby B-type stars located to the west, while the pillar tips are illuminated by two O-stars (see Section \ref{clusters}). However, these B-type stars are very likely not the main sources of feedback, as the pillars in R44 and R45 do not point in their direction but rather towards the two O-stars HD 303316 and HD 305518 marked in Fig.~\ref{orient}. This will be discussed in more detail in Section \ref{clusters}.

\begin{figure*}
\centering
\mbox{
\subfloat[]{\includegraphics[scale=0.43]{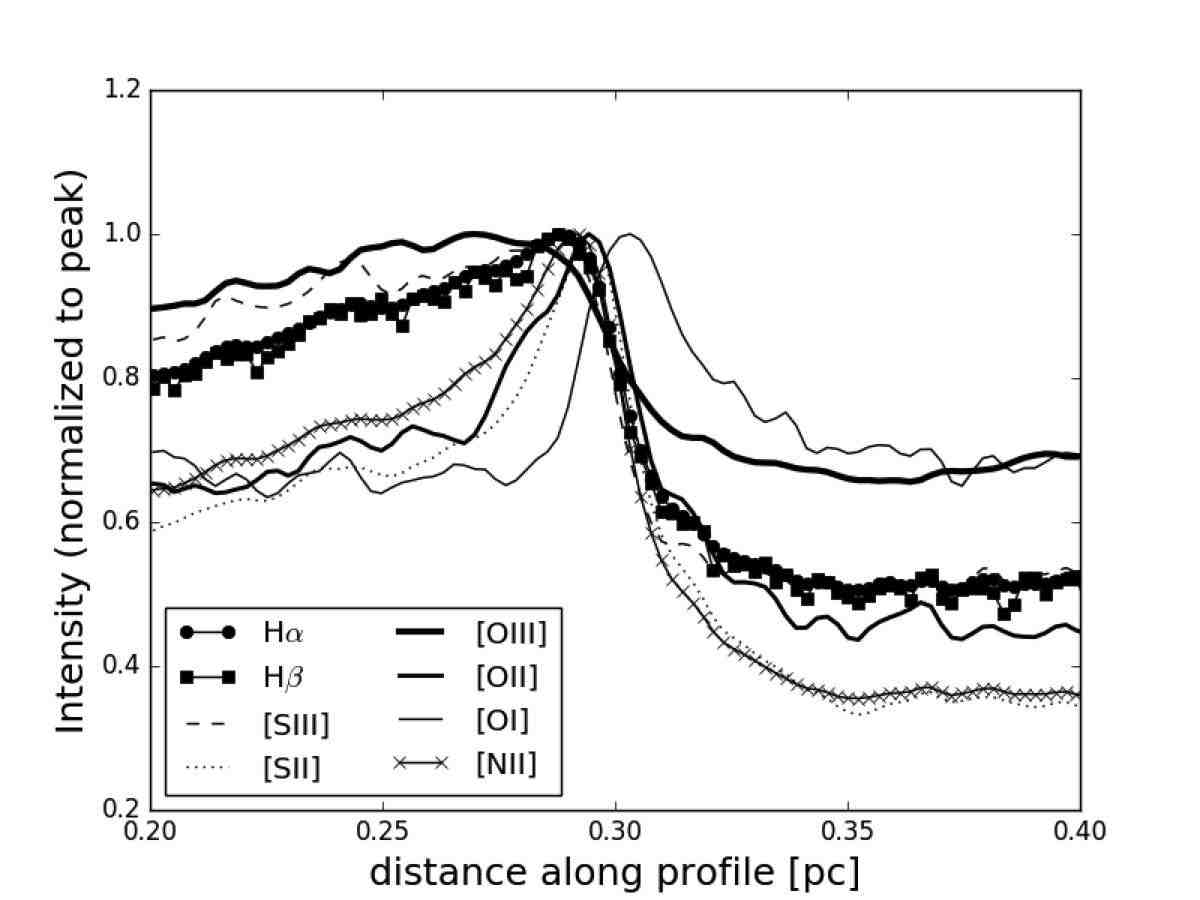}}
\subfloat[]{\includegraphics[scale=0.43]{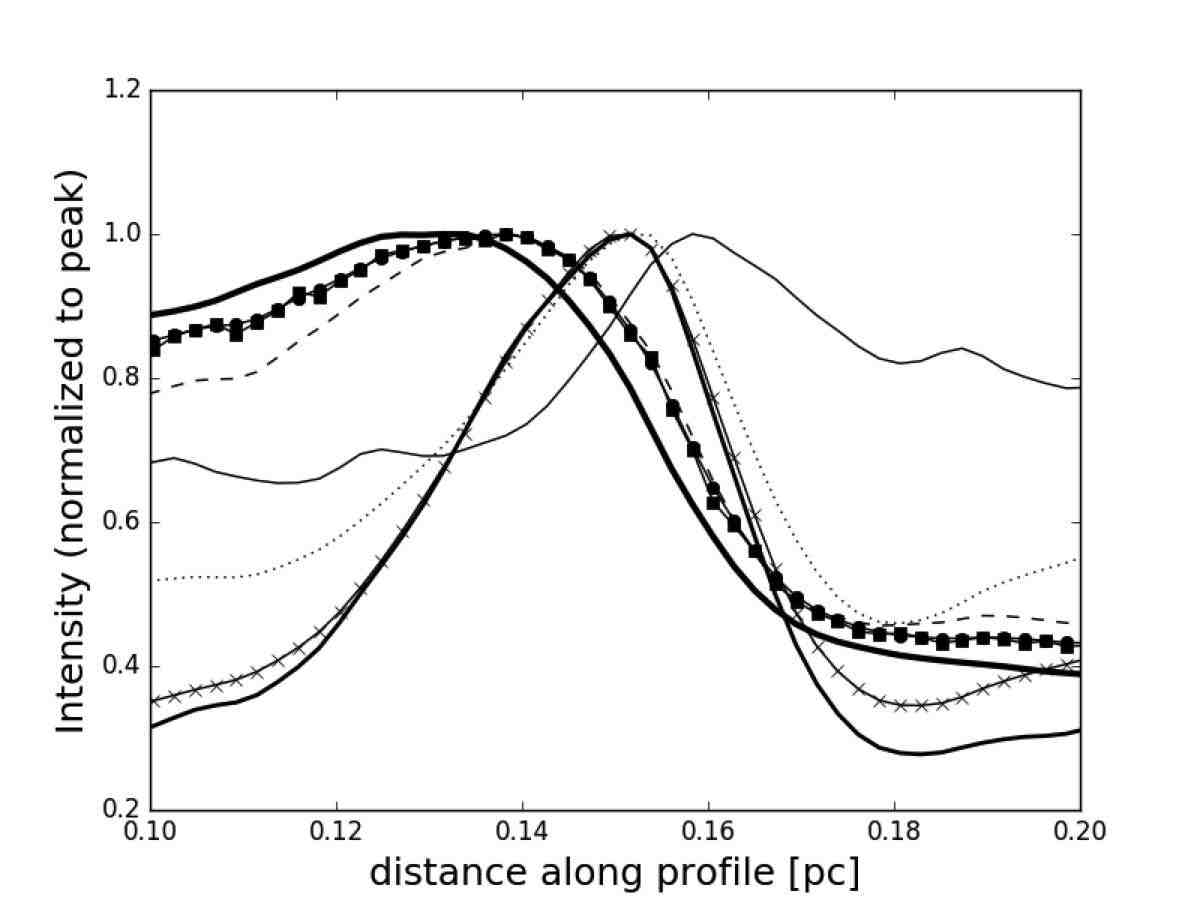}}}
\mbox{
\subfloat[]{\includegraphics[scale=0.43]{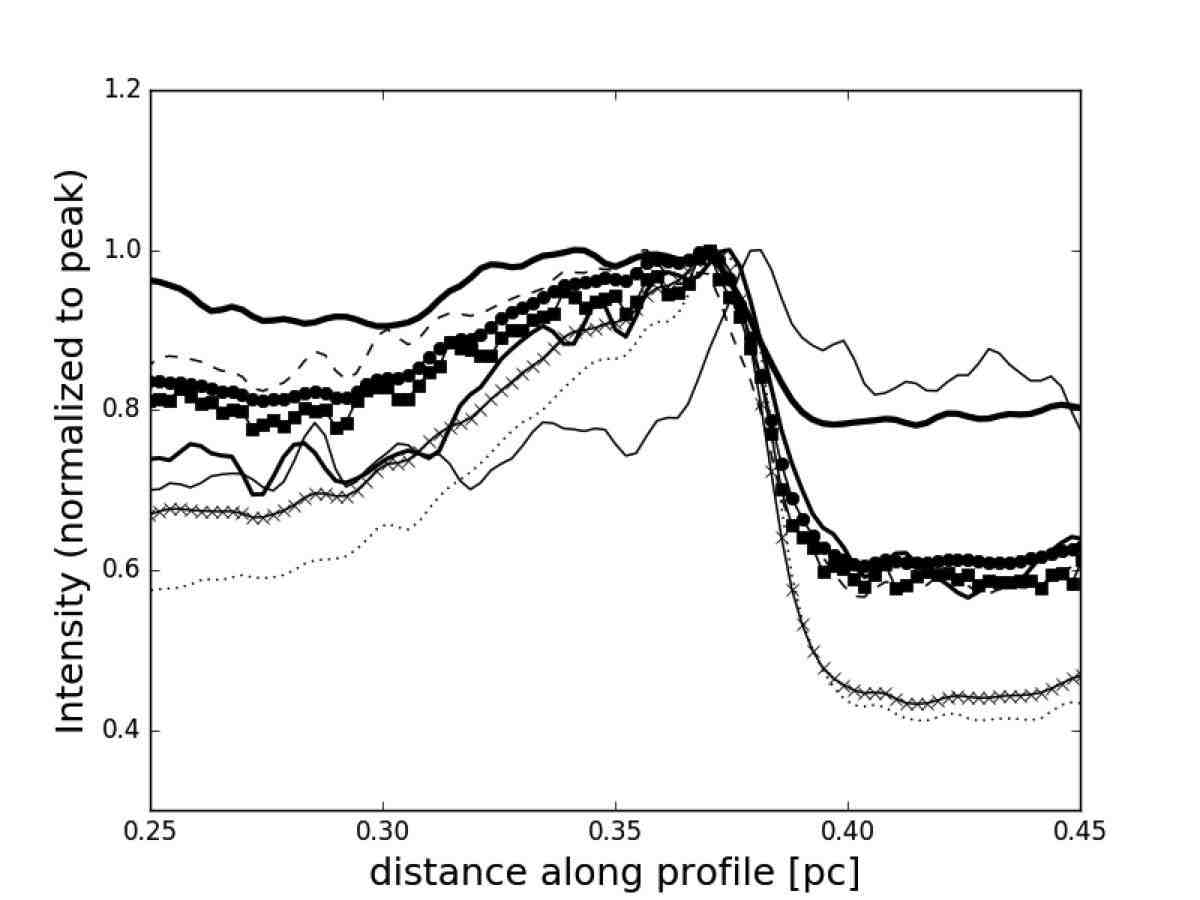}}
\subfloat[]{\includegraphics[scale=0.43]{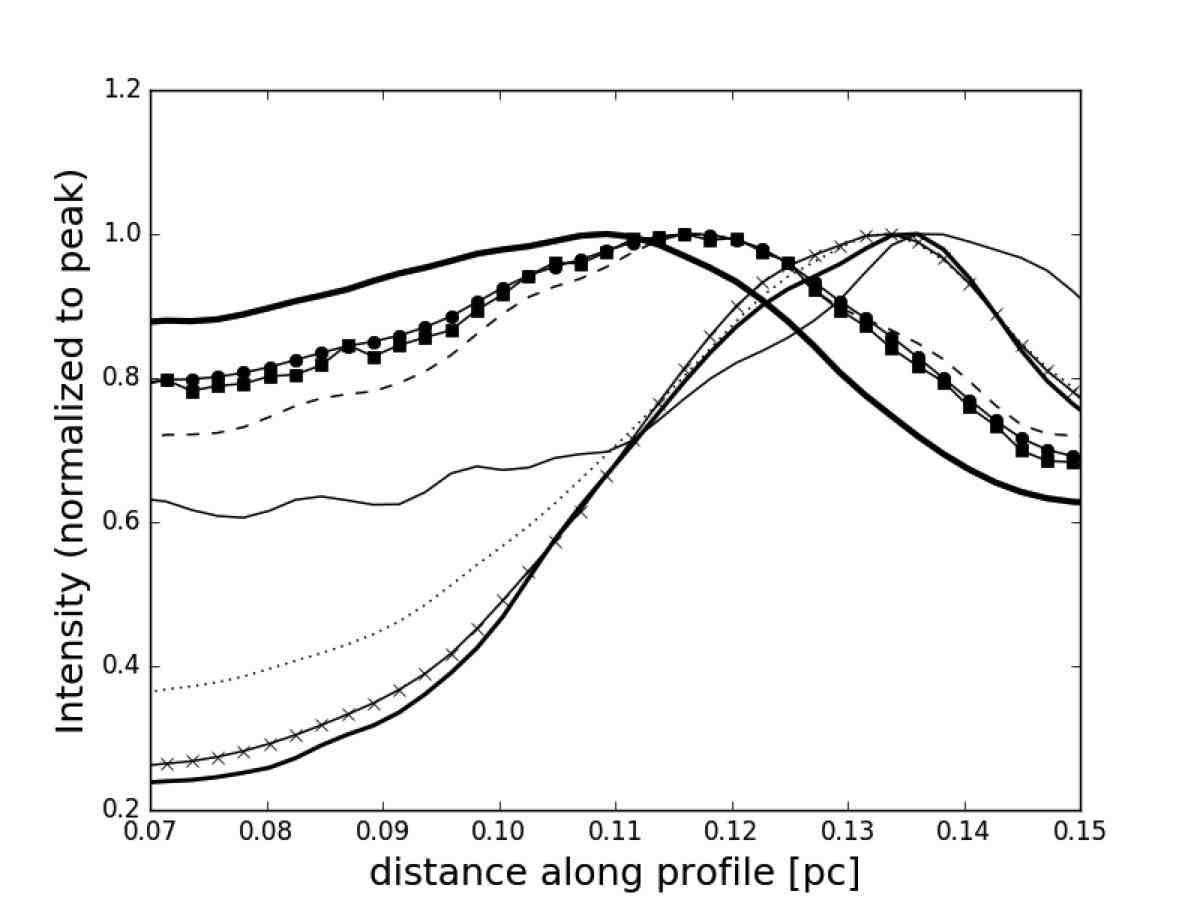}}}
\mbox{
\subfloat[]{\includegraphics[scale=0.43]{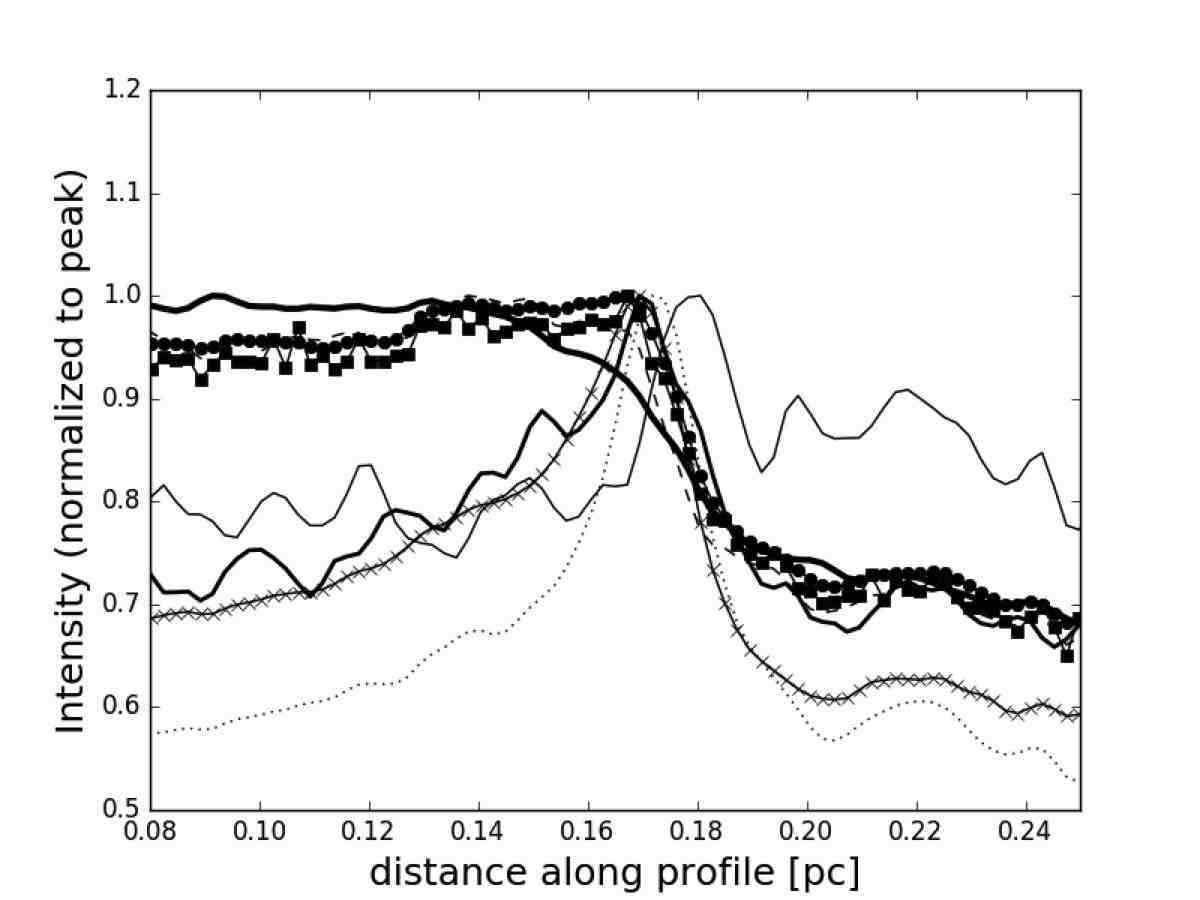}}
\subfloat[]{\includegraphics[scale=0.43]{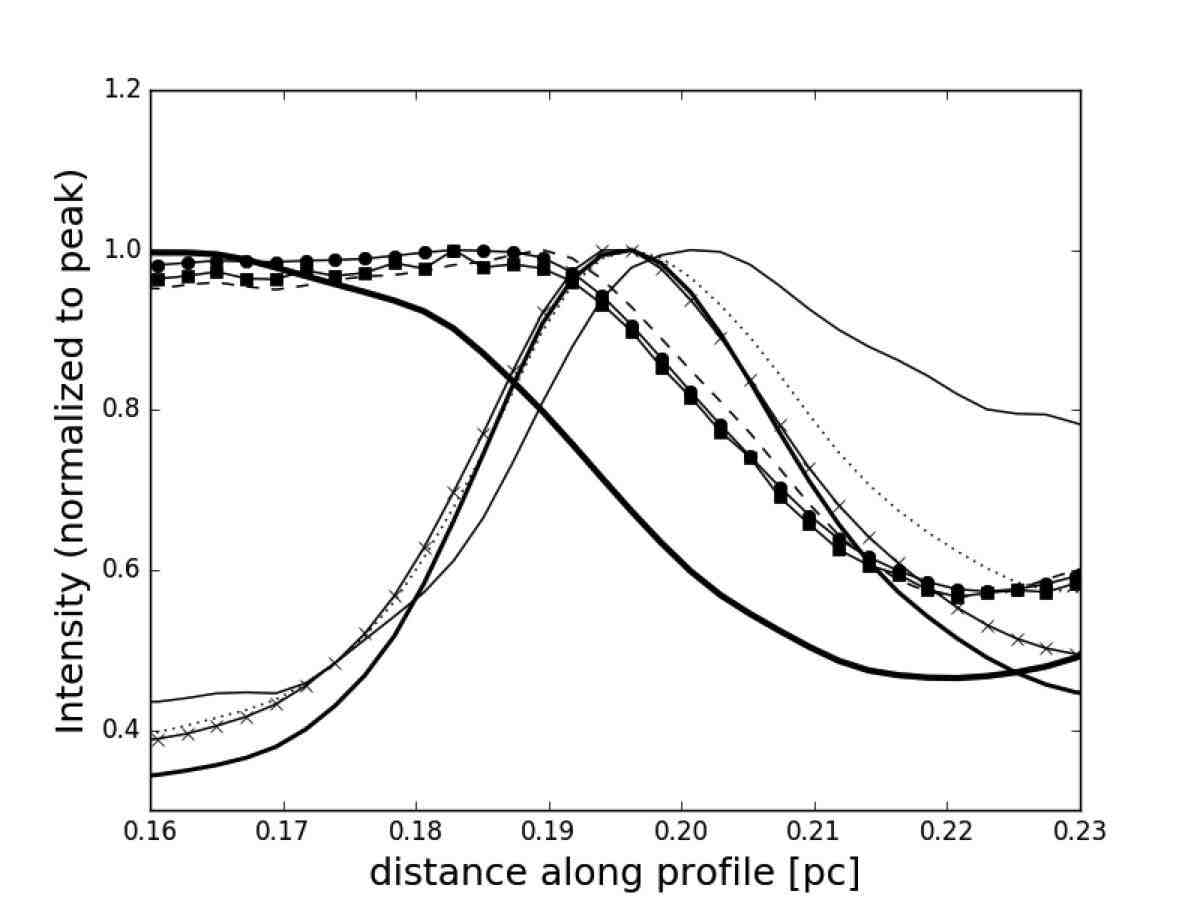}}}
\caption{Emission line intensity profiles along the slits shown in Fig.~\ref{slits} for R44 (left panels, slit 1, 2 and 3 in top, middle and bottom respectively) and R37 (right panels, slit 1, 2 and 3 in top, middle and bottom respectively). The legend in panel (a) applies to all other panels.}
\label{intensities1}
\end{figure*}

\begin{figure*}
\centering
\mbox{
\subfloat[]{\includegraphics[scale=0.43]{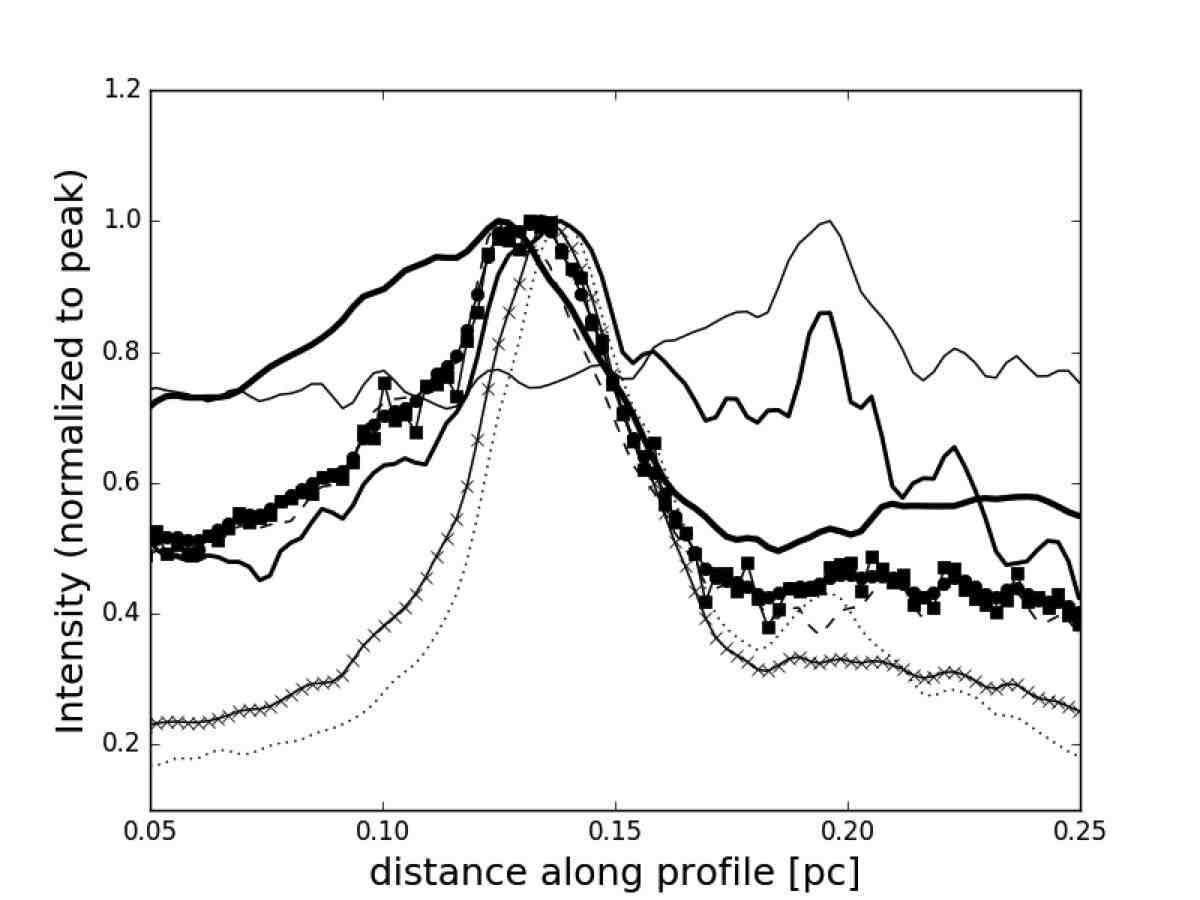}}
\subfloat[]{\includegraphics[scale=0.43]{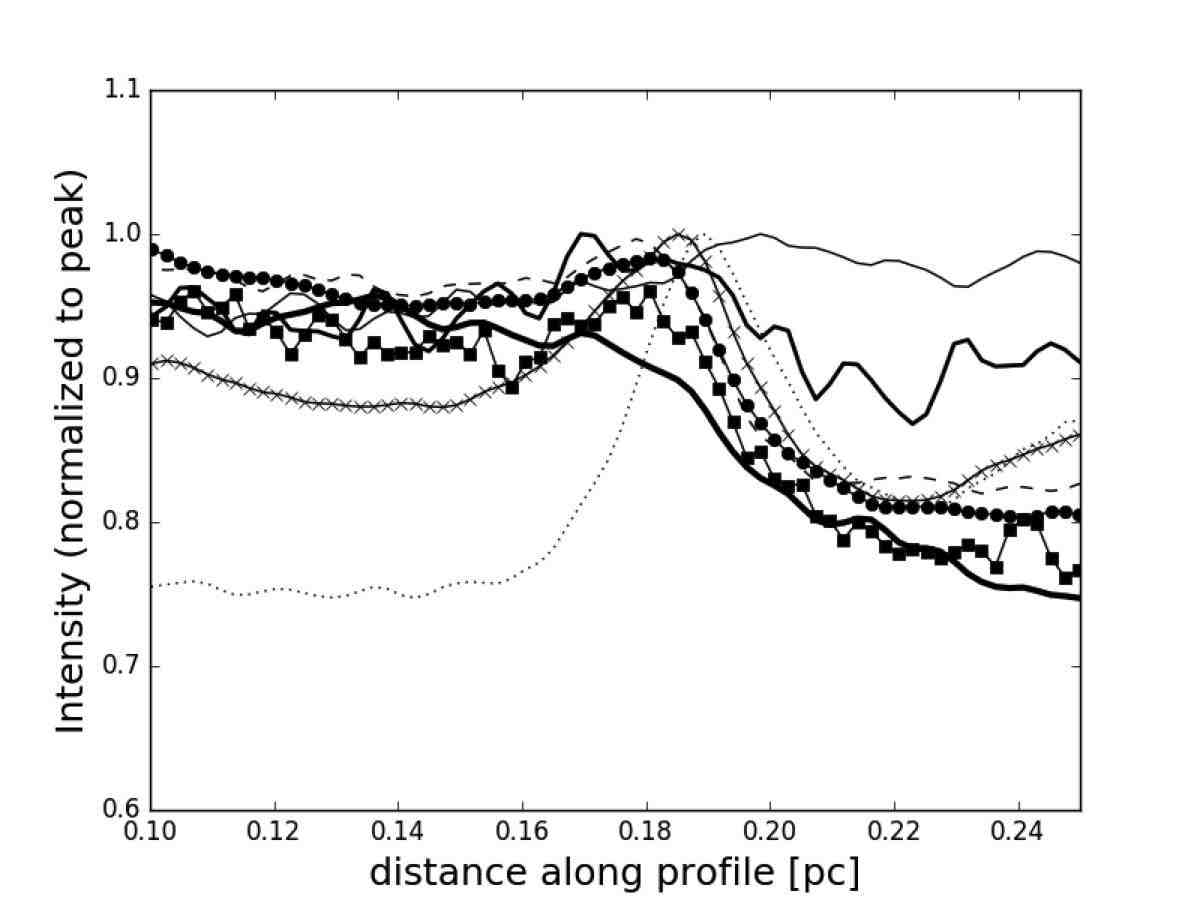}}}
\caption{Emission line intensity profiles along the slits shown in Fig.~\ref{slits} for R18 (panel a) and R45 (panel b). The legend is the same as in Fig.~\ref{intensities1}a.}
\label{intensities2}
\end{figure*}

To analyse the structure of the ionisation fronts at the pillar tips, in Figures \ref{intensities1} and \ref{intensities2} we plot the profiles of the various emission lines along slits positioned as shown in Fig.~\ref{slits} (from left to right, the abscissa in each plot indicates the distance along the slits, moving from the ambient matter towards the pillar tip and into the pillar material). Where more than one emission line for the same ionisation state of a given species is detected, the profile is computed from the mean value of these two (e.~g [SII]$\lambda$6717 and [SII]$\lambda$6731). 

A general trend in order and shape of the profiles is seen in all regions: (i) the emission lines with higher ionisation energies peak first (closer to the ionising sources), followed by the emission lines with progressively lower ionisation energies (as already discussed in MC15); (ii) the emission lines originating from the higher-ionised states, which are less localised at the pillar tips but instead more diffuse ([SIII], [OIII] and H$\alpha,\beta$ and which predominantly originate in the \textsc{HII} region) show a very shallow rise when moving towards the pillar tip (for slit 3 in R44 and R37 the trend is plateau-like), followed by a steep decline after the pillar tip, while the more localised emission lines originating from the lower-ionised states show a steeper rise when moving along the slit toward the pillar tip. We therefore conclude that the ionisation fronts at the pillar tips follow the well-known stratified structure expected for expanding \textsc{HII} regions (e.g. \citealt{hill87}, \citealt{hester96}).

The [OIII] line (corresponding to the mean value of the $\lambda$4959 and the $\lambda$5007 lines) peaks first in all regions except for R18 (where its peak coincides with that of the [SIII] line), and to quantify the structure of the ionisation fronts we report the distance in parsecs of the peak of the other emission lines from the [OIII] peak in Table \ref{peaks}. The spatial resolution of MUSE does not allow a full disentanglement between emission lines from the same ionisation state. However we are able to resolve the peaks between the double ionised, single ionised and neutral lines, e.g. [OIII], [OII] and [OI]. Especially when compared to the values reported in MC15 for M\,16 (comparison possible due to the relatively similar heliocentric distance of 2 and 2.3 kpc respectively\footnote{For this analysis we did not include the NGC 3603 pillar, as the much greater distance to this region (about 6.9 kpc) does not allow us to resolve the emission line peaks.}), it appears that the ionisation fronts in the Carina pillars are thicker (wider), as the emission line peaks appear further apart and the distance $\Delta(\mathrm{[OIII]}-\mathrm{[OI]})$ appears, in general, greater. 

\begin{table}
\begin{center}
\footnotesize
\caption{Location of the peaks (in 10$^{-2}$ pc) of the emission lines with respect to the position of the [OIII] peak (E$_{ion}$ = 35.12 eV) and their ionisation energy. Also reported here are the peak distances for the middle pillar in M\,16 (P2, see MC15). The seeing-limited resolution of the MUSE observations is about 0.2 arcsec, which correspond to 0.002 pc at the distance of Carina.}
\begin{tabular}{lcccccc}
\hline
\hline
 &  [SIII] & H$\alpha,\beta$ & [NII] & [SII] & [OII] & [OI] \\
\hline
R44 slit1 & 1.78 & 1.78 & 2.23 & 2.45 & 2.45 & 3.35 \\
R44 slit2 & 1.56 & 2.89 & 2.89 & 3.12 & 3.35 & 4.01 \\
R44 slit3 & 4.68 & 7.58 & 7.81 & 8.03 & 7.81 & 8.92 \\
\hline 
R37 slit1 & 0.67 & 0.67 & 2.01 & 2.01 & 2.01 & 2.68 \\
R37 slit2 & 0.67 & 0.67 & 2.45 & 2.45 & 2.68 & 2.66 \\
R37 slit3 & 3.245 & 2.68 & 4.01 & 4.01 & 4.01 & 4.46 \\
\hline
R18 & 0.000 & 0.67 & 1.12 & 1.34 & 1.34 & 7.14 \\
\hline
R45 & 1.115 & 0.67 & 11.59 & 11.59 & 11.37 & 14.72 \\
\hline
M\,16 P2 &  0.29 & 0.444 & 0.74 & 0.74 & 0.74 & 1.78 \\
\hline
E$_{ion}$ (eV) &  23.23 & 13.59 & 14.53 & 10.36 & 13.62 & - \\
\hline
\label{peaks}
\end{tabular}
\end{center}
\end{table}

This trend could be the consequence of the difference in ionising flux perceived by each pillar, which in turn is scaled to the distance from the ionising sources. The incident ionising flux is weaker on pillars further form the ionising sources, and pillars located further from the ionising sources are also less dense (see Fig.~\ref{qne} and Section \ref{physparams}). Together, this allows a deeper penetration of the ionising photons in less dense pillars, and therefore increasing the size of the ionisation front, while high-density pillars do not allow the radiation to penetrate as much and therefore show narrower ionisation fronts. A further possible effect is that for denser material the extinction (and thus the shielding) rises faster per unit path length, resulting in a narrower ionisation front. 

\begin{figure*}
\includegraphics[scale=0.6,trim={0 6cm 0 8cm},clip]{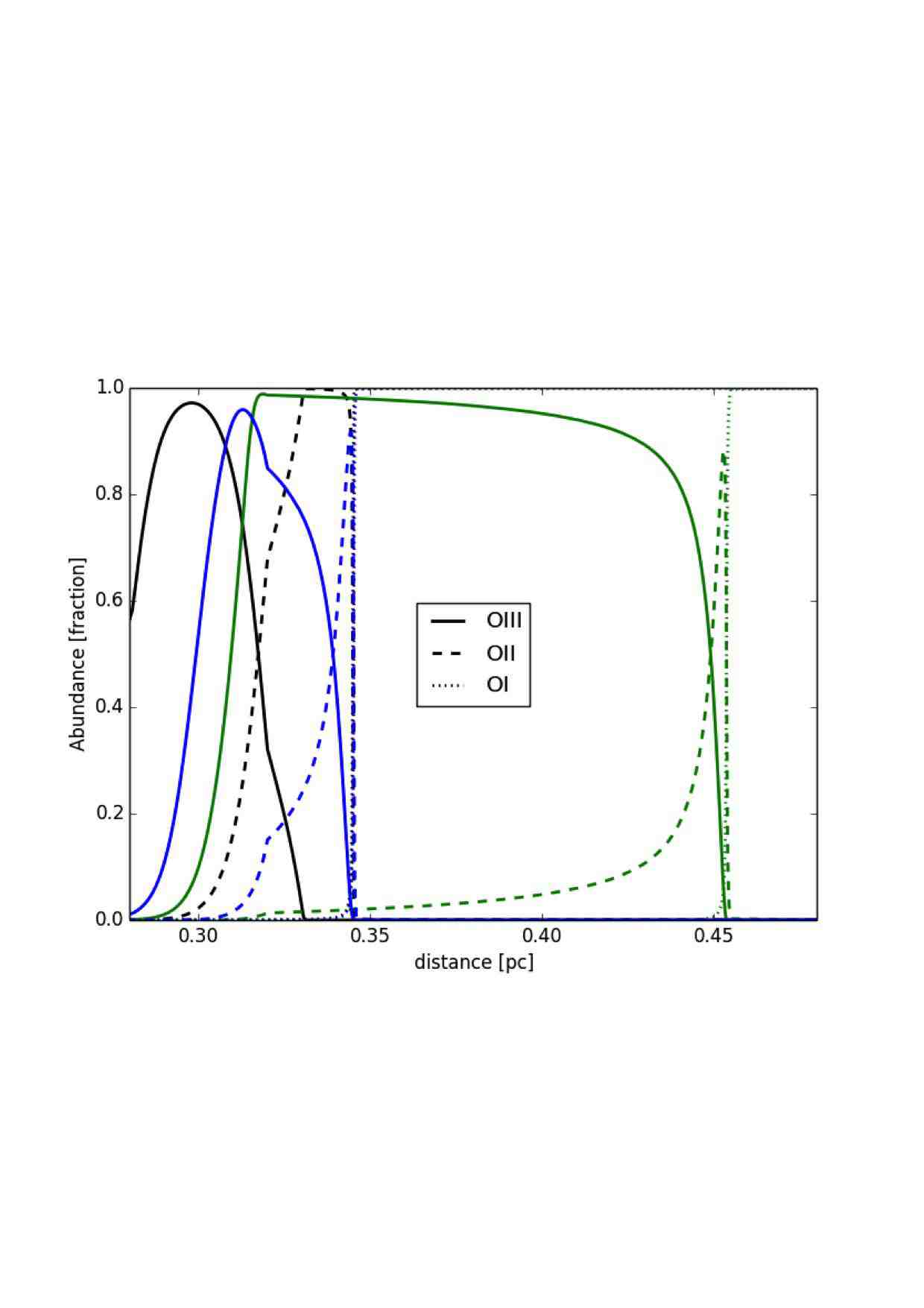}
\caption{Fraction of the total species abundance of [OIII] (solid lines), [OII] (dashed lines) and [OI] (dotted lines) as a function of distance along the pillar/ambient matter interface of simulated pillar-like structures, analogous to Figures \ref{intensities1} and \ref{intensities2}. The simulated pillars were exposed to three different ionising source with fluxes $\log(Q_{0})=$49 photons s$^{-1}$ (T = 45500 K), $\log(Q_{0})=$49 photons s$^{-1}$ (T = 31500 K) and $\log(Q_{0})=$50 photons s$^{-1}$(black, blue and green lines respectively). See text Section \ref{profiles}.}
\label{sims}
\end{figure*}

In MC15 we compared the emission line profiles along the three M\,16 pillar tips to simulations of ionised pillars \citep{gritsch10} which were post-processed with an optical radiative transfer code to obtain the simulated [OIII], H$\alpha$, [NII] and [SII] maps. This first qualitative comparison showed that the ionisation structure was recovered in the simulations. Here, we perform a preliminary analysis of theoretical ionisation front models by varying the initial conditions and processing the radiation with \textsc{cloudy} \citep{ferland13}. We assume the conditions of the Orion photo-dissociation region calculations \citep{ferland13} but replace the ionising source with a black body at a given temperature T. The hydrogen density is assumed to be 1 cm$^{-3}$ and changes rapidly to 5000 cm$^{-3}$ at 0.3 pc. The resulting abundance fractions of the species [OI], [OII] and [OIII] along the pillar/ambient matter interface is shown in Fig.~\ref{sims}. The calculations confirm a general trend in which the ionisation front gets narrower with higher ionising flux, i.e. the separation between the [OIII] and the [OI] line peaks gets smaller with increasing flux. Compared to the simulated emission line profiles shown in MC15, these do not fully recover the shape of the observed lines, as in MC15 we used simulations which where post-processed with \textsc{mocassin} \citep{ercolano03}, a 3D radiation transfer code, while here we use a theoretical one-dimensional density profile. The observations in this work hint at but are not sufficient to confirm the picture in which distance from/type of ionising source determine the thickness of the ionisation front, as no clear trend in $\Delta(\mathrm{[OIII]}-\mathrm{[OI]})$ vs $d_\mathrm{proj}$ and $\Delta(\mathrm{[OIII]]-\mathrm[OI]})$ vs $Q_\mathrm{0,pil}$ is observed. To further test this,
observations with higher angular resolution are needed, in combination with dedicated simulations in which the distance, the ionising photon flux, the pillar density as well as the geometry and the illumination of the pillar can be varied and the emission line intensity profiles carefully analysed.

\subsection{Physical parameters}\label{physparams}

For all observed regions, the electron density and temperature was determined as in MC15. For this, a continuum subtraction was performed on every data cube, followed by an extinction correction based on the H$\alpha$/H$\beta$ ratio and assuming R$_\mathrm{v}$=3.2 \citep{turner79} using the nebular analysis tool \textsc{pyneb} \citep{pyneb}. The line ratios used to compute the electron density $n_\mathrm{e}$ and temperature $T_\mathrm{e}$ are [SII]$\lambda$6731/[SII]$\lambda$6717 and [NII]$\lambda$6548+6584/[NII]$\lambda$5755 respectively, and the corresponding maps are shown in Figures \ref{nete1} to \ref{nete3}. These figures also show circular regions at the pillar tips (marked with black or white circles) from which mean $n_\mathrm{e}$ and $T_\mathrm{e}$ values were extracted and reported in Table \ref{params}. 

\begin{figure*}
\mbox{
\subfloat[]{\includegraphics[scale=0.45]{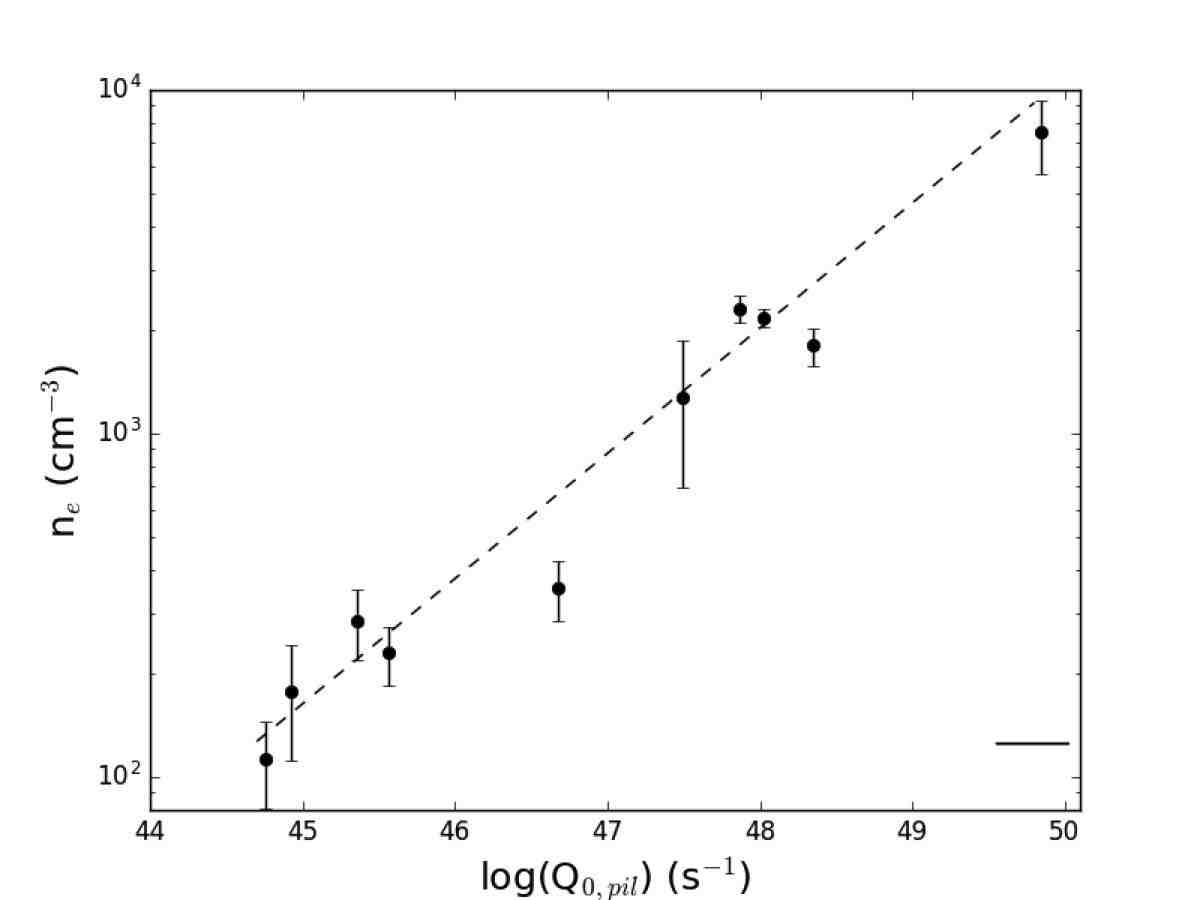}}
\subfloat[]{\includegraphics[scale=0.225]{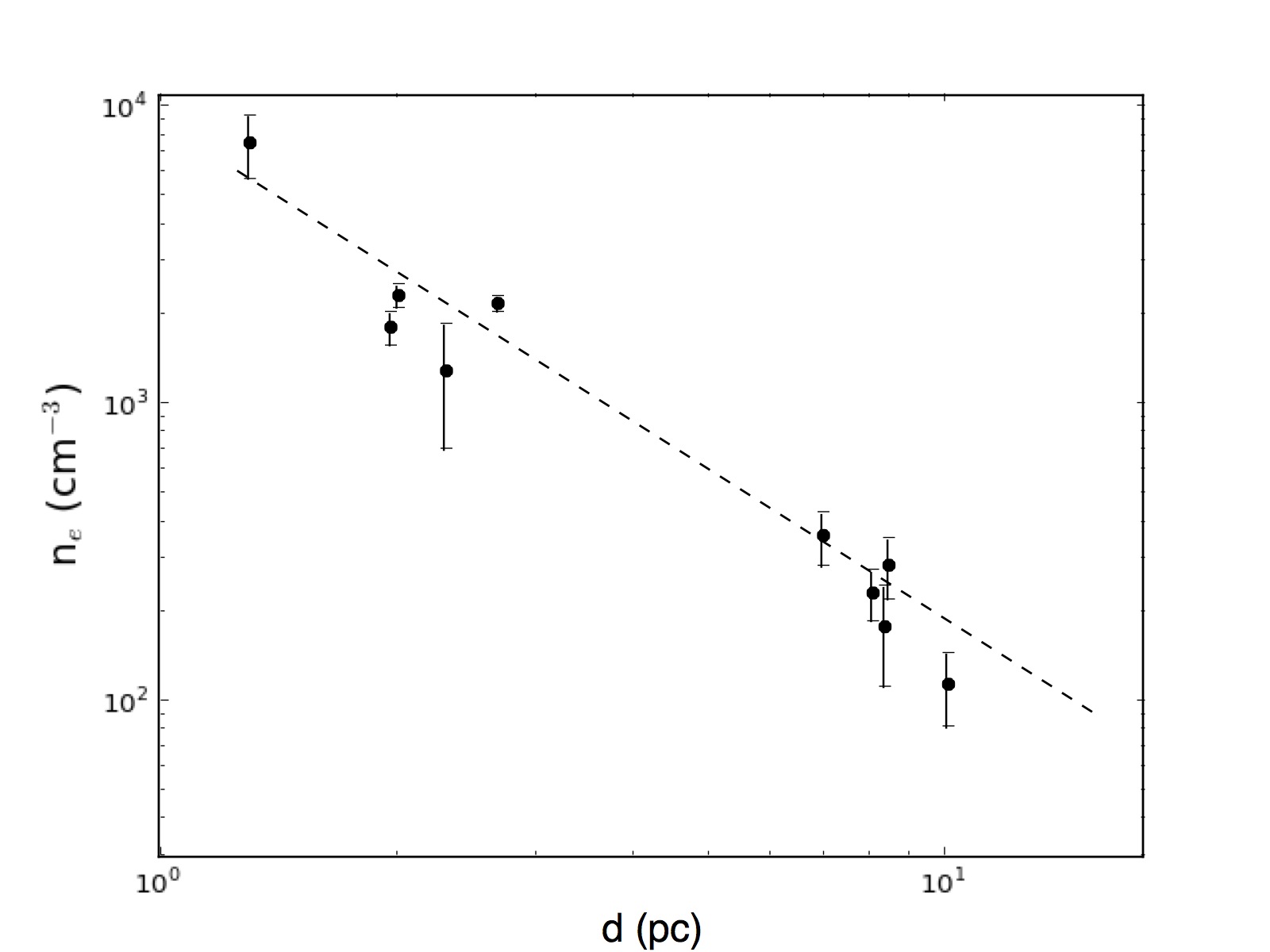}}}
\caption{Panel (a): the electron density $n_\mathrm{e}$ against the ionising photon flux at the pillar tips, $Q_\mathrm{0,pil}$ (circles) and the best fit power law (dashed) with index $p=0.36\pm0.04$ (see text Section \ref{physparams}). The horizontal line indicates the average error of $\log(Q_\mathrm{0,pil})$. Panel (b): $n_\mathrm{e}$ against the projected distance from the ionising sources (circles) and the best fit power law (dashed) with index $p=-1.67\pm0.19$.}
\label{qne}
\end{figure*}

When plotted against the ionising photon flux perceived at each pillar tip ($Q_\mathrm{0,pil}$, computed by scaling the photon flux $Q_{0}$ to the subtended solid angle between pillar and ionising source, see Section \ref{clusters}), we find a tight correlation between the electron density and $Q_\mathrm{0,pil}$. This is shown in Fig.~\ref{qne}a, where the uncertainties in $n_\mathrm{e}$ are the standard deviation of each circular extraction region, while for the uncertainty in $Q_\mathrm{0,pil}$ (indicated by the horizontal line in the lower right corner of the figure) we assume that the pillars are projected to within $\pm37^{\circ}$ and use this to provide a 25\% error on the true distance. This correlation shows that the observed electron densities are tightly correlated with the perceived photon flux at the pillar tips. This relation is linked to the initial conditions under which pillar-like structures form, and, together with the tight correlation between the electron density and the distance shown in Fig.~\ref{qne}b (which is related to the feedback-driving stars rather than the initial conditions), a possible argument is that if low-density pillars had been further in (closer to the ionising stars), they would have been eroded much faster. The combination of these two relations could represent a threshold density pillars need to have in order to withstand a given ionising photon flux at a given distance and on a given timescale. These findings are of great importance when considering the destructive effect of ionising feedback on the pillars, as the rate at which pillars are photo-evaporated therefore not only depends on the strength of the impinging ionising radiation, but also and crucially so, on the density of the pillar material, as well as the distance from the ionising source. We will discuss this further in Section \ref{mdot} when analysing the mass-loss rate of each pillar. 

\begin{figure*}
\centering
\mbox{
\subfloat[]{\includegraphics[scale=0.35]{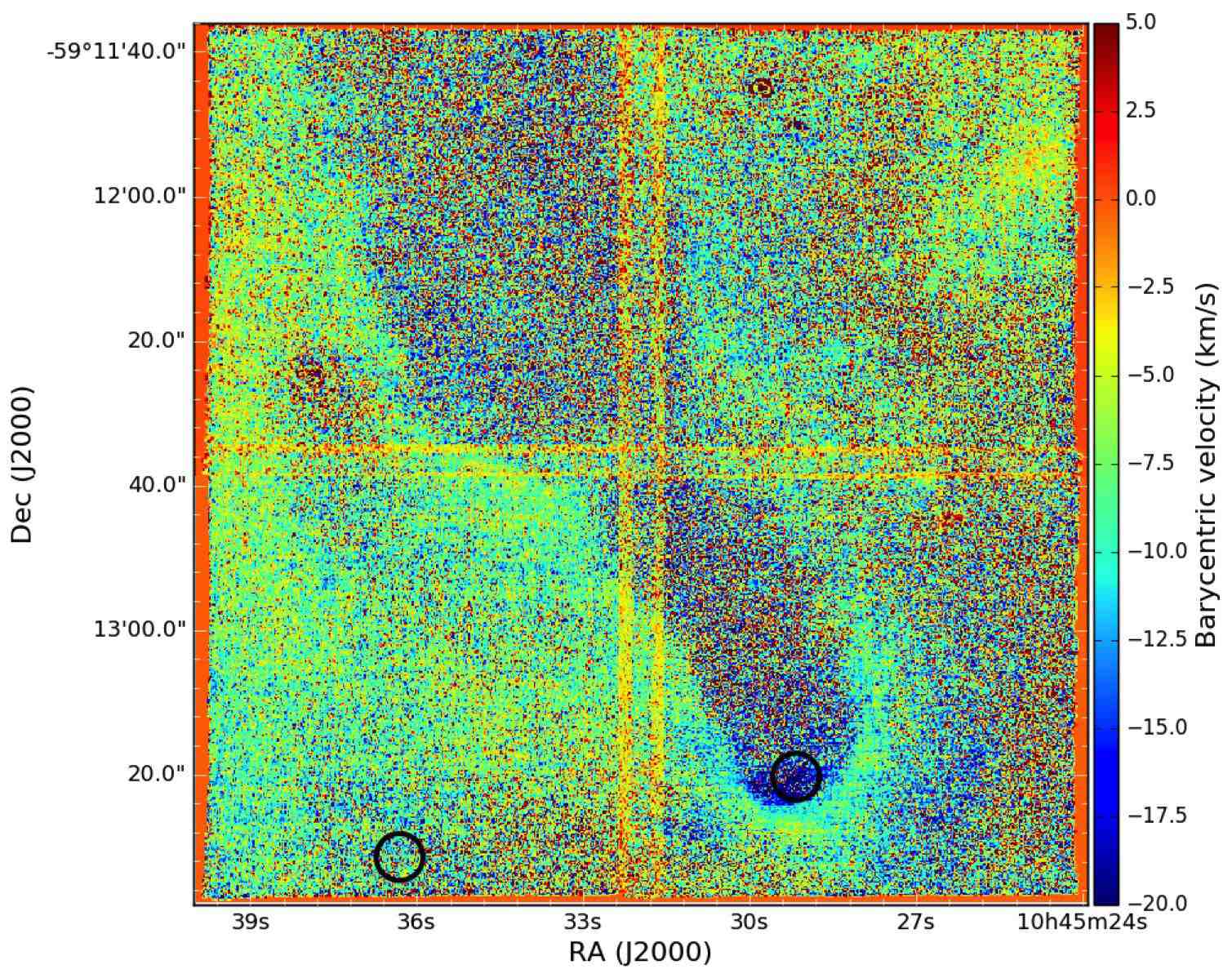}}
\subfloat[]{\includegraphics[scale=0.35]{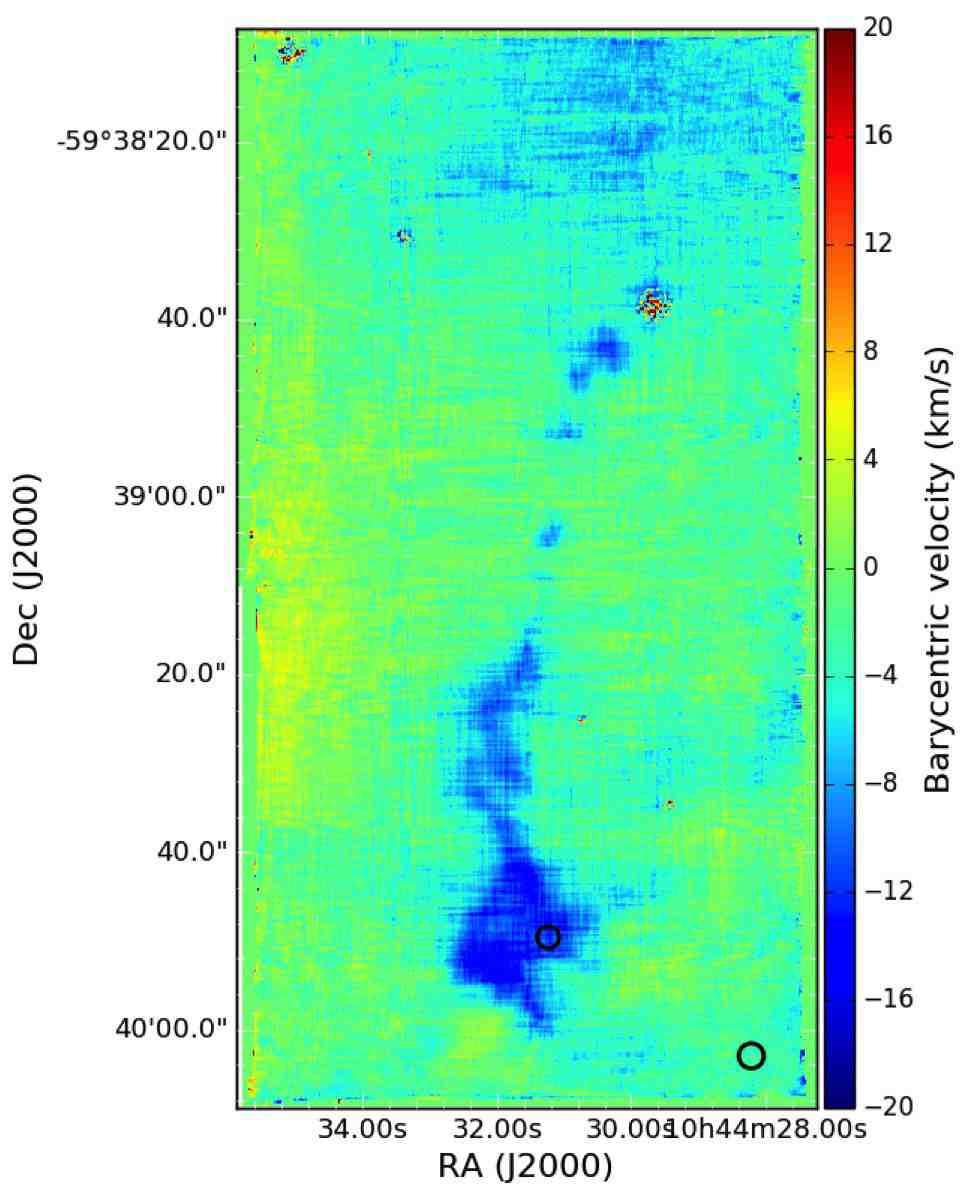}}}
\mbox{
\subfloat[]{\includegraphics[scale=0.45]{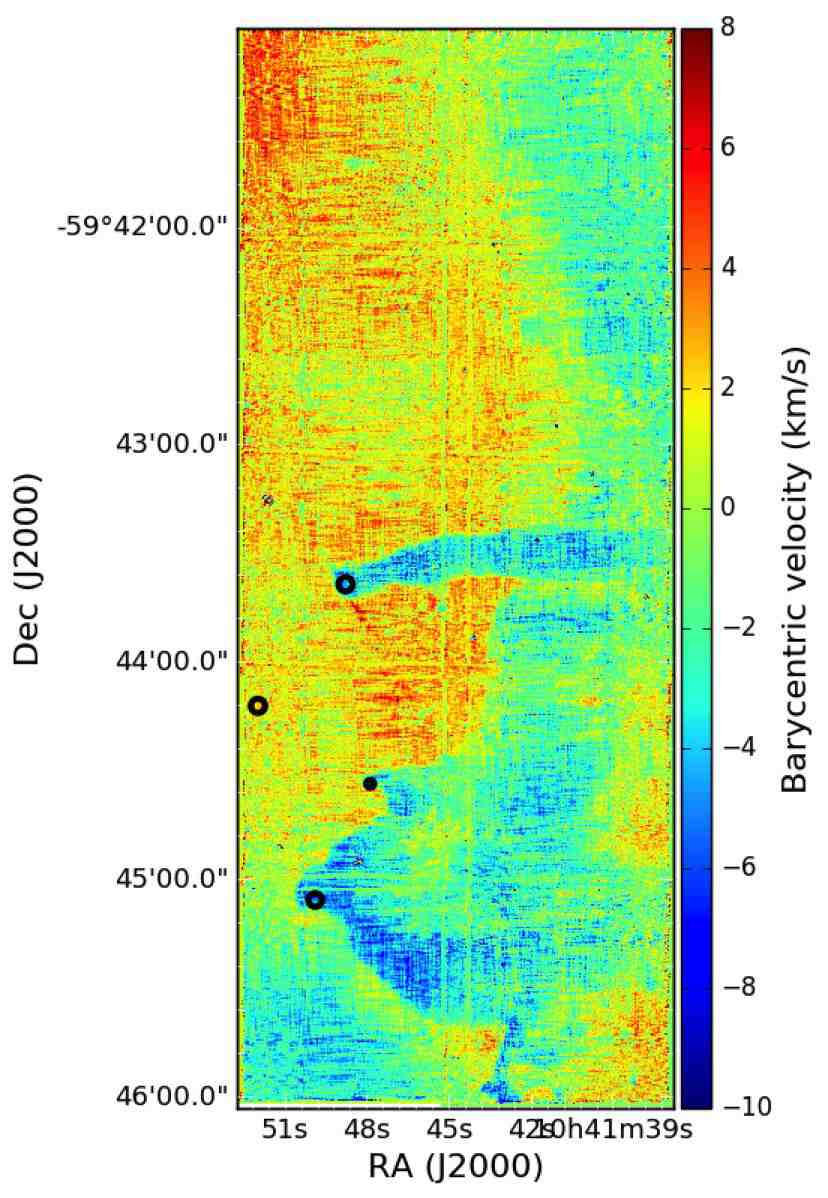}}
\subfloat[]{\includegraphics[scale=0.35]{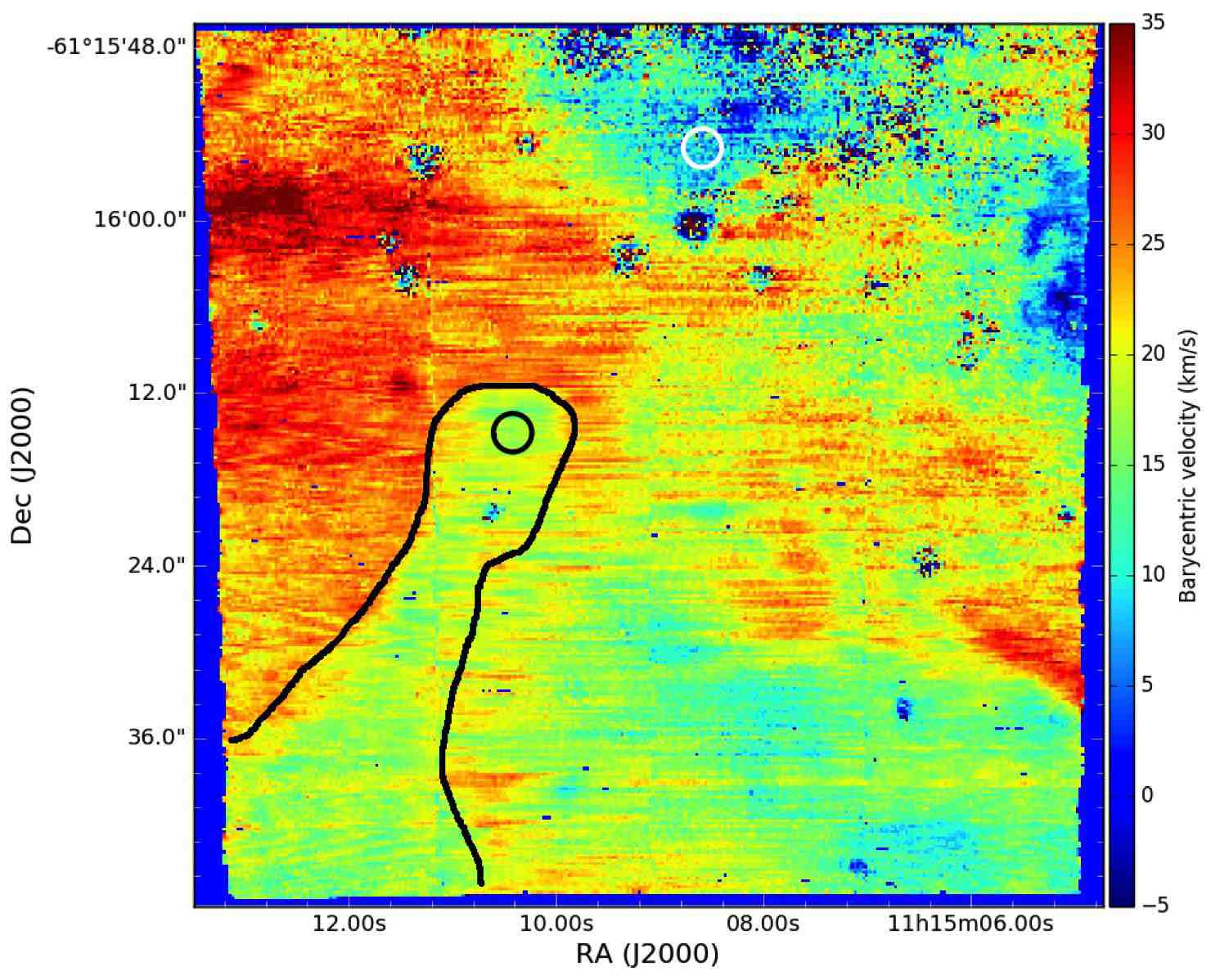}}}
\mbox{
\subfloat[]{\includegraphics[scale=0.35]{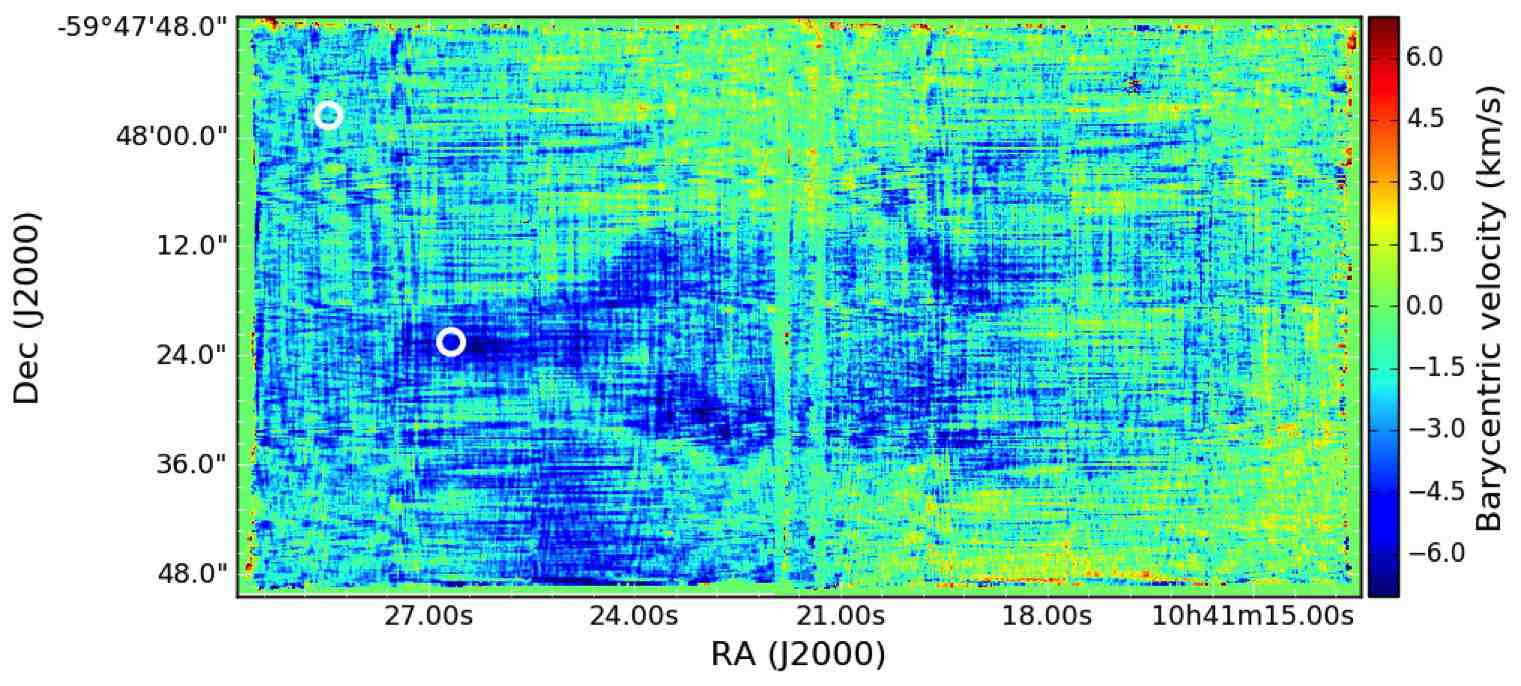}}}
\caption{Velocity maps of the four Carina regions (R18, R37, R44 and R45 in panels a, b, c and e respectively) and NGC 3603 (panel d). Circles (black or white depending on visibility) mark the regions used to extract velocity values of the pillar material and the surrounding interstellar medium (see text Section \ref{physparams} and Table \ref{params}). To better highlight the pillar in NGC 3603, a black contour tracing the pillar was added.}
\label{velmaps}
\end{figure*}

As in MC15, we normalise and stack several emission lines in each single pixel (thus assuming that all the lines in one pixel originate under one set of excitation conditions) and then fit the resulting velocity spectrum in each pixel with a gaussian function to produce a velocity map for each region\footnote{The \textsc{python} packages \textsc{spectral$\_$cube} (spectral-cube.readthedocs.org) and \textsc{pyspeckit} \citep{2011ascl.soft09001G} were used for the fitting routine.}. The lines used for the stacked spectrum are H$\alpha$, the two [NII] lines, the two [SII] lines, [OI]$\lambda$6300,6363 and HeI$\lambda$6678. This procedure is of great advantage for the medium spectral resolution of MUSE ($\Delta$v = 150 km s$^{-1}$ at 4600 \AA\ and $\Delta$v = 75 km s$^{-1}$ at 9300 \AA), as the line spread function is undersampled, and by applying this technique a better-sampled velocity spectrum is obtained (see MC15 for details). The velocity maps are shown in Fig.~\ref{velmaps}: all velocity maps show barycentric velocity, automatically corrected for by the MUSE data reduction pipeline\footnote{For the molecular gas in the Carina Complex, \cite{yonekura05} find LSR (local standard of rest) velocities ranging from about -26 km s$^{-1}$ to about -12 km s$^{-1}$. By converting the barycentric velocities to LSR values, the MUSE data is in good agreement with these values. As an example, a barycentric velocity value of 0 km s$^{-1}$ corresponds to approximately -12 km s$^{-1}$ in the LSR frame.}. The black or white circles in each panel show the circular apertures used to extract the velocities of the pillar material and the surrounding ambient matter and determine the velocity of the photo-evaporative flow (values reported in Table \ref{params}). All maps show a certain degree of either image/instrument artefacts (such as the checked pattern visible in all panels except for (a)\footnote{The checked pattern is particularly visible in the velocity map of NGC 3603 (panel (d)), as the 90$^{\circ}$ rotation dither pattern implemented in the Carina observations was not used.}, or the overlap regions resulting from mosaicking the various pointings for each region), low S/N (e.~g. R18), or both (e.~g. R45). However, each pillar is clearly distinguishable from the surrounding ambient matter (making a confident estimate of the photo-evaporation flow velocity possible), being typically blueshifted with respect to the latter. The blueshift of the pillar material traces the photo-evaporative flow, as this flows out perpendicular to the pillar surface (\citealt{hester96}, MC15) and results in the pillar surfaces pointing towards the observer being blueshifted along the line of sight. The velocity of the photo-evaporative flow is smaller than $c_\mathrm{s}$ ($\sim$10 km s$^{-1}$, the sound speed of ionised gas) in all regions. This fact reflects that the matter is being accelerated along a density-gradient, the densest gas is slower than the less dense gas, and the matter already moving at the sound speed of ionised gas is already merging with the \textsc{HII} region.

\subsection{Quantifying ionising feedback: relating the ionising photon flux to the photo-evaporation rate}\label{main}

\subsubsection{Ionisation from the nearby massive star clusters}\label{clusters}
To determine the impact of ionising feedback from the nearby massive clusters, a first assumption of which cluster is acting on which region needs to be made. The cases of NGC 3603 and M\,16 are relatively simple, as in both regions there is only one massive nearby star cluster (the HD 97950 star cluster and NGC 6611 respectively). The case of Carina however is less simple, due to the complex nature of this massive and large star-forming region. We therefore compute the projected distances from the observed pillars to the three most massive clusters, Tr 14, 15 and 16, and relate this to the direction the pillars point back to, therefore assuming that the direction of the pillars is indicative for the origin of the dominating impinging radiation. We use the \textsc{simbad} coordinates for each of these clusters: 10:43:56.0 -59:33:00 for Tr 14, 10:44:43.0 -59:22:00 for Tr 15 and 10:45:10.0 -59:43:00 for Tr 16 (all coordinates are in J2000). The location of the pillars with respect to the central coordinates of the three star clusters, as well as the known O- and B-stars in the Carina region \citep{gagne11} are shown in Fig.~\ref{orient}. Based on the orientation of the pillars and their distance to the nearby clusters, we assume the following for each region:

\begin{itemize}
\item R18: the pillar is located at $\sim$ 10.5' (7.0 pc) north-east of Tr 15, and points directly toward the central coordinates of this cluster; we therefore assume that Tr 15 is the feedback-driving cluster acting on this region.
\item R37: this ionised globule lies at $\sim$ 6.1' (4.1 pc) south-east of Tr 16. \cite{smith04} propose that the dominant ionising sources might be either the O4 I star Tr16-244, the WNL star WR25 (HD 93162), or both, since the main ionisation fronts point in that direction. Indeed (as shown by the black arrow in Fig.~\ref{slits}), the globule and its main ionisation front do not point back towards the central coordinates of Tr 16, but rather towards Tr16-244 and HD 93162, and together with the fact that it is also closer in projection to these two stars (2.7 pc) we therefore assume that these are the main ionising sources. 
\item R44 and R45: these pillars lie closer in projection to Tr 14 than Tr 16 (12.9 vs. 16.9 pc and 16.1 pc vs. 19.1 pc respectively). However, they do not point back towards the central coordinates of these two clusters, but rather towards an O6- and an O9.5-star (HD303316 and HD305518) at about 2.7 pc north-east. Given the large projected distance and the weak emission from these pillars, we therefore assume that it is the combined flux of the two above mentioned O-type stars acting on these regions, rather than Tr 16 or Tr 14. To simplify the discussion, we will refer to the combination of these two O-stars as HD30 henceforth.
\end{itemize}

For NGC 3603 there is no full census of B-type stars, and for Carina we see that the B-type stars only contribute a few percent of the ionising flux  (see below), and to estimate the energy input in terms of photons s$^{-1}$ from the massive clusters, we assume that the O-type stellar population of each cluster is dominating the energy input. We convert the spectral type of the known O stars to a flux $Q_{0}$ by using the values given in \cite{martins05}. For Tr 14, 15 and 16 this was done in \cite{smith06}, however these authors also include B-type stars in their estimate of the total flux for each cluster. Here, we therefore use the list of O-stars in \cite{smith06} as well as their reported values for $\log(Q_{0})$ (also taken from \citealt{martins05}) and find $\log(Q_{0})$=49.49 s$^{-1}$ for Tr 15 and $\log(Q_{0})$=50.92 s$^{-1}$ for Tr 16. The comparison between these two values for $\log(Q_{0})$ and those obtained by including the B-stars reveals that the contribution to the ionising flux of the B-stars is only 15\% for Tr 15 and 8\% for Tr 16. 

As mentioned above, we assume that R37 is mainly affected by Tr16-244 and HD 93162, which correspond to an O4 I star and a WN6 star respectively, and their ionising fluxes add up to $\log(Q_{0})$=48.02 s$^{-1}$ (henceforth we will refer to the combination of these two stars as O4/WNL). For R44 and R45 the combined fluxes of HD303316 and HD305518 (HD30) amount to $\log(Q_{0})$=49.02 s$^{-1}$.

For NGC 3603, \cite{moffat83} and \cite{drissen95} derived spectra for 13 and 14 objects respectively, of which 11 were classified as early-type O-stars. However, neither of these studies could resolve the central part of the cluster, which was classified as a Trapezium-like WR system \citep{walborn73}. It was not until \cite{melena08} and a combination of optical spectroscopy and high-resolution $HST$ imaging that the classification of a total number of 38 cluster stars was possible. These authors find a large number of O- and B-type stars with spectral types as early as O3, and high-mass objects with M $\ga$ 120 M$_{\odot}$ and Wolf-Rayet features. They discuss that the highest mass stars seem to be slightly younger than the 1-4 Myr old population with M $\la$ 40 M$_{\odot}$, having ages between 1 and 2 Myr. We derive the integrated ionising flux $Q_{0}$ of each O-type star in Table 1 of \cite{melena08} by comparing their spectral type to the one derived from simulations in \cite{martins05}. For the 33 O-type stars in this table, we find a total ionising photon flux of log(Q$_{0}$)$\sim$ 50.98 s$^{-1}$. For the number and spectral type of O-stars in M\,16 we refer to Table 11 in \cite{evans05}, and compute  log(Q$_{0}$)$\sim$ 49.87 s$^{-1}$ for the 13 O-stars of this cluster.

We then scale the integrated ionising fluxes $Q_{0}$ of each cluster to the distance of each pillar to determine the ionising fluxes impinging on each pillar tip, $Q_\mathrm{0, pil}$. For this we measure the radius of the pillar tips to determine the exposed pillar area (assuming that the pillar tip is a half sphere with radius $r$) and derive the solid angle which each pillar subtends with the respective cluster as $\Omega=2\pi r^{2}/d_\mathrm{proj}^{2}$, where $d_\mathrm{proj}$ is the projected pillar-cluster distance. All values are reported in Table \ref{params}.


\begin{sidewaystable*}
\vspace*{-17cm}
\centering
\footnotesize
\begin{tabular}{lcccccccccc}
\hline
\hline
Pillar & Cluster & d$_\mathrm{proj}$ & r$_{surf}$ & $\Omega$ & log(Q$_{0}$) & log(Q$_\mathrm{0,pil}$) & $n_\mathrm{e}$  & T$_\mathrm{e}$  & $v$  & $\dot{M}$  \\
 & & (pc) & (pc) & (sr) &(photons s$^{-1}$) & (photons s$^{-1}$) & (cm$^{-3}$) & (K) & (km s$^{-1}$) & (M$_{\odot}$ Myr$^{-1}$) \\
\hline
R18 & Tr 15 & 7.0 & 0.11 & 1.55$\times$10$^{-3}$ & 49.49 & 46.68 & 355$\pm$73 & 9832$\pm$2663 & 6.3$\pm$4.8 & 3.1$\pm$0.8 \\
R37 & O4/WNL & 2.7 & 0.09 & 6.98$\times$10$^{-3}$ & 50.18 & 48.02 &  2171$\pm$133 & 9738$\pm$274 & 8.9$\pm$0.1 & 17.7$\pm$0.1\\
R44 P1 & HD30 & 8.5 & 0.05 & 2.17$\times$10$^{-4}$ & 49.02 & 45.36 & 285$\pm$66 & 8418$\pm$1224 & 6.4$\pm$0.3 & 0.5$\pm$0.2\\
R44 P2 & HD30 & 8.4 & 0.03 & 8.01$\times$10$^{-5}$ & 49.02 & 44.92 & 177$\pm$66 & 7746$\pm$1422 & 3.3$\pm$0.1 & 0.07$\pm$0.37 \\
R44 P3 & HD30 & 8.1 & 0.06 & 3.45$\times$10$^{-4}$ & 49.02 & 45.56 & 229$\pm$44 & 8189$\pm$1209 & 6.6$\pm$0.1 & 0.6$\pm$0.2 \\
R45 & HD30 & 10.1 & 0.03 & 5.54$\times$10$^{-5}$ & 49.02 & 44.76 & 113$\pm$32 & 7453$\pm$1262 & 3.4$\pm$0.1 & 0.04$\pm$0.28 \\
M\,16 P1 & NGC 6611 & 1.97 & 0.14 & 31.73$\times$10$^{-3}$ & 49.87 & 48.35  & 1800$\pm$230 & 8777$\pm$749 & 4.5$\pm$0.3 & 17.9$\pm$0.1\\
M\,16 P2 & NGC 6611 & 2.01 & 0.08 & 9.95$\times$10$^{-3}$ & 49.87 & 47.87 & 2306$\pm$205 & 8792$\pm$601 & 5.3$\pm$0.3 & 8.9$\pm$0.1 \\
M\,16 P3 & NGC 6611 & 2.31 & 0.06 & 4.24$\times$10$^{-3}$ & 49.87 & 47.49 & 1277$\pm$579 & 8823$\pm$1905 & 8.8$\pm$11.8 & 4.6$\pm$1.4\\
NGC 3603 & HD97950 & 1.30 & 0.14 & 72.87$\times$10$^{-3}$ & 50.92 & 49.78 & 7536$\pm$1824 & 11347$\pm$809 & 9.2$\pm$2.4 & 150.0$\pm$0.4\\
\hline
\end{tabular}
\\[1.5pt]
\caption{Properties and physical parameters of the considered pillars (column 1): feedback-driving cluster (column 2), projected distance from cluster (column 3), radius of pillar tip (column 4), subtended solid angle pillar-cluster (column 5), integrated ionising photon flux of the cluster (column 6), photon flux at pillar tip (column 7), electron density (column 8), electron temperature (column 9), velocity of the photo-evaporative flow (column 10) and mass loss rate of the pillar tip (column 11). HD30 stands for the combination of HD303316 and HD305518, while O4/WNL stands for the combination of Tr16-244 and HD 93162 (see text Section \ref{clusters}).}
\label{params}
\end{sidewaystable*}

\subsubsection{Mass-loss rate due to photo-evaporation}\label{mdot}

In MC15 the mass-loss rate was computed with the optically thin [SII]$\lambda$6731 line. In this method, the luminosity of the [SII] line is used to compute the mass of the line-emitting matter, which is then used to compute the mass-loss rate as $\dot{M}=Mv/l$ (where $v$ is the velocity of the photo-evaporative flow and $l$ the size of the line-emitting region). However, this method is only valid if the density of the line-emitting matter is $N\gtrsim N_\mathrm{c}$, with $N_\mathrm{c}$ the critical density of the [SII]$\lambda$6731 line (N$_\mathrm{c}\sim3.9\times10^{3}$ cm$^{-3}$). As the densities computed for the Carina pillars are in general much lower than this (see Table \ref{params}), we use the expression given in \cite{smith04} to compute the mass loss rate,

\begin{equation}
\dot{M}\simeq\pi r^{2}m_\mathrm{H}n_\mathrm{H}v\qquad \, \mathrm{(kg/s)}
\label{mdoteq}
\end{equation}

\noindent where $r$ is the curvature radius of the pillar tip, $v$ the velocity of the photo-evaporative flow and $n_\mathrm{H}$ the matter density. As in \cite{smith04}, we adopt $n_\mathrm{e}\simeq0.7n_\mathrm{H}$ \citep{sankrit00}, and we position circular apertures at the ionisation fronts, i.e. the pillar tips, to extract values for $v$ and $n_\mathrm{e}$. The positions of the circular extraction regions are marked in Figures \ref{velmaps}, \ref{nete1} and \ref{nete2}, and the extracted values are reported in Table \ref{params}. The errors on the $v$, $n_\mathrm{e}$ and $T_\mathrm{e}$ measurements correspond to the standard deviation of the values of included pixel values. The uncertainty for $\dot{M}$ is then computed by propagating the uncertainties in velocity and electron density. For R37 we find $\dot{M}\sim18$ M$_{\odot}$ Myr$^{-1}$, which agrees with the value of $\dot{M}\sim20$ M$_{\odot}$ Myr$^{-1}$ found by \cite{smith04}. For M\,16, in MC15 we report a mass-loss rate of $\dot{M}\sim70$ M$_{\odot}$ Myr$^{-1}$, while here the combined mass-loss rate of the three pillars (Table \ref{params}) is about $\dot{M}\sim20$ M$_{\odot}$ Myr$^{-1}$. The difference of these two values comes from the fact that in MC15 we determined the mass-loss rate of the the combined three pillars by computing the total mass of the [SII]-emitting matter of the three pillars, while here we only consider the ionisation fronts of the pillar tips.

\begin{figure}
\includegraphics[scale=0.45]{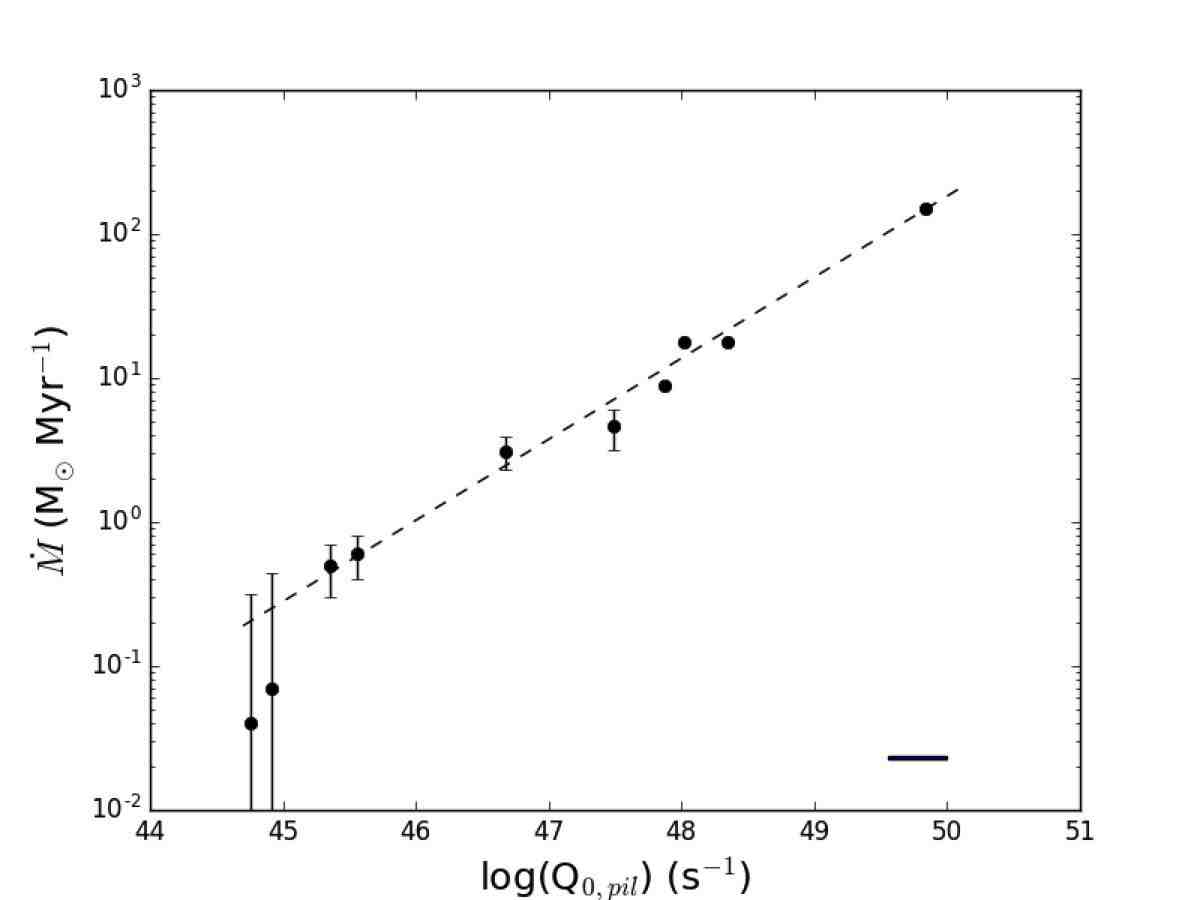}
\caption{The mass-loss rate $\dot{M}$ against the ionising photon flux at the pillar tips, Q$_\mathrm{0,pil}$ (circles) and the best fit linear (dashed) relation. The power law index is $p=0.56\pm0.02$. The horizontal line indicates the average uncertainty of $\log(Q_\mathrm{0,pil})$. See text Section \ref{mdot}.}
\label{qmd}
\end{figure}

Since the mass-loss rate of the pillars is due to photo-evaporation caused by the impinging ionising photons, we investigate the presence of a correlation between the ionising photon flux perceived by the pillar tips, $\log(Q_{0})$, and the mass-loss rate $\dot{M}$. The resulting plot is shown in Fig.~\ref{qmd}, where the data points are best fit with a power law of index $p=0.56$. Together with the correlation between $n_\mathrm{e}$ and $\log(Q_\mathrm{0,pil})$, and $n_\mathrm{e}$ and the projected distance, we find as expected that pillars with lower densities are further away from the ionising stars/star clusters, and these also show lower mass-loss rates. The photo-evaporative effect of ionising radiation therefore depends on the relative flux perceived at the pillar tip, the density, as well as the distance of the pillar from the massive stars.

There are a number of theoretical works where the dependency between $\dot{M}$ and $Q_\mathrm{0,pil}$ has been quantified for various different structures and environments, such as interstellar clouds subject to the ionising radiation of a young star (e.g. \citealt{bertoldi89}, \citealt{bertoldi90}) or the photo-evaporation of cloud clumps in the vicinity of planetary nebulae (e.g. \citealt{mellema98}). To test the theoretical understanding of photo-evaporation, we compare our empirically determined $\dot{M}$-$Q_\mathrm{0,pil}$ relation to the 2-d hydrodynamical simulations of the dynamical evolution of a neutral globule exposed to the ionising radiation of OB stars of \cite{lefloch94}. Specifically, these authors assume an initially spherical neutral cloud with uniform density, a distance $d$ to the ionising star(s) which is large compared to the size of the cloud, and a incident ionising flux parametrised by $\Phi=Q_\mathrm{0}/4\pi d^{2}$. In their model, the ionised cloud undergoes a 2-stage evolution consisting of an initial (short, $\sim10^{5}$ yr) collapse phase, followed by a (longer, a few $10^{5}$ yr) cometary phase. The latter terminates with the cloud being rapidly disrupted due to small-scale instabilities. \cite{lefloch94} give two $\dot{M}$-flux relations, depending on whether the incident ionising flux is (a) processed to balance the recombination occurring in the photo-evaporative flow, or is (b) processed in the ionisation front. For the two cases, corresponding to the insulating boundary layer (IBL) regime and the ionisation front (IF) regime respectively, the $\dot{M}$-flux relations are of the form (see equations 37 and 38 in \citealt{lefloch94})

\begin{equation}
\dot{M}_\mathrm{IBL} \propto \mu \Big( \frac{\Phi}{10^{7}\, \mathrm{cm}^{-2}\mathrm{s}^{-1}} \Big)^{1/2} \, \Big( \frac{r}{1\,\mathrm{pc}} \Big)^{3/2} \, M_{\odot}\,\mathrm{Myr}
\label{coll}
\end{equation}

\begin{equation}
\dot{M}_\mathrm{IF} \propto \mu \Big( \frac{\Phi}{10^{7}\, \mathrm{cm}^{-2}\mathrm{s}^{-1}} \Big) \, \Big( \frac{r}{1\,\mathrm{pc}} \Big)^{2} \, M_{\odot}\,\mathrm{Myr}
\label{if}
\end{equation}

\begin{figure*}
\includegraphics[scale=0.35]{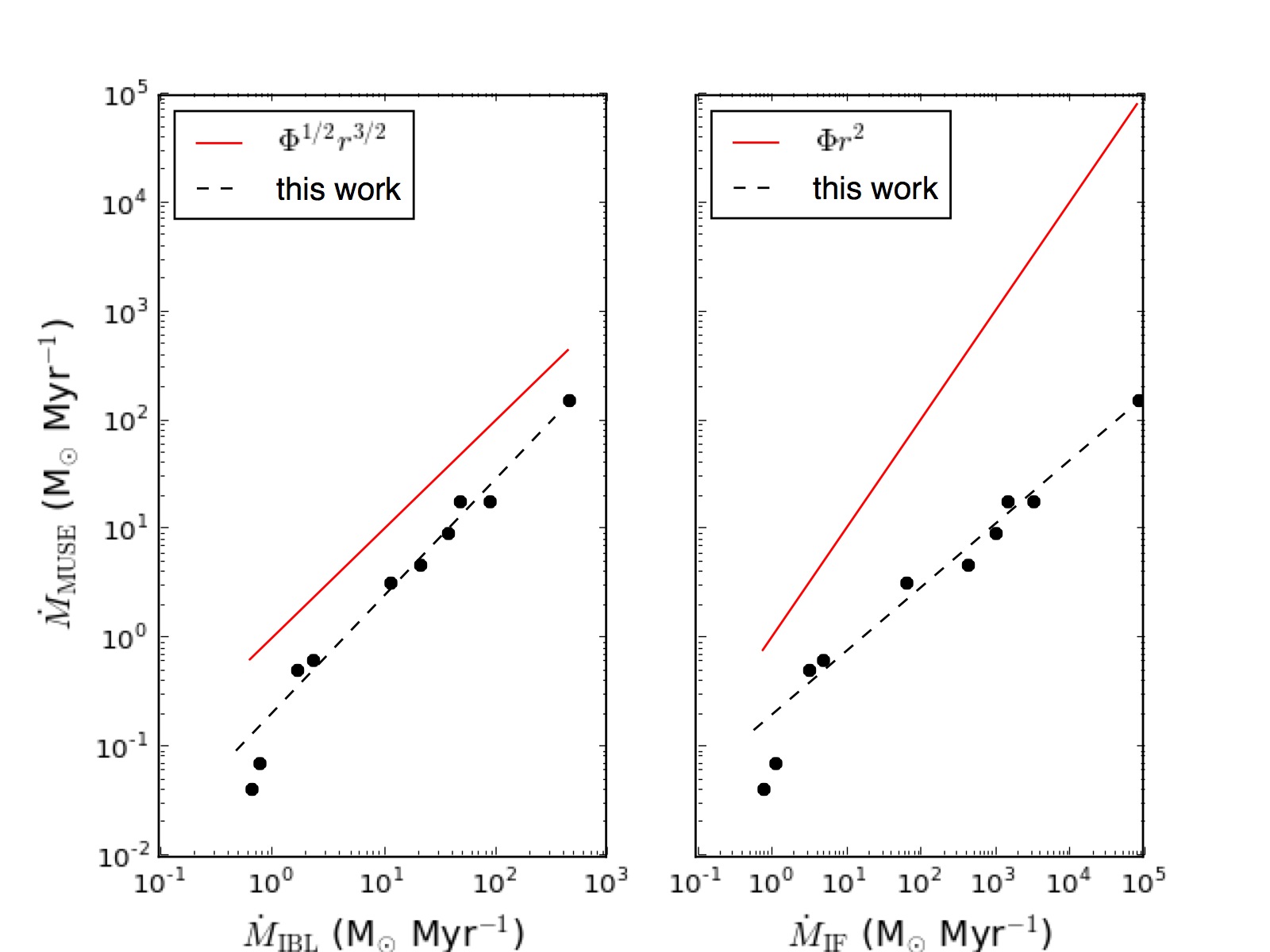}
\caption{The mass-loss rate $\dot{M}_\mathrm{MUSE}$ compared to the mass-loss rate derived from the \protect\cite{lefloch94} model (black circles) and the best fit to the data points (dashed line). The red line in the left panel shows the case for which $\dot{M}_\mathrm{MUSE}$=$\dot{M}_\mathrm{collapse}$, while that in the right panel corresponds to $\dot{M}_\mathrm{MUSE}$=$\dot{M}_\mathrm{cometary}$. See text Section \ref{mdot}.}
\label{model}
\end{figure*}

\begin{figure*}
\includegraphics[scale=0.3]{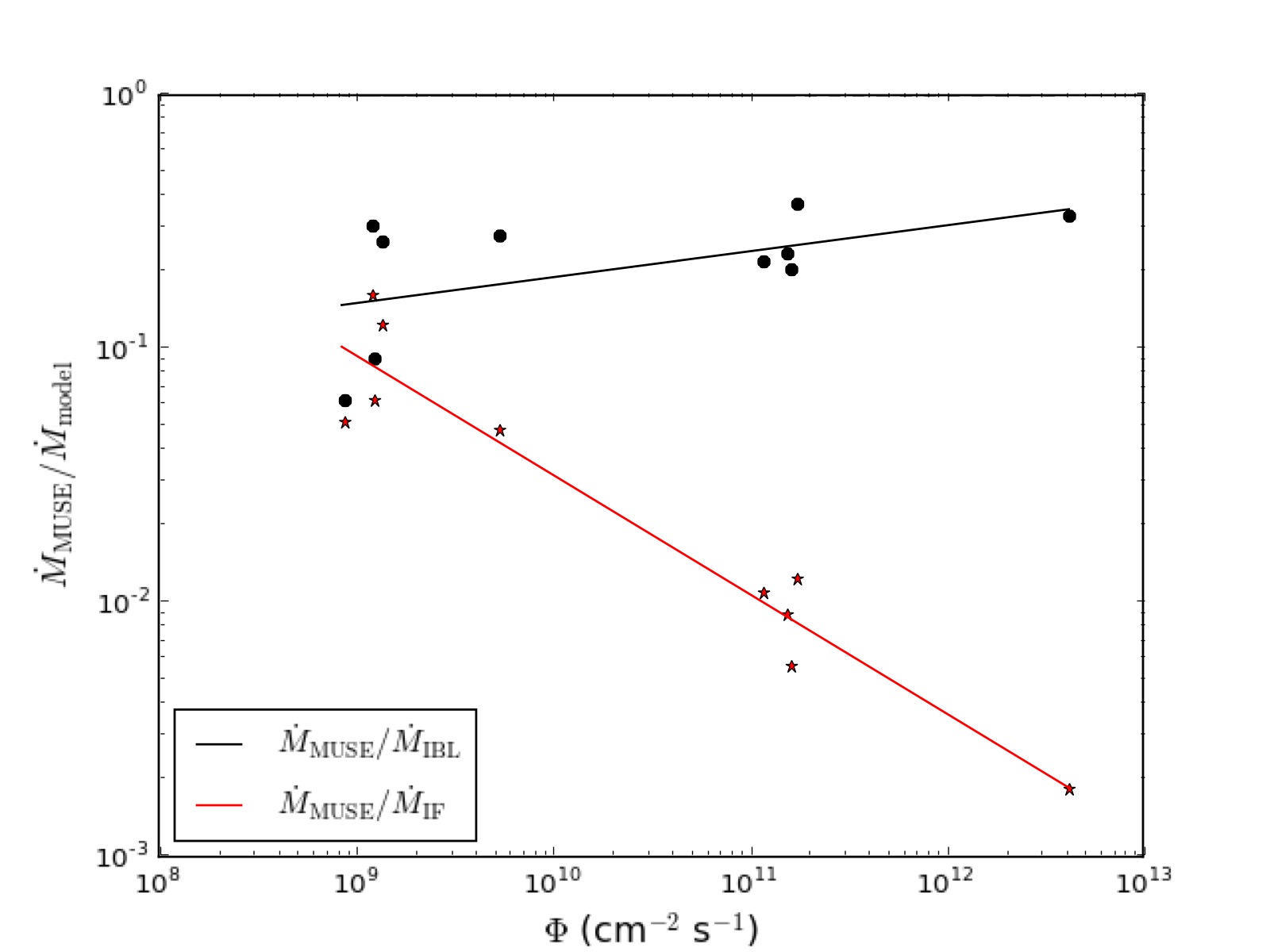}
\caption{The ratio of the measured mass-loss rate to that derived from the \protect\cite{lefloch94} model, $\dot{M}_\mathrm{model}/\dot{M}_\mathrm{MUSE}$ versus the photon flux $\Phi=Q_{0}/4\pi d^{2}$ for the IBL regime (black dots) and the IF regime (red stars). See text Section \ref{mdot}.}
\label{model2}
\end{figure*}

Where $\mu$ is the mean molecular weight (for which we assume $\mu=1.3$). The comparison is shown in figures \ref{model} and \ref{model2}. The first shows the mass-loss rate derived in this work compared to that derived from \cite{lefloch94} via the above equations by using the $Q_\mathrm{0}$ (parametrised as $\Phi$) and $r$ values from Table~\ref{params}. The comparison is shown with a red line that indicates $\dot{M}_\mathrm{MUSE}$=$\dot{M}_\mathrm{model}$, demonstrating that the data are in good agreement with the IBL regime of \cite{lefloch94}, as well as a clear discrepancy with the IF regime. This is further demonstrated in Fig.~\ref{model2}, which shows the ratio $\dot{M}_\mathrm{MUSE}/\dot{M}_\mathrm{model}$ as a function of the ionising flux.

These results show that in the pillars observed with MUSE, the incident photons are not processed in the ionisation front, but rather compensate the recombination occurring in the gas that streams away from the pillar-surfaces and therefore absorbs part of the ionising photons. The incident photon flux is attenuated by the photo-evaporative flow absorbing part of the photons, the mass-loss rate is reduced, and for a given mass the lifetime is consequently prolonged (compared to a scenario where the entire photon budget goes into ionisation). However, the observationally-derived mass-loss rates are systematically lower by a factor $\sim2.21$ than the predicted model values. The small difference between the empirically-derived mass-loss rates and those derived via the \cite{lefloch94} IBL model regime can be the result of small differences in the parameters, e.g. the assumed mean molecular weight, the  ionised sound speed, geometrical factors of the models or the observations. For example, if we consider the pillar tips being half spheres, there would be an extra factor 2 in Eq. \ref{mdoteq}.

The model-data comparison demonstrates that our understanding of the physics of photo-evaporation is correct, confirming a scenario in which the ionising radiation heats and ionises the cloud material, producing an ionised photo-evaporative flow which then partly absorbs further incoming photons. However, the connection between electron density, projected distance, flux and mass-loss rate and their dependencies on each other are neither trivial nor understood. In fact, whether and how the $n_\mathrm{e}$-$\log(Q_\mathrm{0,pil})$, $n_\mathrm{e}$-$d_\mathrm{proj}$ and $\dot{M}$-$\log(Q_\mathrm{0,pil})$ relations derived in this work are connected, needs to be thoroughly analysed with detailed models that exploit the full parameter space of the flux, the distance to the ionised structures, size of the structures, and density profiles. 

This result might be challenged because of the assumptions made for the ionising sources of R37, R44 and R45. However, even if both R37 and R44/45 are under the full influence of Tr 16 instead of O4/WNL and HD30, the correlations would still hold, as in that case $\log(Q_\mathrm{0,pil})=48.40$ for R37, $\log(Q_\mathrm{0,pil})=46.11$ for R45 and $\log(Q_\mathrm{0,pil})=46.67,46.22,46.82$ for the three pillars in R44. Furthermore, selection and geometrical effects need to be considered when interpreting this correlation, as it depends on projected distances instead of real distances. However, pillars are only identified as such with a favourable spatial orientation and illumination (i.~e. if viewed edge-on), and the consequence of this is that for projection angles up to 60 degrees, the true distance is no more than a factor of 2 greater than the projected distance to the ionising sources. A further note on this is that photo-evaporation is certainly not the only mechanism able to destroy pillar-like structures in these kind of regions, as the massive stellar content implies strong stellar winds, as well as supernova events which will occur at some point during the evolution of the star-forming region. These events will contribute to a more rapid disruption of the pillars. 

A natural expansion of this analysis would be to compute the (remaining) lifetime of each pillar and analyse the presence of a correlation with $\log(Q_\mathrm{0,pil})$. With a CO-derived mass of 10-20 $M_{\odot}$ from \cite{cox95}, \cite{smith04} determined a remaining lifetime of about $10^{5.3-6}$ yr for the globule in R37, while for the M\,16 pillars a remaining lifetime of about  $10^{6.47}$ yr is computed (MC15). With molecular line data (e.~g. $^{12}CO$), or alternative via SED fitting of in-house \emph{Herschel} images available for the CNC \citep{preibisch12}, one could attempt a mass computation of all the pillars in the MUSE Carina data set and consequently determine their lifetimes. 


\subsection{Ionised jets}\label{jetsection}
We further exploit the MUSE Carina data set to study feedback in massive star-forming regions by analysing the presence, orientation and morphology of two ionised jets originating from two of the observed pillars.

In MC15, we developed a novel method to detect jets from young stars that have not yet had time to fully emerge from the embedding material. This method uses the the so called S$_{23}$ parameter ($(\mathrm{[SII]}+\mathrm{[SIII]})/\mathrm{H}\beta$, \citealt{vilchez96}) to detect a jet that is only now starting to emerge from the pillar. The S$_{23}$ parameter detection needs to be complemented with a line-of-sight velocity map, derived via a pixel-by-pixel guassian fitting routine to the stacked spectrum of several lines in the 600-700 nm range, as was done to compute the velocity maps of the pillars discussed here. In the case of M\,16, the well studied pillars were not previously known to host jets, and it was only due to a careful analysis of the MUSE data that such a detection could be made. We inspected the Carina pillars in terms of the S$_{23}$ parameter, but no jet could be identified in the pillars with this parameter. However, we identified two jets in the Carina data set nonetheless: these are jets protruding for several arcseconds from the pillar tips that host their sources, and they are detected as ionised jets predominantly visible in the H$\alpha$ and [SII] lines. In \cite{mcleod16}, we propose the utility of the S$_{23}$ parameter as an indicator of the shock contribution to the excitation of line-emitting atoms, meaning that if a jet is traced by this parameter, it is because it is interacting with the host pillar material from which it is emerging by creating a layer of shocked material at the ionised pillar surface. From the fact that the two jets are not identified with the S$_{23}$ parameter map, we conclude that these jets are more evolved than the M\,16 case, as in the latter the jet is not yet detected beyond the pillar boundaries. The R44 and the R18 jets however are seen to extend well beyond the pillars which host their sources. In the following two subsections, we will discuss the R18 and R44 jets separately. 

The two detected jets are in regions R18 and R44, and are comparable to the $HST$-detected jets in \cite{smith10}. In the case of R44, the jet corresponds to the known Herbig-Haro object HH 1010. With this data set however, because of the spectral and spatial coverage of MUSE, we are not only able to identify the ionised jets in emission lines such as H$\alpha$ and [SII], but we are also able to distinguish their kinematics. This is because the two detected jets are only seen in the emission line wings, and we can therefore split them into a red- and a blue-shifted component respectively. In the case of R18, this is the first reported detection of the ionised jet.

\subsubsection{HH 1124 in R18}\label{R18text}
\begin{figure}
\hspace{-0.5cm}
\includegraphics[scale=0.38]{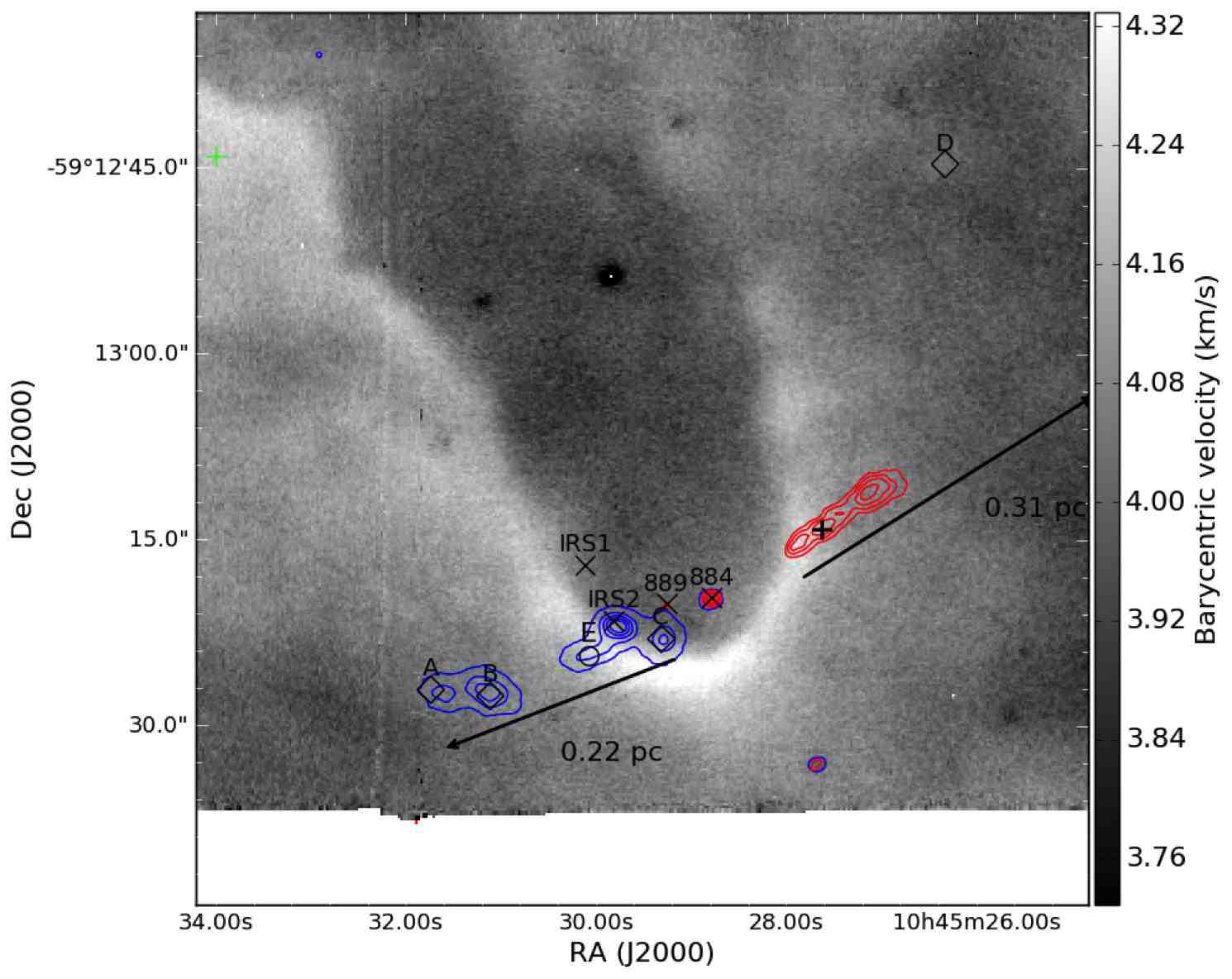}
\caption{Continuum-subtracted H$\alpha$ intensity map of the tip of pillar R18 and HH 1124. Blue and red contours mark the location of the red and blue lobes of the bipolar jet originating from the pillar tip. The contours are extracted from the corresponding two slices of the data cube at red = 6566.43 \AA\ and blue = 6558.93 \AA. The emission line knots A-D as well as IRS1/2, 884 and 889 are the sources discussed in \protect\cite{hartigan15}, while the region marked with E is only detected here as part of the blue counterpart of the jet (see text Section \ref{R18text}). The black arrows indicate the extent of the jet, best compared with the RGB composite of the pillar tip shown in fig.~\ref{R18rgb}, and the black cross indicated the position of aperture used to determine the jet velocity of the red lobe (see text Section \ref{R18text}).}
\label{r18jet}
\end{figure}

The R18 pillar and the sources at its tip are discussed in \cite{hartigan15}, and the main features presented by these authors are reported in Fig. \ref{r18jet}: four emission line knots (A to D), two infrared sources (IRS1 and IRS2), and two young stellar objects (PCYC 884 and 889, for simplicity referred to as 884 and 889). These authors attribute all emission line knots to a jet (HH 1124) originating from the tip of the pillar that is being bent north by the O-stars responsible for the formation of the pillar. In Fig.~\ref{r18jet} a fifth knot is marked with $E$, which (as will be discussed below), is a faint knot belonging to the blue lobe of the jet and was not detected by \cite{hartigan15} but only in this work.

The presence of a jet becomes clear when inspecting the wavelength slices of the data cube around the bright single-ionised emission lines such as H$\alpha$, [SII] and [NII]: when moving through the cube around the central wavelength of these lines, the two jet lobes become visible in the blue and red slices before and after the central frame (see coloured contours in Fig.~\ref{r18jet}). 
The ionised jet does not, however, follow the direction of the jet proposed in \cite{hartigan15}, where emission line knot D is discussed to be part of it and it is proposed that the jet is being bent north by the feedback of the nearby massive stars of Tr 15. Here, we find a P.A. of $\sim$123$^{\circ}$ for the red counterpart, and a P.A. of $\sim$113$^{\circ}$ for the blue counterpart. The pillar itself is at P.A. $\sim$210$^{\circ}$, meaning that the jet is approximately perpendicular to the pillar (the angle of the pillar being 97$^{\circ}$ and 87$^{\circ}$ with the red and the blue counterparts respectively), and that there is an indication for the red jet-lobe being bent $\sim10^{\circ}$ north with respect to the blue lobe due to the high-mass stellar feedback (in the form of stellar winds or outflowing material from the inner parts of the nebula, \citealt{bally01}). When compared to the conclusion reached in \citeauthor{hartigan15} that there is evidence for strong jet bending due to stellar winds (because these authors include emission knot D as being part of the jet), the jet bending seen in this work is minimal.


Compared to the blue, the red lobe of the jet is more continuous, and it can be traced up to $\sim$ 0.46' ($\sim$ 0.31 pc) from the pillar surface into the surrounding medium. The blue lobe on the other hand consists of at least 5 knots, out of which 4 correspond to emission line knots A, B, C and IRS2, as is shown in Fig.~\ref{r18jet} . If measured from knot C to knot A, the spatial extent of the blue lobe is $\sim$ 0.33' ($\sim$ 0.22 pc). From this data we conclude that emission line knot D is not part of the jet, but there is a fifth detectable knot, marked as E in Fig.~\ref{r18jet}.

\begin{figure*}
\centering
\includegraphics[scale=0.45]{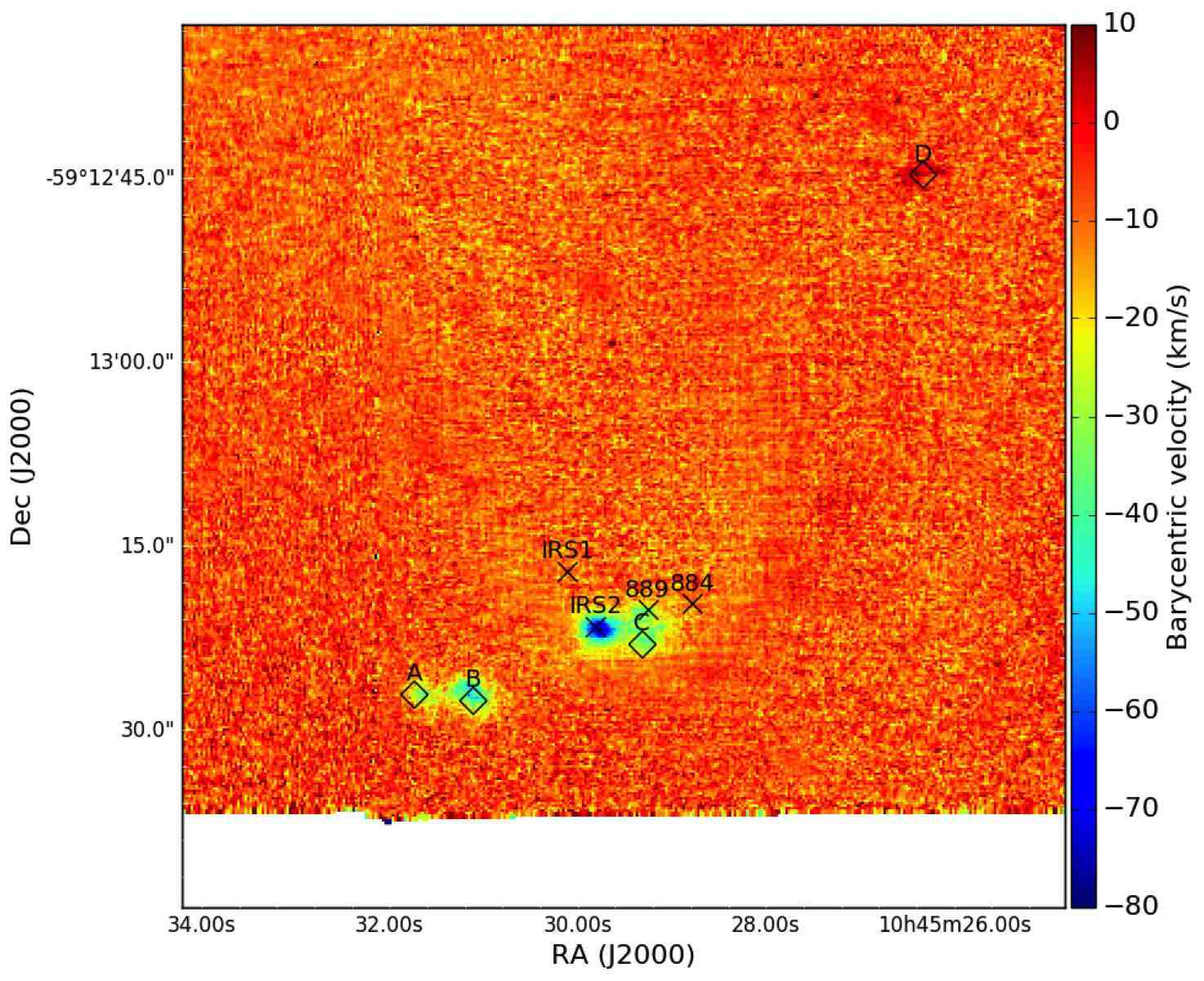}
\caption{Velocity map of the [SII]$\lambda$6717 line zoomed on the pillar tip and the emission line knots/sources located at the tip (see text Section \ref{R18text}).}
\label{R18_SII}
\end{figure*}

In terms of velocity, emission line knots A, B, C, as well as IRS2 and 889 are blueshifted with respect to both the pillar and the ambient matter. This is shown in Fig.~\ref{R18_SII}, where the discussed emission line knots and young stellar objects are overlaid on the [SII]$\lambda$6717 velocity map\footnote{We show the [SII]$\lambda$6717 velocity map instead of that of the H$\alpha$ line, because the latter is contaminated by instrument artefacts in the form of a checked pattern as is shown in Fig.~\ref{velmaps}. Other than being unpleasant to the eye, this pattern does not influence our analysis.}. However, by inspecting the emission line profiles for the sources/knots in Fig. \ref{r18jet}, we detect double components in the low-ionization emission lines ([OI], [SII], [NII], and H$\alpha$) towards knots B and E as well as IRS2, but not towards knots A, C, D and 889. In both IRS2 and knot B, the blue component is $\sim$50\% weaker in intensity than the red component, the two are separated by about 200 km s$^{-1}$ (as is shown in Fig.~\ref{R18_SII} and Table \ref{specfit_table} for the [SII]$\lambda$6717 line), the blue component being at $\sim$ -200 km s$^{-1}$ and the red at $\sim$ -35 km s$^{-1}$. In emission line knot E the brightness of the blue component is only about 15\% that of the red component, so that a two-component gaussian fit was not possible. One possible scenario for this peculiar configuration is that the blue component of the double-peaked line profile is tracing the $\sim$ 200 km s$^{-1}$ jet, while the red component is tracing a $\sim$ 30 km s$^{-1}$ stellar wind.

To determine the dynamical age $t_{dyn}$ of the jet, we assume that the jet velocity of the blue lobe is $\sim$ 200 km s$^{-1}$, while a velocity of $\sim$ -8 km s$^{-1}$ for the red lobe is obtained by fitting the spectrum of a 3-pixel aperture centred on the black cross marked in Fig.~\ref{r18jet}. With a total spatial extent of $\sim$ 0.5 pc, we obtain $t_{dyn}\approx10^{3.4}$ yr. This age is much younger than the age of Tr 15, indicating that the jet-driving source must have formed after the formation of the cluster. Whether the formation of the jet source was triggered by the feedback of Tr 15 cannot be determined from this data.

The determination of the source of the jet and wind is not trivial. In \cite{hartigan15}, IRS2 is mentioned to be a possible sub-arcsecond binary, however, the MUSE data does not allow a verification of this. A source at the approximate location of YSO 884 is discussed and analysed in \cite{ohlen12}: with multi-wavelength photometric data ranging from the near- to the far-infrared, these authors obtain an estimate of 2.6 $M_{\odot}$ for the stellar mass of this source. However, the angular resolution of the data used by these authors does not allow them to distinguish between YSOs 884 and 889, and indeed it is treated as a single compact green object. The question whether IRS2 or one of the two YSOs is responsible for driving the jet could be tackled with high angular resolution molecular data, e.g. with CO or a shock tracer like SiO, as to trace the jet directly back to its source.

\begin{table}
\footnotesize
\centering
\caption{Best fit parameters of the 2-component gaussian fitting performed on the [SII]$\lambda$6717 spectra of emission line knot B and IRS2 (see Fig.~\ref{r18jet}). All values are in km s$^{-1}$.}
\begin{tabular}{lcccc}
\hline
\hline
Source & Centroid & Width & Centroid & Width \\
 & (blue) & (blue) & (red) & (red) \\
\hline
IRS2 & -228.7$\pm$10.8 & 86.1$\pm$10.2 & -31.7$\pm$3.6 & 60.8$\pm$2.9 \\
Knot B & -208.5$\pm$28.0 & 102.6$\pm$26.9 & -2.5$\pm$8.3 & 69.4$\pm$7.1 \\
\hline
\end{tabular}
\label{specfit_table}
\end{table}

\subsubsection{HH 1010 in R44}\label{jetR44}
\begin{figure}
\hspace{-0.5cm}
\includegraphics[scale=0.38]{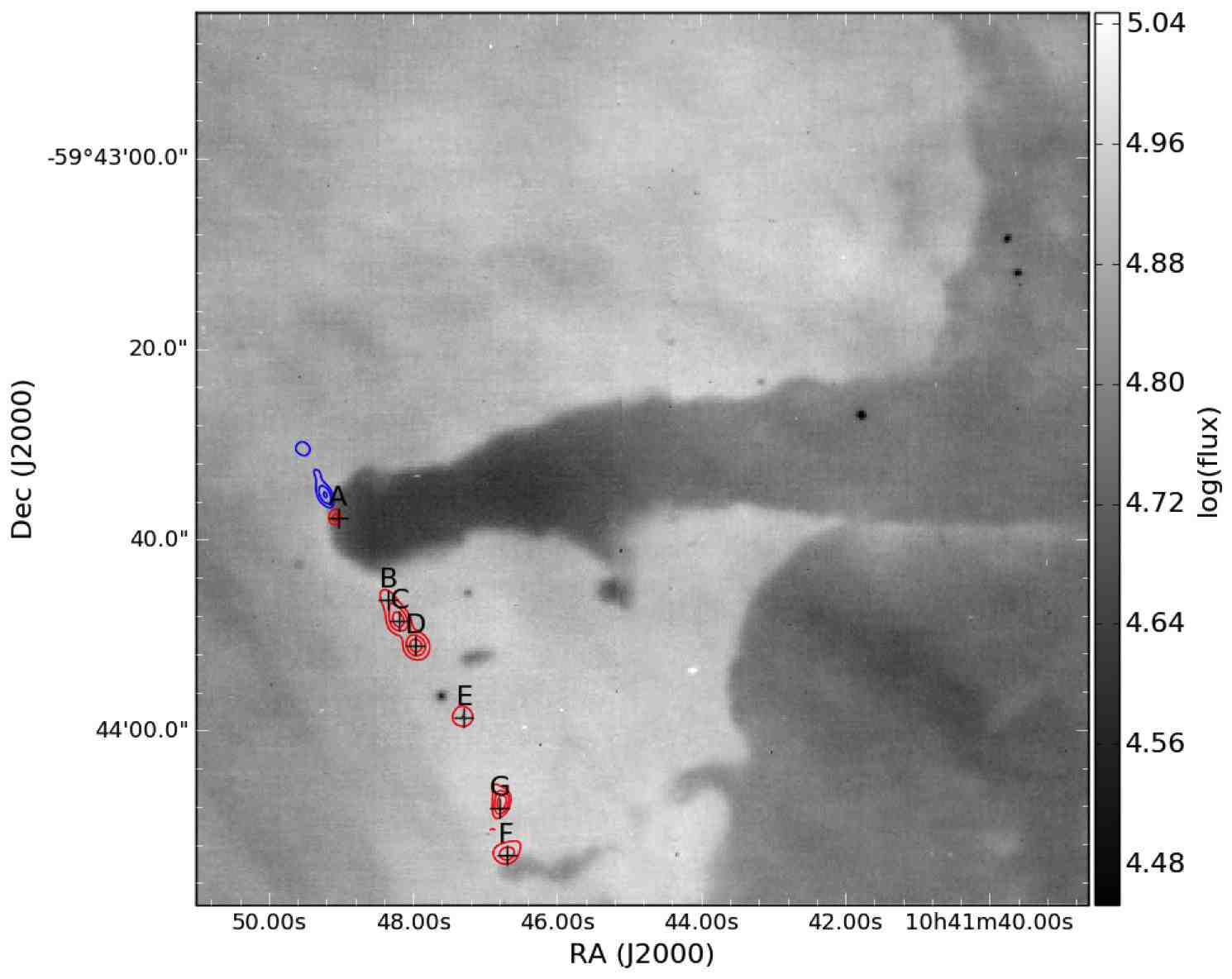}
\caption{Continuum-subtracted H$\alpha$ intensity map of the tip of pillar R44. Blue and red contours mark the location of the red and blue lobes of the bipolar jet originating from the pillar tip. The contours are extracted from the corresponding two slices of the data cube at red = 6566.43 \AA\ and blue = 6558.93 \AA. The knots of the red lobe are marked with the white diamonds and letters from $A$ to $G$. See text Section \ref{jetR44}.}
\label{jetR44_fig}
\end{figure}

\begin{figure}
\hspace{-0.5cm}
\includegraphics[scale=0.38]{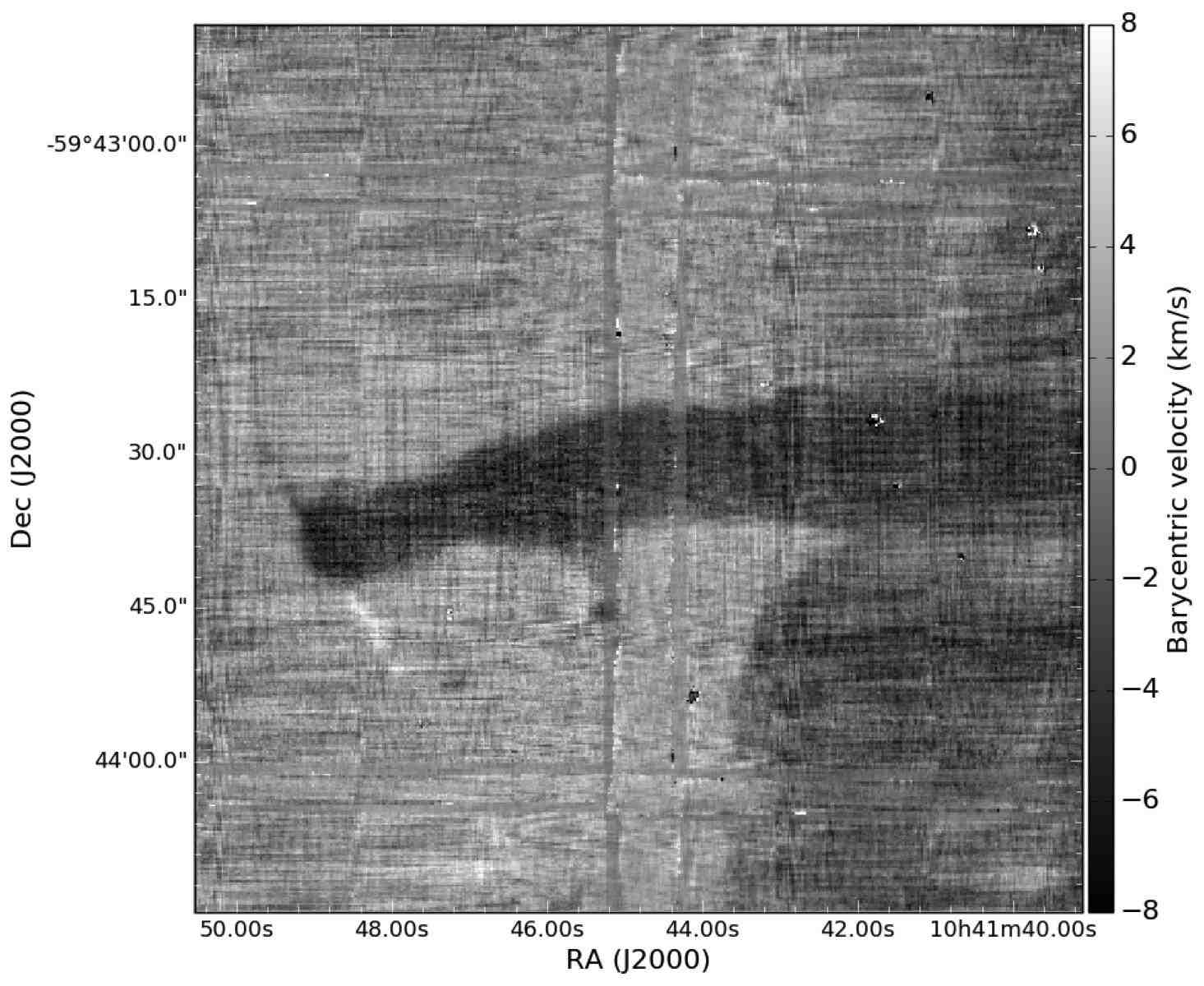}
\caption{Velocity map of P1 in R44 showing the red and blue lobes of the jet (see text Section \ref{jetR44}).}
\label{jetR44_vel}
\end{figure}

The highly collimated jet HH 1010 at the tip of pillar 44-P1 was identified with $HST$ H$\alpha$ imaging in \cite{smith10}, where a PA of about 216$^{\circ}$ and an extent of 60 arcseconds are reported. These values for extent and position angle are confirmed with the MUSE data, where the jet can again be separated into a red and a blue component by extracting slices on either side of the central H$\alpha$ wavelength from the data cube. Fig.~\ref{jetR44_fig} shows the continuum-subtracted H$\alpha$ map with contours extracted from the 6566.43 \AA\ and 6558.93 \AA\ slices for red and blue respectively. The red lobe of the jet is composed of a series of knots ($A$ to $G$ in Fig.~\ref{jetR44_fig}) which extend over almost 40" (0.45 pc), knot $A$ being at the tip of the dark host pillar approximately where \cite{smith10} report the presence of a faint optical source, and knot $G$ being close to the rim of a small globule south-west of the pillar. There is a very shallow S-shaped wiggle in the red lobe, which might be an indication for jet precession \citep{reipurth01}. The blue lobe is fainter than the red one, more collimated, and spans over $\sim$ 10" (0.11 pc) north-east. Unlike HH 1124 in R18, none of the knots of the red lobe show remarkable line profiles. However, the velocity of the two lobes can be determined from the stacked velocity map shown in Fig.~\ref{velmaps}c, of which a close-up is shown in Fig.~\ref{jetR44_vel}: the blue lobe emerges from the pillar with $v\approx-2$ km s$^{-1}$, and the red lobe with $v\approx6$ km s$^{-1}$, yielding a jet (radial) velocity of $\sim$ 8 km s$^{-1}$. Together with a total spatial extent of $\sim$ 0.7 pc, this yield a dynamical age of  $\sim10^{4.9}$ yr. As for HH 1124, this age is much smaller than the age of the nearby clusters, indicating that the driving source formed after the formation of the clusters. As for R18, there is no indication for a feedback-bent jet, as the lobes (except for the slight S-shape of the red lobe) are not seen to bend away from the direction of the incident radiation. This could be an indication either for the fact that the jet has enough mass to resist being bent by the feedback from the nearby massive stars, or for the fact that, compared to HH 1124 in R18, HH 1010 is exposed to a weaker radiation field, which is less likely to bend jets.

With multi-wavelength near-infrared data, \cite{povich11} report the presence of a 1.8 M$_{\odot}$ YSO at J2000 10:41:48.670 -59:43:38.10 (PCYC55), which is just south of knot $A$. We conclude that this object is most likely the source of the jet.

\section{Conclusions}\label{conclusions}

In this paper, we presented MUSE optical integral field data of pillar-like structures in the Carina Nebula Complex. We exploit the unprecedented spectral and spatial coverage of this instrument to analyse the morphology and kinematics of the ionised gas component in the observed regions by producing integrated line maps of the main ionised and neutral emission lines covered by MUSE (H$\alpha$, H$\beta$, [OI], [OII], [OIII], [SII], [SIII] and [NII]), as well as velocity maps and maps of the electron temperature and density. The analysis performed in this paper is oriented towards understanding the link between the feedback-driving massive stars and the pillar-like structures in their vicinity. The main results are:

$\bullet$ The behaviour of the various emission lines along slits positioned across the pillar/ambient matter interface was derived. The pillar tips present the expected ionisation stratification, where the line-emitting species with higher ionisation energy peak first and further away from the pillar, followed by the species with lower and lower ionisation energy. By comparing the separation between the [OIII] and the [OI] lines ([OIII] having the highest ionisation energy, and [OI] tracing the neutral material), we find a slight trend in which the separation between the highly ionised and the neutral material gets larger with distance from/ weaker intensity of the ionising sources. We include a preliminary comparison with analytical density profiles under the influence of three different types of ionising sources, and find that the the ionisation front gets narrower with increasing photon flux. This however needs to be thoroughly tested with higher angular resolution imaging, as well as dedicated simulations in which a full parameter space of pillars under varying ionising conditions can performed.

$\bullet$ For each region, we compute the electron density, electron temperature and velocity maps. A tight correlation between the electron density, $n_\mathrm{e}$ and the ionising photon flux at each pillar tip, $\log(Q_\mathrm{0,pil})$, is found, as well as a correlation between the electron density and the projected distance from the ionising sources, yielding a lower limit for the density a pillar must have at a certain distance not to be destroyed in a region with given ionising conditions. 

$\bullet$ We compute the mass-loss rate $\dot{M}$ due to the photo-evaporative effect of the nearby ionising stars, and find that $\dot{M}$ strongly depends on the perceived photon flux $\log(Q_\mathrm{0,pil})$. This empirical correlation is the first observational quantification of cloud evaporation covering a significant dynamic range in impinging ionising flux. We perform a comparison with the mass-loss rate as described in the models of \cite{lefloch94}, and find a good agreement between the observations and the theory. Specifically, we find that the observed mass-loss rate of pillars follows a square root proportionality to the impinging photon flux, rather than a linear proportionality, therefore favouring the IBL regime of the \cite{lefloch94} models. We conclude that, as expected, not all of the impinging photons lead to ionisation of the pillar material, as the photo-evaporative flow absorbs part of the radiation to balance the effect of recombination. However, the observed mass-loss rates are systematically lower than the model prediction by a factor of about 2.2, which could be partly explained by the assumed geometry when computing the mass-loss rate.
To understand the connection between the density/flux, density/distance and mass-loss rate/flux relations, specific numerical simulations exploring a full range of parameters are needed. A very interesting perspective for future analyses is to test the presence of a correlation between the lifetime of pillars and the ionising photon flux acting on them. This could be achieved with either molecular line data or via multi-wavelength SED fitting to dust-emission maps in order to obtain a mass estimate for each pillar.

$\bullet$ With the spectrally resolved MUSE data, we identify two jets at the tips of R18 and R44 P1, where HH 1010 in R44 is a previously detected and discussed jet, while the morphology and extent of HH 1124 in R18 (previously only identified but not analysed) is discussed for the first time in this work. Only HH 1124 shows signs of being bent away from the direction of the incoming radiation from the massive stars, while this is not the case for HH 1010, where no jet-bending is observed, which could indicate either that it has sufficient mass to resist the jet bending, or that the radiation field is not strong enough to cause jet-bending. HH 1124 presents a rather peculiar morphology in terms of velocity, which we attribute to the presence of both a jet and a wind, possibly originating from the same source. The red lobe of HH 1010 displays a slight S-shape, hinting at jet precession. Also, both jets are younger than the clusters in their vicinity, indicating that the formation of their driving sources occurred after the formation of the massive clusters.

To summarise, we demonstrate that MUSE optical integral field data can yield valuable information when analysing the effect of ionising feedback from massive stars and star clusters, as it offers the possibility of performing a uniform survey of feedback-driven structures across different star-forming regions, and derive the kind of relations presented in this paper in an unbiased way across these different regions. The survey-capability of MUSE is further augmented by the fact that this instrument offers a unique combination of simultaneous large spatial and spectral coverage, not obtainable via traditional long-slit or Fabry-P\'{e}rot spectroscopy, with a single set of observations.Covering entire structures (not possible with long-slit spectroscopy) in all the various emission lines (expensive with Fabry-P\'{e}rot instruments) is particularly useful, not only because the measured physical parameters can be traced across entire pillar rims and therefore yield a better estimate of these parameters where the S/N values are low and an averaging over several pixels is needed, but also because it allows the analysis of variations perpendicular to the angle of incidence (which is beyond the large-scale scope of this paper, but certainly is an interesting potential follow-up analysis).

As mentioned in Section 1, feedback from massive stars is one of the main uncertainties in galaxy evolution models, which are only able to reproduce observations if feedback is included. An observational quantification of the effect of feedback is therefore crucial in order to better constrain these models. One of the main open questions in the field of massive star formation feedback is its dependence on the characteristics and properties of the different environments the massive stars form in. In the case of ionisation feedback, surveys of ionisation fronts and ionised pillar-like structures are needed, not only throughout different environments in the Milky Way, but also throughout e.g. the Magellanic Clouds, different galactic environments with lower metallicities where single stars and ionisation fronts can still be resolved. In terms of galaxy evolution models and the various massive star formation feedback mechanisms, observations  of more distant environments are needed in order to get large samples of differing geometries of feedback-affected regions and luminosities of feedback drivers. For the reasons listed above, and combined with the fact that this can be achieved within the framework of single observations, MUSE is an ideal instrument for the analysis of massive star formation feedback.

Not only was the effect of ionising feedback analysed through a correlation between the ionising photon flux (originating from nearby massive stars) and the mass-loss rate due to photo-evaporation of pillar-like structures, but the same data set was used to understand the effect of ionisation on the size (width) of the ionisation fronts at the pillar tips, but also to look for signs of jet bending via the detection of two ionised jets originating from the pillar tips. These results set the scene for further investigations to better understand feedback from massive stars, which can be achieved by combining the results from more IFU data sets and simulations of star-forming regions which include feedback recipes.

\section*{Acknowledgments}
We would like to thank Bob O'Dell for the very useful discussions. This work is based on MUSE commissioning data obtained with ESO telescopes at the Paranal Observatory under the programmes 096.C-0574(A) and 60.A-9309(A). The data reduction and analysis make use of \textsc{aplpy} (http:$//$aplpy.github.io), \textsc{spectral$\_$cube} (spectral-cube.readthedocs.org), \textsc{pyspeckit} (pyspeckit.bitbucket.org, \cite{2011ascl.soft09001G}) and \textsc{glue} (glueviz.org). This research was supported by the Christiane N{\"u}sslein-Volhard Foundation (AFMCL) and the DFG cluster of excellence \textit{Origin and Structures of the Universe} (JED). This work was partly supported (LT) by the Italian Ministero dell'Istruzione, Universit{'a} e Ricerca through the grant \textit{Progetti Premiali 2012 -- iALMA} (CUP C52I13000140001).

\nocite{*}
\bibliographystyle{mn2e}
\bibliography{references}

\begin{thebibliography}{}

\bibitem[\protect\citeauthoryear{{Arthur}, {Henney}, {Mellema}, {de Colle} \&
  {V{\'a}zquez-Semadeni}}{{Arthur} et~al.}{2011}]{arthur11}
{Arthur} S.~J.,  {Henney} W.~J.,  {Mellema} G.,  {de Colle} F.,
  {V{\'a}zquez-Semadeni} E.,  2011, \mnras, 414, 1747

\bibitem[\protect\citeauthoryear{{Bally} \& {Reipurth}}{{Bally} \&
  {Reipurth}}{2001}]{bally01}
{Bally} J.,  {Reipurth} B.,  2001, \apj, 546, 299

\bibitem[\protect\citeauthoryear{{Bertoldi}}{{Bertoldi}}{1989}]{bertoldi89}
{Bertoldi} F.,  1989, \apj, 346, 735

\bibitem[\protect\citeauthoryear{{Bertoldi} \& {McKee}}{{Bertoldi} \&
  {McKee}}{1990}]{bertoldi90}
{Bertoldi} F.,  {McKee} C.~F.,  1990, \apj, 354, 529

\bibitem[\protect\citeauthoryear{{Beuther}, {Churchwell}, {McKee} \&
  {Tan}}{{Beuther} et~al.}{2007}]{beuther07}
{Beuther} H.,  {Churchwell} E.~B.,  {McKee} C.~F.,    {Tan} J.~C.,  2007,
  Protostars and Planets V, pp 165--180

\bibitem[\protect\citeauthoryear{{Bressert}, {Bastian}, {Gutermuth}, {Megeath},
  {Allen}, {Evans} II, {Rebull}, {Hatchell}, {Johnstone}, {Bourke}, {Cieza},
  {Harvey}, {Merin}, {Ray} \& {Tothill}}{{Bressert} et~al.}{2010}]{bressert10}
{Bressert} E.,  {Bastian} N.,  {Gutermuth} R.,  {Megeath} S.~T.,  {Allen} L.,
  {Evans} II N.~J.,  {Rebull} L.~M.,  {Hatchell} J.,  {Johnstone} D.,  {Bourke}
  T.~L.,  {Cieza} L.~A.,  {Harvey} P.~M.,  {Merin} B.,  {Ray} T.~P.,
  {Tothill} N.~F.~H.,  2010, \mnras, 409, L54

\bibitem[\protect\citeauthoryear{{Cox} \& {Bronfman}}{{Cox} \&
  {Bronfman}}{1995}]{cox95}
{Cox} P.,  {Bronfman} L.,  1995, \aap, 299, 583

\bibitem[\protect\citeauthoryear{{Dale}, {Ercolano} \& {Bonnell}}{{Dale}
  et~al.}{2012}]{dale12}
{Dale} J.~E.,  {Ercolano} B.,    {Bonnell} I.~A.,  2012, \mnras, 427, 2852

\bibitem[\protect\citeauthoryear{{Drissen}, {Moffat}, {Walborn} \&
  {Shara}}{{Drissen} et~al.}{1995}]{drissen95}
{Drissen} L.,  {Moffat} A.~F.~J.,  {Walborn} N.~R.,    {Shara} M.~M.,  1995,
  \aj, 110, 2235

\bibitem[\protect\citeauthoryear{{Ercolano}, {Barlow}, {Storey} \&
  {Liu}}{{Ercolano} et~al.}{2003}]{ercolano03}
{Ercolano} B.,  {Barlow} M.~J.,  {Storey} P.~J.,    {Liu} X.-W.,  2003, \mnras,
  340, 1136

\bibitem[\protect\citeauthoryear{{Evans} et~al.,}{{Evans}
  et~al.}{2005}]{evans05}
{Evans} C.~J.,  et~al., 2005, \aap, 437, 467

\bibitem[\protect\citeauthoryear{{Feigelson}, {Getman}, {Townsley}, {Broos},
  {Povich}, {Garmire}, {King}, {Montmerle}, {Preibisch}, {Smith}, {Stassun},
  {Wang}, {Wolk} \& {Zinnecker}}{{Feigelson} et~al.}{2011}]{feigelson11}
{Feigelson} E.~D.,  {Getman} K.~V.,  {Townsley} L.~K.,  {Broos} P.~S.,
  {Povich} M.~S.,  {Garmire} G.~P.,  {King} R.~R.,  {Montmerle} T.,
  {Preibisch} T.,  {Smith} N.,  {Stassun} K.~G.,  {Wang} J.,  {Wolk} S.,
  {Zinnecker} H.,  2011, \apjs, 194, 9

\bibitem[\protect\citeauthoryear{{Ferland}, {Porter}, {van Hoof}, {Williams},
  {Abel}, {Lykins}, {Shaw}, {Henney} \& {Stancil}}{{Ferland}
  et~al.}{2013}]{ferland13}
{Ferland} G.~J.,  {Porter} R.~L.,  {van Hoof} P.~A.~M.,  {Williams} R.~J.~R.,
  {Abel} N.~P.,  {Lykins} M.~L.,  {Shaw} G.,  {Henney} W.~J.,    {Stancil}
  P.~C.,  2013, \rmxaa, 49, 137

\bibitem[\protect\citeauthoryear{{Gagne}, {Fehon}, {Savoy}, {Cohen},
  {Townsley}, {Broos}, {Povich}, {Corcoran}, {Walborn}, {Evans}, {Moffat},
  {Naze} \& {Oskinova}}{{Gagne} et~al.}{2011}]{gagne11}
{Gagne} M.,  {Fehon} G.,  {Savoy} M.~R.,  {Cohen} D.~H.,  {Townsley} L.~K.,
  {Broos} P.~S.,  {Povich} M.~S.,  {Corcoran} M.~F.,  {Walborn} N.~R.,  {Evans}
  N.~R.,  {Moffat} A.~F.~J.,  {Naze} Y.,    {Oskinova} L.~M.,  2011, VizieR
  Online Data Catalog, 219

\bibitem[\protect\citeauthoryear{{Ginsburg} \& {Mirocha}}{{Ginsburg} \&
  {Mirocha}}{2011}]{2011ascl.soft09001G}
{Ginsburg} A.,  {Mirocha} J., , 2011, {PySpecKit: Python Spectroscopic
  Toolkit}, Astrophysics Source Code Library

\bibitem[\protect\citeauthoryear{{Gritschneder}, {Burkert}, {Naab} \&
  {Walch}}{{Gritschneder} et~al.}{2010}]{gritsch10}
{Gritschneder} M.,  {Burkert} A.,  {Naab} T.,    {Walch} S.,  2010, \apj, 723,
  971

\bibitem[\protect\citeauthoryear{{Harayama}, {Eisenhauer} \&
  {Martins}}{{Harayama} et~al.}{2008}]{harayama08}
{Harayama} Y.,  {Eisenhauer} F.,    {Martins} F.,  2008, \apj, 675, 1319

\bibitem[\protect\citeauthoryear{{Hartigan}, {Reiter}, {Smith} \&
  {Bally}}{{Hartigan} et~al.}{2015}]{hartigan15}
{Hartigan} P.,  {Reiter} M.,  {Smith} N.,    {Bally} J.,  2015, \aj, 149, 101

\bibitem[\protect\citeauthoryear{{Hester} et~al.,}{{Hester}
  et~al.}{1996}]{hester96}
{Hester} J.~J.,  et~al., 1996, \aj, 111, 2349

\bibitem[\protect\citeauthoryear{{Hill} \& {Hollenbach}}{{Hill} \&
  {Hollenbach}}{1978}]{hill87}
{Hill} J.~K.,  {Hollenbach} D.~J.,  1978, \apj, 225, 390

\bibitem[\protect\citeauthoryear{{Hur}, {Sung} \& {Bessell}}{{Hur}
  et~al.}{2012}]{hur12}
{Hur} H.,  {Sung} H.,    {Bessell} M.~S.,  2012, \aj, 143, 41

\bibitem[\protect\citeauthoryear{{Klaassen}, {Mottram}, {Dale} \&
  {Juhasz}}{{Klaassen} et~al.}{2014}]{klaassen14}
{Klaassen} P.~D.,  {Mottram} J.~C.,  {Dale} J.~E.,    {Juhasz} A.,  2014,
  \mnras, 441, 656

\bibitem[\protect\citeauthoryear{{Lebouteiller}, {Bernard-Salas}, {Brandl},
  {Whelan}, {Wu}, {Charmandaris}, {Devost} \& {Houck}}{{Lebouteiller}
  et~al.}{2008}]{lebout08}
{Lebouteiller} V.,  {Bernard-Salas} J.,  {Brandl} B.,  {Whelan} D.~G.,  {Wu}
  Y.,  {Charmandaris} V.,  {Devost} D.,    {Houck} J.~R.,  2008, \apj, 680, 398

\bibitem[\protect\citeauthoryear{{Lefloch} \& {Lazareff}}{{Lefloch} \&
  {Lazareff}}{1994}]{lefloch94}
{Lefloch} B.,  {Lazareff} B.,  1994, \aap, 289, 559

\bibitem[\protect\citeauthoryear{{Luridiana}, {Morisset} \& {Shaw}}{{Luridiana}
  et~al.}{2015}]{pyneb}
{Luridiana} V.,  {Morisset} C.,    {Shaw} R.~A.,  2015, \aap, 573, A42

\bibitem[\protect\citeauthoryear{{Martins}, {Schaerer} \& {Hillier}}{{Martins}
  et~al.}{2005}]{martins05}
{Martins} F.,  {Schaerer} D.,    {Hillier} D.~J.,  2005, \aap, 436, 1049

\bibitem[\protect\citeauthoryear{{McLeod}, {Dale}, {Ginsburg}, {Ercolano},
  {Gritschneder}, {Ramsay} \& {Testi}}{{McLeod} et~al.}{2015}]{M16}
{McLeod} A.~F.,  {Dale} J.~E.,  {Ginsburg} A.,  {Ercolano} B.,  {Gritschneder}
  M.,  {Ramsay} S.,    {Testi} L.,  2015, \mnras, 450, 1057

\bibitem[\protect\citeauthoryear{{McLeod}, {Weilbacher}, {Ginsburg}, {Dale},
  {Ramsay} \& {Testi}}{{McLeod} et~al.}{2016}]{mcleod16}
{McLeod} A.~F.,  {Weilbacher} P.~M.,  {Ginsburg} A.,  {Dale} J.~E.,  {Ramsay}
  S.,    {Testi} L.,  2016, \mnras, 455, 4057

\bibitem[\protect\citeauthoryear{{Melena}, {Massey}, {Morrell} \&
  {Zangari}}{{Melena} et~al.}{2008}]{melena08}
{Melena} N.~W.,  {Massey} P.,  {Morrell} N.~I.,    {Zangari} A.~M.,  2008, \aj,
  135, 878

\bibitem[\protect\citeauthoryear{{Mellema}, {Arthur}, {Henney}, {Iliev} \&
  {Shapiro}}{{Mellema} et~al.}{2006}]{mellema06}
{Mellema} G.,  {Arthur} S.~J.,  {Henney} W.~J.,  {Iliev} I.~T.,    {Shapiro}
  P.~R.,  2006, \apj, 647, 397

\bibitem[\protect\citeauthoryear{{Mellema}, {Raga}, {Canto}, {Lundqvist},
  {Balick}, {Steffen} \& {Noriega-Crespo}}{{Mellema} et~al.}{1998}]{mellema98}
{Mellema} G.,  {Raga} A.~C.,  {Canto} J.,  {Lundqvist} P.,  {Balick} B.,
  {Steffen} W.,    {Noriega-Crespo} A.,  1998, \aap, 331, 335

\bibitem[\protect\citeauthoryear{{Moffat}}{{Moffat}}{1983}]{moffat83}
{Moffat} A.~F.~J.,  1983, \aap, 124, 273

\bibitem[\protect\citeauthoryear{{Molinari} et~al.,}{{Molinari}
  et~al.}{2010}]{molinari10}
{Molinari} S.,  et~al., 2010, \aap, 518, L100

\bibitem[\protect\citeauthoryear{{Ohlendorf}, {Preibisch}, {Gaczkowski},
  {Ratzka}, {Grellmann} \& {McLeod}}{{Ohlendorf} et~al.}{2012}]{ohlen12}
{Ohlendorf} H.,  {Preibisch} T.,  {Gaczkowski} B.,  {Ratzka} T.,  {Grellmann}
  R.,    {McLeod} A.~F.,  2012, \aap, 540, A81

\bibitem[\protect\citeauthoryear{{Peimbert}, {Rayo} \&
  {Torres-Peimbert}}{{Peimbert} et~al.}{1978}]{peimbert78}
{Peimbert} M.,  {Rayo} J.~F.,    {Torres-Peimbert} S.,  1978, \apj, 220, 516

\bibitem[\protect\citeauthoryear{{Povich}, {Smith}, {Majewski}, {Getman},
  {Townsley}, {Babler}, {Broos}, {Indebetouw}, {Meade}, {Robitaille},
  {Stassun}, {Whitney}, {Yonekura} \& {Fukui}}{{Povich}
  et~al.}{2011}]{povich11}
{Povich} M.~S.,  {Smith} N.,  {Majewski} S.~R.,  {Getman} K.~V.,  {Townsley}
  L.~K.,  {Babler} B.~L.,  {Broos} P.~S.,  {Indebetouw} R.,  {Meade} M.~R.,
  {Robitaille} T.~P.,  {Stassun} K.~G.,  {Whitney} B.~A.,  {Yonekura} Y.,
  {Fukui} Y.,  2011, \apjs, 194, 14

\bibitem[\protect\citeauthoryear{{Preibisch}}{{Preibisch}}{2011}]{preibisch11}
{Preibisch} T.,  2011, in {von Berlepsch} R.,  ed., Reviews in Modern Astronomy
  Vol.~23 of Reviews in Modern Astronomy, {Star formation at High Resolution,
  Zooming into the Carina nebula, the nearest laboratory of massive Star
  feedback}.
p.~223

\bibitem[\protect\citeauthoryear{{Preibisch}, {Roccatagliata}, {Gaczkowski} \&
  {Ratzka}}{{Preibisch} et~al.}{2012a}]{preibisch12}
{Preibisch} T.,  {Roccatagliata} V.,  {Gaczkowski} B.,    {Ratzka} T.,  2012a,
  \aap, 541, A132

\bibitem[\protect\citeauthoryear{{Preibisch}, {Roccatagliata}, {Gaczkowski} \&
  {Ratzka}}{{Preibisch} et~al.}{2012b}]{CNC-Herschel1}
{Preibisch} T.,  {Roccatagliata} V.,  {Gaczkowski} B.,    {Ratzka} T.,  2012b,
  \aap, 541, A132

\bibitem[\protect\citeauthoryear{{Preibisch}, {Schuller}, {Ohlendorf},
  {Pekruhl}, {Menten} \& {Zinnecker}}{{Preibisch} et~al.}{2011}]{CNC-Laboca}
{Preibisch} T.,  {Schuller} F.,  {Ohlendorf} H.,  {Pekruhl} S.,  {Menten}
  K.~M.,    {Zinnecker} H.,  2011, \aap, 525, A92

\bibitem[\protect\citeauthoryear{{Reipurth} \& {Bally}}{{Reipurth} \&
  {Bally}}{2001}]{reipurth01}
{Reipurth} B.,  {Bally} J.,  2001, \araa, 39, 403

\bibitem[\protect\citeauthoryear{{Reiter} \& {Smith}}{{Reiter} \&
  {Smith}}{2013}]{reiter13}
{Reiter} M.,  {Smith} N.,  2013, \mnras, 433, 2226

\bibitem[\protect\citeauthoryear{{Reiter} \& {Smith}}{{Reiter} \&
  {Smith}}{2014}]{reiter14}
{Reiter} M.,  {Smith} N.,  2014, \mnras, 445, 3939

\bibitem[\protect\citeauthoryear{{Roccatagliata}, {Preibisch}, {Ratzka} \&
  {Gaczkowski}}{{Roccatagliata} et~al.}{2013}]{CNC-Herschel2}
{Roccatagliata} V.,  {Preibisch} T.,  {Ratzka} T.,    {Gaczkowski} B.,  2013,
  \aap, 554, A6

\bibitem[\protect\citeauthoryear{{Rochau}, {Brandner}, {Stolte}, {Gennaro},
  {Gouliermis}, {Da Rio}, {Dzyurkevich} \& {Henning}}{{Rochau}
  et~al.}{2010}]{rochau10}
{Rochau} B.,  {Brandner} W.,  {Stolte} A.,  {Gennaro} M.,  {Gouliermis} D.,
  {Da Rio} N.,  {Dzyurkevich} N.,    {Henning} T.,  2010, \apjl, 716, L90

\bibitem[\protect\citeauthoryear{{Sankrit} \& {Hester}}{{Sankrit} \&
  {Hester}}{2000}]{sankrit00}
{Sankrit} R.,  {Hester} J.~J.,  2000, \apj, 535, 847

\bibitem[\protect\citeauthoryear{{Schaye} et~al.,}{{Schaye}
  et~al.}{2015}]{schaye15}
{Schaye} J.,  et~al., 2015, \mnras, 446, 521

\bibitem[\protect\citeauthoryear{{Smith}}{{Smith}}{2006a}]{smith06}
{Smith} N.,  2006a, \mnras, 367, 763

\bibitem[\protect\citeauthoryear{{Smith}}{{Smith}}{2006b}]{smith06b}
{Smith} N.,  2006b, \apj, 644, 1151

\bibitem[\protect\citeauthoryear{{Smith}, {Barb{\'a}} \& {Walborn}}{{Smith}
  et~al.}{2004}]{smith04}
{Smith} N.,  {Barb{\'a}} R.~H.,    {Walborn} N.~R.,  2004, \mnras, 351, 1457

\bibitem[\protect\citeauthoryear{{Smith} \& {Brooks}}{{Smith} \&
  {Brooks}}{2008}]{smith08}
{Smith} N.,  {Brooks} K.~J.,  2008, {The Carina Nebula: A Laboratory for
  Feedback and Triggered Star Formation}.
p.~138

\bibitem[\protect\citeauthoryear{{Smith} et~al.,}{{Smith}
  et~al.}{2010a}]{smith10}
{Smith} N.,  et~al., 2010a, \mnras, 405, 1153

\bibitem[\protect\citeauthoryear{{Smith} et~al.,}{{Smith}
  et~al.}{2010b}]{smith10b}
{Smith} N.,  et~al., 2010b, \mnras, 406, 952

\bibitem[\protect\citeauthoryear{{Sommer-Larsen}, {G{\"o}tz} \&
  {Portinari}}{{Sommer-Larsen} et~al.}{2003}]{sommerlarsen03}
{Sommer-Larsen} J.,  {G{\"o}tz} M.,    {Portinari} L.,  2003, \apj, 596, 47

\bibitem[\protect\citeauthoryear{{Sung} \& {Bessell}}{{Sung} \&
  {Bessell}}{2004}]{sung04}
{Sung} H.,  {Bessell} M.~S.,  2004, \aj, 127, 1014

\bibitem[\protect\citeauthoryear{{Tremblin}, {Audit}, {Minier} \&
  {Schneider}}{{Tremblin} et~al.}{2012}]{tremblin12}
{Tremblin} P.,  {Audit} E.,  {Minier} V.,    {Schneider} N.,  2012, \aap, 538,
  A31

\bibitem[\protect\citeauthoryear{{Turner} \& {Moffat}}{{Turner} \&
  {Moffat}}{1979}]{turner79}
{Turner} D.~G.,  {Moffat} A.~F.~J.,  1979, \jrasc, 73, 301

\bibitem[\protect\citeauthoryear{{Vilchez} \& {Esteban}}{{Vilchez} \&
  {Esteban}}{1996}]{vilchez96}
{Vilchez} J.~M.,  {Esteban} C.,  1996, \mnras, 280, 720

\bibitem[\protect\citeauthoryear{{Vogelsberger}, {Genel}, {Springel}, {Torrey},
  {Sijacki}, {Xu}, {Snyder}, {Bird}, {Nelson} \& {Hernquist}}{{Vogelsberger}
  et~al.}{2014}]{vogel14}
{Vogelsberger} M.,  {Genel} S.,  {Springel} V.,  {Torrey} P.,  {Sijacki} D.,
  {Xu} D.,  {Snyder} G.,  {Bird} S.,  {Nelson} D.,    {Hernquist} L.,  2014,
  \nat, 509, 177

\bibitem[\protect\citeauthoryear{{Walborn}}{{Walborn}}{1973}]{walborn73}
{Walborn} N.~R.,  1973, \apjl, 182, L21

\bibitem[\protect\citeauthoryear{{Walch}}{{Walch}}{2014}]{walch14}
{Walch} S.~K.,  2014, Astrophysics and Space Science Proceedings, 36, 173

\bibitem[\protect\citeauthoryear{{Wang}, {Feigelson}, {Townsley}, {Broos},
  {Getman}, {Wolk}, {Preibisch}, {Stassun}, {Moffat}, {Garmire}, {King},
  {McCaughrean} \& {Zinnecker}}{{Wang} et~al.}{2011}]{wang11}
{Wang} J.,  {Feigelson} E.~D.,  {Townsley} L.~K.,  {Broos} P.~S.,  {Getman}
  K.~V.,  {Wolk} S.~J.,  {Preibisch} T.,  {Stassun} K.~G.,  {Moffat} A.~F.~J.,
  {Garmire} G.,  {King} R.~R.,  {McCaughrean} M.~J.,    {Zinnecker} H.,  2011,
  \apjs, 194, 11

\bibitem[\protect\citeauthoryear{{Weilbacher}, {Streicher}, {Urrutia}, {Jarno},
  {P{\'e}contal-Rousset}, {Bacon} \& {B{\"o}hm}}{{Weilbacher}
  et~al.}{2012}]{pipeline}
{Weilbacher} P.~M.,  {Streicher} O.,  {Urrutia} T.,  {Jarno} A.,
  {P{\'e}contal-Rousset} A.,  {Bacon} R.,    {B{\"o}hm} P.,  2012, in Society
  of Photo-Optical Instrumentation Engineers (SPIE) Conference Series Vol.~8451
  of Society of Photo-Optical Instrumentation Engineers (SPIE) Conference
  Series, {Design and capabilities of the MUSE data reduction software and
  pipeline}.
p.~0

\bibitem[\protect\citeauthoryear{{Westmoquette}, {Dale}, {Ercolano} \&
  {Smith}}{{Westmoquette} et~al.}{2013}]{west13}
{Westmoquette} M.~S.,  {Dale} J.~E.,  {Ercolano} B.,    {Smith} L.~J.,  2013,
  \mnras, 435, 30

\bibitem[\protect\citeauthoryear{{White}, {Nelson}, {Holland}, {Robson},
  {Greaves}, {McCaughrean}, {Pilbratt}, {Balser}, {Oka}, {Sakamoto},
  {Hasegawa}, {McCutcheon}, {Matthews}, {Fridlund}, {Tothill}, {Huldtgren} \&
  {Deane}}{{White} et~al.}{1999}]{white99}
{White} G.~J.,  {Nelson} R.~P.,  {Holland} W.~S.,  {Robson} E.~I.,  {Greaves}
  J.~S.,  {McCaughrean} M.~J.,  {Pilbratt} G.~L.,  {Balser} D.~S.,  {Oka} T.,
  {Sakamoto} S.,  {Hasegawa} T.,  {McCutcheon} W.~H.,  {Matthews} H.~E.,
  {Fridlund} C.~V.~M.,  {Tothill} N.~F.~H.,  {Huldtgren} M.,    {Deane} J.~R.,
  1999, \aap, 342, 233

\bibitem[\protect\citeauthoryear{{Wolk}, {Broos}, {Getman}, {Feigelson},
  {Preibisch}, {Townsley}, {Wang}, {Stassun}, {King}, {McCaughrean}, {Moffat}
  \& {Zinnecker}}{{Wolk} et~al.}{2011}]{wolk11}
{Wolk} S.~J.,  {Broos} P.~S.,  {Getman} K.~V.,  {Feigelson} E.~D.,  {Preibisch}
  T.,  {Townsley} L.~K.,  {Wang} J.,  {Stassun} K.~G.,  {King} R.~R.,
  {McCaughrean} M.~J.,  {Moffat} A.~F.~J.,    {Zinnecker} H.,  2011, \apjs,
  194, 12

\bibitem[\protect\citeauthoryear{{Yonekura}, {Asayama}, {Kimura}, {Ogawa},
  {Kanai}, {Yamaguchi}, {Barnes} \& {Fukui}}{{Yonekura}
  et~al.}{2005}]{yonekura05}
{Yonekura} Y.,  {Asayama} S.,  {Kimura} K.,  {Ogawa} H.,  {Kanai} Y.,
  {Yamaguchi} N.,  {Barnes} P.~J.,    {Fukui} Y.,  2005, \apj, 634, 476

\end{thebibliography}


\appendix
\section[]{}
\begin{figure*}
\mbox{
\subfloat[]{\includegraphics[scale=0.35]{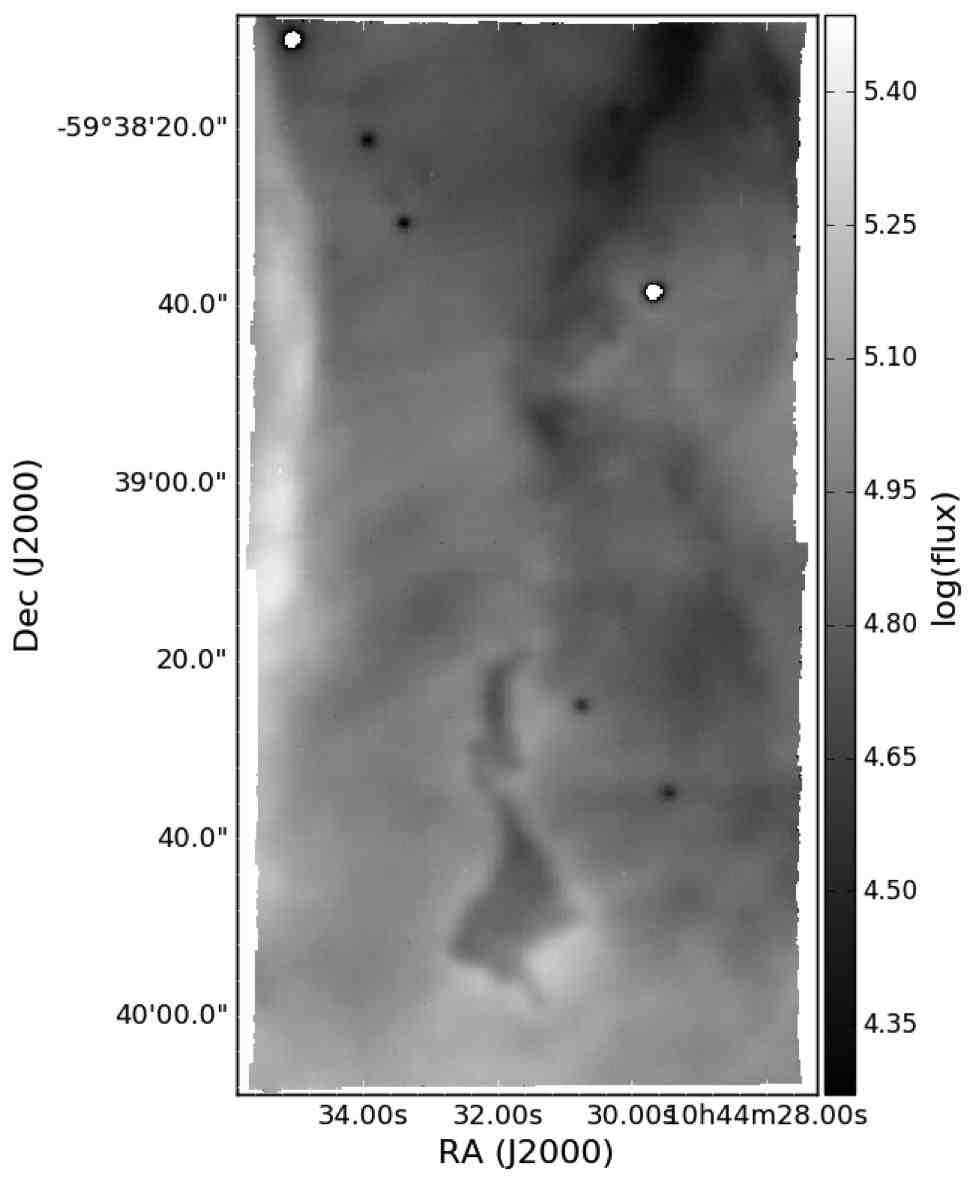}}
\subfloat[]{\includegraphics[scale=0.35]{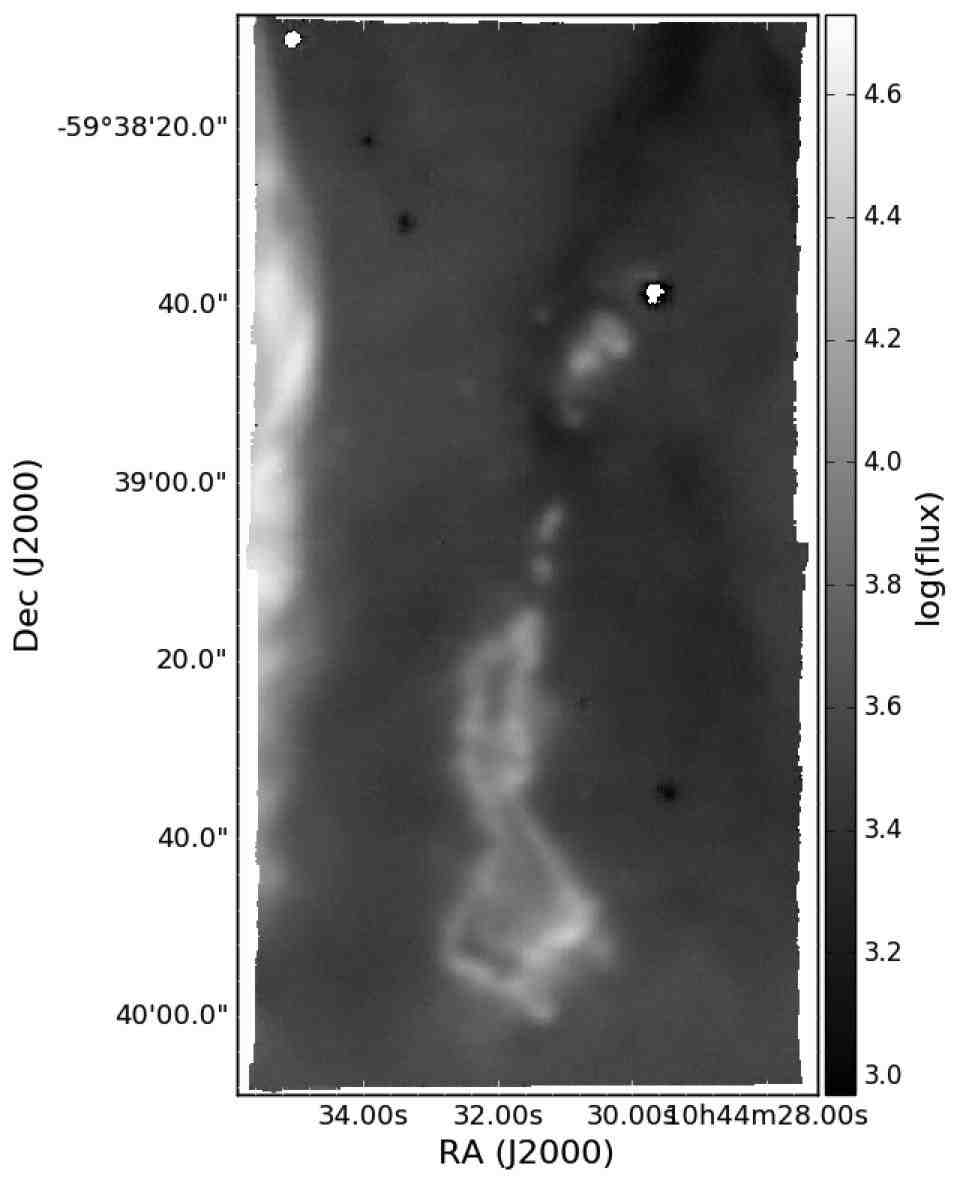}}}
\mbox{
\subfloat[]{\includegraphics[scale=0.35]{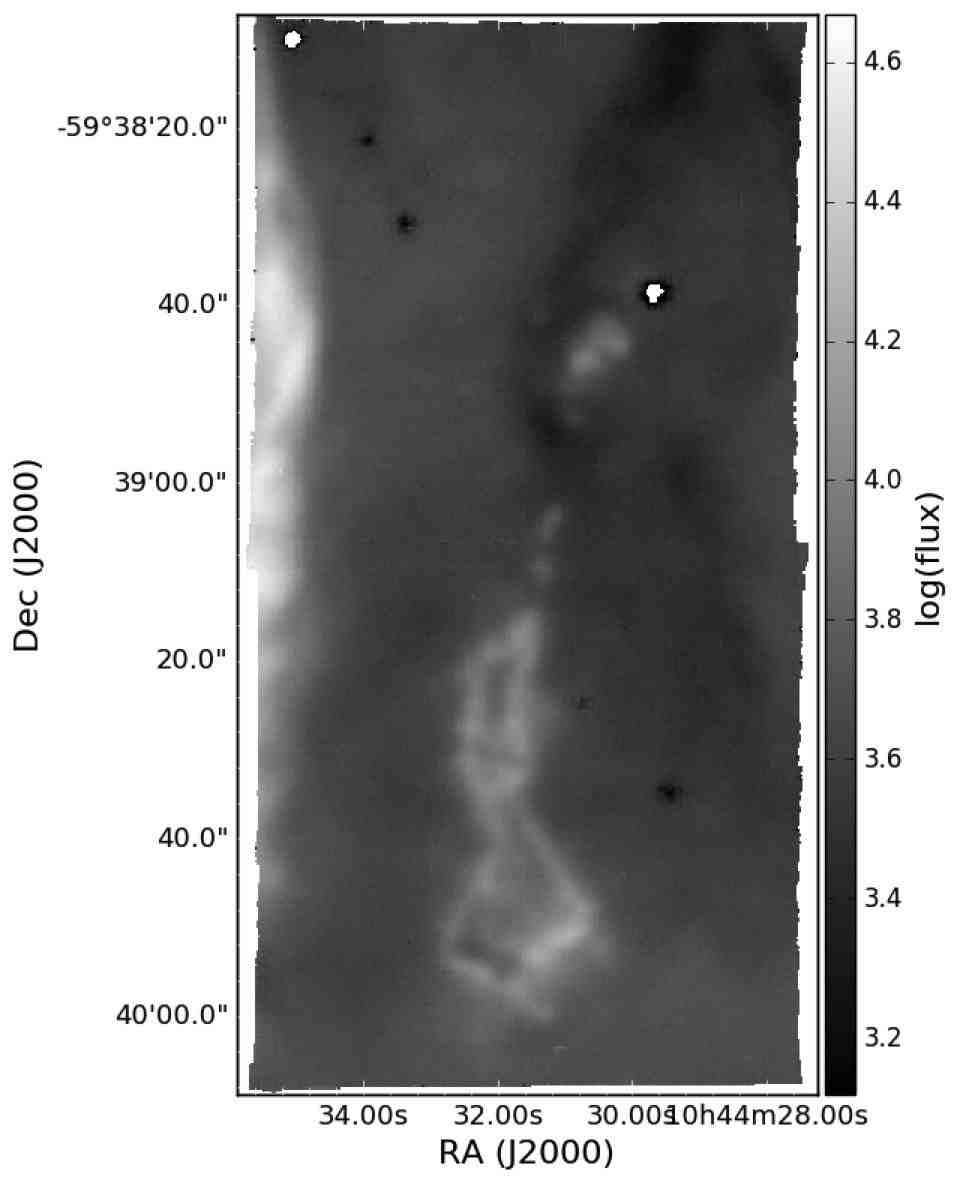}}
\subfloat[]{\includegraphics[scale=0.35]{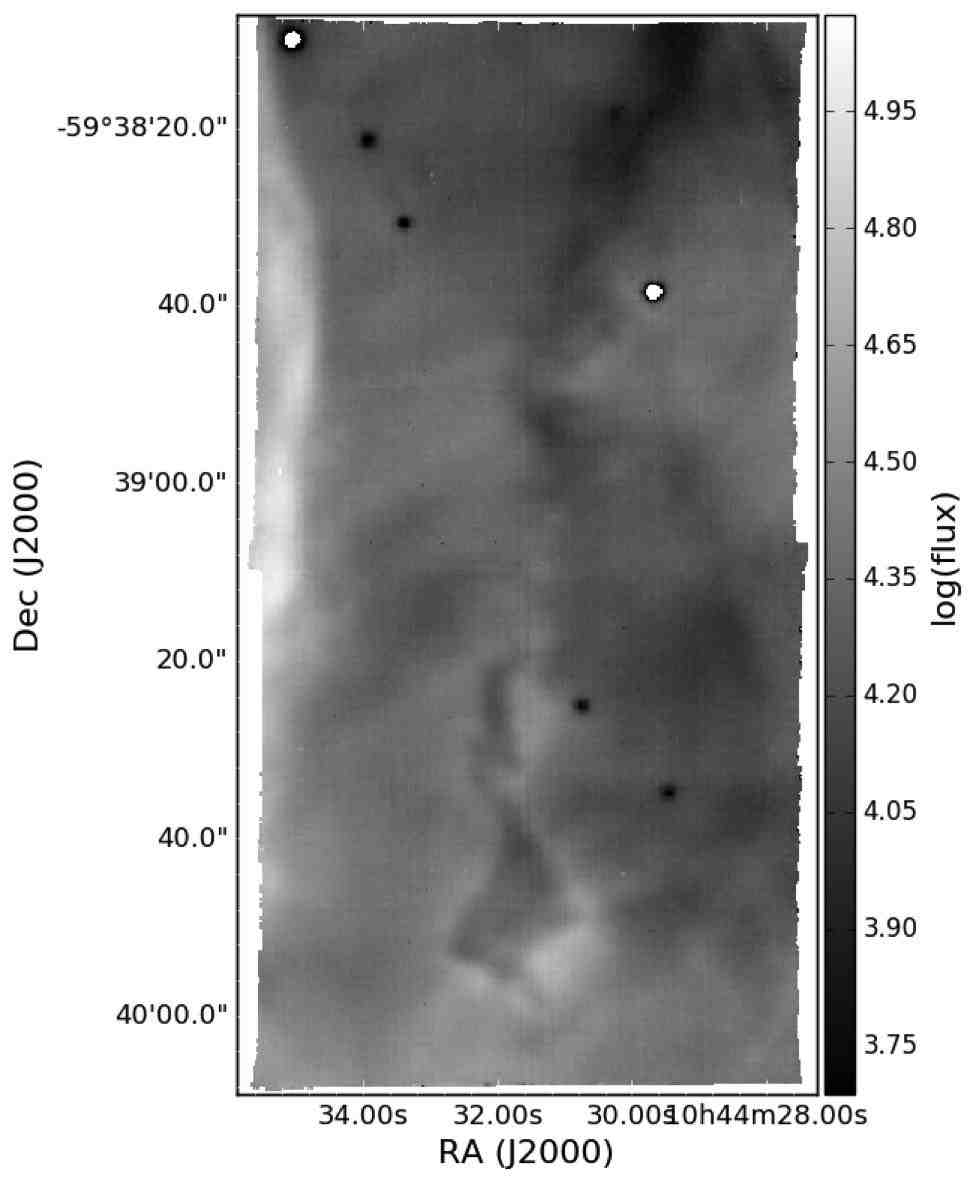}}}
\mbox{
\subfloat[]{\includegraphics[scale=0.35]{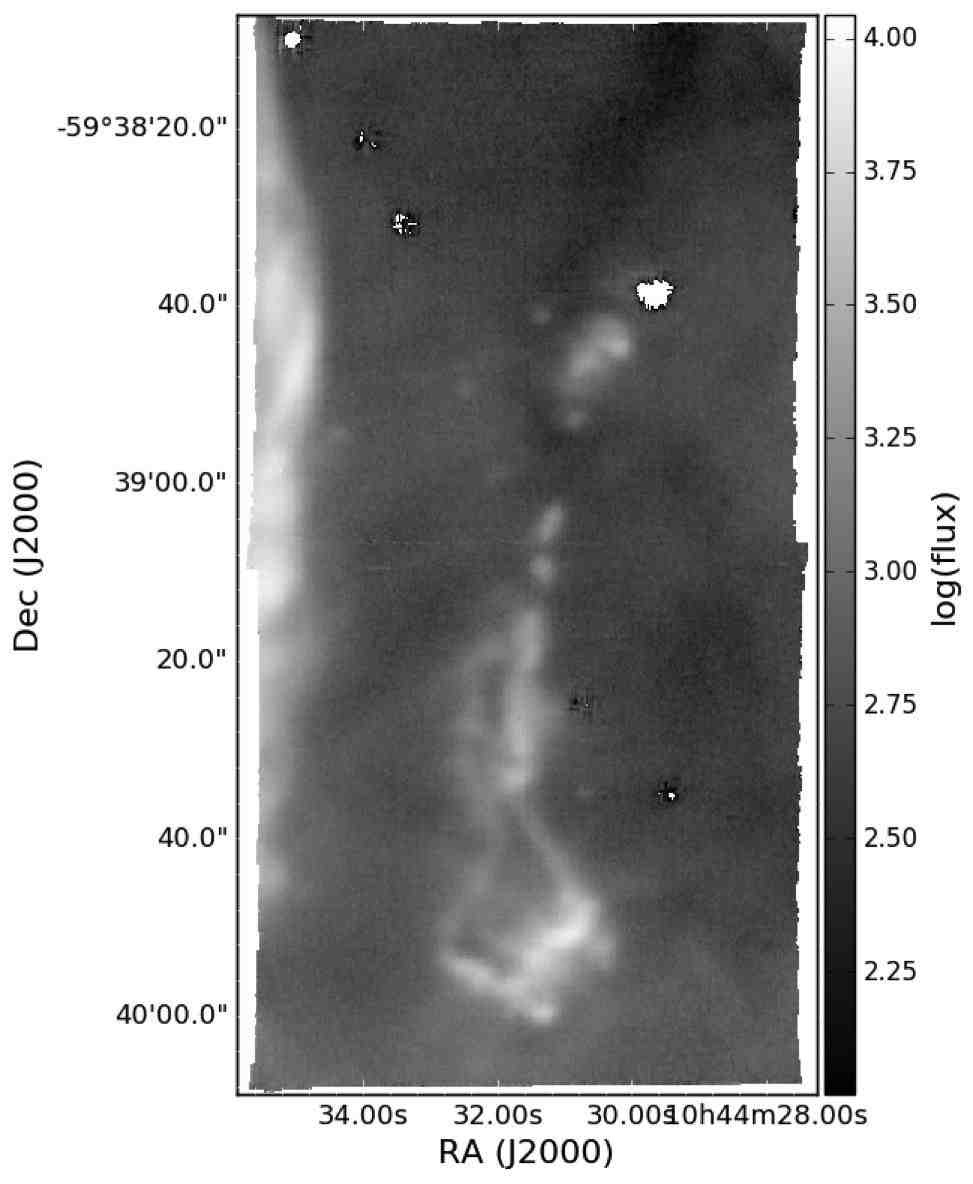}}
\subfloat[]{\includegraphics[scale=0.35]{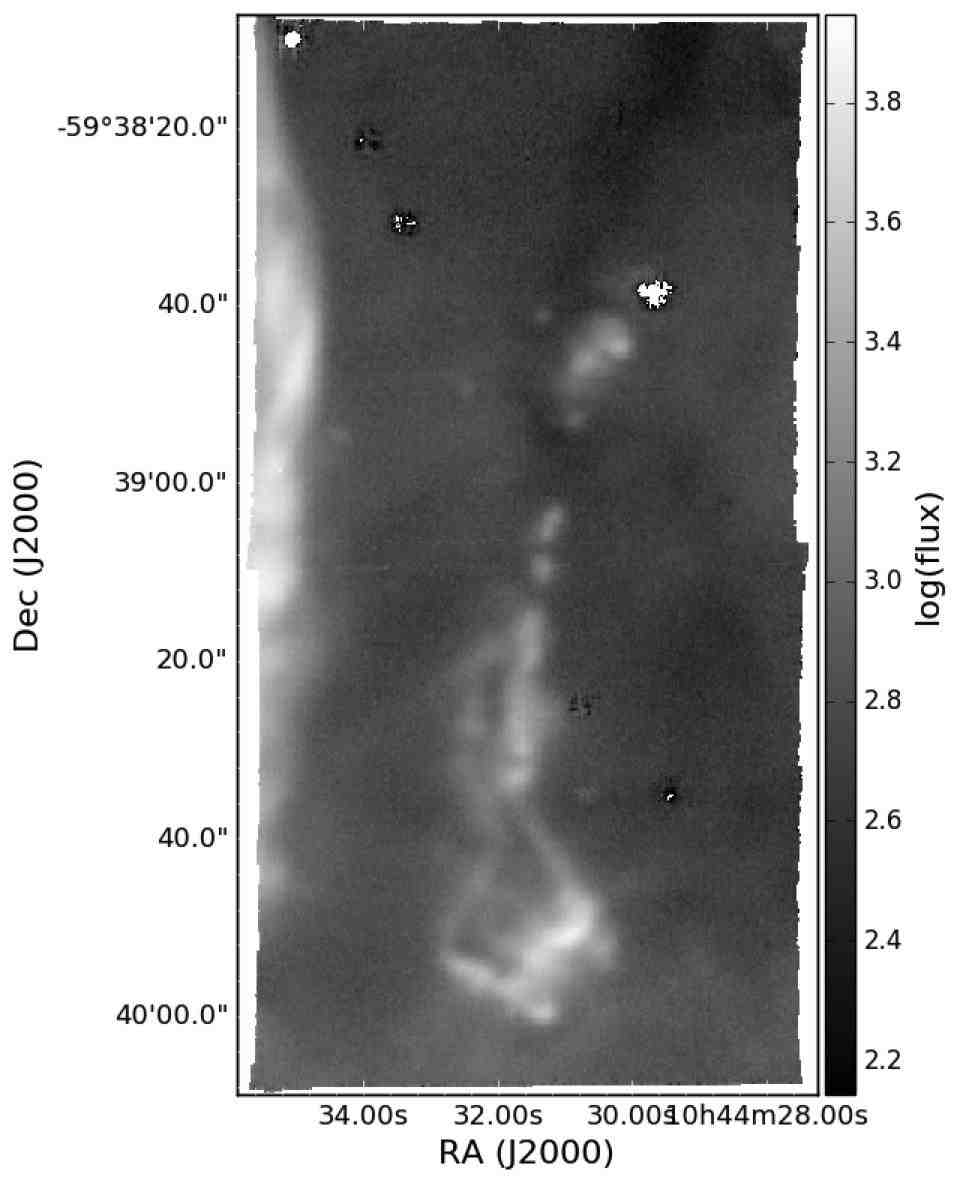}}}
\caption{Continuum-subtracted integrated emission line intensity maps of R37, from (a) to (f): H$\beta$, [SII]$\lambda$6731, [SII]$\lambda$6717, [SIII]$\lambda$9068, [OII]$\lambda$7320, [OII]$\lambda$7330.}
\label{maps1}
\end{figure*}

\begin{figure*}
\mbox{
\subfloat[]{\includegraphics[scale=0.35]{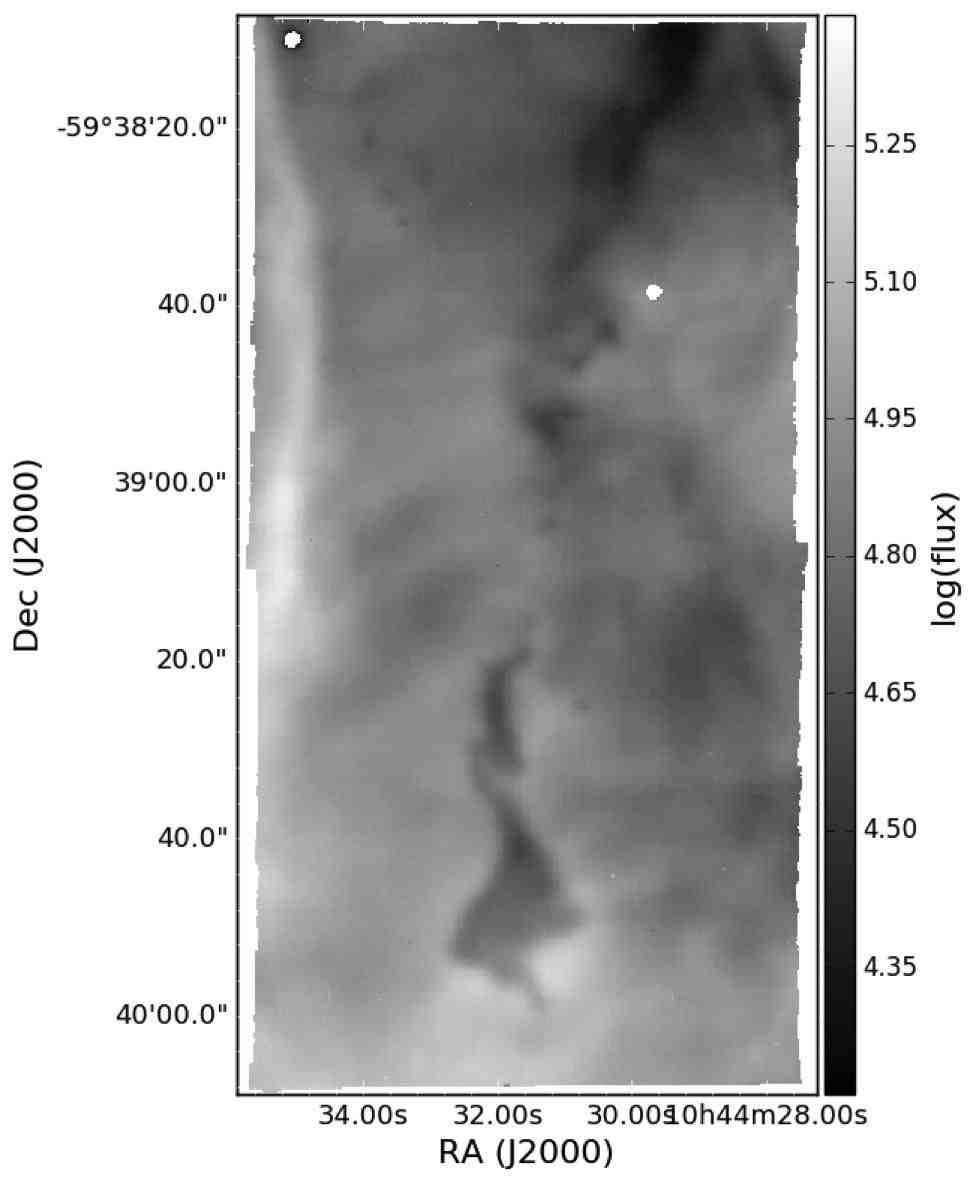}}
\subfloat[]{\includegraphics[scale=0.35]{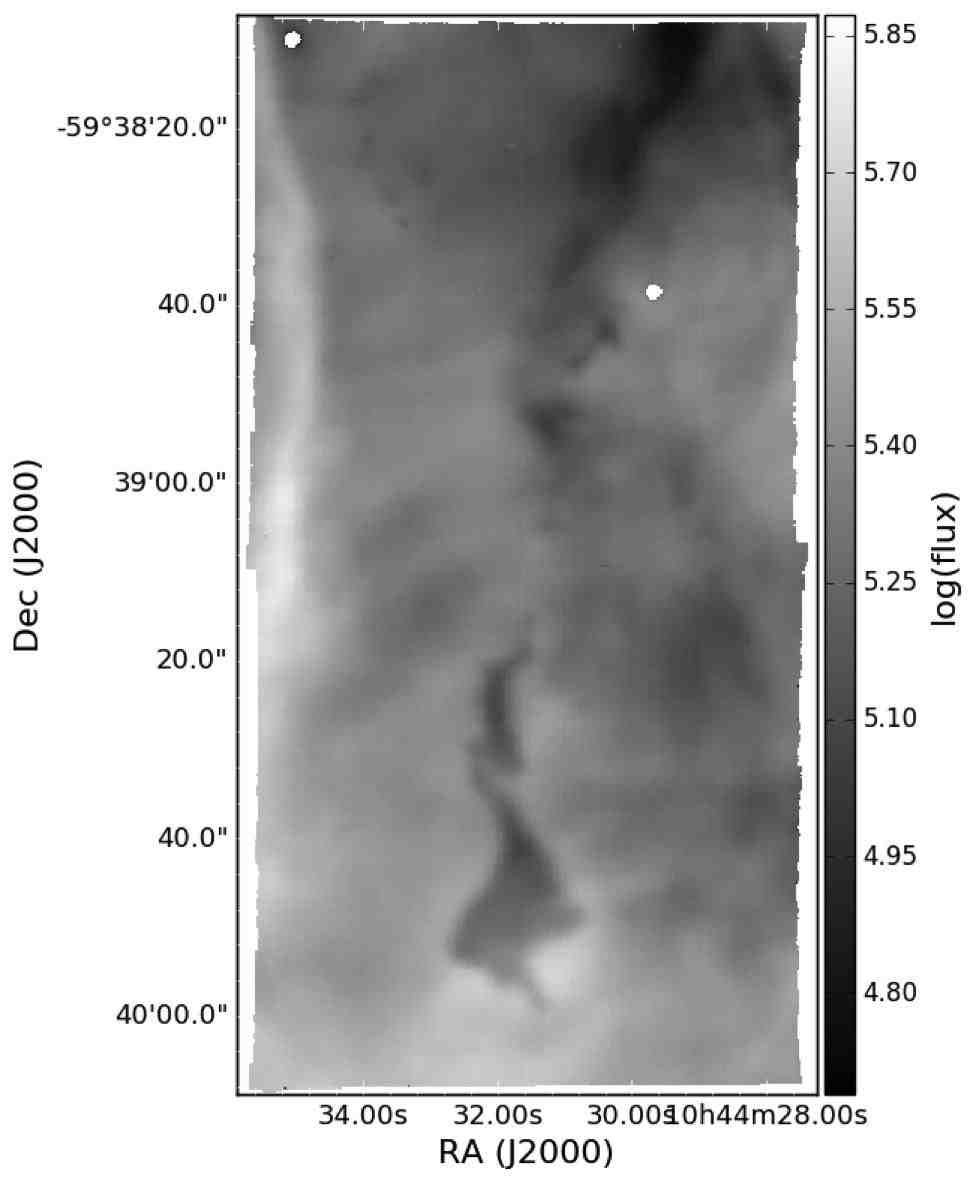}}}
\mbox{
\subfloat[]{\includegraphics[scale=0.35]{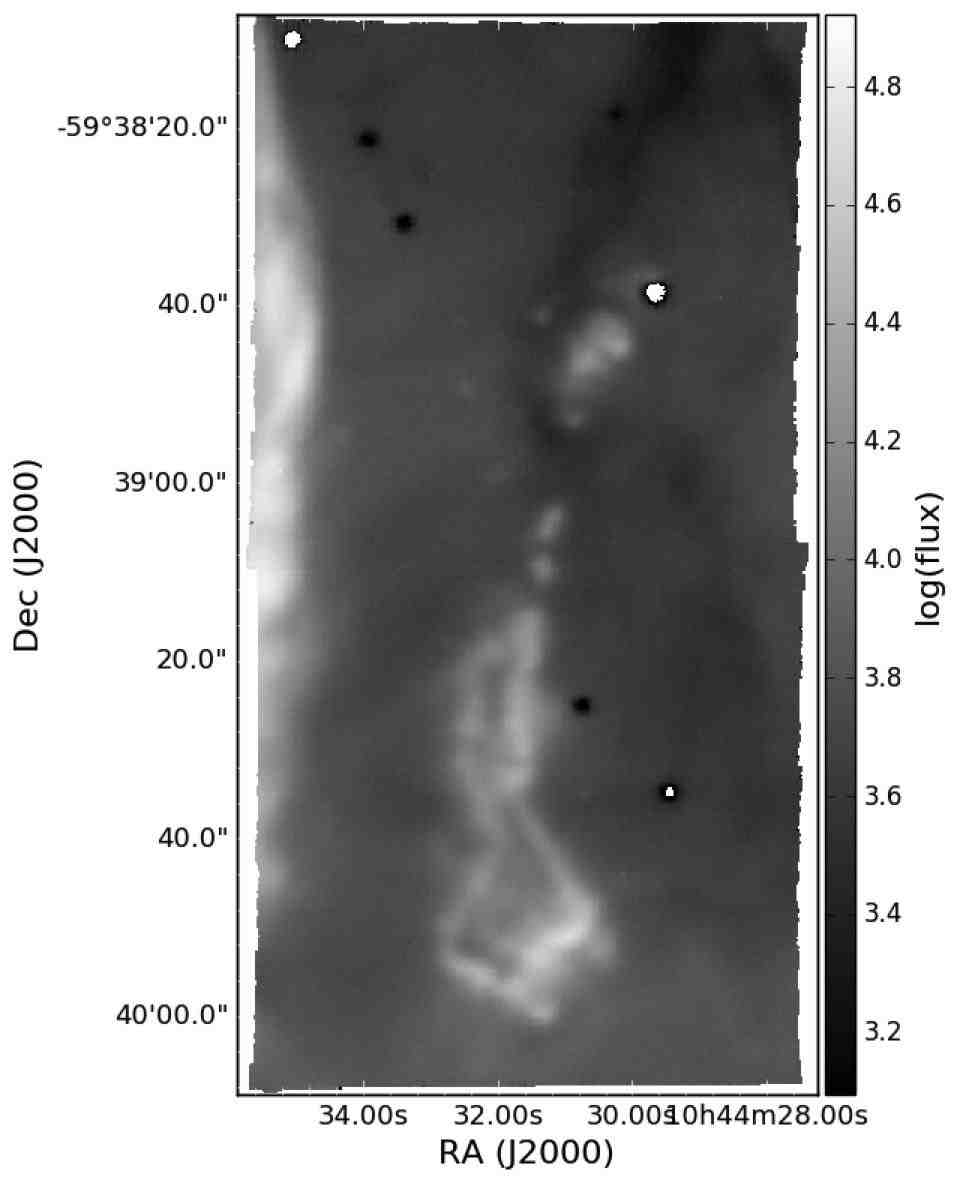}}
\subfloat[]{\includegraphics[scale=0.35]{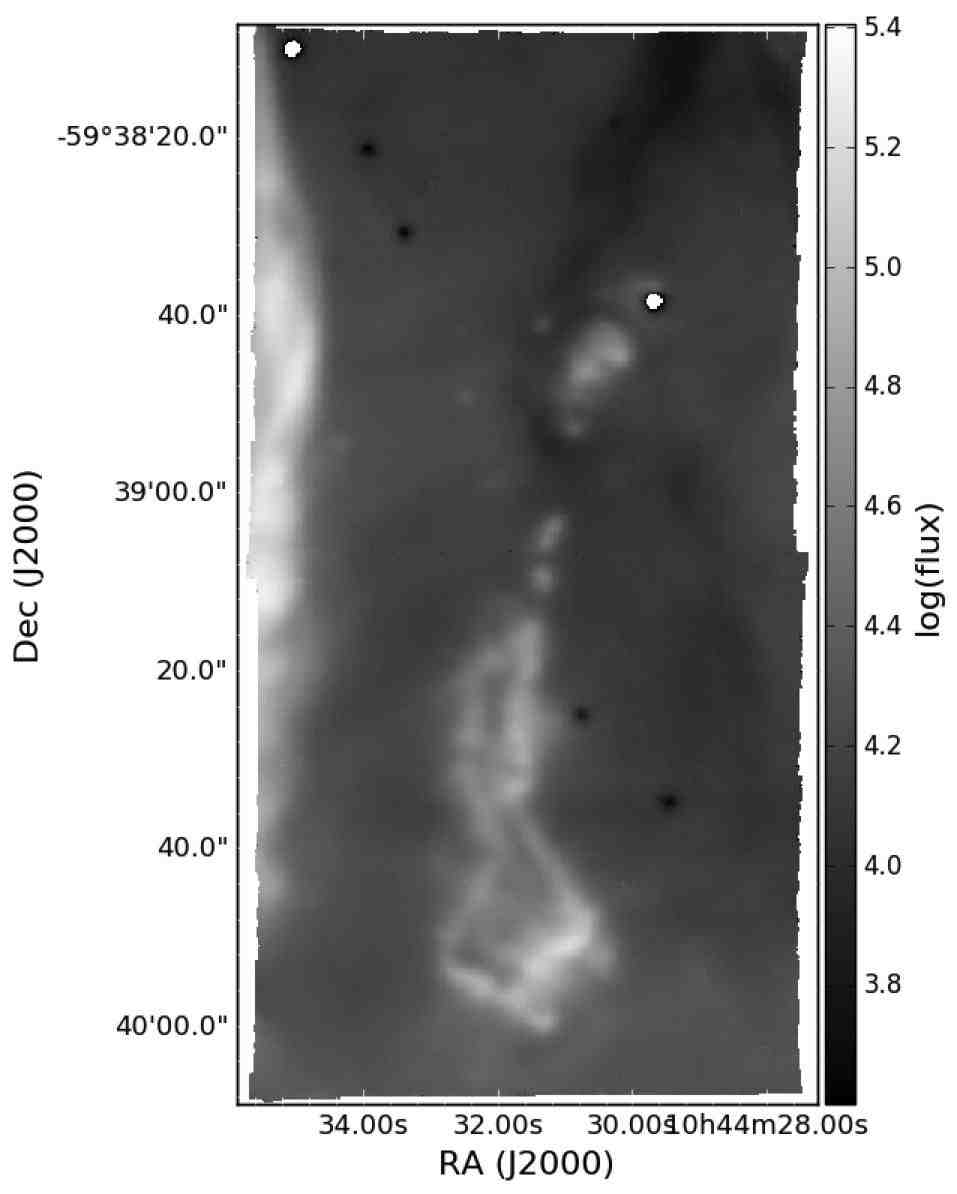}}}
\mbox{
\subfloat[]{\includegraphics[scale=0.35]{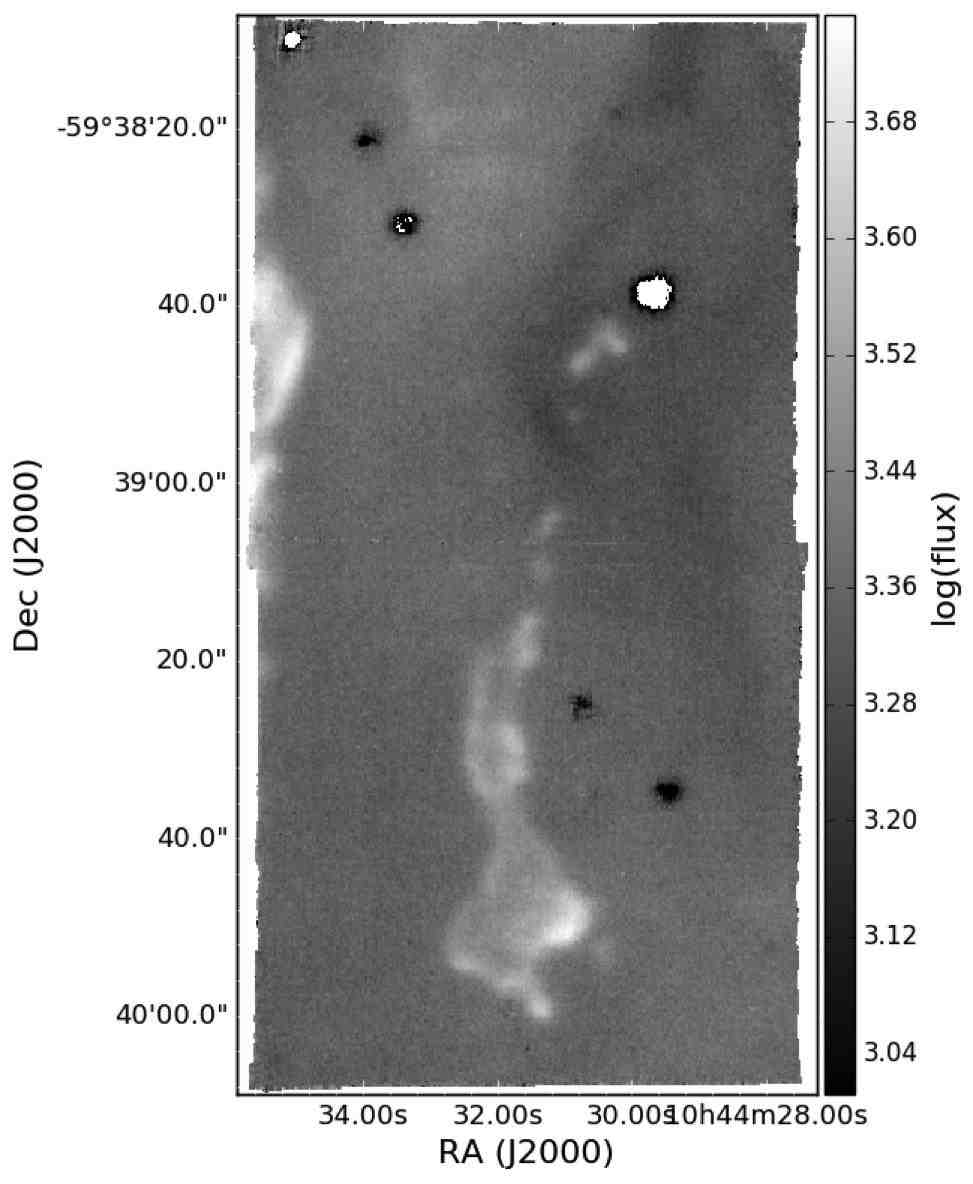}}
\subfloat[]{\includegraphics[scale=0.35]{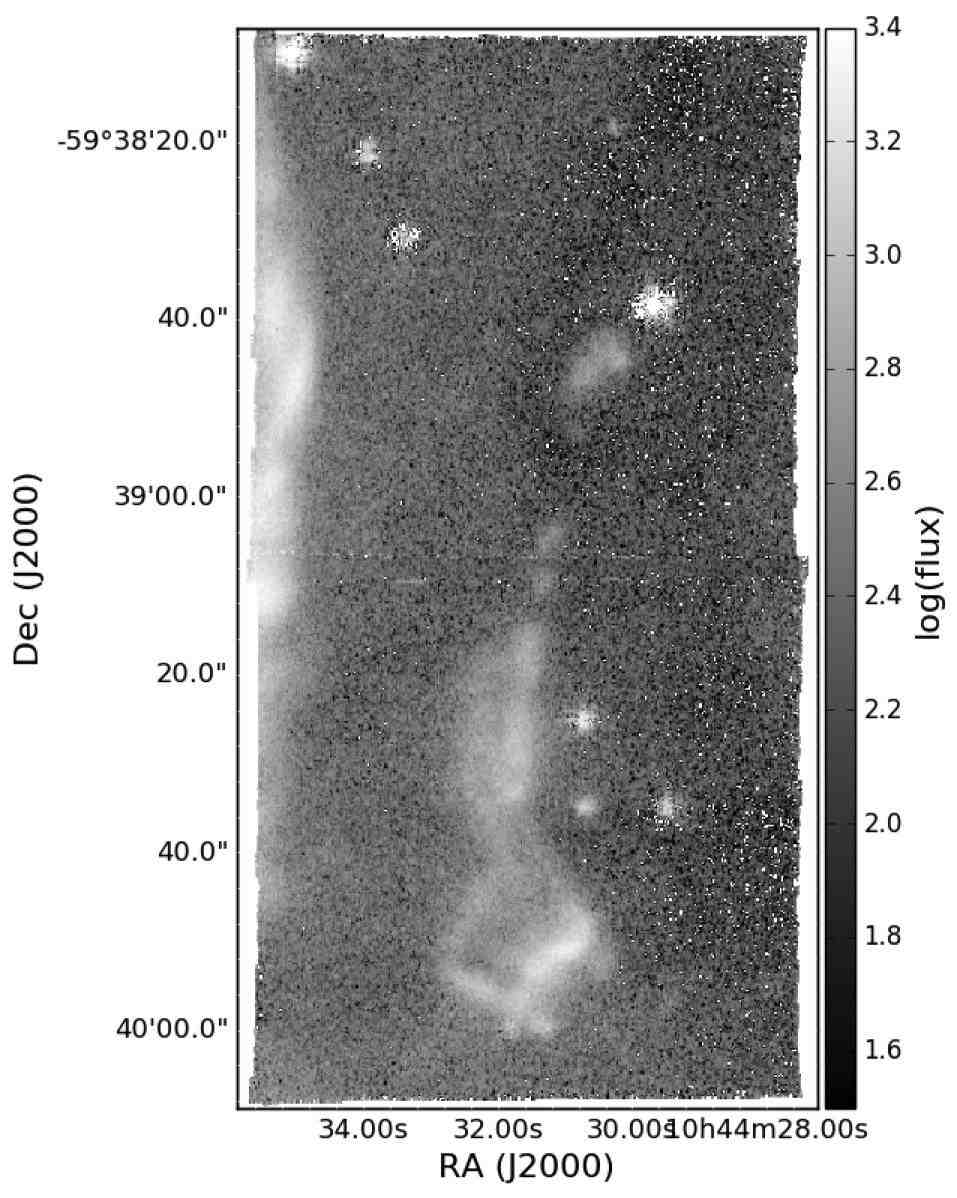}}}
\caption{Same as Fig.~A1 , (a) to (e): [OIII]$\lambda$4959, [OIII]$\lambda$5007, [NII]$\lambda$6548, [NII]$\lambda$6584, [OI]$\lambda$6300, [NII]$\lambda$5755.}
\label{maps2}
\end{figure*}

\begin{figure*}
\mbox{
\subfloat[]{\includegraphics[scale=0.35]{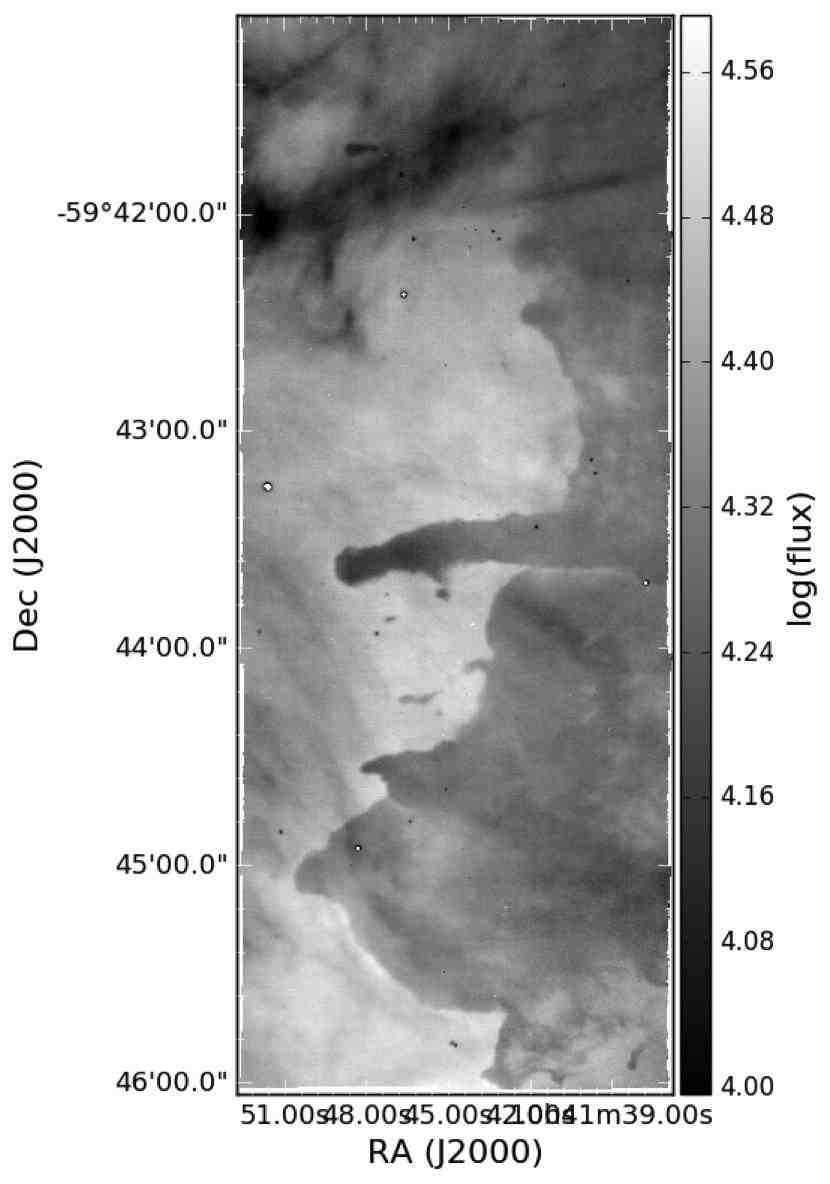}}
\subfloat[]{\includegraphics[scale=0.35]{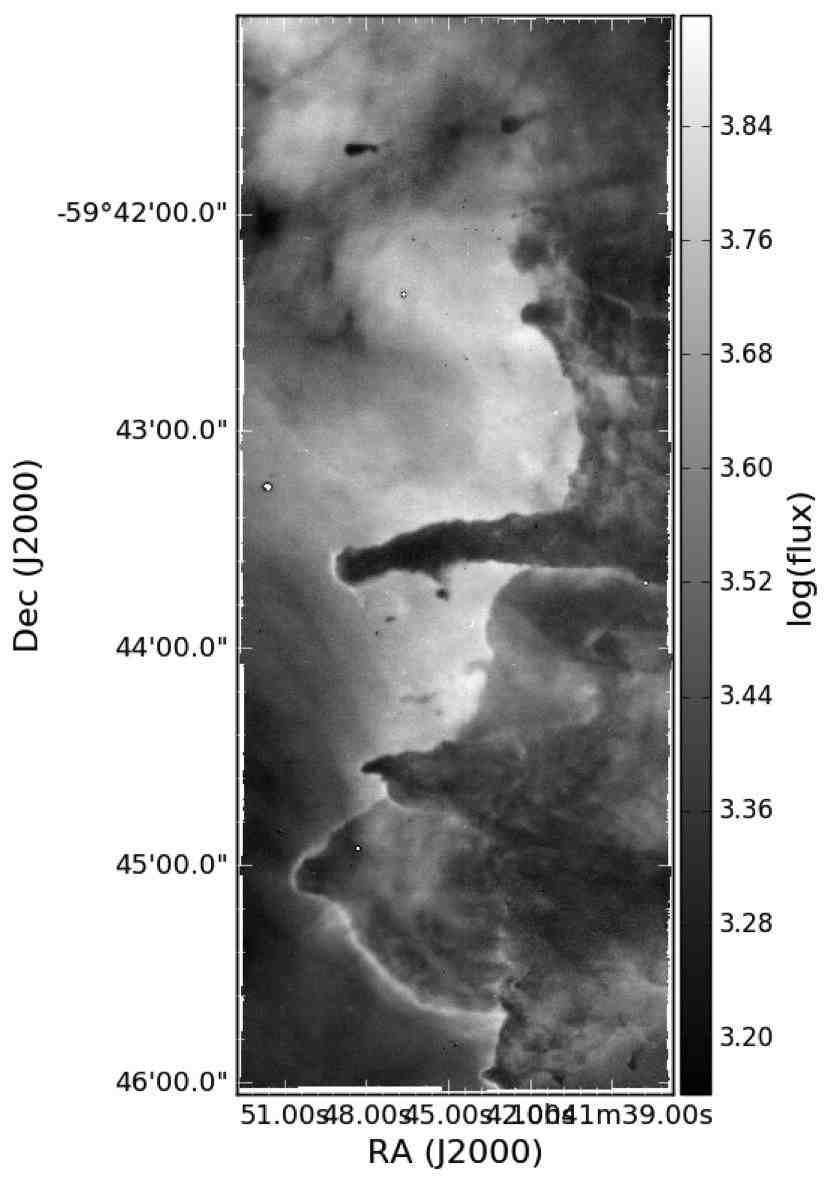}}}
\mbox{
\subfloat[]{\includegraphics[scale=0.35]{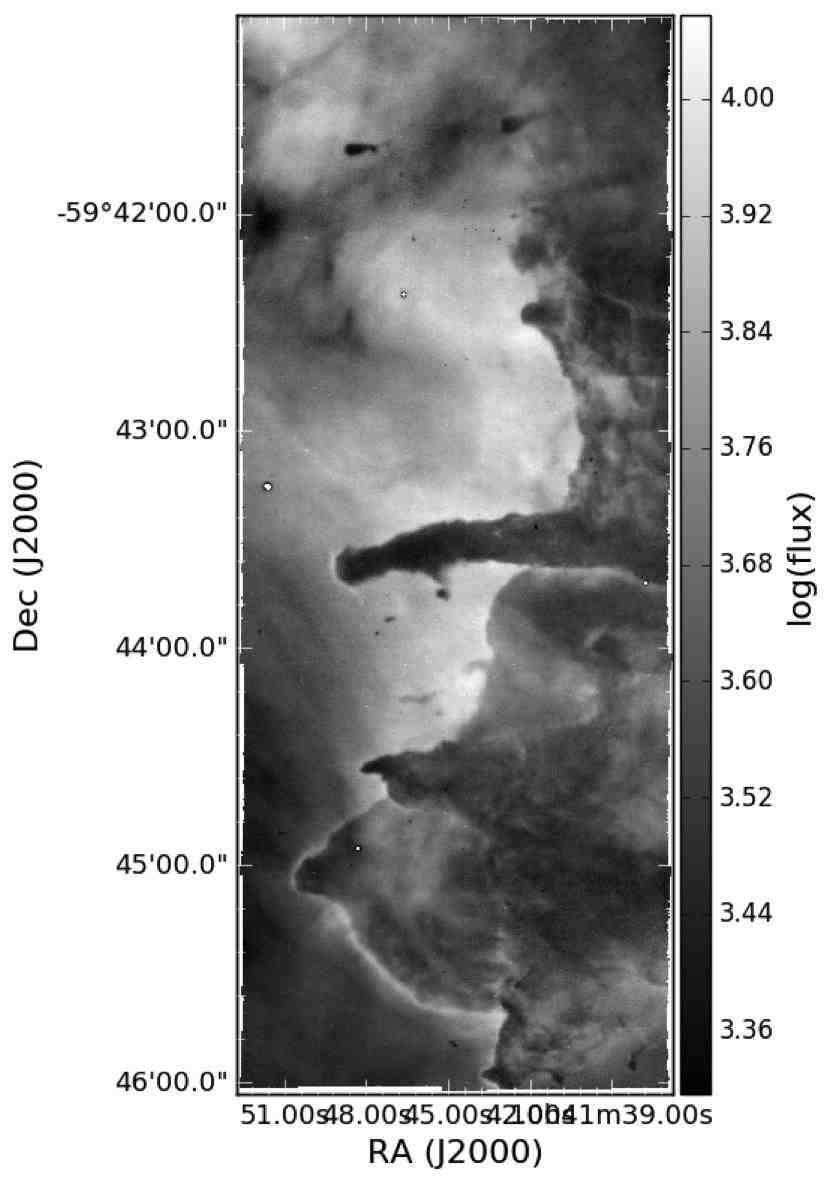}}
\subfloat[]{\includegraphics[scale=0.35]{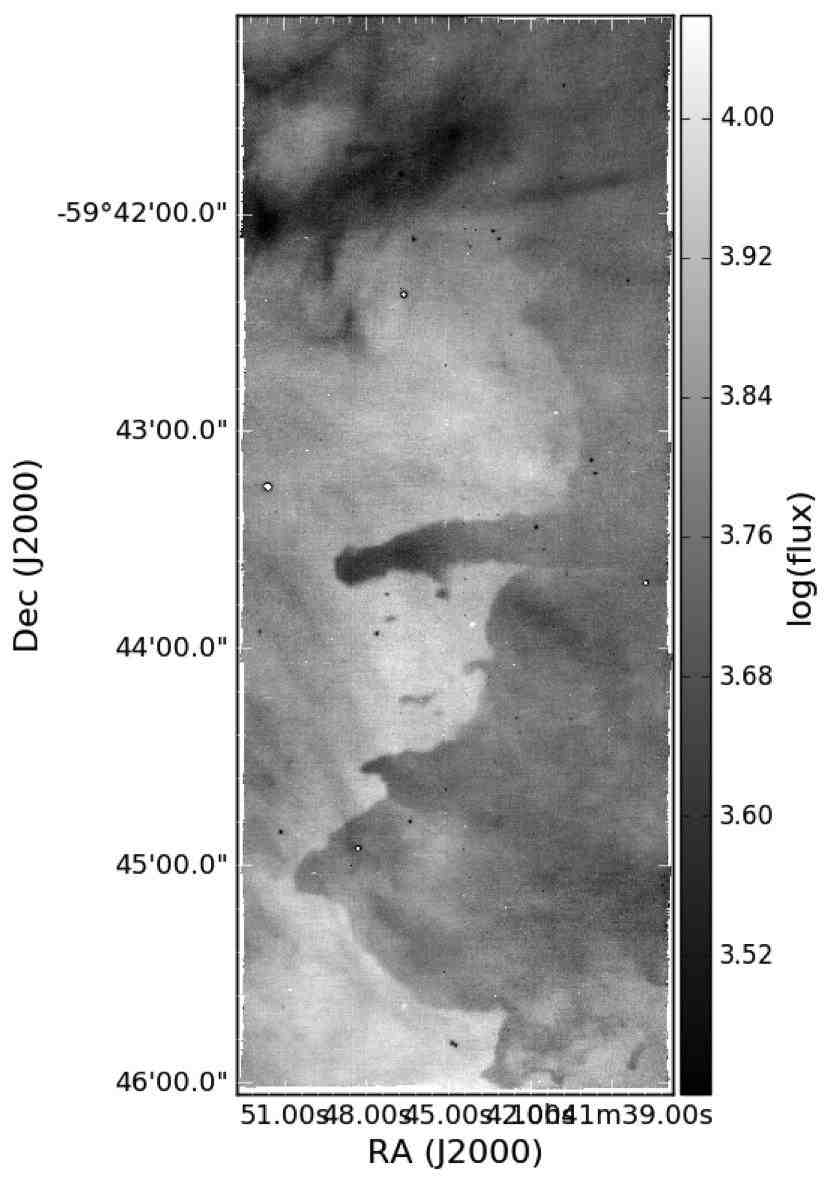}}}
\mbox{
\subfloat[]{\includegraphics[scale=0.35]{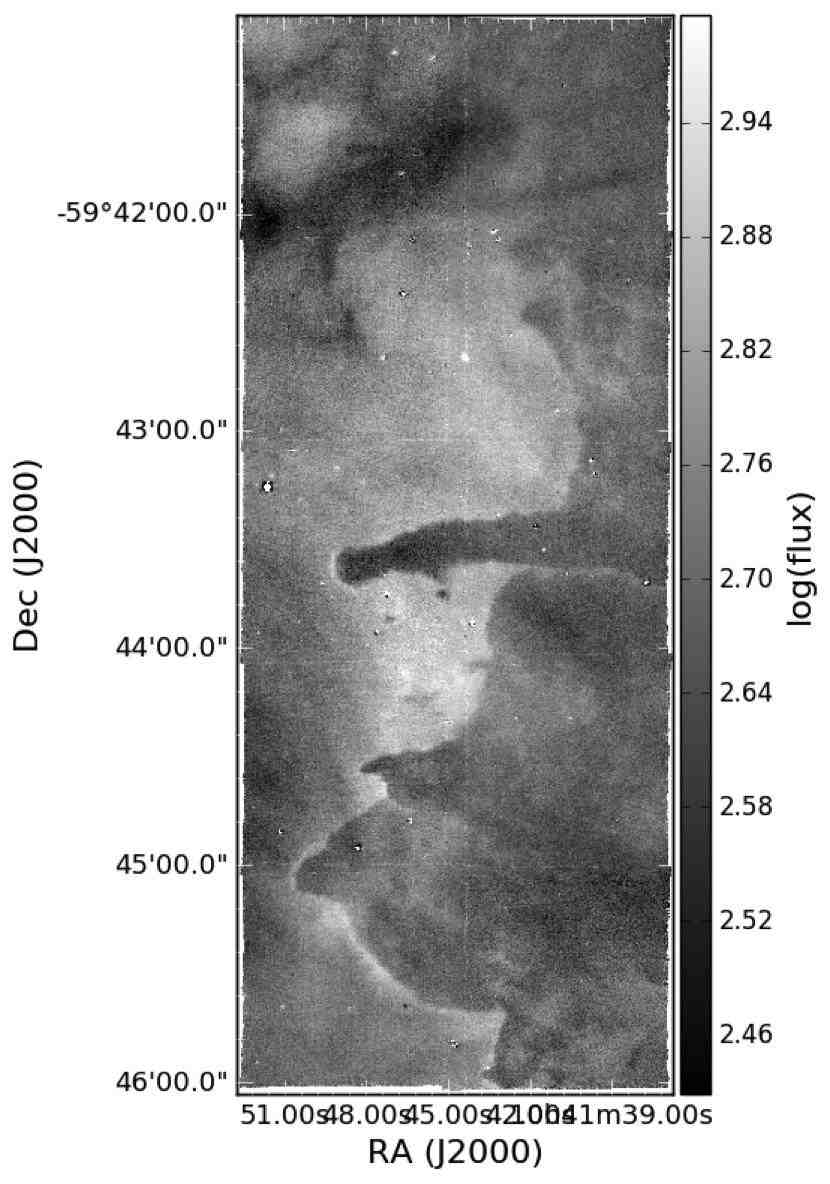}}
\subfloat[]{\includegraphics[scale=0.35]{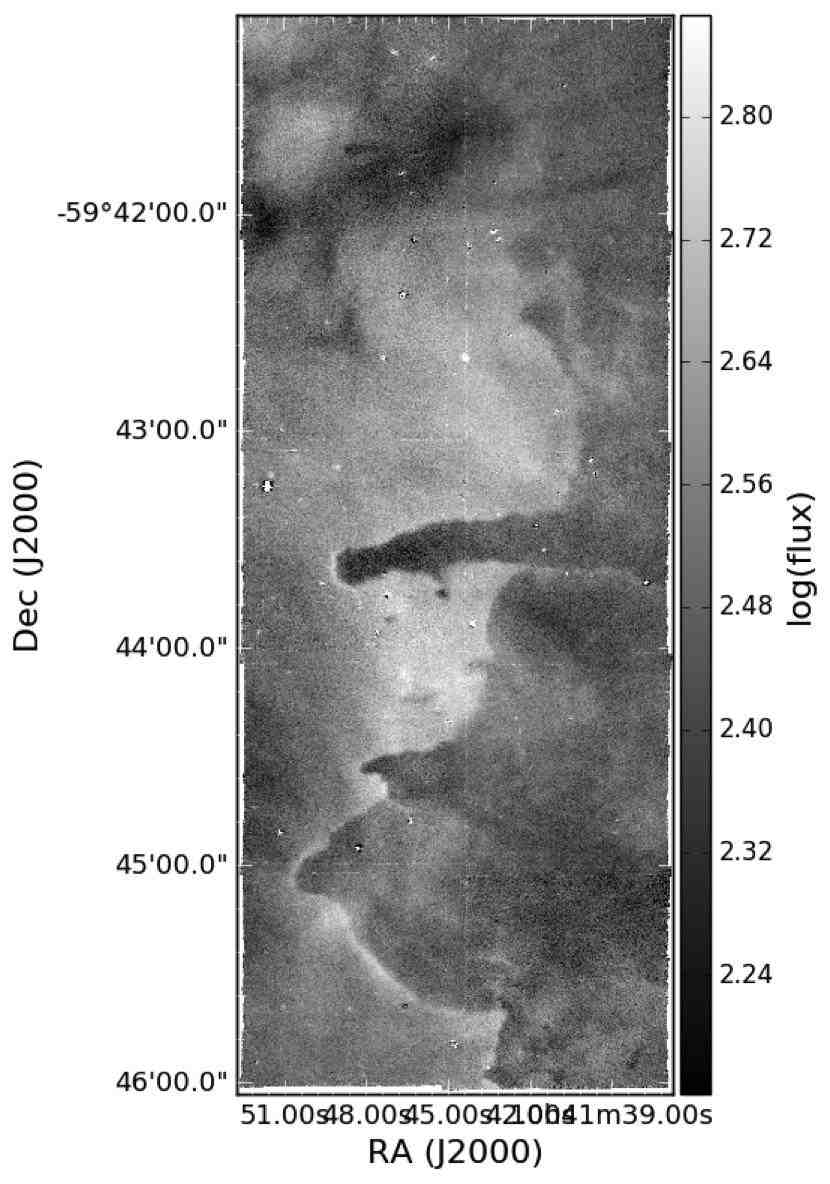}}}
\caption{Same as Fig. A1 for R44.}
\label{maps3}
\end{figure*}

\begin{figure*}
\mbox{
\subfloat[]{\includegraphics[scale=0.35]{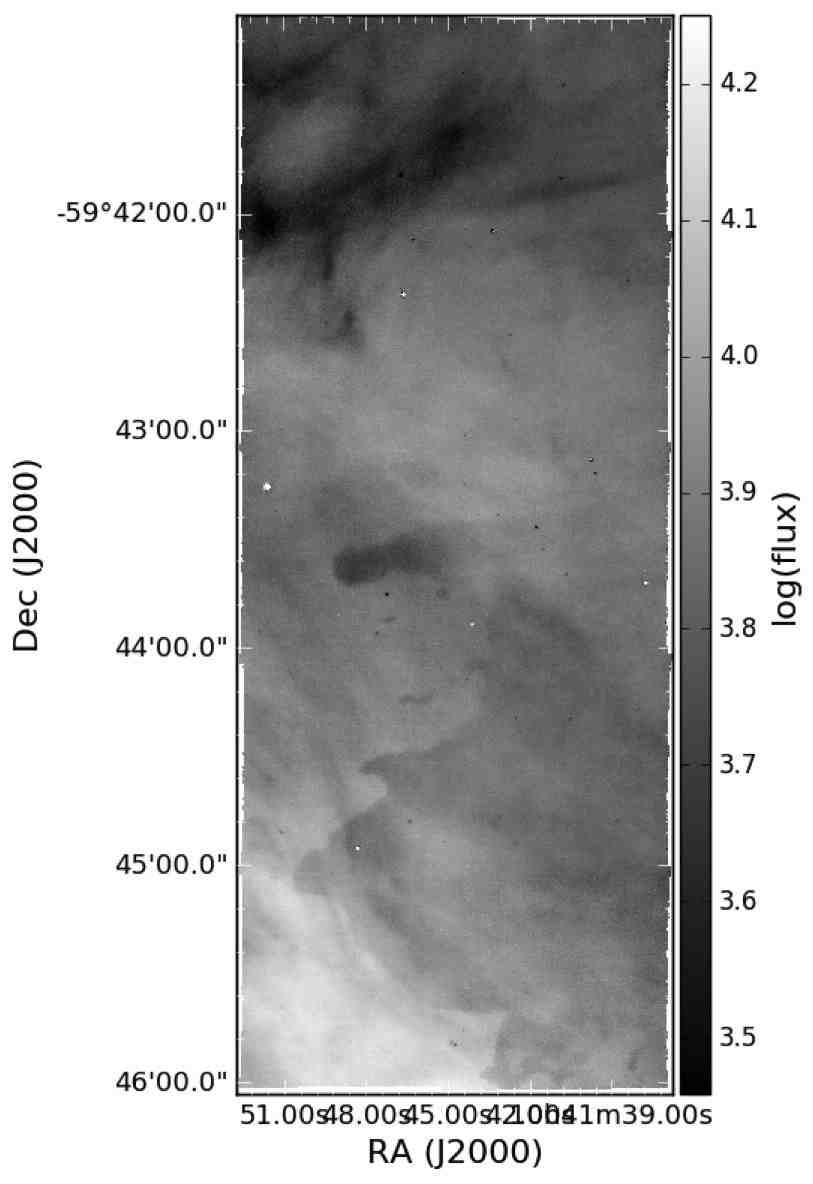}}
\subfloat[]{\includegraphics[scale=0.35]{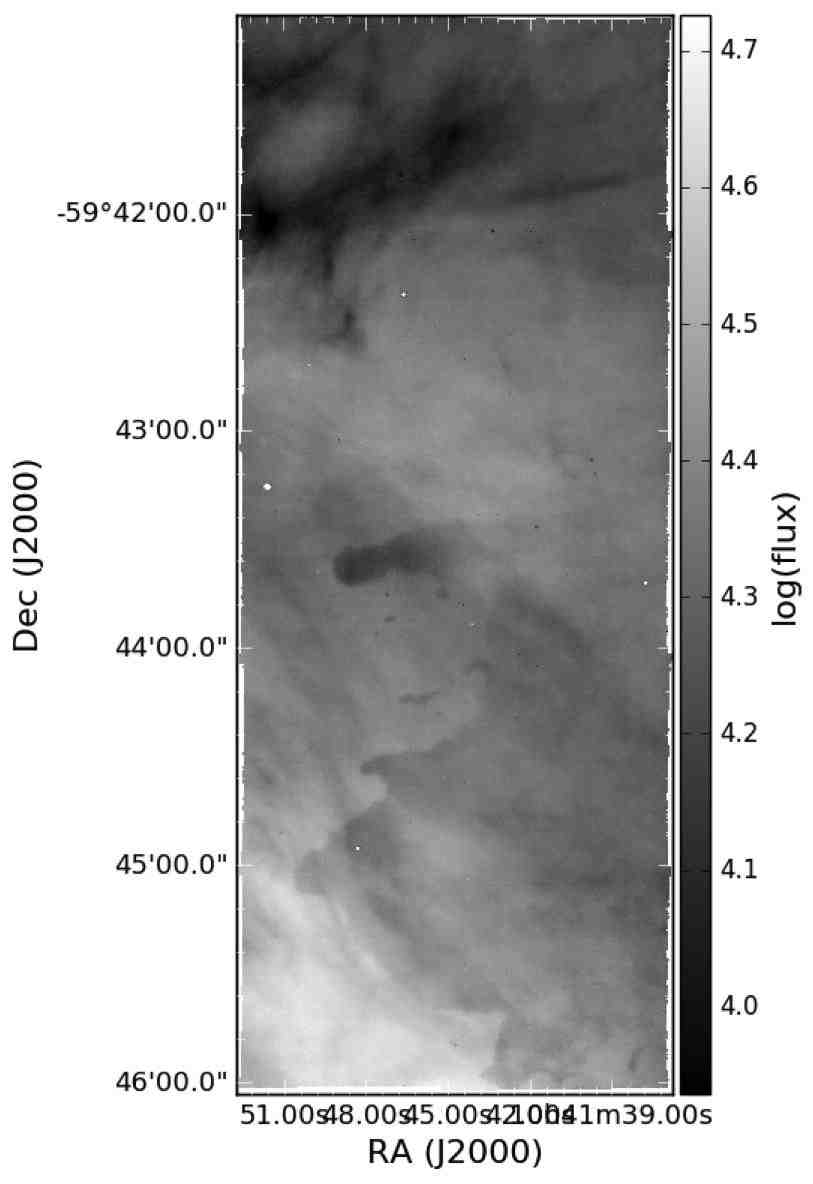}}}
\mbox{
\subfloat[]{\includegraphics[scale=0.35]{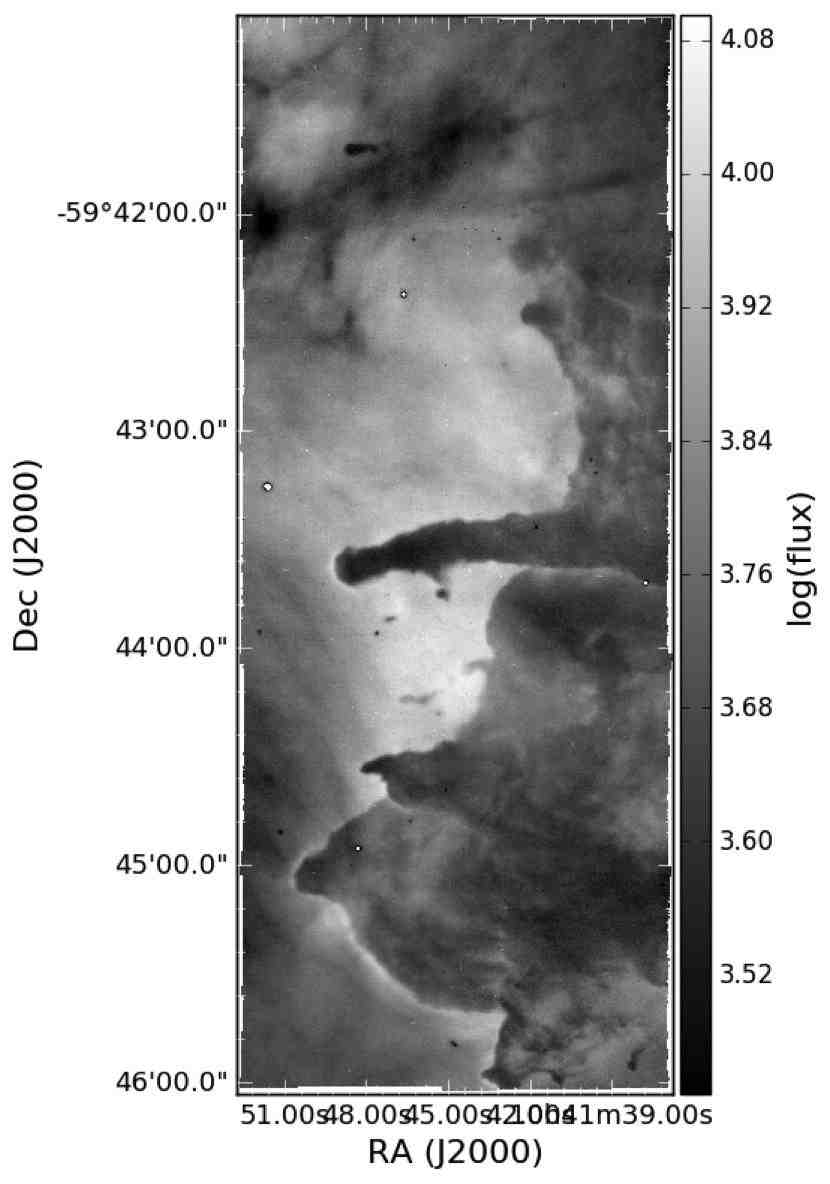}}
\subfloat[]{\includegraphics[scale=0.35]{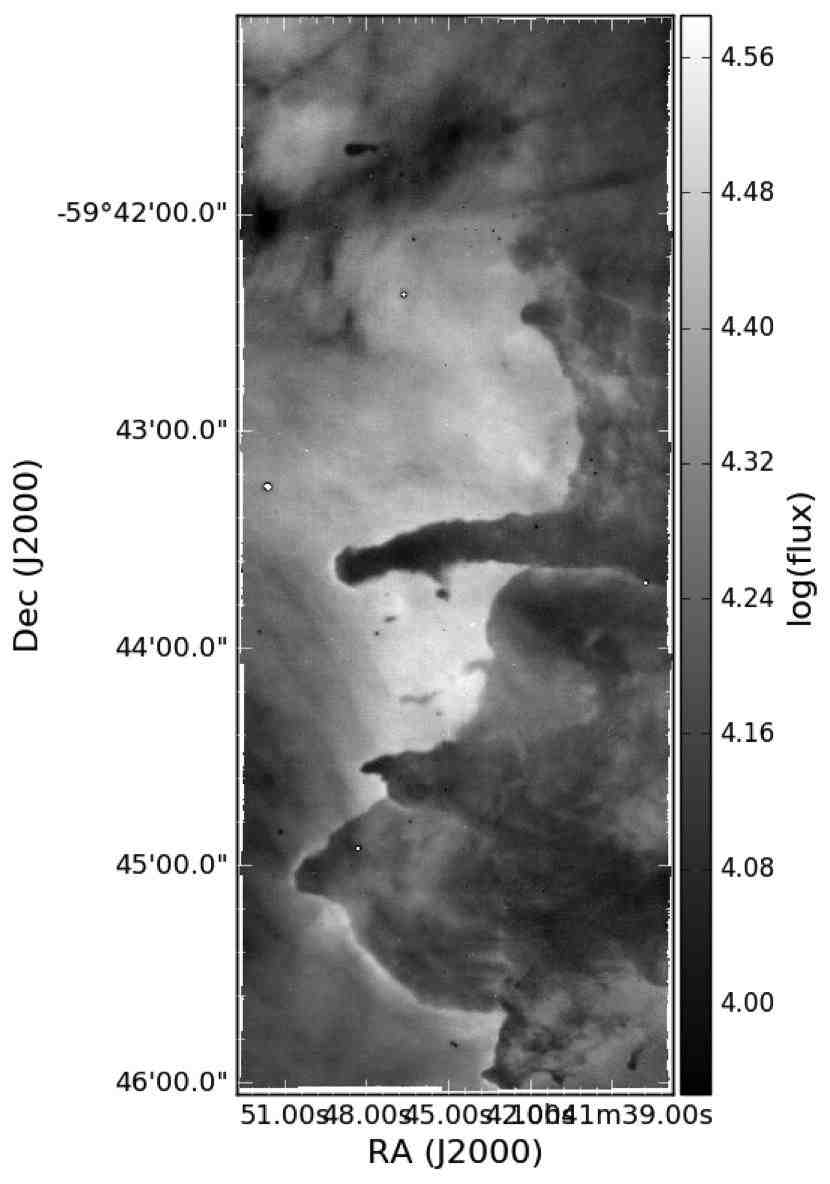}}}
\mbox{
\subfloat[]{\includegraphics[scale=0.35]{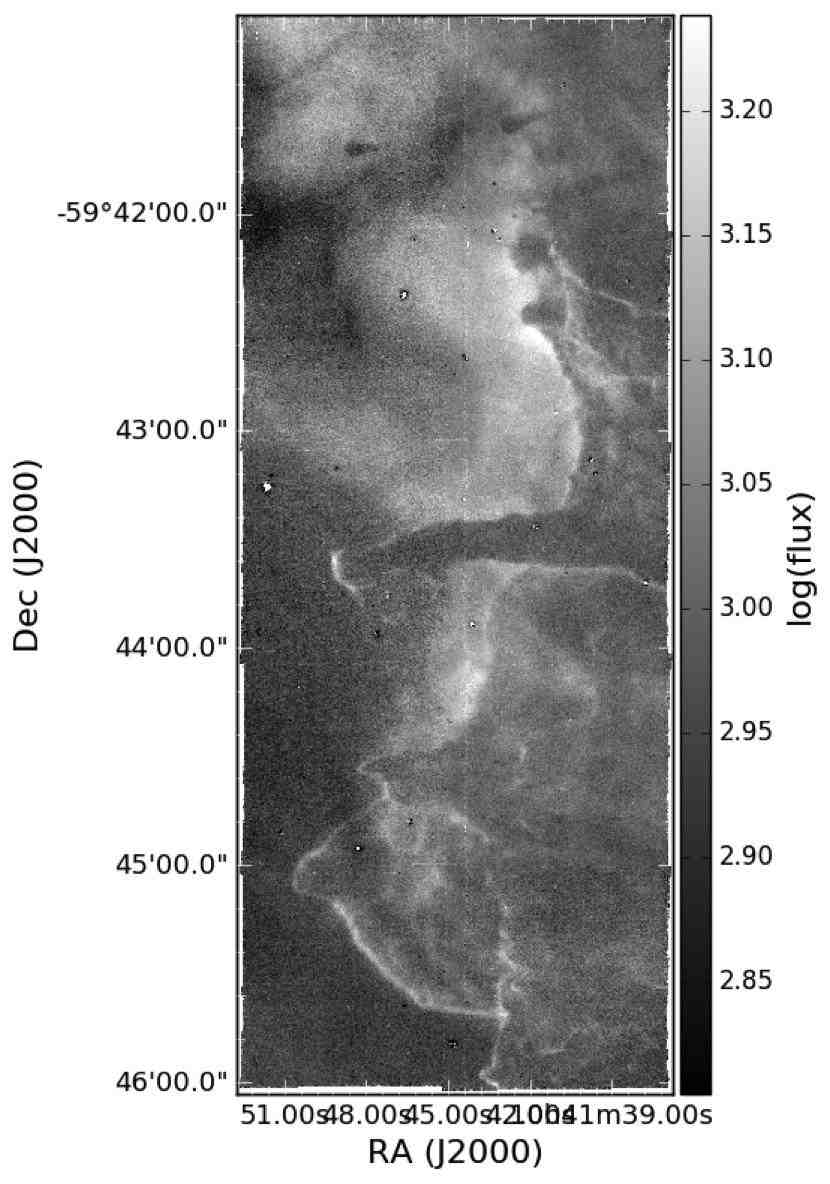}}
\subfloat[]{\includegraphics[scale=0.35]{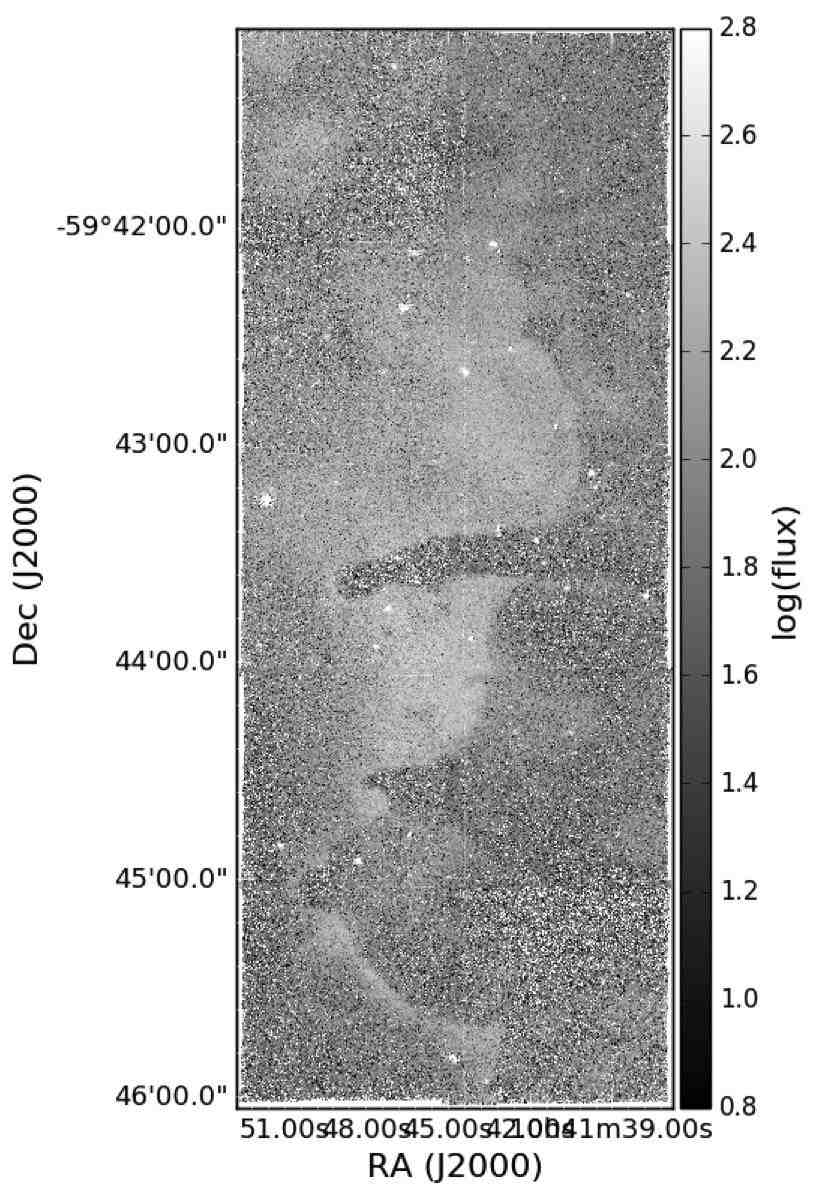}}}
\caption{Same as Fig. A2 for R44.}
\label{maps4}
\end{figure*}

\begin{figure*}
\mbox{
\subfloat[]{\includegraphics[scale=0.35]{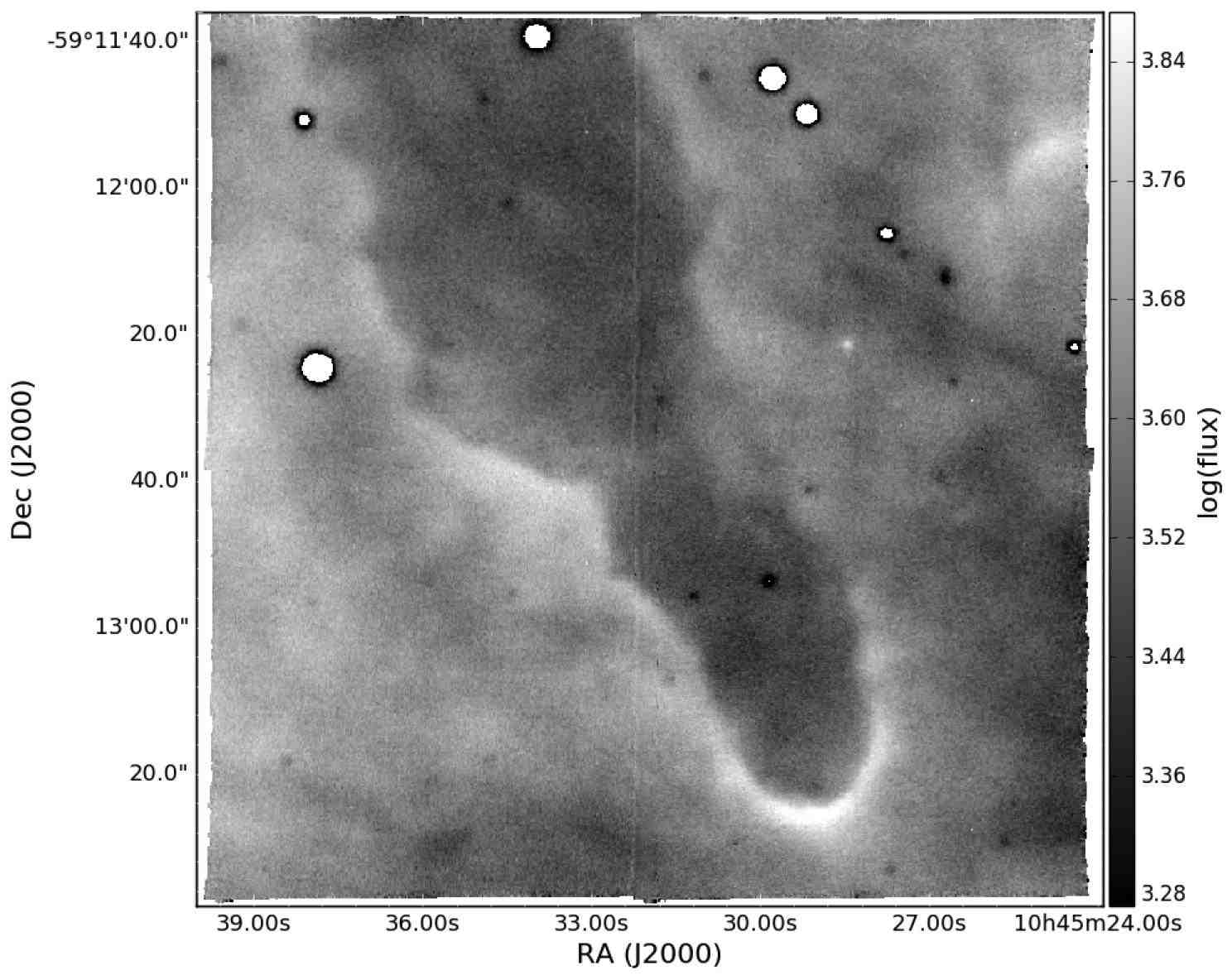}}
\subfloat[]{\includegraphics[scale=0.35]{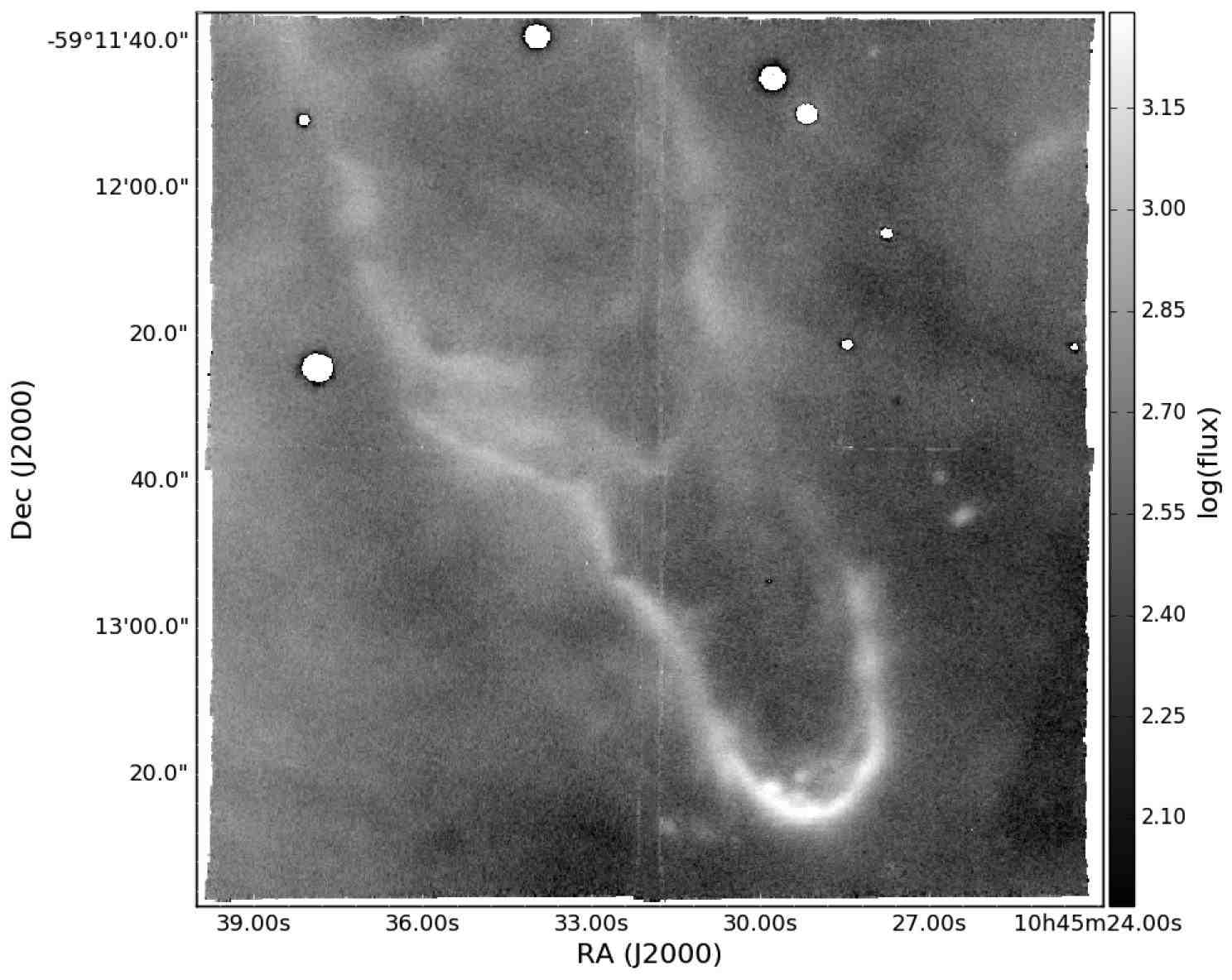}}}
\mbox{
\subfloat[]{\includegraphics[scale=0.35]{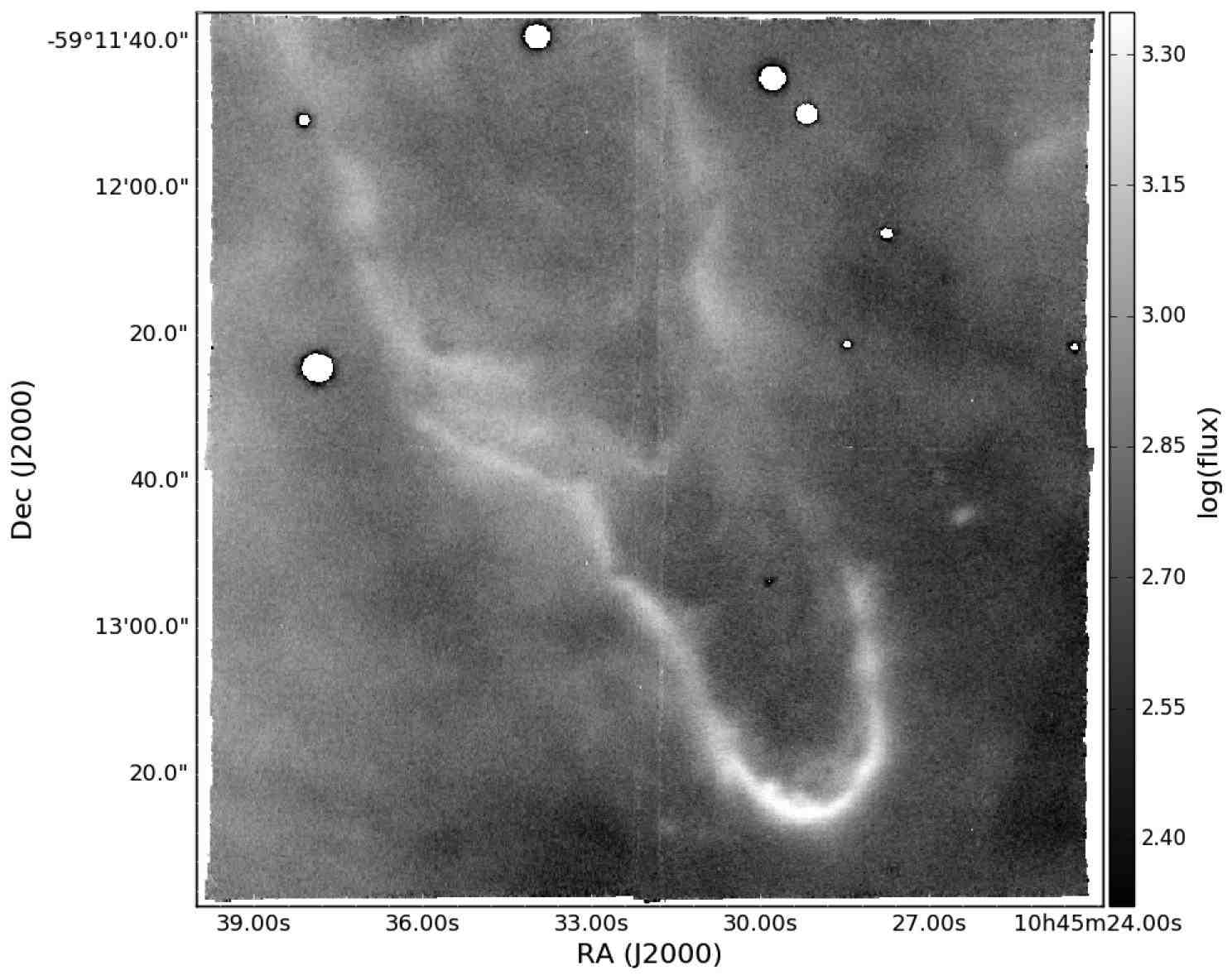}}
\subfloat[]{\includegraphics[scale=0.35]{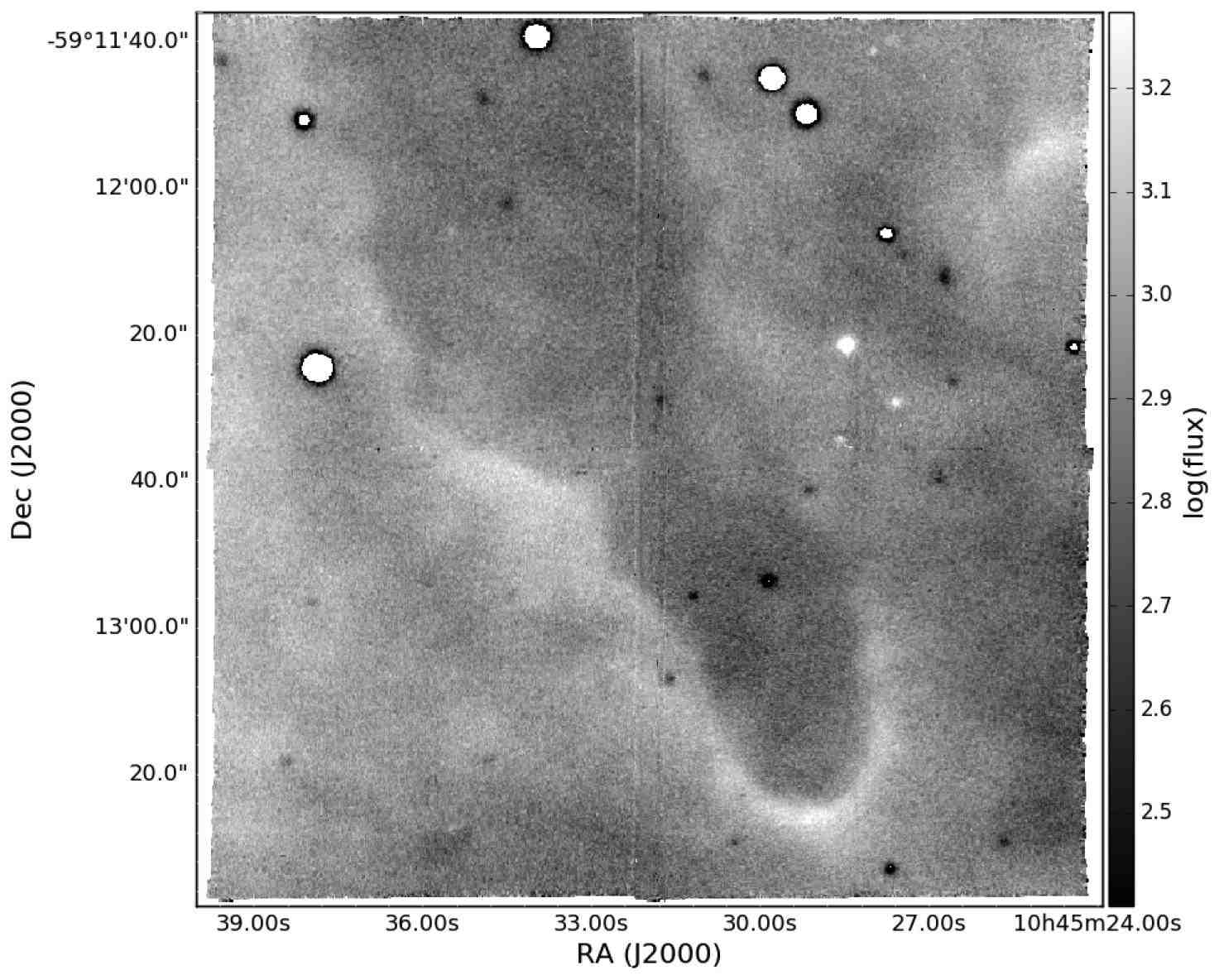}}}
\mbox{
\subfloat[]{\includegraphics[scale=0.35]{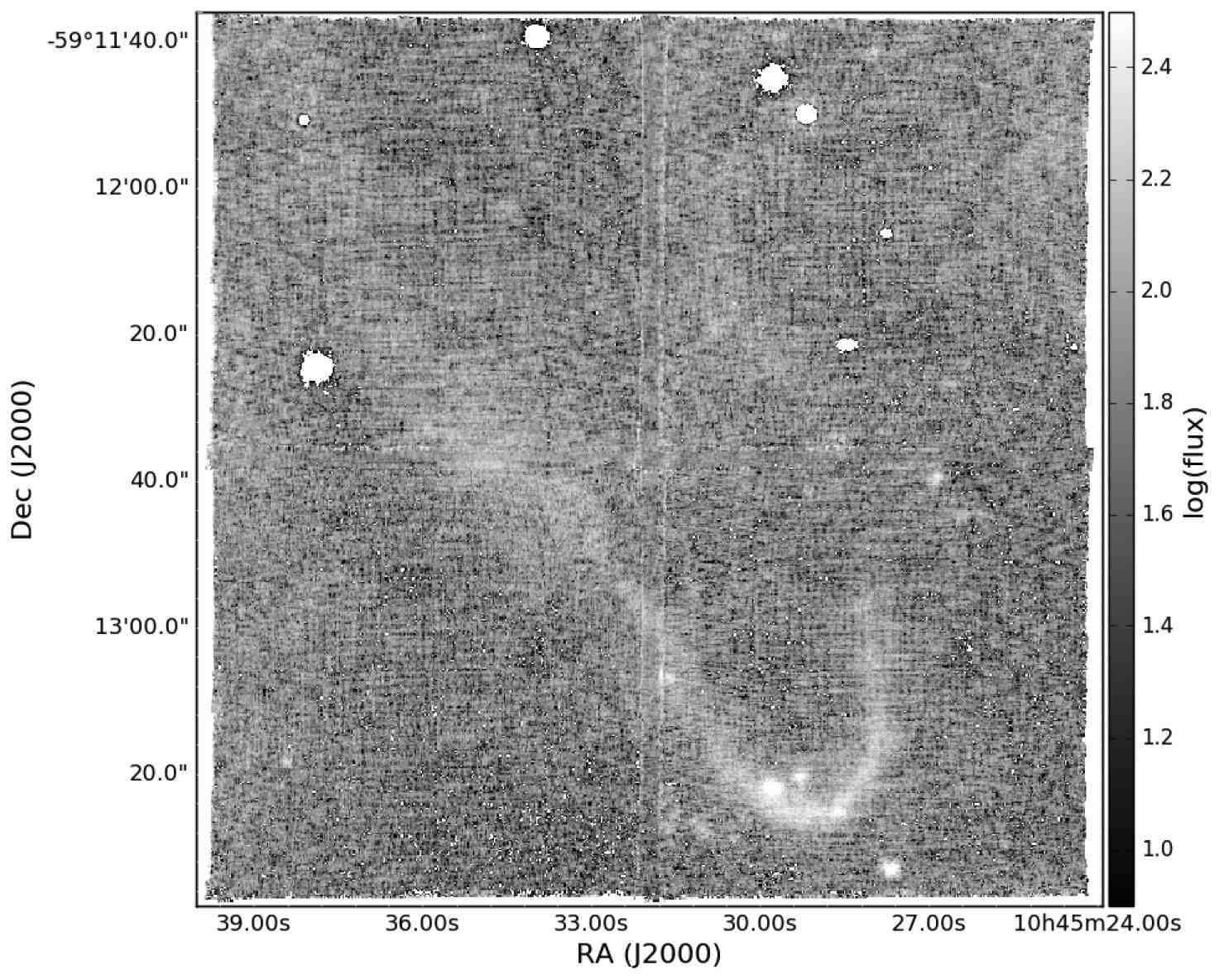}}
\subfloat[]{\includegraphics[scale=0.35]{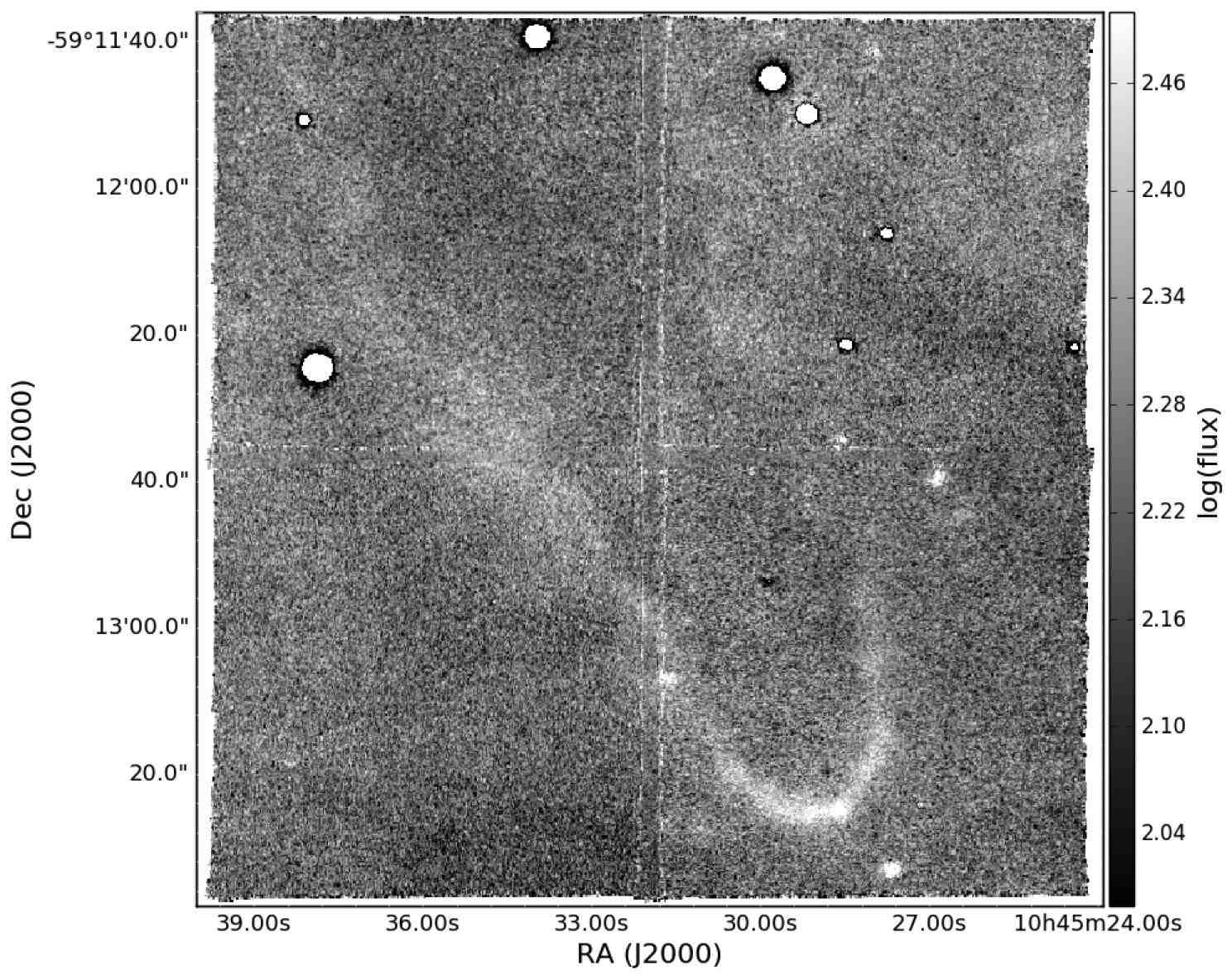}}}
\caption{Same as Fig. A1 for R18.}
\label{maps5}
\end{figure*}

\begin{figure*}
\mbox{
\subfloat[]{\includegraphics[scale=0.35]{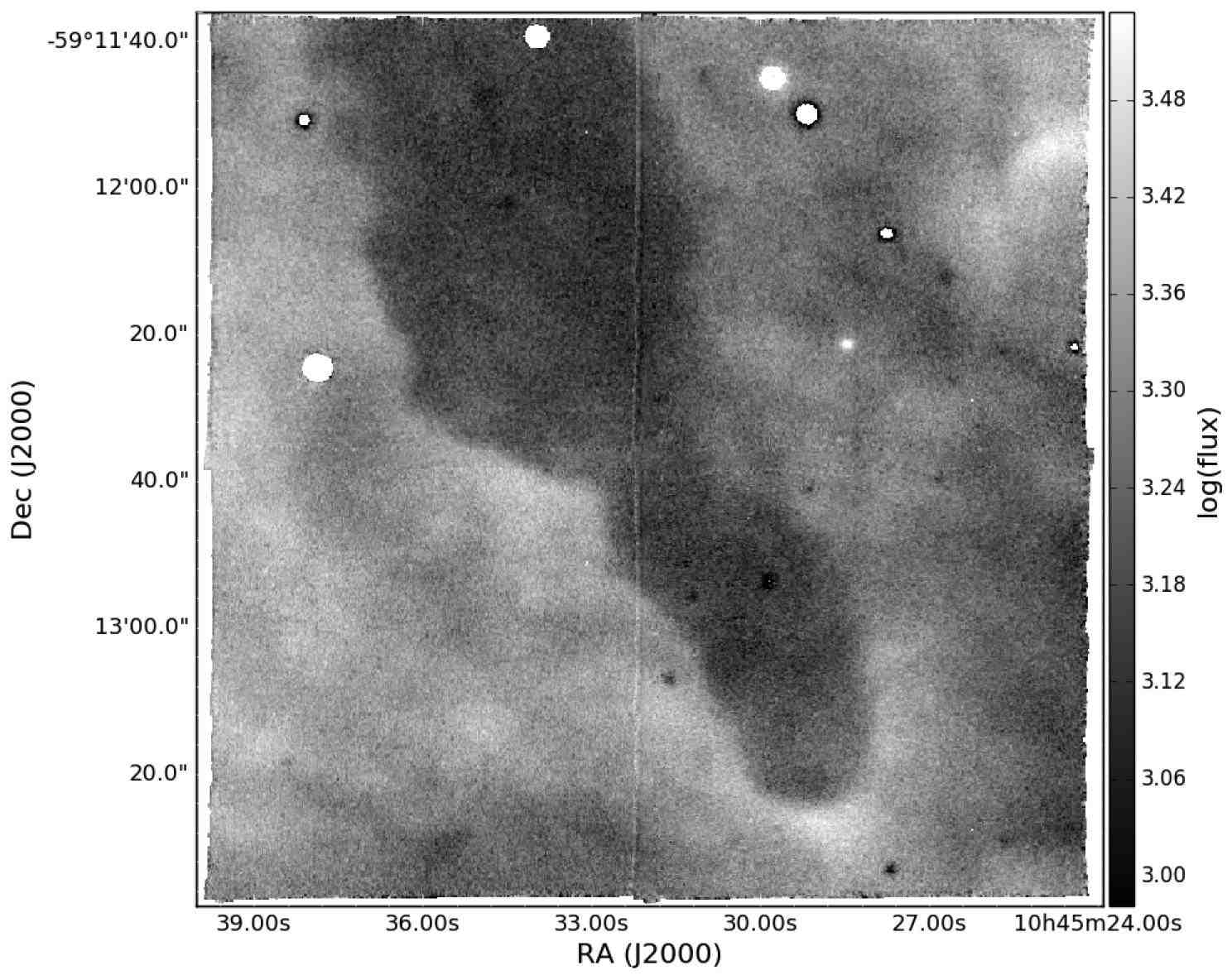}}
\subfloat[]{\includegraphics[scale=0.35]{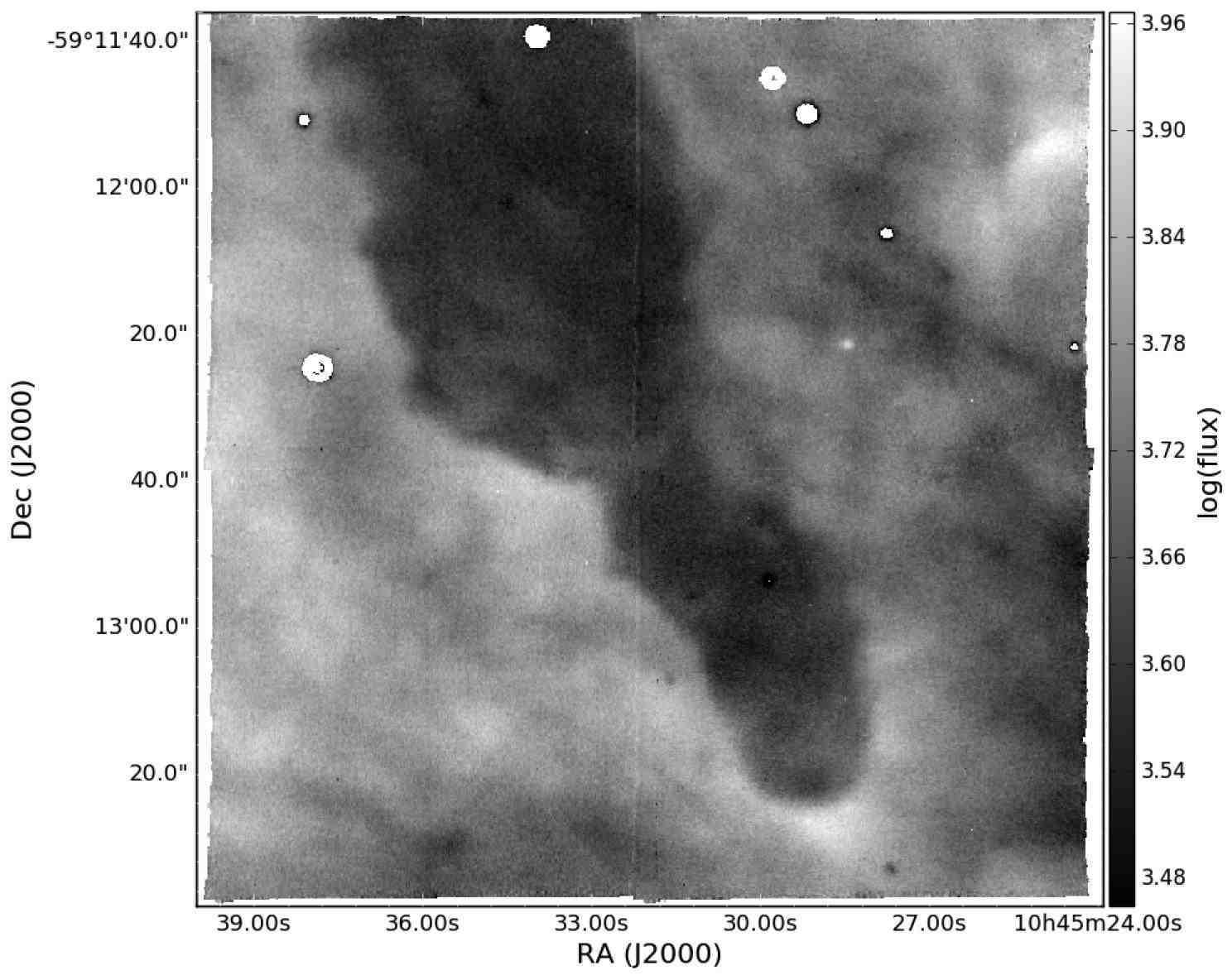}}}
\mbox{
\subfloat[]{\includegraphics[scale=0.35]{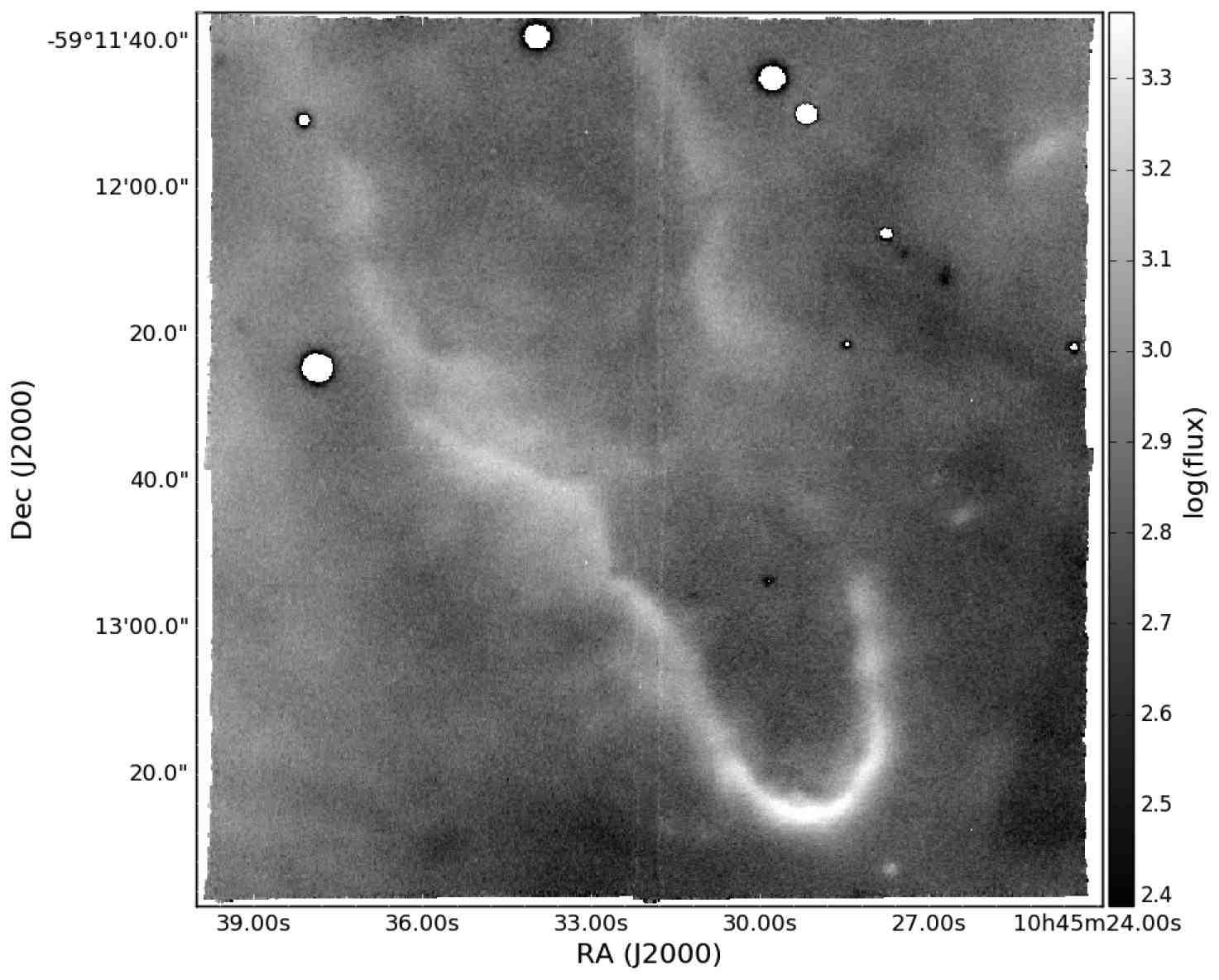}}
\subfloat[]{\includegraphics[scale=0.35]{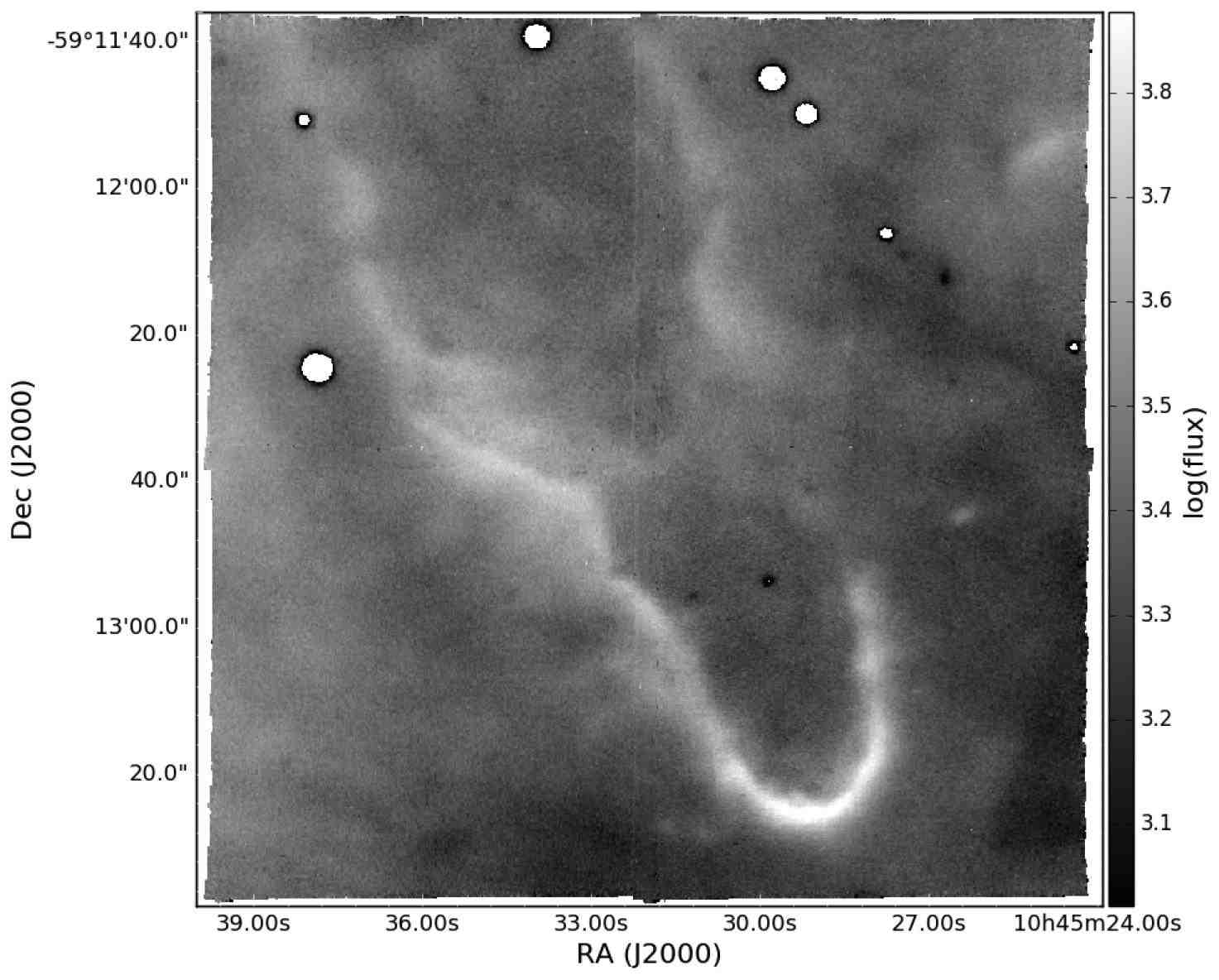}}}
\mbox{
\subfloat[]{\includegraphics[scale=0.35]{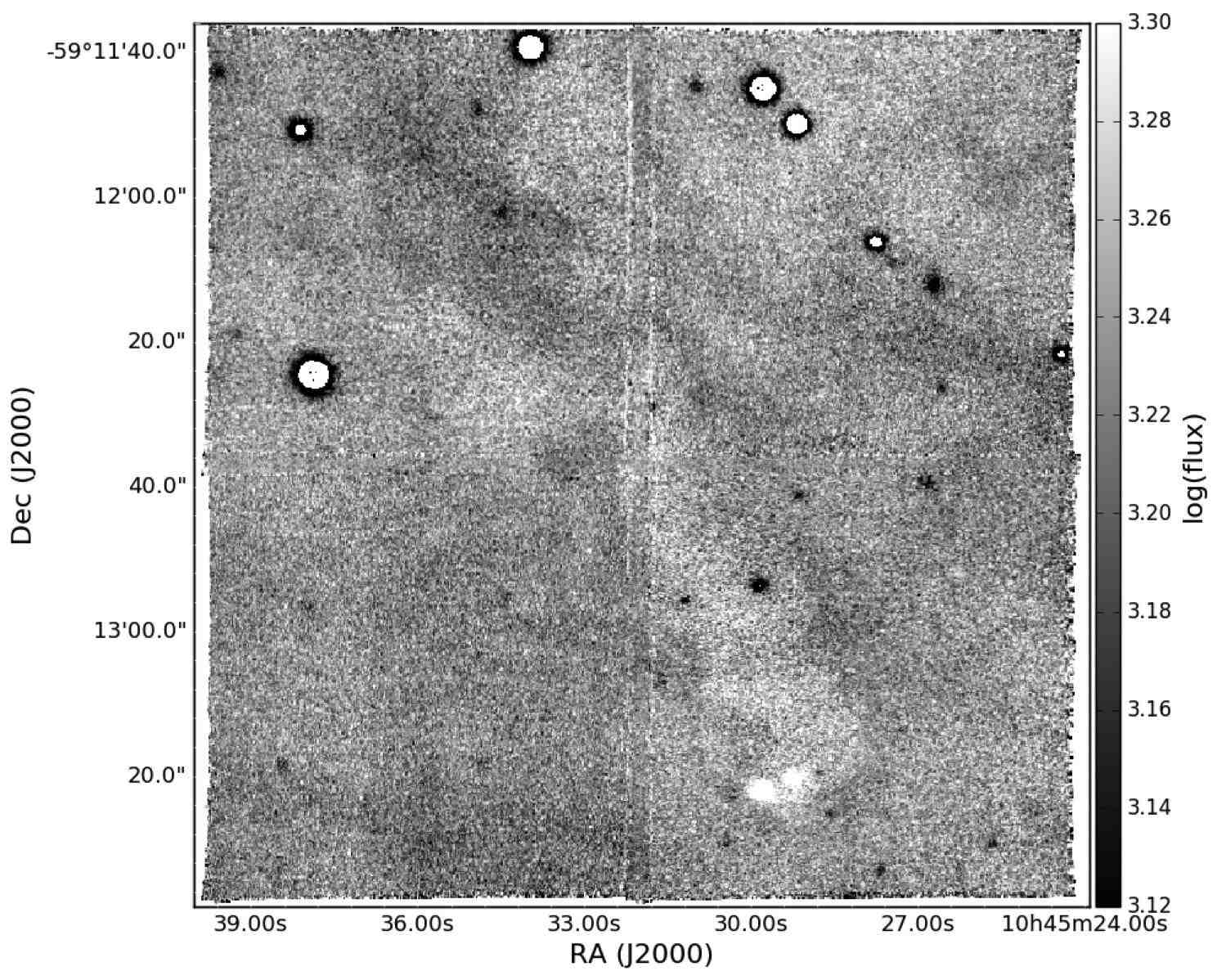}}
\subfloat[]{\includegraphics[scale=0.35]{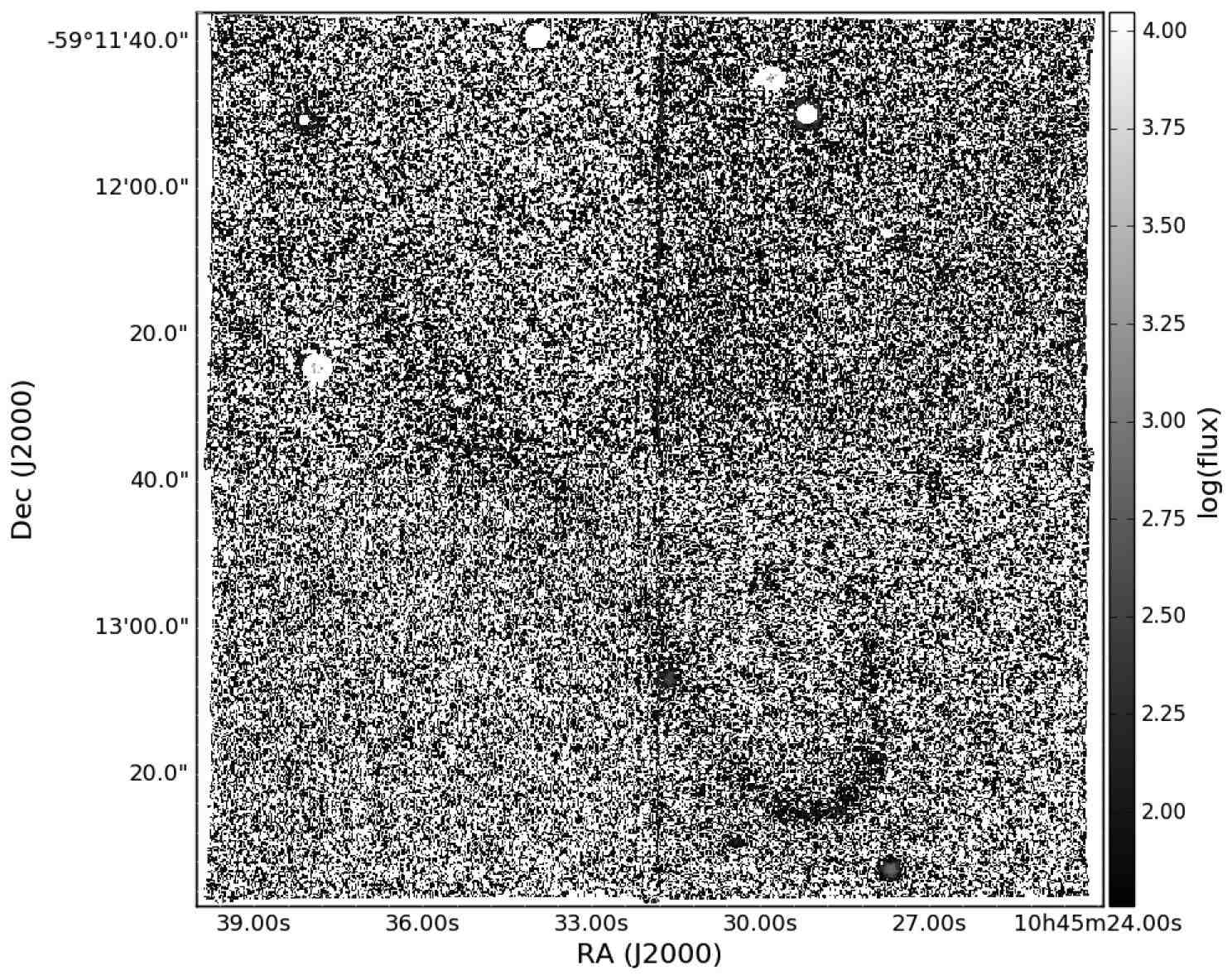}}}
\caption{Same as Fig. A2 for R18.}
\label{maps6}
\end{figure*}

\begin{figure*}
\mbox{
\subfloat[]{\includegraphics[scale=0.35]{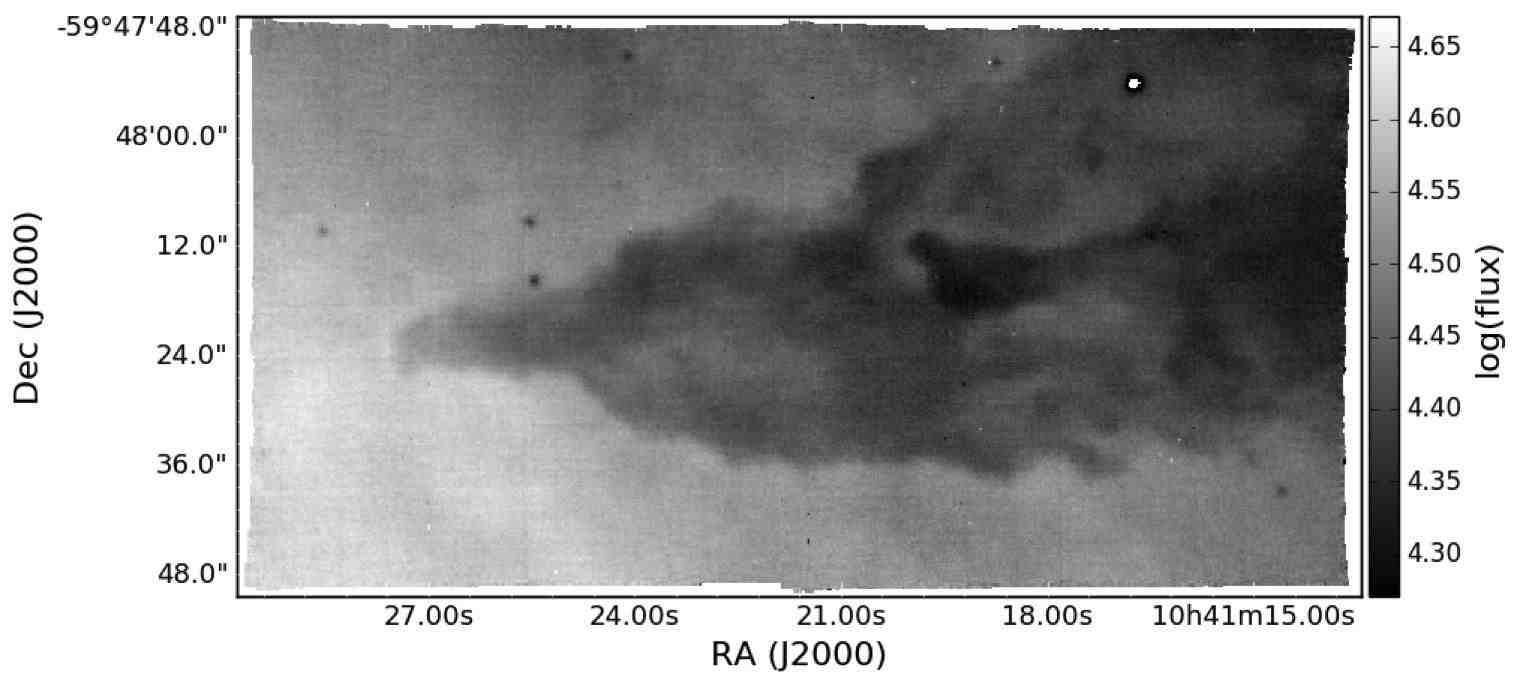}}
\subfloat[]{\includegraphics[scale=0.35]{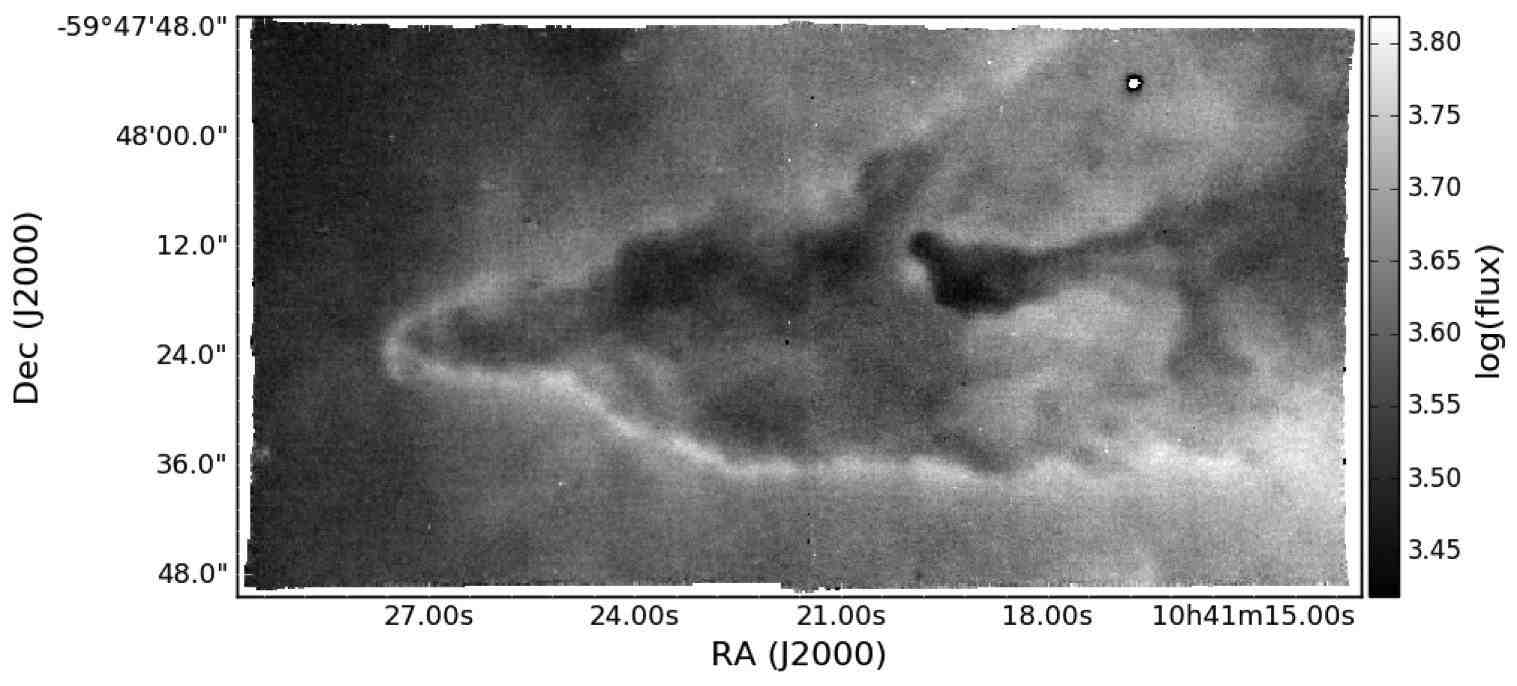}}}
\mbox{
\subfloat[]{\includegraphics[scale=0.35]{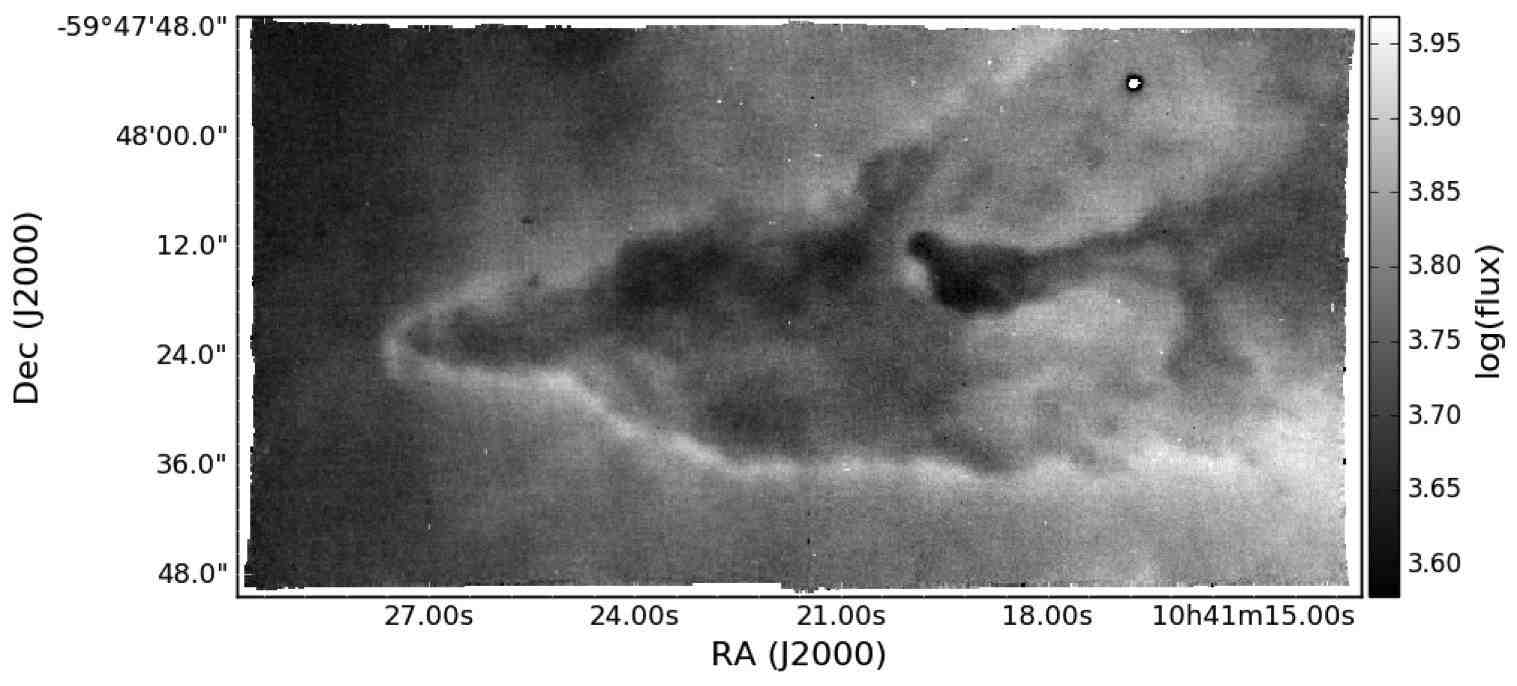}}
\subfloat[]{\includegraphics[scale=0.35]{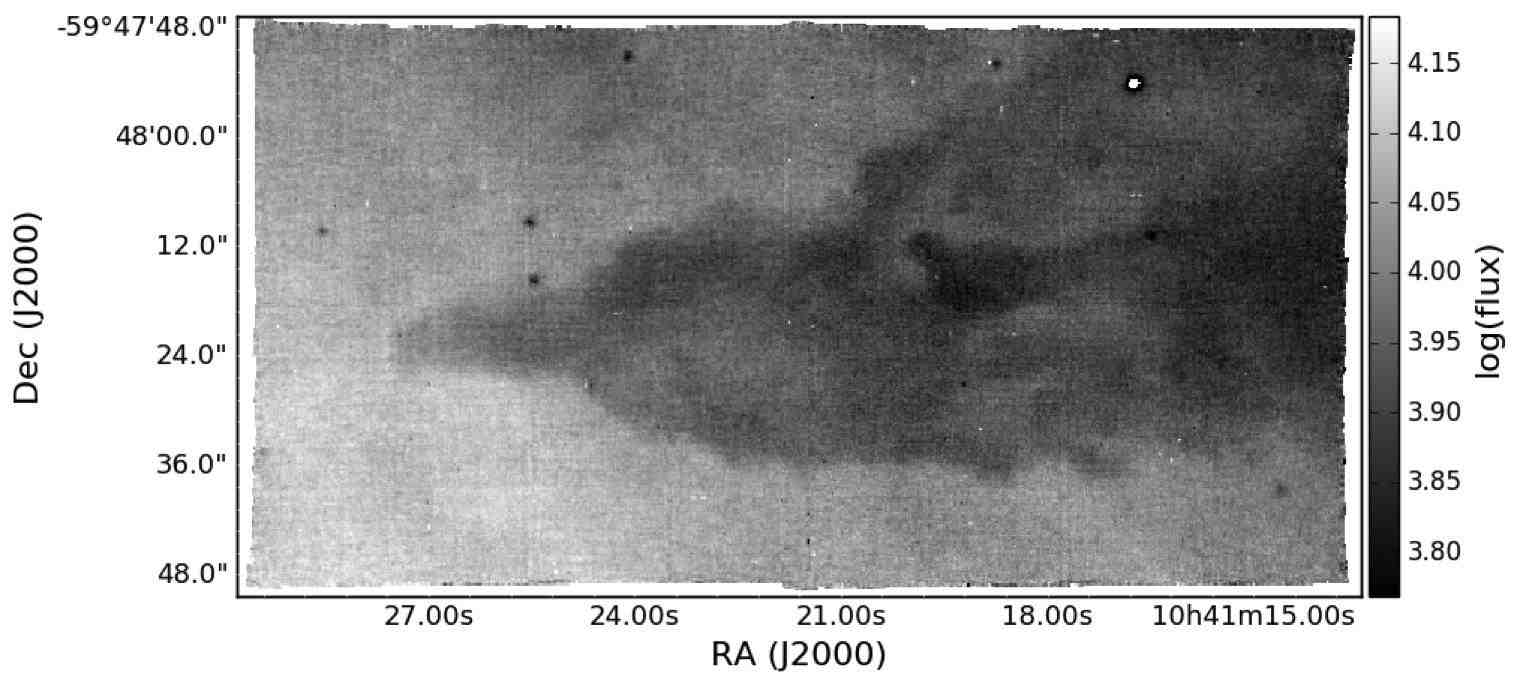}}}
\mbox{
\subfloat[]{\includegraphics[scale=0.35]{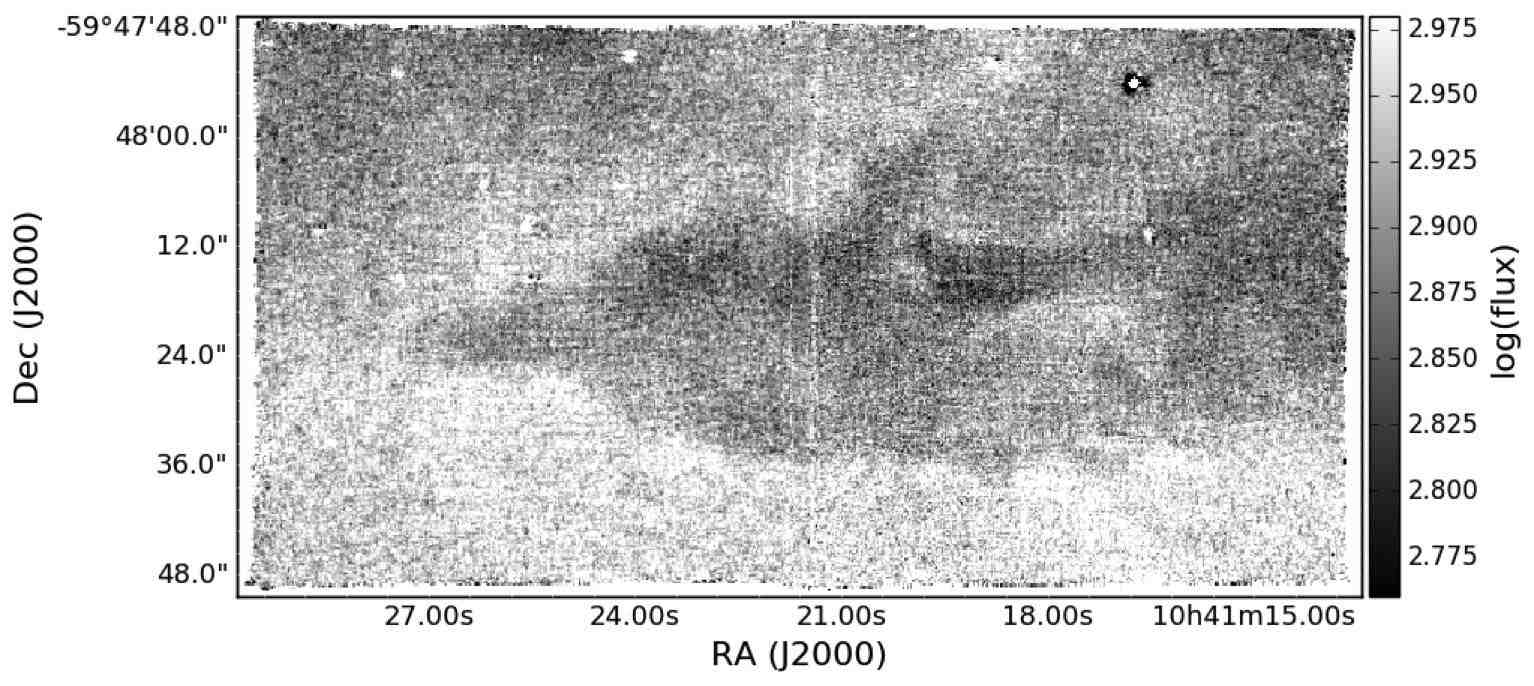}}
\subfloat[]{\includegraphics[scale=0.35]{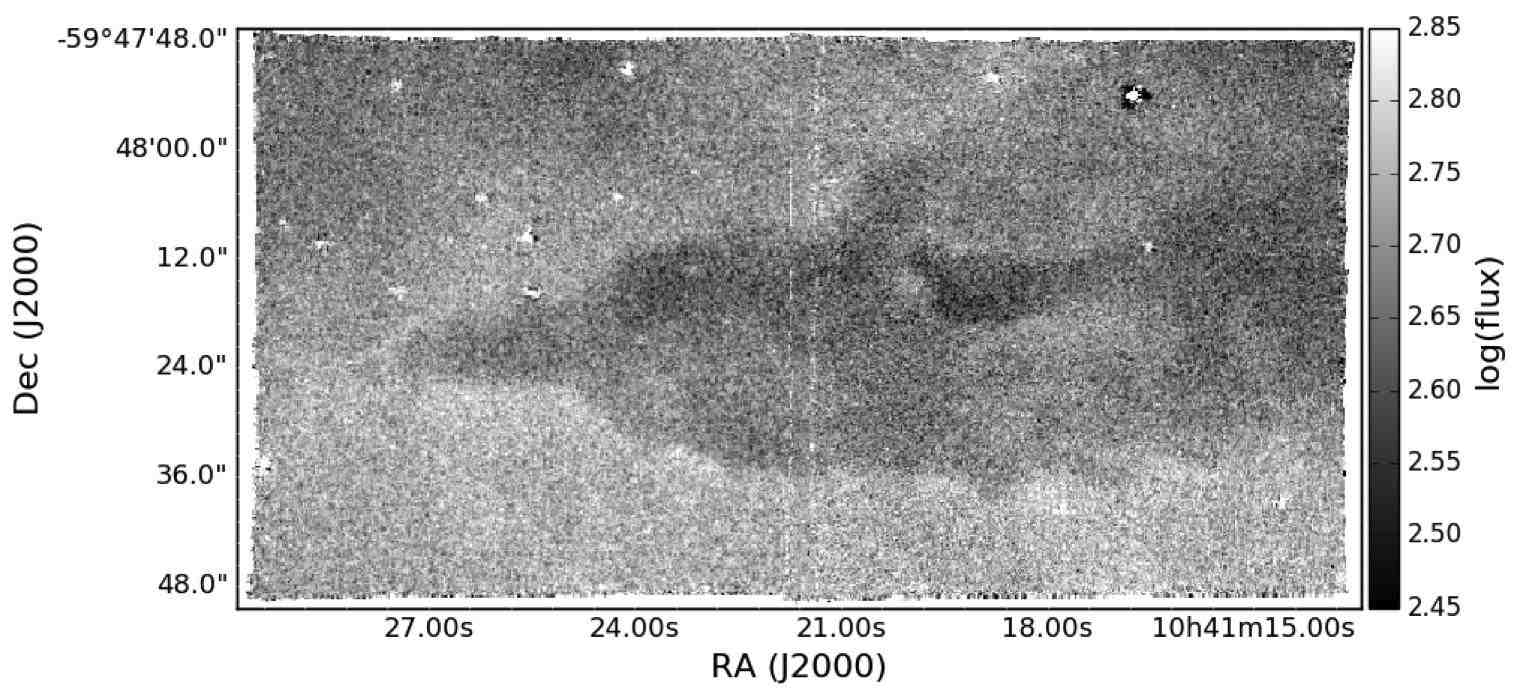}}}
\caption{Same as Fig. A1 for R45.}
\label{maps7}
\end{figure*}

\begin{figure*}
\mbox{
\subfloat[]{\includegraphics[scale=0.35]{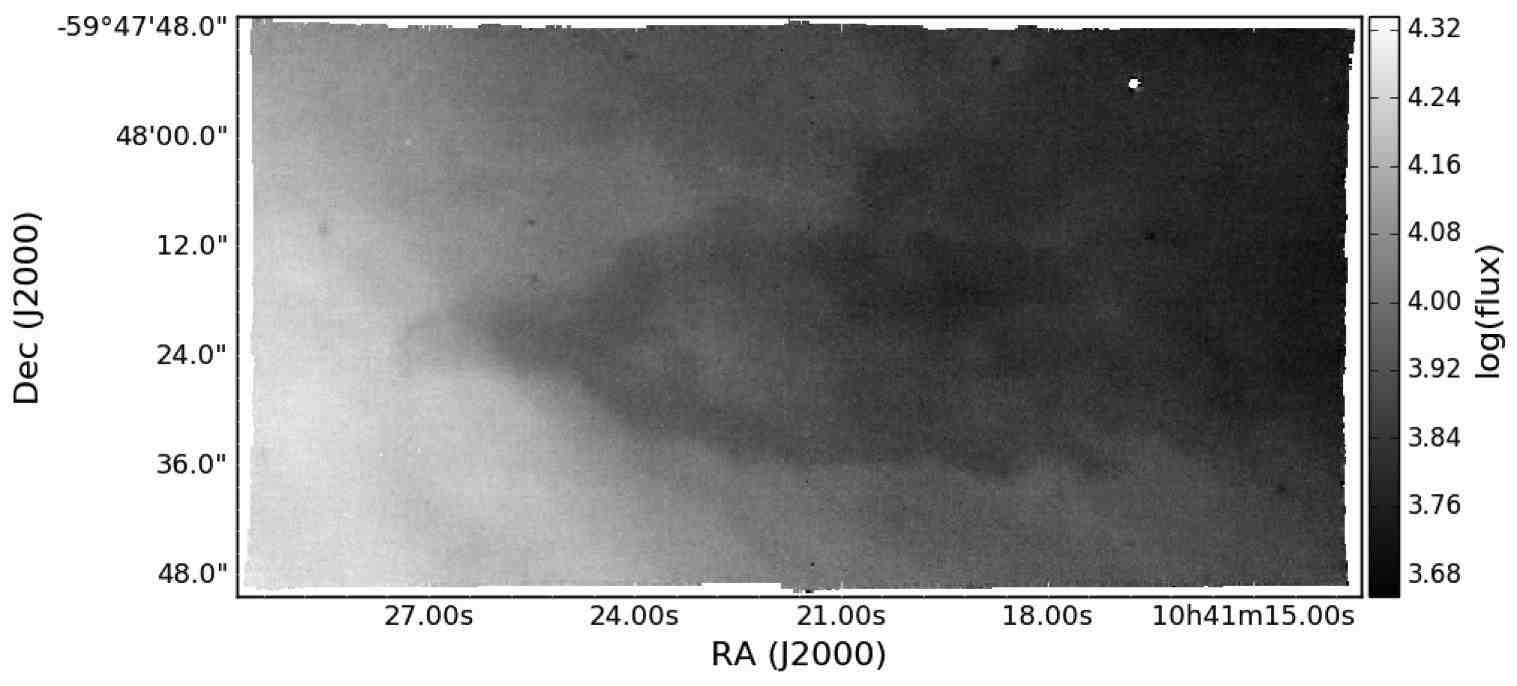}}
\subfloat[]{\includegraphics[scale=0.35]{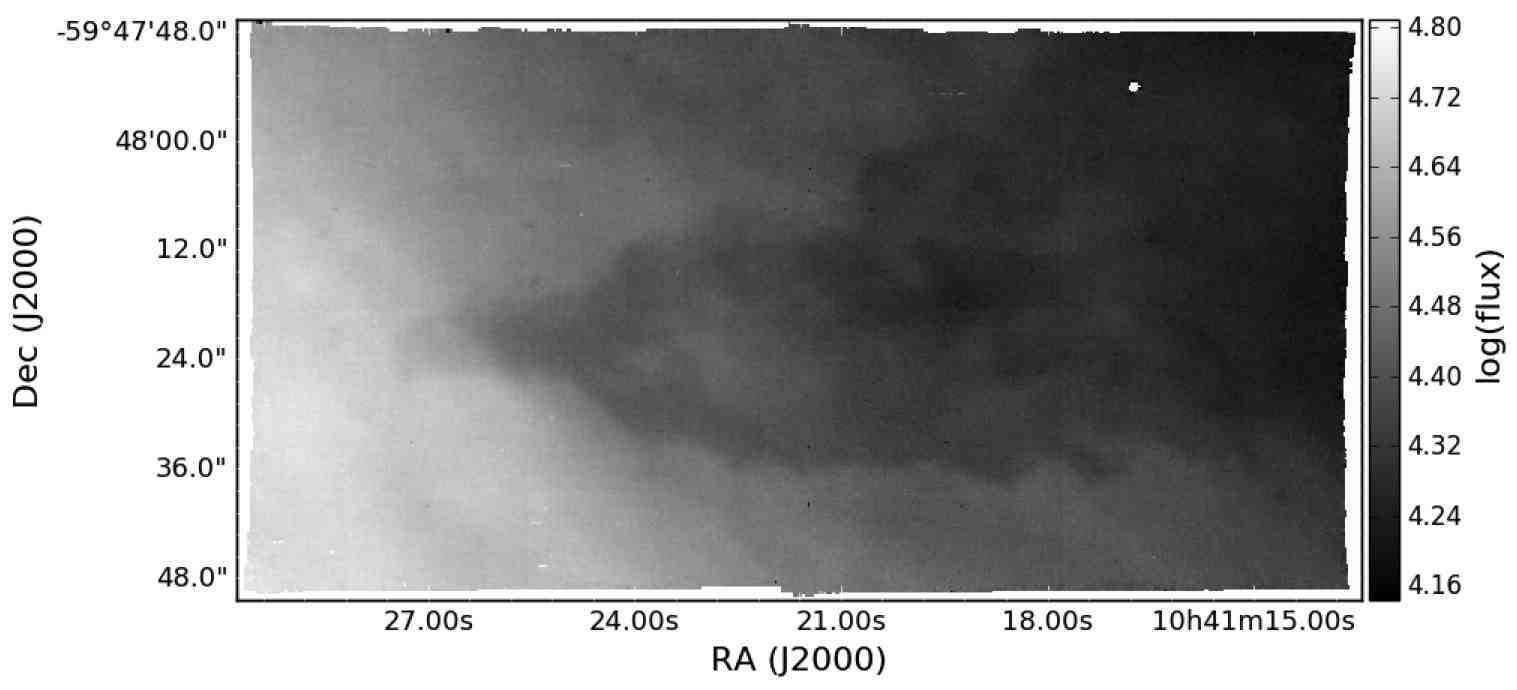}}}
\mbox{
\subfloat[]{\includegraphics[scale=0.35]{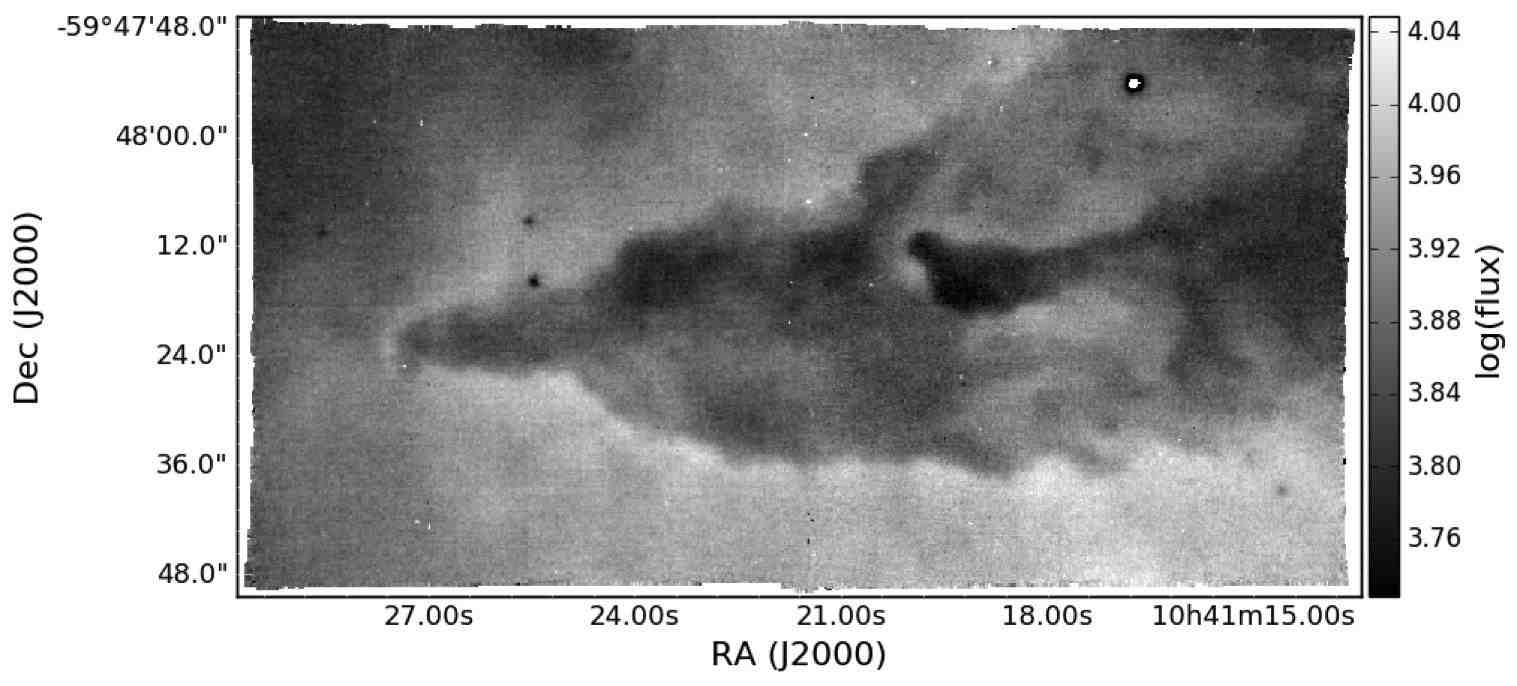}}
\subfloat[]{\includegraphics[scale=0.35]{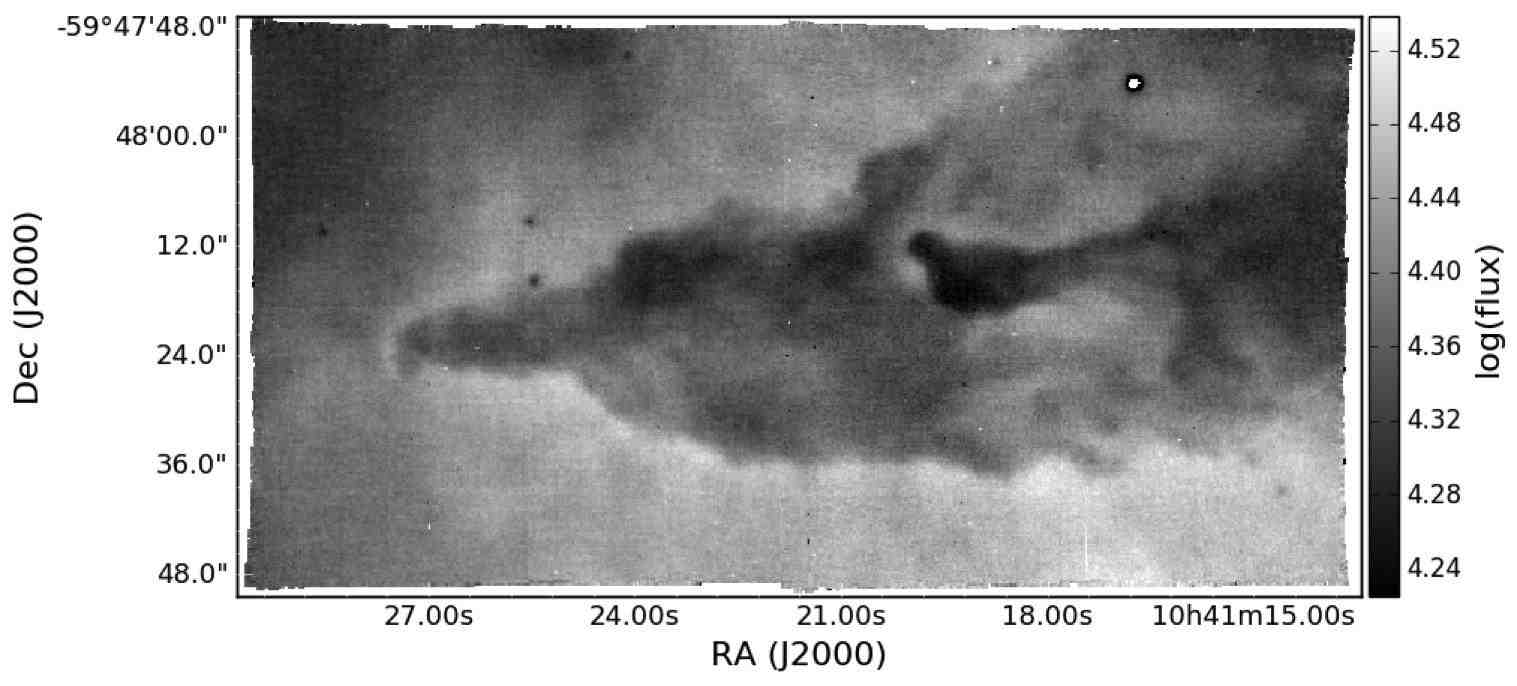}}}
\mbox{
\subfloat[]{\includegraphics[scale=0.35]{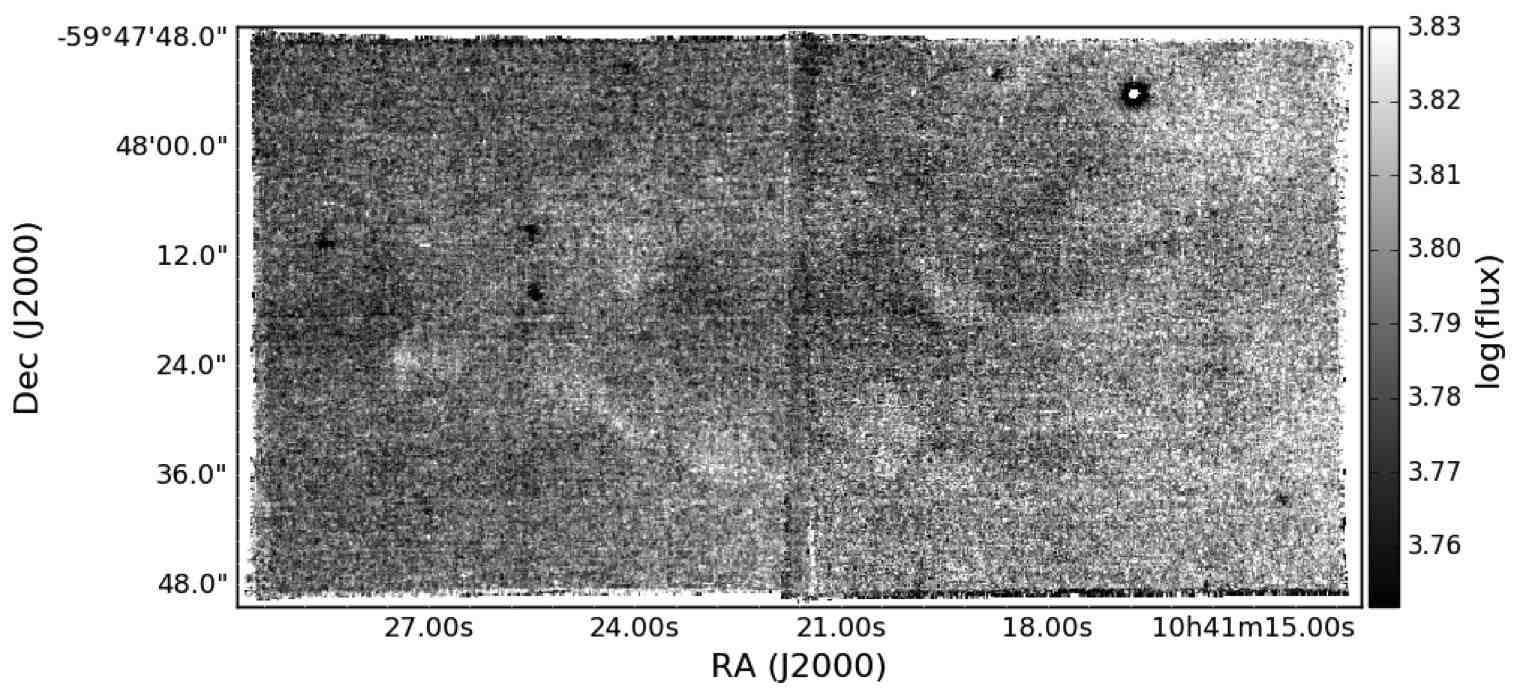}}
\subfloat[]{\includegraphics[scale=0.35]{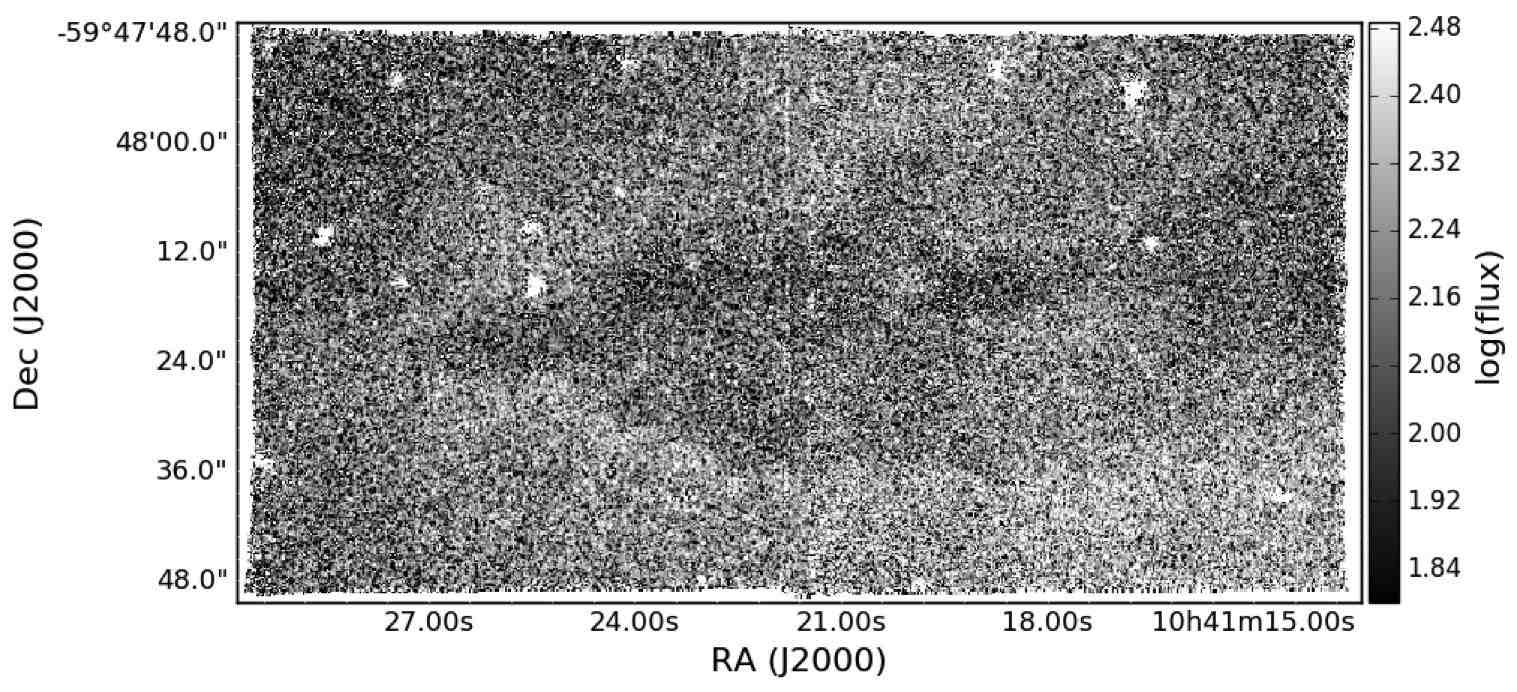}}}
\caption{Same as Fig. A2 for R45.}
\label{maps8}
\end{figure*}

\begin{figure*}
\mbox{
\subfloat[]{\includegraphics[scale=0.35]{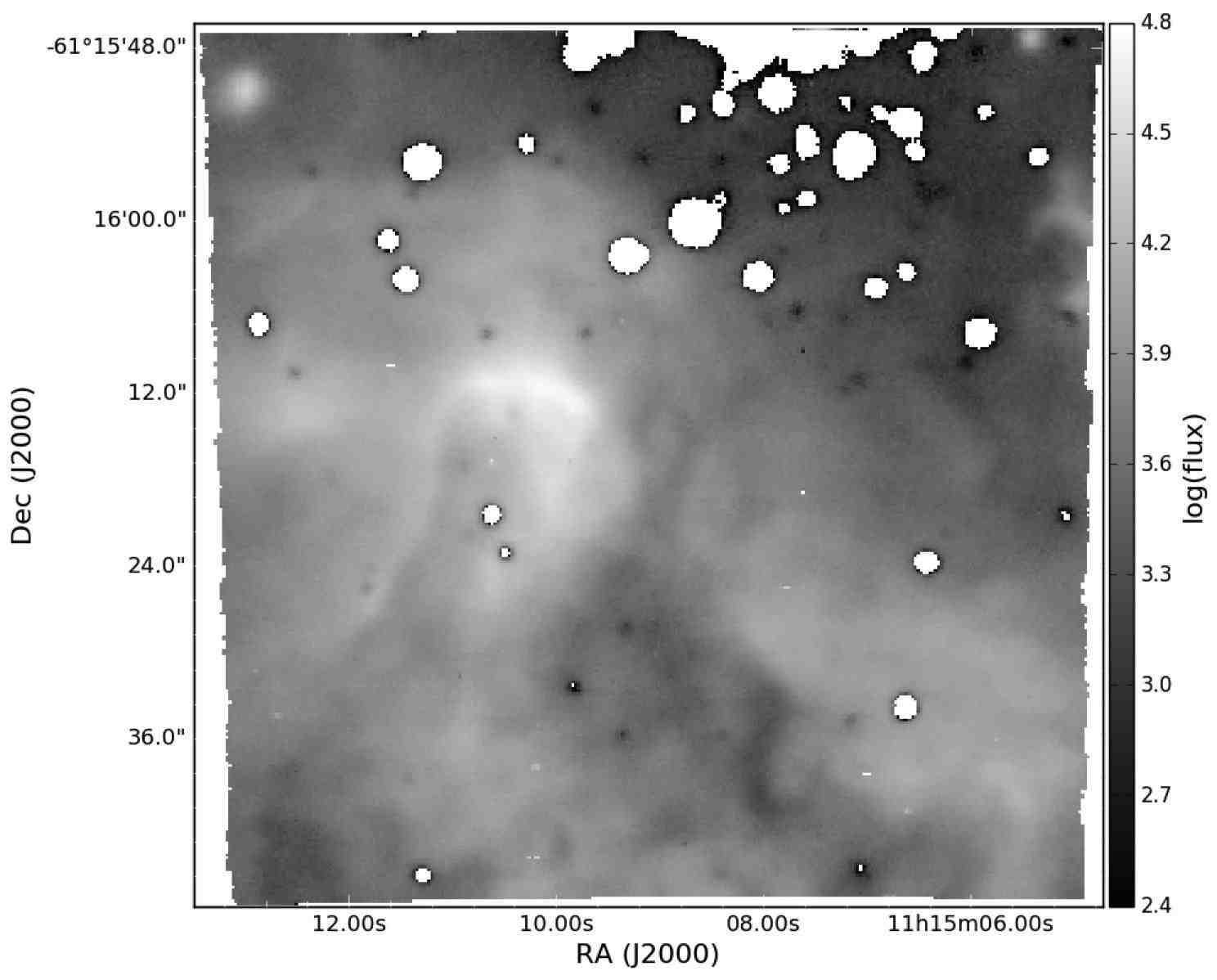}}
\subfloat[]{\includegraphics[scale=0.35]{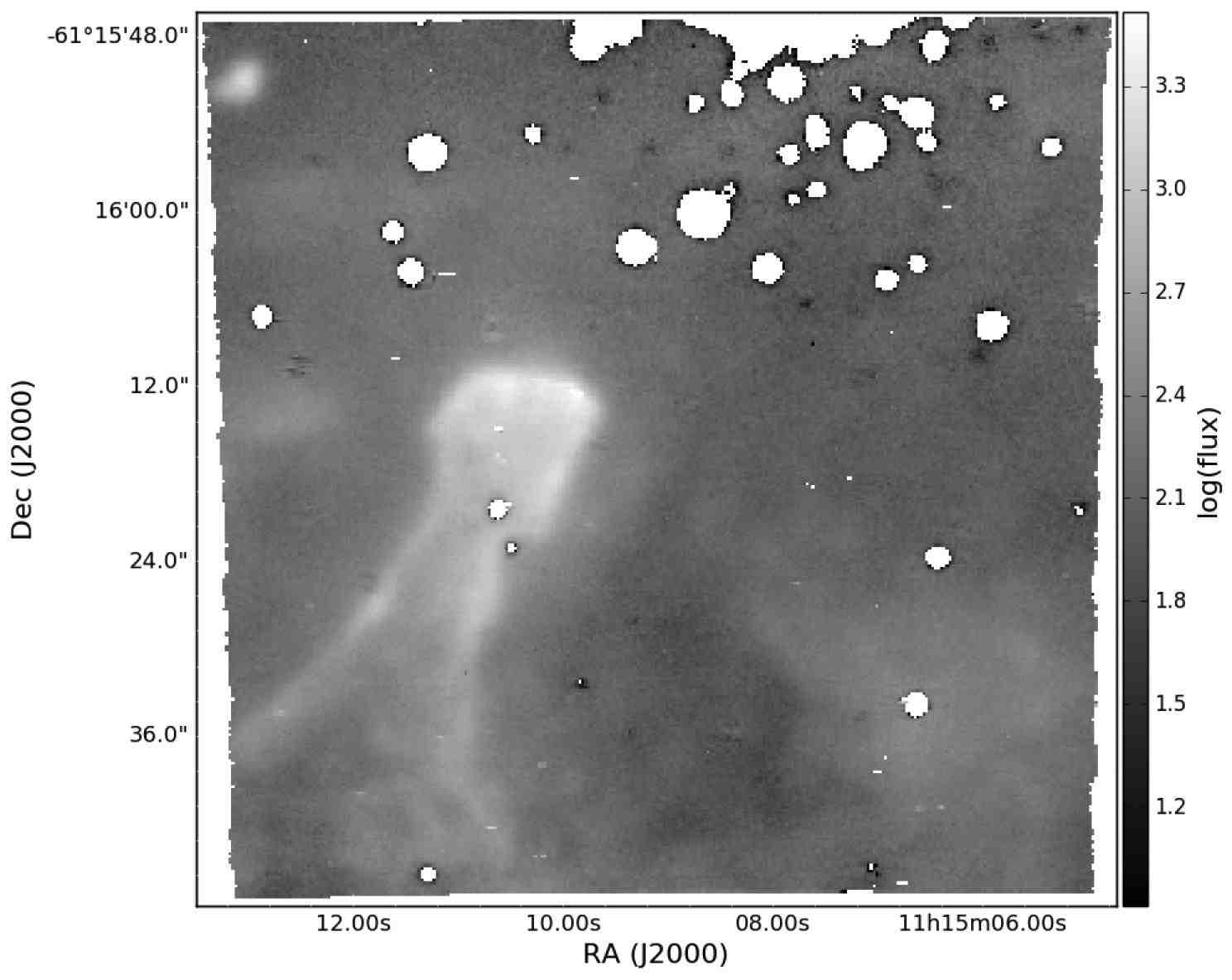}}}
\mbox{
\subfloat[]{\includegraphics[scale=0.35]{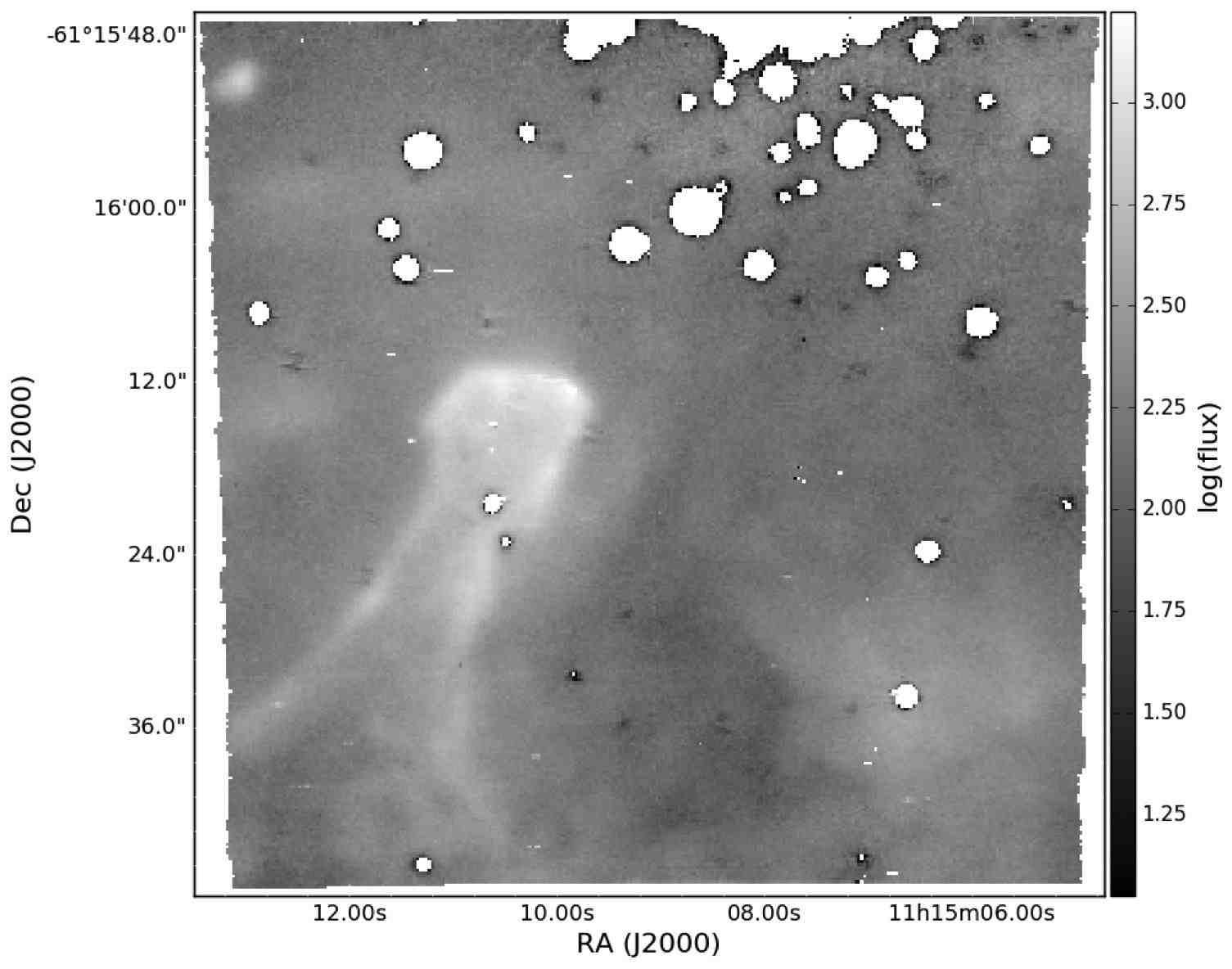}}
\subfloat[]{\includegraphics[scale=0.35]{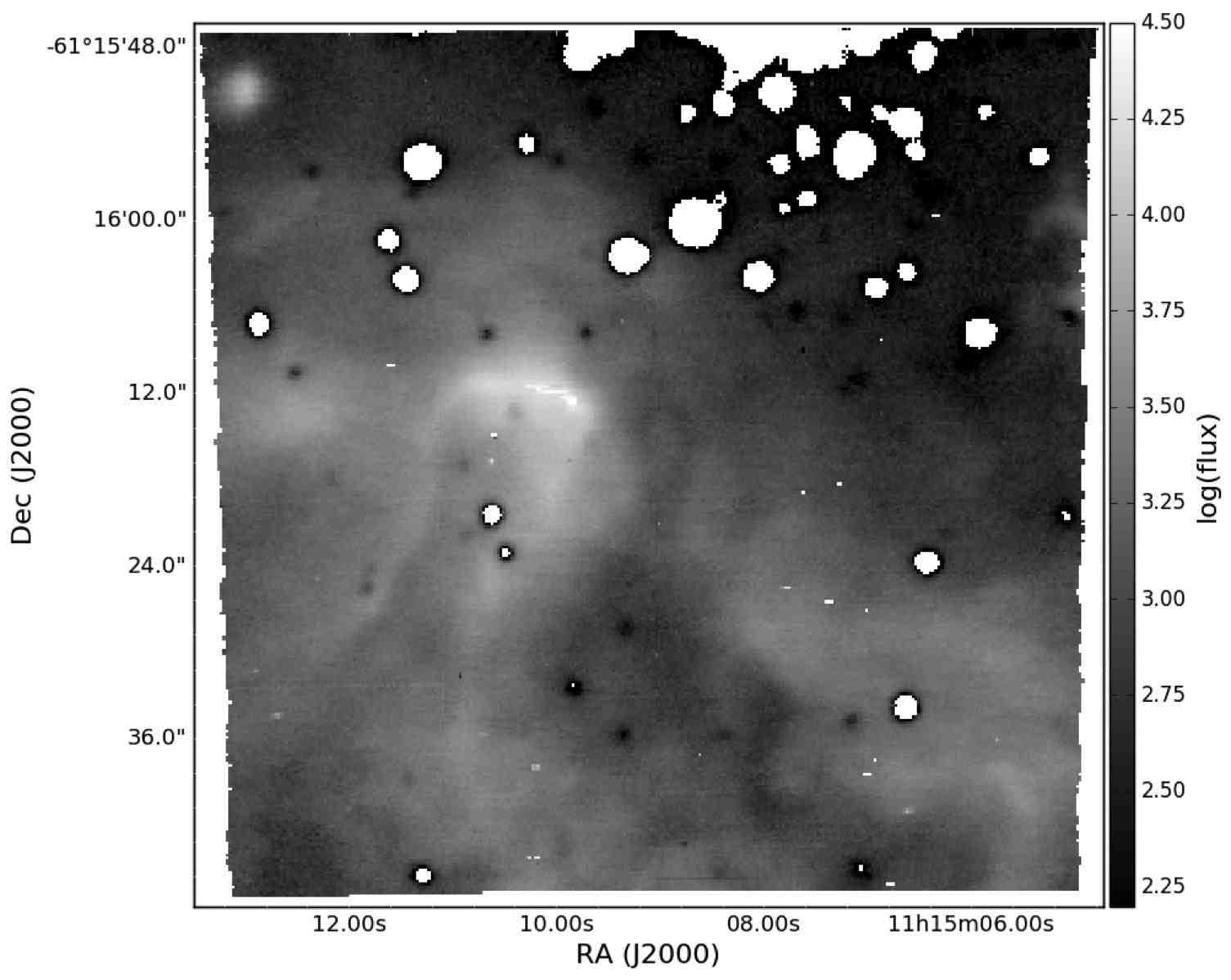}}}
\mbox{
\subfloat[]{\includegraphics[scale=0.35]{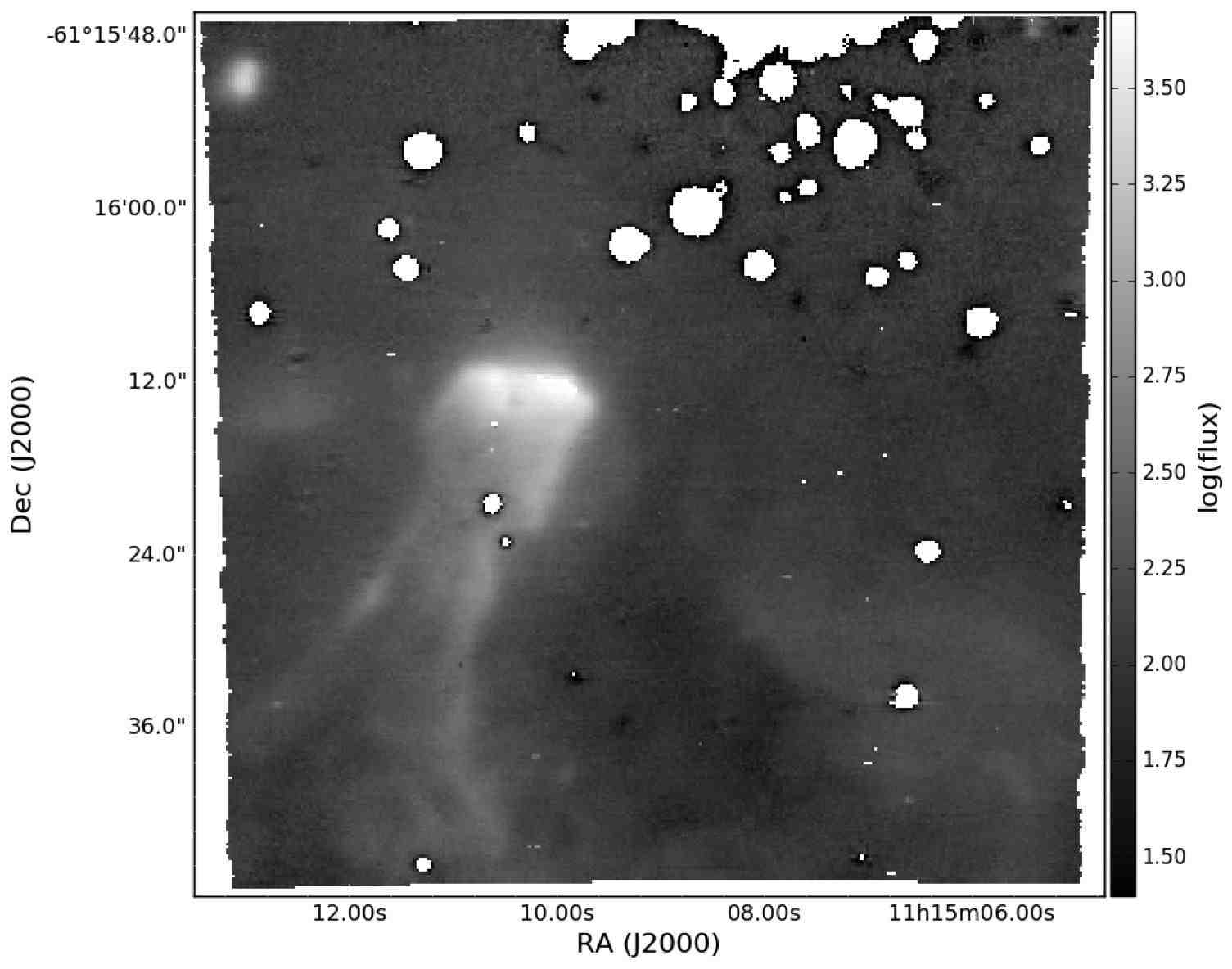}}
\subfloat[]{\includegraphics[scale=0.35]{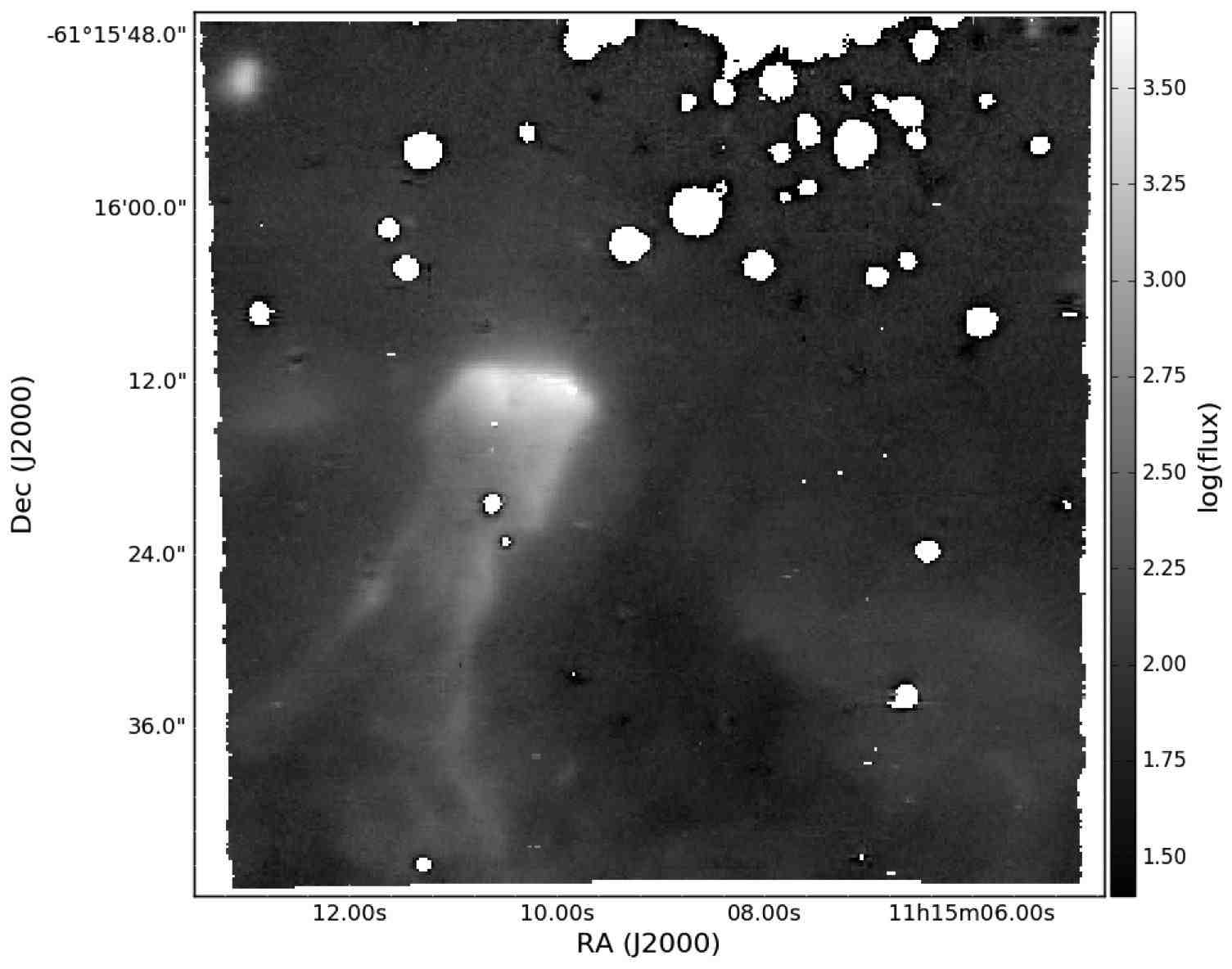}}}
\caption{Same as Fig. A1 for NGC 3603.}
\label{maps9}
\end{figure*}

\begin{figure*}
\mbox{
\subfloat[]{\includegraphics[scale=0.35]{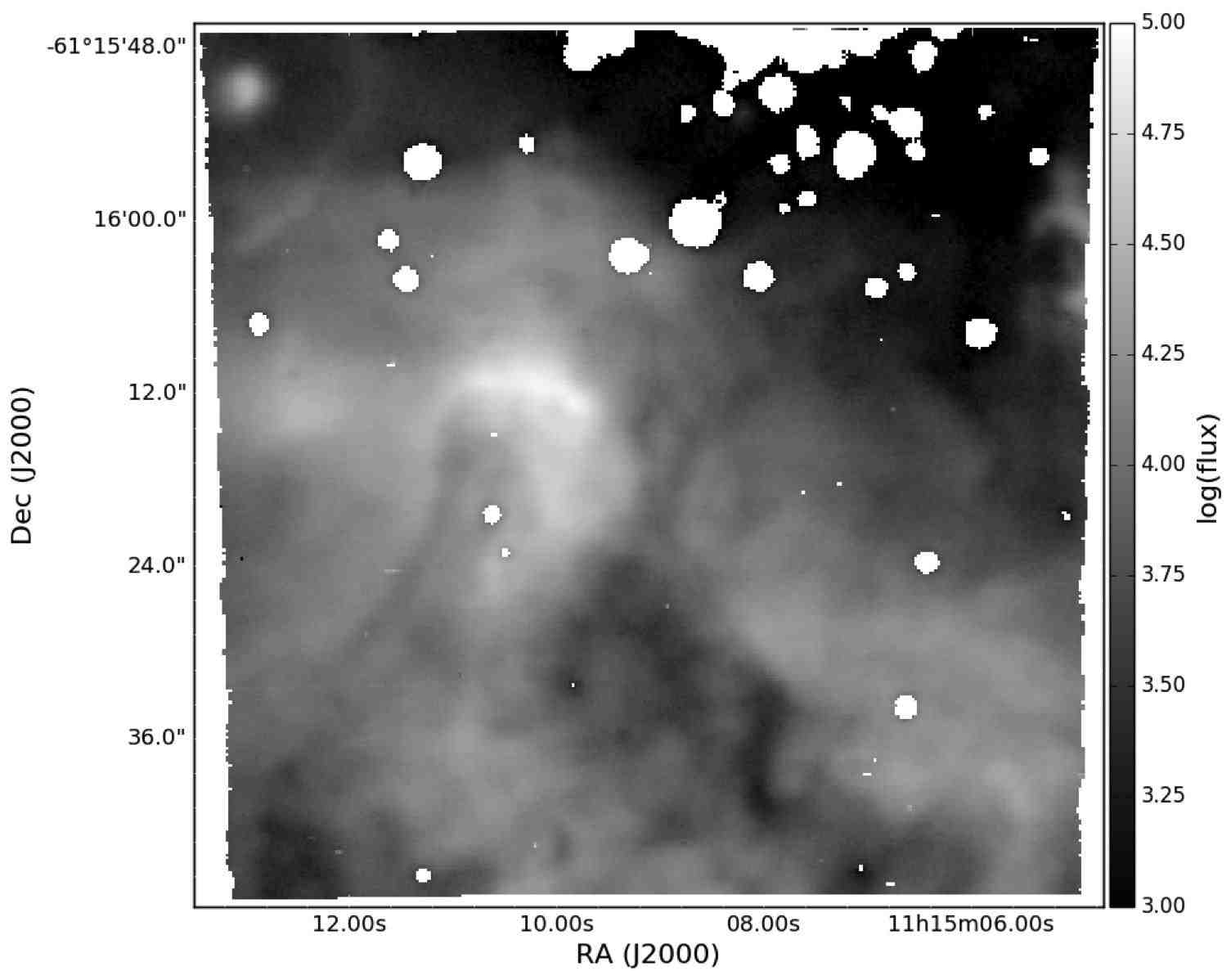}}
\subfloat[]{\includegraphics[scale=0.35]{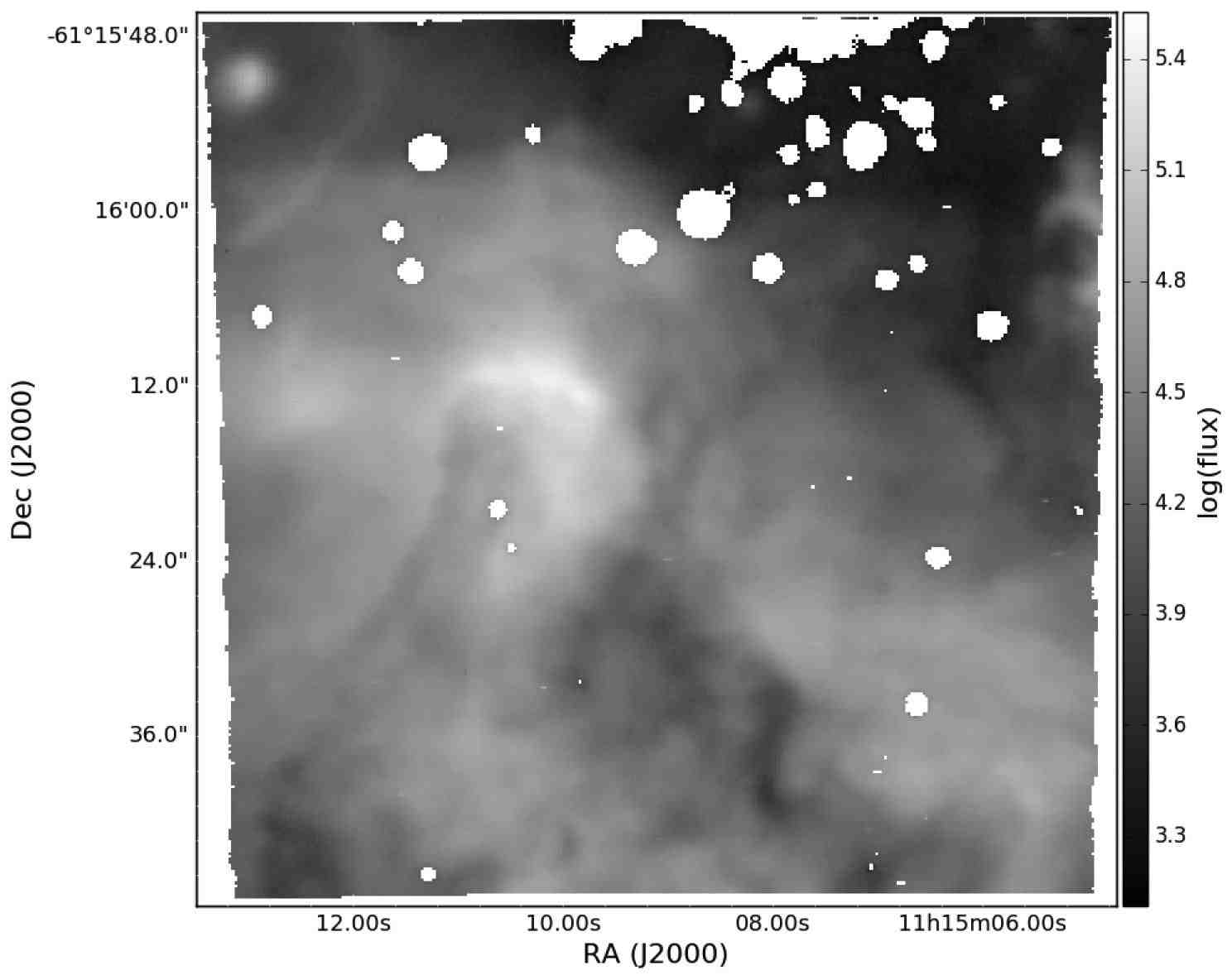}}}
\mbox{
\subfloat[]{\includegraphics[scale=0.35]{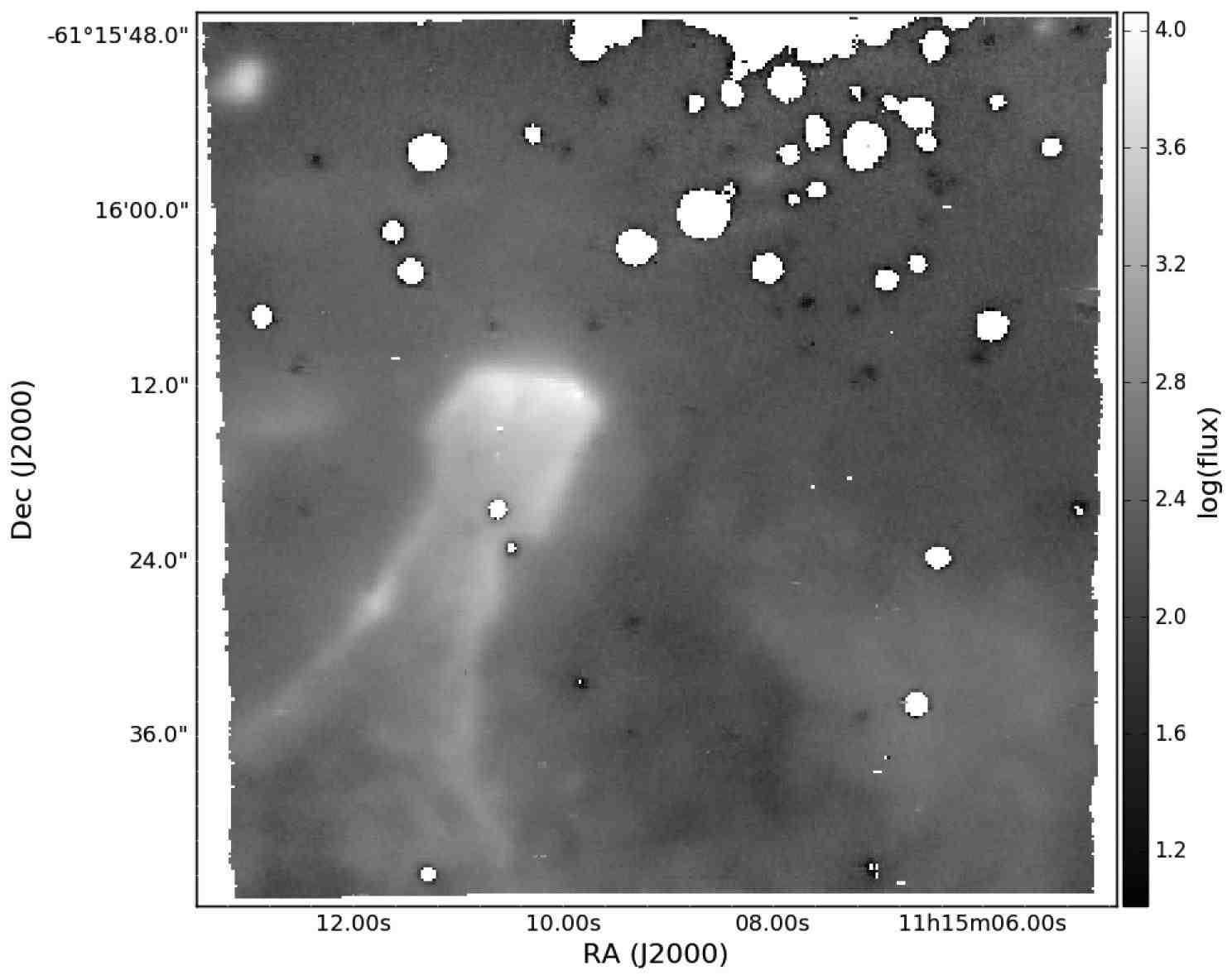}}
\subfloat[]{\includegraphics[scale=0.35]{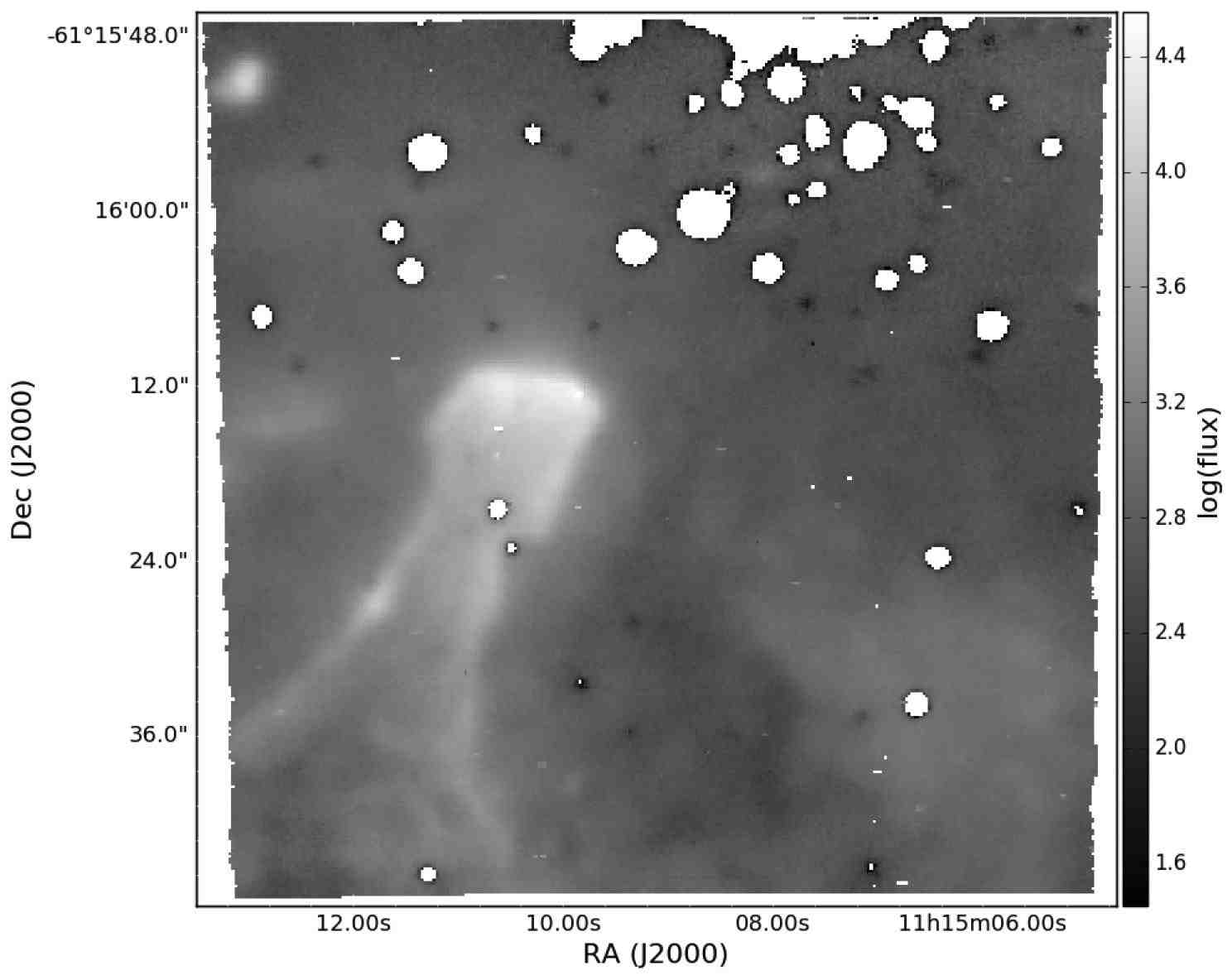}}}
\mbox{
\subfloat[]{\includegraphics[scale=0.35]{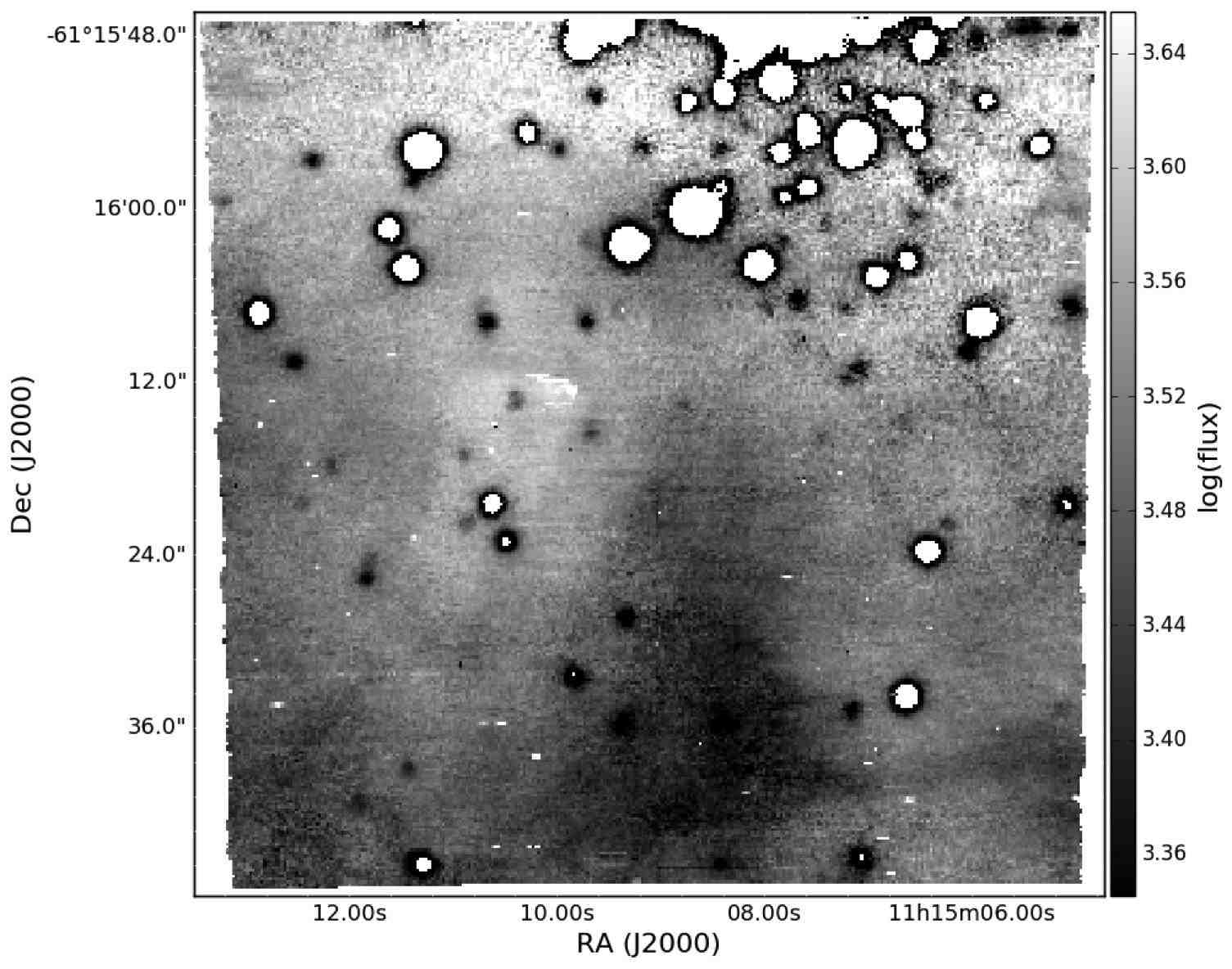}}
\subfloat[]{\includegraphics[scale=0.35]{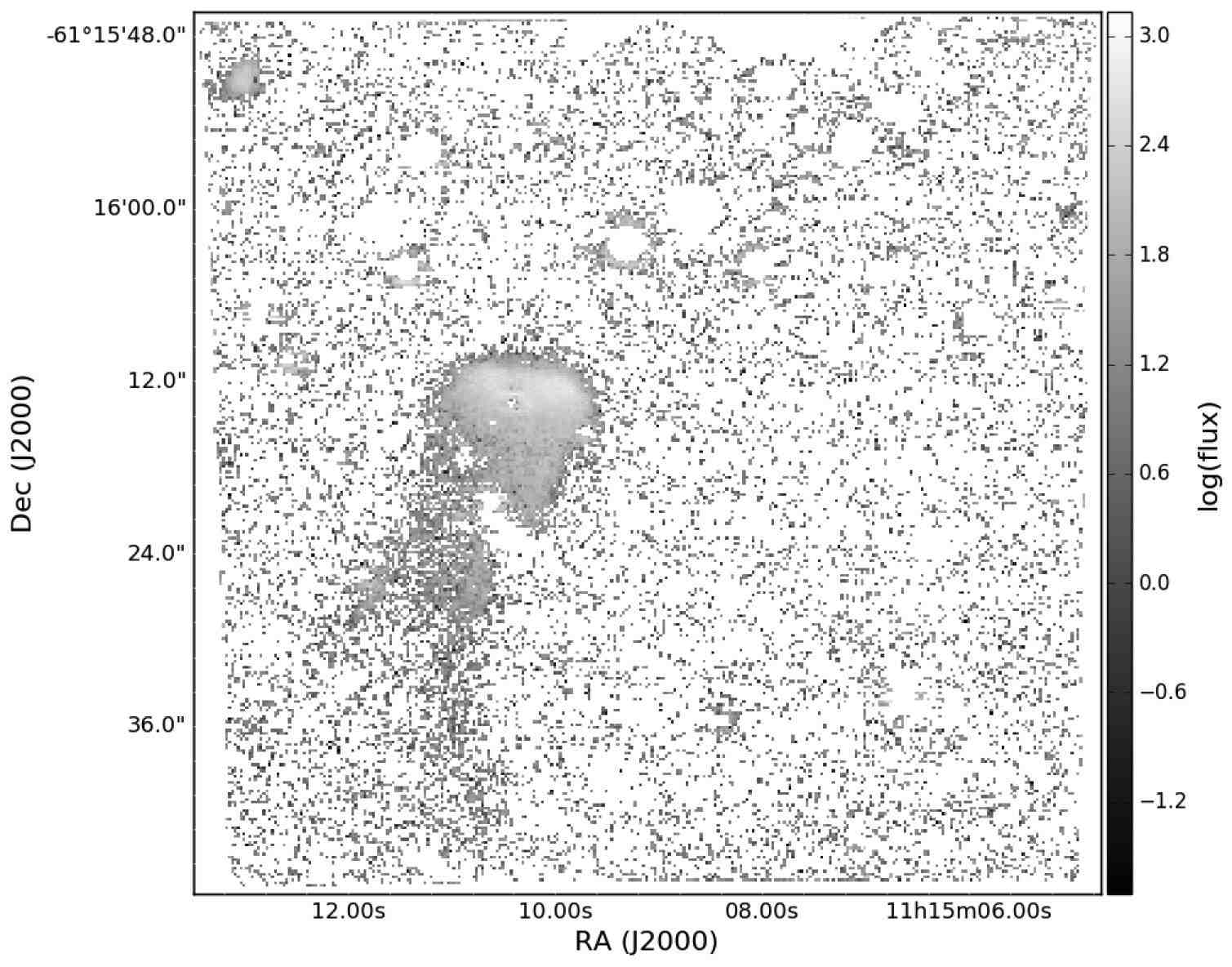}}}
\caption{Same as Fig. A2 for NGC 3603.}
\label{maps10}
\end{figure*}

\begin{figure*}
\centering
\mbox{
\subfloat[]{\includegraphics[scale=0.35]{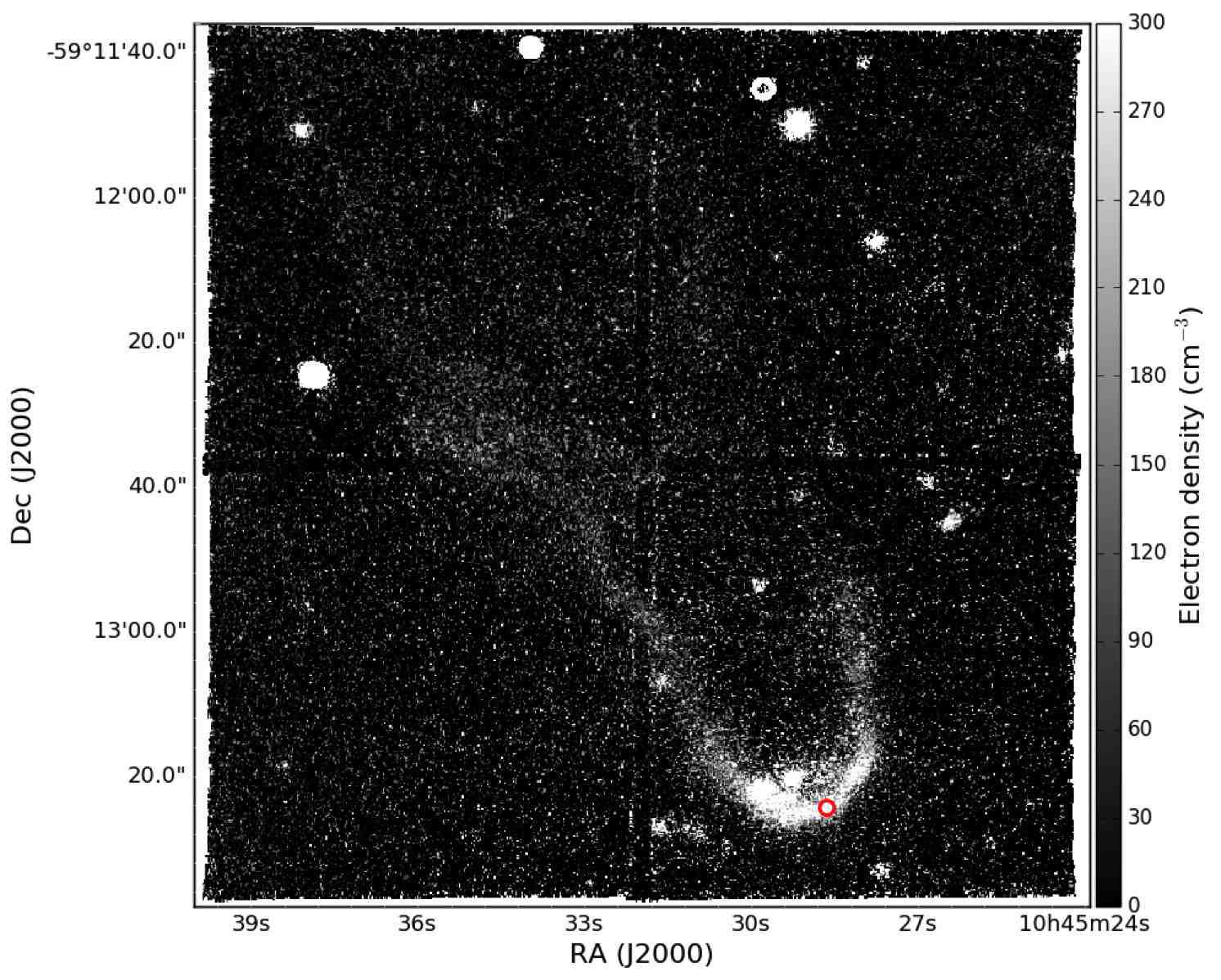}}
\subfloat[]{\includegraphics[scale=0.35]{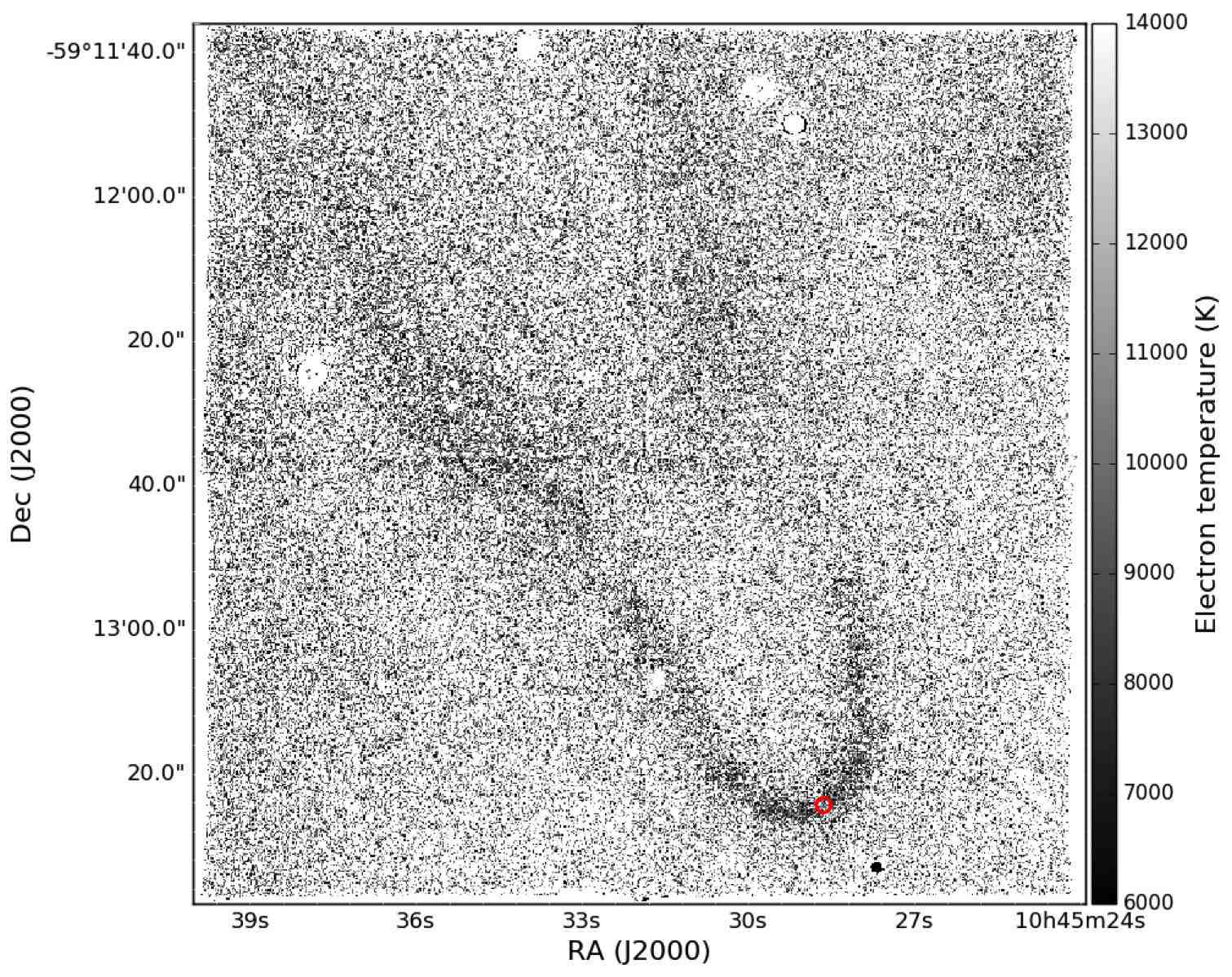}}}
\mbox{
\subfloat[]{\includegraphics[scale=0.45]{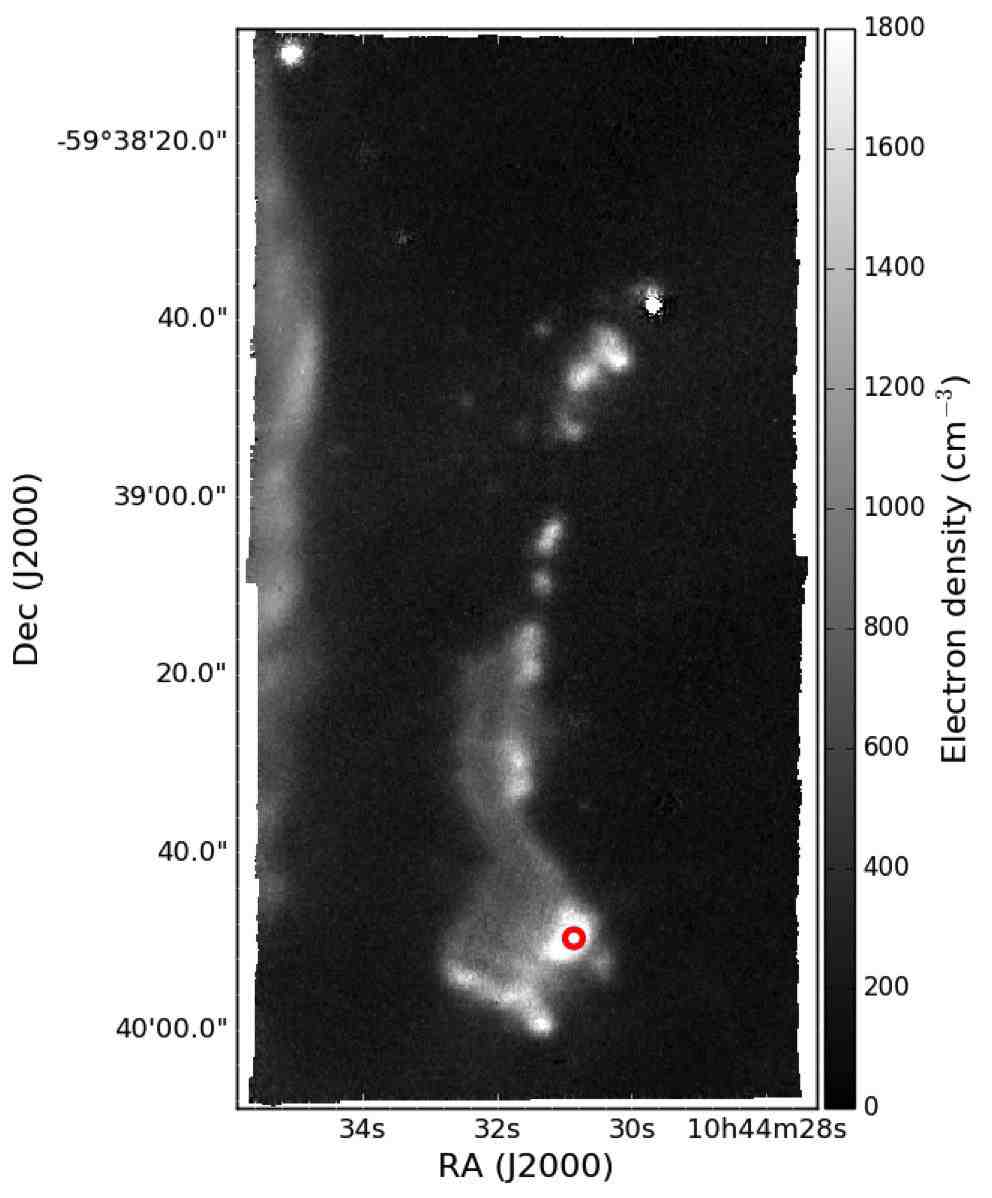}}
\subfloat[]{\includegraphics[scale=0.45]{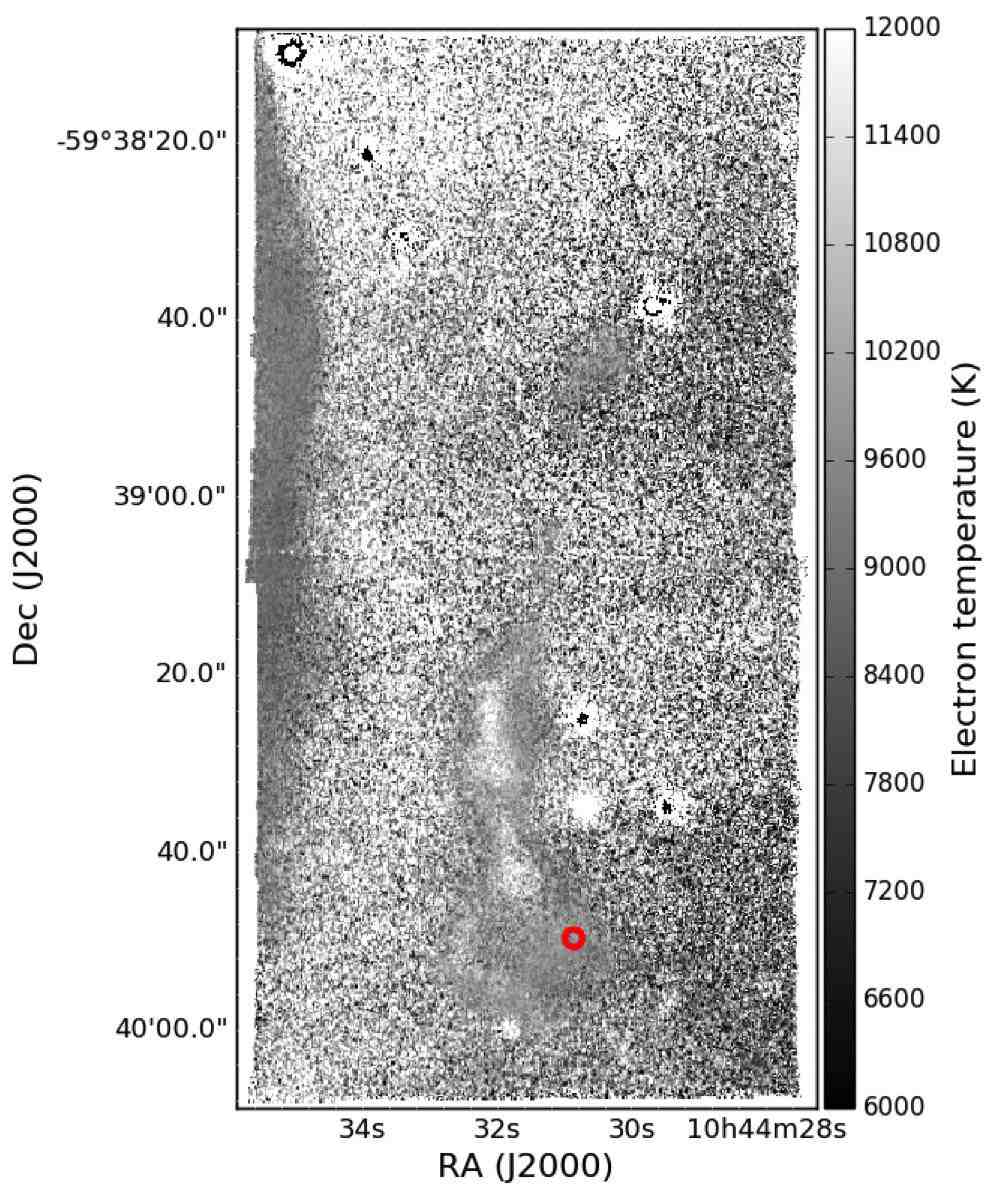}}}
\caption{Electron density and temperature maps of R18 (top panels) and R37 (bottom panels). Red circles indicate the regions used to extract values reported in Table \ref{params}.}
\label{nete1}
\end{figure*}

\begin{figure*}
\centering
\mbox{
\subfloat[]{\includegraphics[scale=0.45]{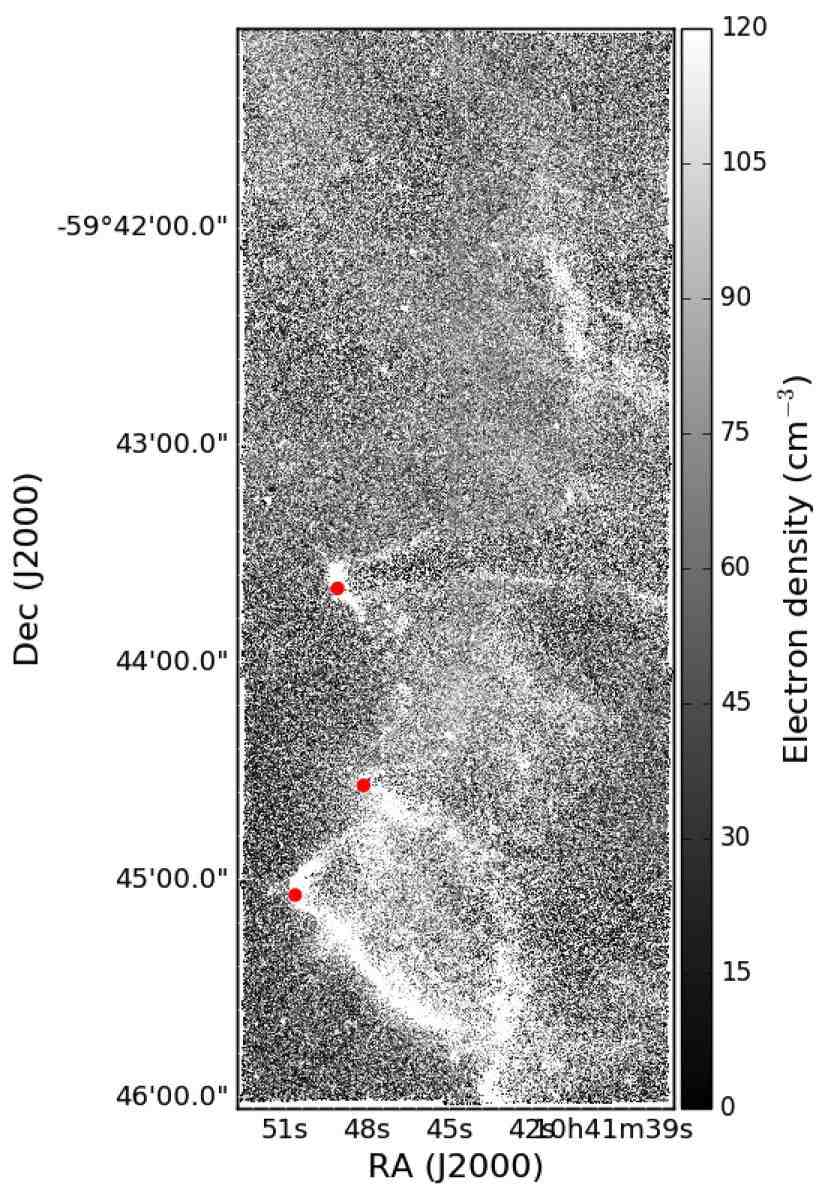}}
\subfloat[]{\includegraphics[scale=0.45]{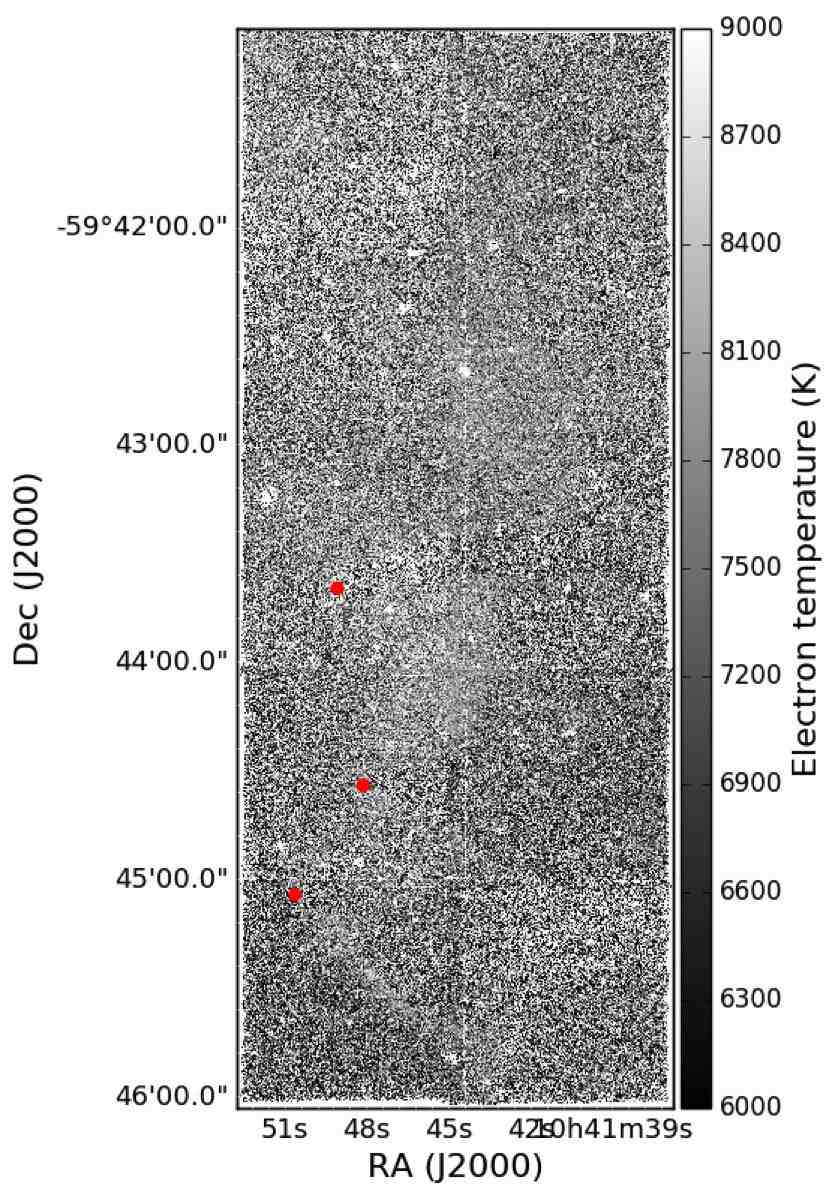}}}
\mbox{
\subfloat[]{\includegraphics[scale=0.45]{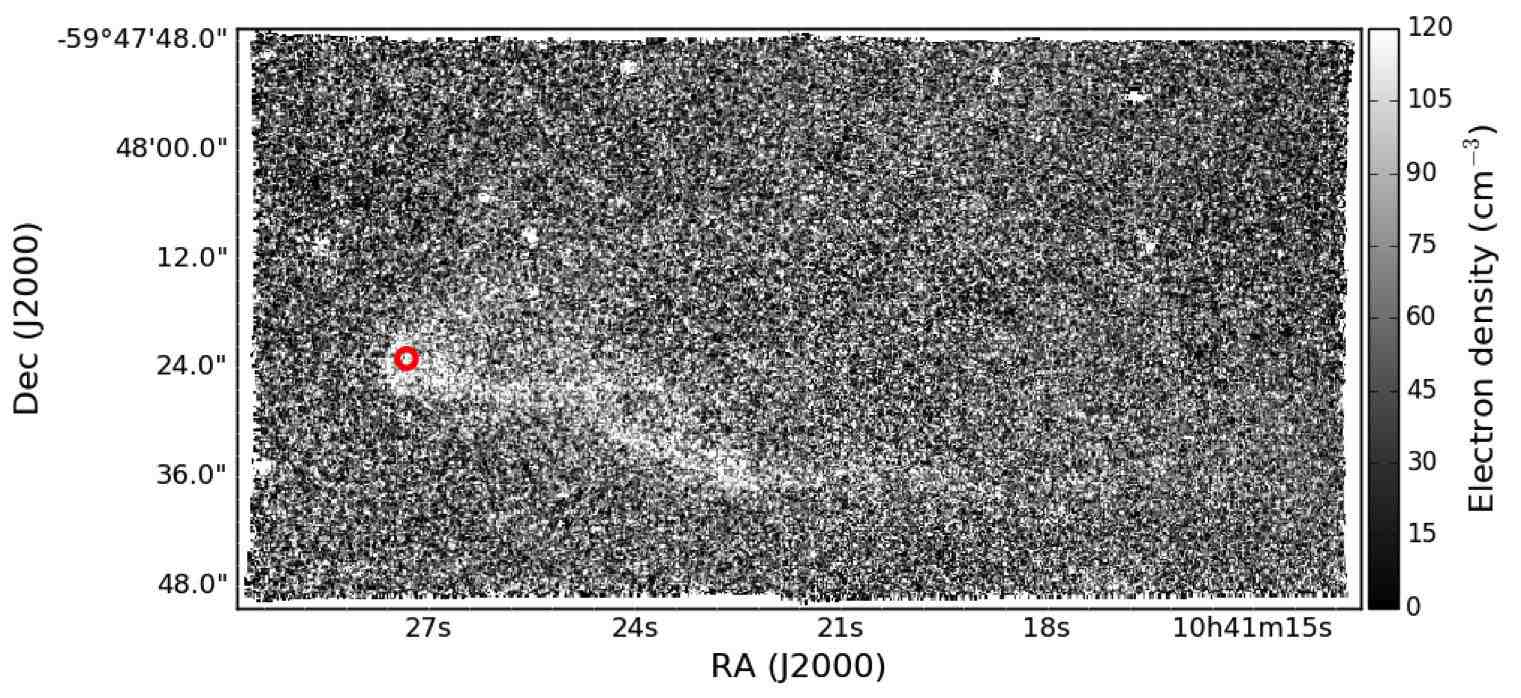}}}
\mbox{
\subfloat[]{\includegraphics[scale=0.45]{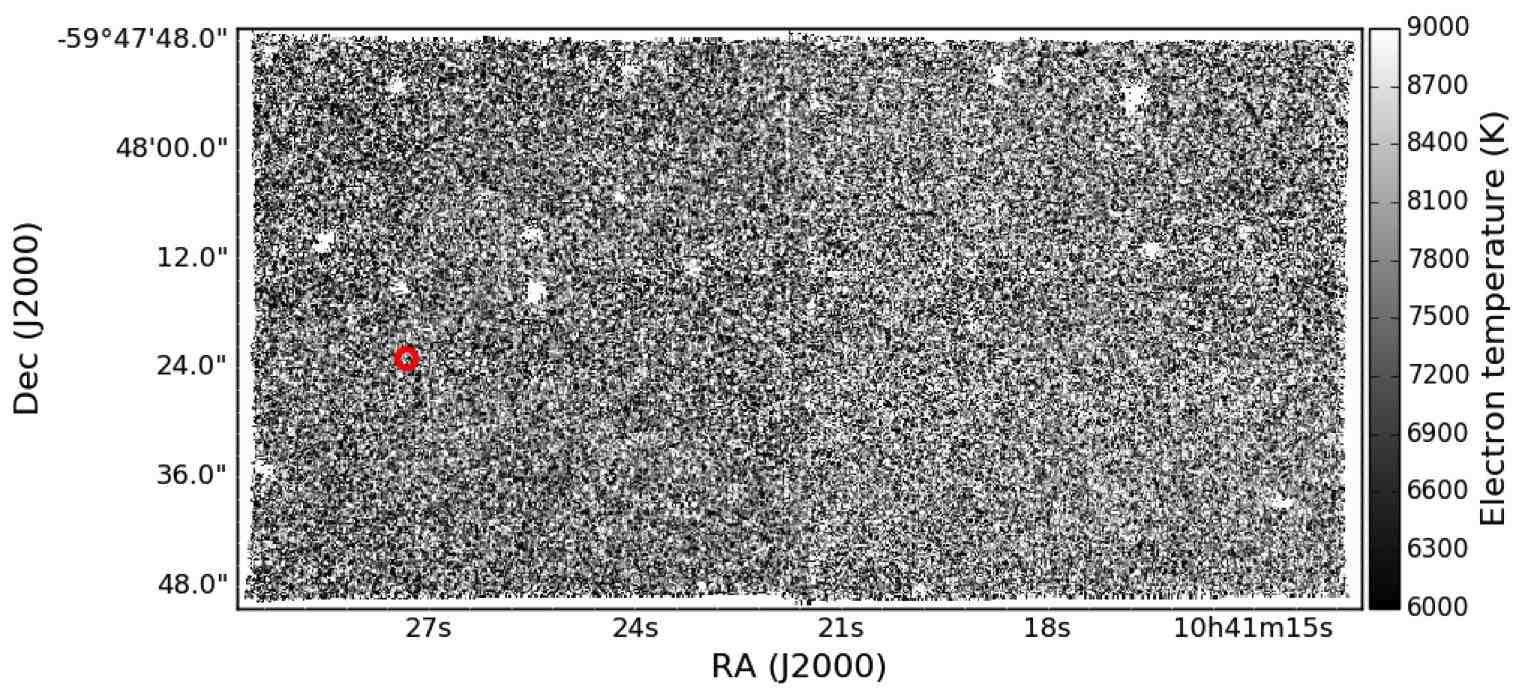}}}
\caption{Same as Fig.~\ref{nete1}, but for R44 (top panels), and R45 (bottom panels).}
\label{nete2}
\end{figure*}

\begin{figure}
\includegraphics[scale=0.45]{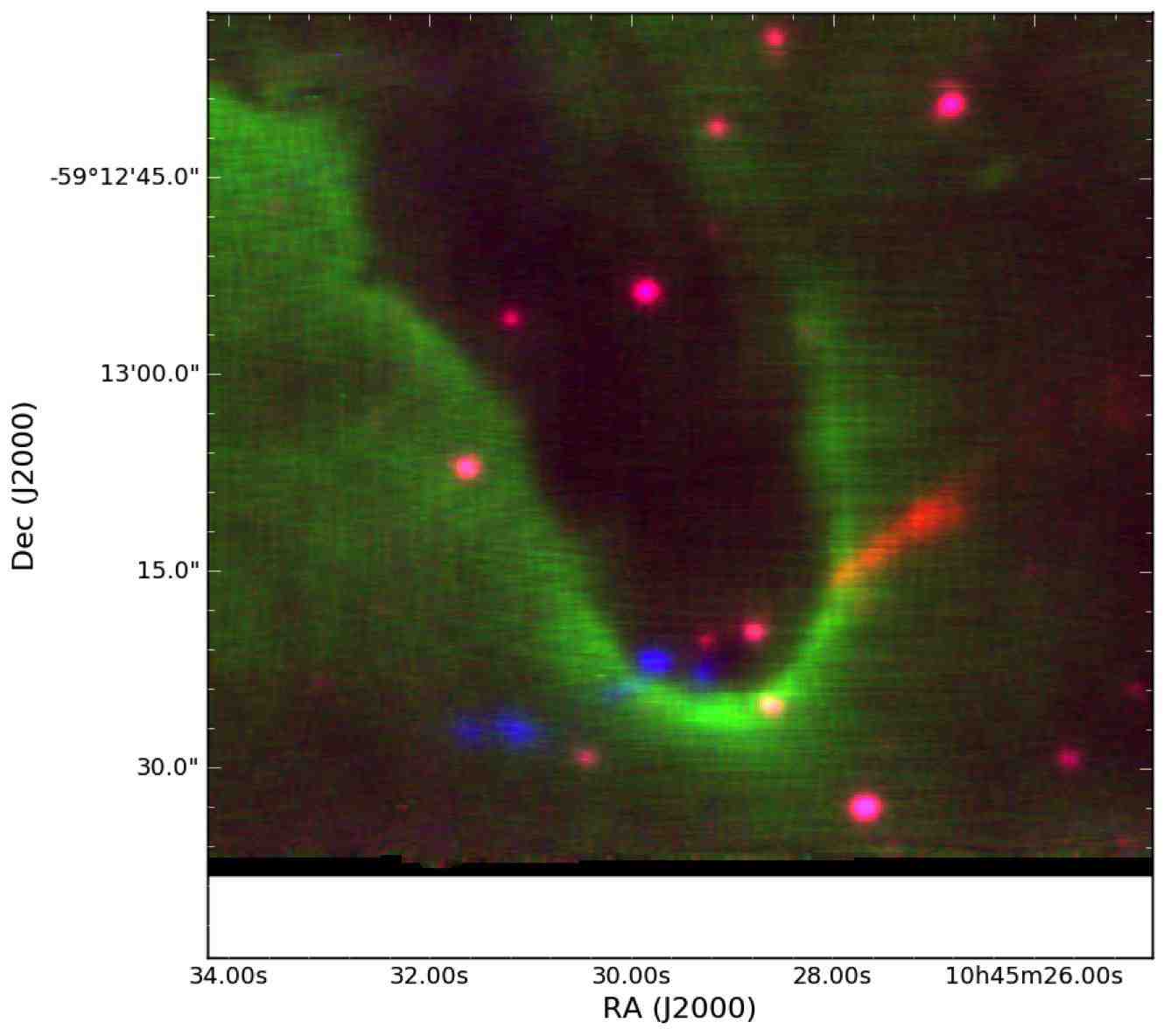}
\caption{RGB composite of three slices around the H$\alpha$ line showing the jet HH 1124 in R18. See text Section \ref{jetsection}.}
\label{R18rgb}
\end{figure}

Figures \ref{maps1} to \ref{maps10} show continuum-subtracted integrated line maps of the main emission lines analysed in this work. Figures \ref{nete1} to \ref{nete2} show the electron temperature and density maps of the various regions, while Figure \ref{R18rgb} shows a three-colour composite of the jet HH 1124 in R18.

\begin{figure*}
\centering
\mbox{
\subfloat[]{\includegraphics[scale=0.45]{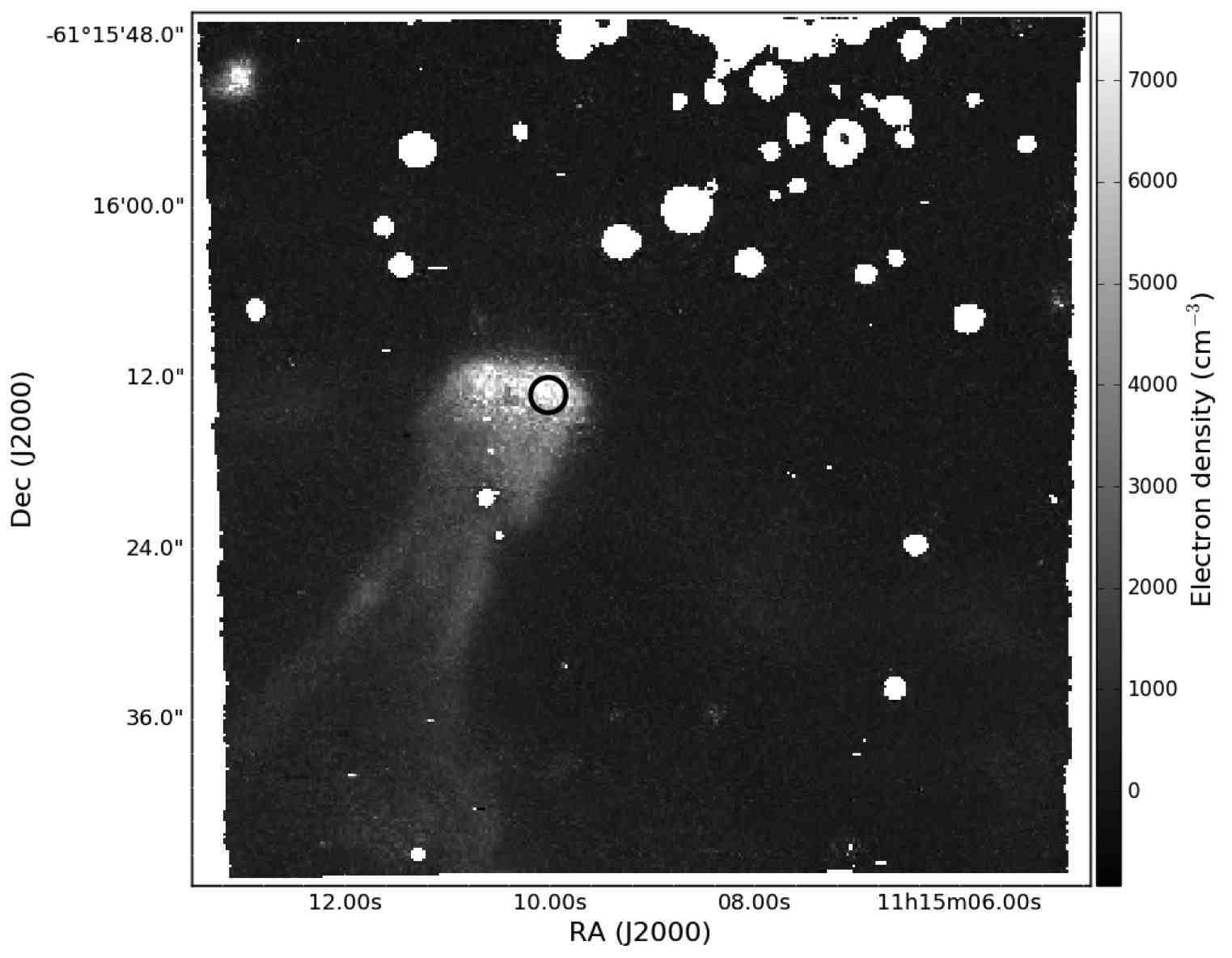}}}
\mbox{
\subfloat[]{\includegraphics[scale=0.45]{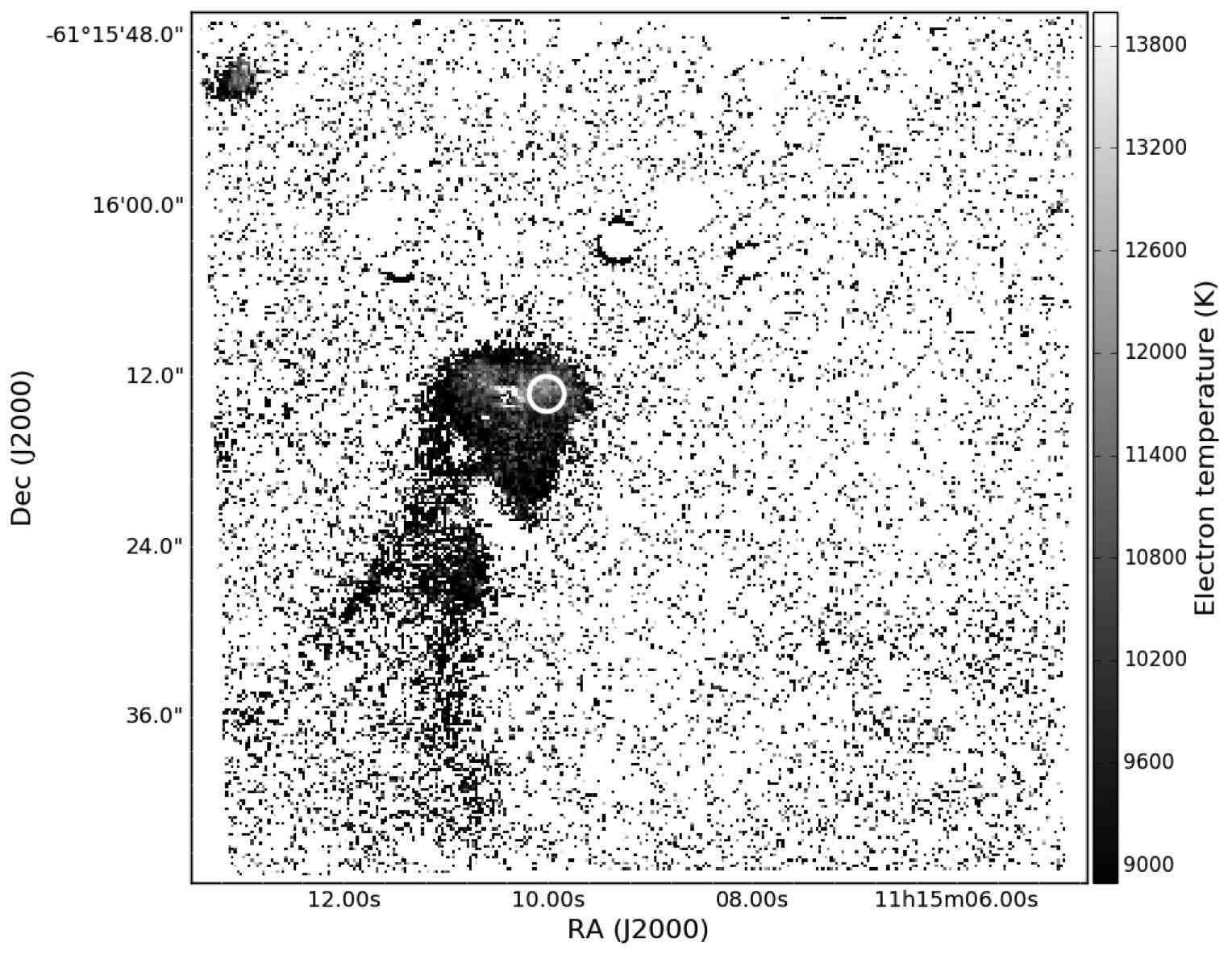}}}
\caption{Same as Fig.~\ref{nete1}, but for NGC 3603.}
\label{nete3}
\end{figure*}

\bsp
\label{lastpage}

\end{document}